\documentclass[fleqn,]{wlscirep}
\usepackage[utf8]{inputenc}
\usepackage{algorithm}
\usepackage{algpseudocode}
\usepackage[T1]{fontenc}
\usepackage{bm}
\usepackage{float}
\usepackage{multirow}
\usepackage{wrapfig} 
\usepackage{makecell} 
\usepackage{quantikz}
\usepackage{subcaption}
\usepackage{tabularx}
\usepackage{graphicx}
\usepackage{pdflscape}
\usepackage{caption}
\usepackage{xcolor}
\usepackage{longtable}
\usepackage[normalem]{ulem}

\definecolor{revisionblue}{RGB}{0,102,204}
\definecolor{revisionred}{RGB}{192,0,0}

\title{Toward chemically accurate quantum simulations on quantum hardware: A quantum-HPC hybrid approach}

\author[$\dagger$, 1]{\small Anurag K. S. V.}
\author[$\dagger$, 1]{\small Ashish Kumar Patra}
\author[2]{\small Manas Mukherjee}
\author[3]{\small Alok Shukla}
\author[1]{\small Sai Shankar P.}
\author[1, 4]{\small Ruchika Bhat}
\author[5]{\\ \small Radhika T. S. L.}
\author[, 1]{\small Jaiganesh G.\thanks{(Corresponding Author) email: jaiganesh@qclairvoyance.in, drjaiganesh15@gmail.com}}

\affil[1]{\small Qclairvoyance Quantum Labs, Secunderabad, TG 500094, India.}
\affil[2]{\small Centre for Quantum Technologies, National University of Singapore, Singapore 117543, Singapore.}
\affil[3]{\small School of Arts and Sciences, Ahmedabad University, Ahmedabad, GJ 380009, India.}
\affil[4]{\small The University of Arizona, Tucson, AZ 85721, USA.}
\affil[5]{\small Department of Mathematics, BITS Pilani, Hyderabad Campus, Hyderabad, TG 500078, India.}

\affil[$\dagger$]{\small These authors contributed equally to this work}

\begin{abstract}
We report one of the most comprehensive experimental demonstrations to date of quantum-computing-based molecular simulation on IQM’s Sirius 24-qubit superconducting quantum processor, using up to 16 operational qubits. We employ Sample-based Quantum Diagonalization (SQD) with the Local Unitary Cluster Jastrow (LUCJ) ansatz to compute ground-state energies of benchmark molecular systems, including H$_2$, LiH, BeH$_2$, H$_2$O, and NH$_3$. We further introduce the Linear-CNOT Unitary Coupled-Cluster Singles and Doubles (LCNot-UCCSD) ansatz within the SQD framework, reducing the classical pre-computation overhead at the cost of increased quantum circuit depth, and demonstrate its effectiveness by accurately recovering the ground-state energies of H$_2$, LiH, and BeH$_2$. A comparative analysis of the two ansätze is presented, highlighting their respective merits, demerits, and practical use cases in near-term quantum simulations. In addition, we perform one-dimensional potential energy surface (1D-PES) scans for H$_2$ and HeH$^+$ using STO-3G and 6-31G basis sets, and for LiH and BeH$_2$ using STO-3G. We also report the first experimental generation of a full two-dimensional potential energy surface (2D-PES) map, consisting of a $32 \times 32$ grid in bond length and bond angle, for the water molecule in the STO-3G basis on IQM quantum hardware. Beyond small-molecule benchmarks, we integrate Density Matrix Embedding Theory (DMET) with SQD(LUCJ) to obtain chemically accurate 4-electron-in-4-orbital active-space energies for eight ligand-like molecules, as well as for the pharmacologically relevant amantadine molecular system, representing the first of such experimental demonstrations. Across all studies, the majority of quantum-computed energies agree with reference full configuration interaction (FCI) results and with DMET-CASCI energies for embedded systems, to within chemical accuracy for the chosen basis sets. These results demonstrate the robustness of quantum sampling-based classical diagonalization approaches while highlighting the promise of hybrid embedding strategies combined with near-term ansätze for advancing quantum chemistry toward classically intractable molecular systems and establishing their practicality on current quantum hardware.

\end{abstract}

\begin{document}

\flushbottom
\maketitle

\thispagestyle{empty}


\vspace{1.0em}
\tableofcontents

\newpage

\section{Introduction} \label{sec:introduction}

    \subsection{The Electronic-Structure Challenge}

    One of the quintessential objectives of computational quantum chemistry is the rigorous description of the electronic structure of matter~\cite{Szabo1996, Helgaker2000, Levine2014, McArdle2020}. This endeavor is the bedrock upon which modern computational materials science~\cite{Hohenberg1964, Kohn1965, Martin2004, Hafner2008, Burke2012, Jain2013}, pharmacology~\cite{Jorgensen2004, Friesner2005, Raha2007, Cournia2017, Heifetz2020, Niazi2026}, and chemical engineering rest~\cite{Bell2011, Norskov2014, Deglmann2015, Peters2017}. The ability to predict the behaviour of electrons, how they distribute themselves around nuclei, how they form bonds, and how they respond to external stimuli dictates our capacity to design novel catalysts for carbon capture~\cite{Norskov2002, Gagliardi2017, Seh2017, Keith2018, Nitopi2019}, synthesize life-saving therapeutics~\cite{Jorgensen2004, Raha2007, Molani2024, Goullieux2025, Niazi2026}, and engineer high-temperature superconductors~\cite{Anderson1987, Pickett1989, Keimer2015}. At the theoretical core of this pursuit lies the non-relativistic time-independent Schrödinger equation~\cite{Schrodinger1926}. In principle, the exact solution to this equation for a given electronic Hamiltonian of $N$ electrons and $M$ nuclei yields the ground-state energy of the provided Hamiltonian with a precision limited only by relativistic effects\cite{Pyykko1978, Pyykko1988, Dyall2007, Pyykko2012, Pyykko2025} and the breakdown of the Born-Oppenheimer approximation\cite{Born1927, Yarkony2012, Curchod2018, Pavosevic2020}.

    However, the practical realization of this predictive power is hindered by a formidable computational barrier known as the curse of dimensionality~\cite{Bellman1961, Kohn1999}. The Hilbert space in which the electronic wavefunction resides grows exponentially with the number of active electrons and orbitals~\cite{Helgaker2000}. In the Full Configuration Interaction (FCI) formalism, the exact wavefunction is expressed as a linear combination of all Slater determinants consistent with particle number and spin symmetries~\cite{Szabo1996}. For a molecular system described by $N_{\mathrm{orb}}$ spatial orbitals and $N_{\mathrm{ele}}$ electrons (where $N_{\alpha}$ and $N_{\beta}$ denote the number of electrons with $\alpha$-spin and $\beta$-spin, respectively), the dimension of this space grows combinatorially. This scaling follows $\binom{2N_{\mathrm{orb}}}{N_\mathrm{ele}}$ for general open-shell systems and $\binom{N_{\mathrm{orb}}}{N_{\alpha}} \binom{N_{\mathrm{orb}}}{N_{\beta}}$ for closed-shell systems, rapidly exceeding classical memory and compute limits~\cite{Helgaker2000, Knowles1984}. Historically, this scaling confined exact diagonalization methods to systems containing roughly 10--14 correlated electrons, with dimension sizes just exceeding one billion ($10^9$) determinants~\cite{Olsen1990, Sherrill1999}. Recent algorithmic breakthroughs in distributed computing and tensor compression have significantly expanded this horizon. Notably, massively parallel implementations have enabled CASCI calculations on active spaces as large as 24 electrons in 24 orbitals; while formally spanning $7.31 \times 10^{12}$ determinants, the exploitation of $D_{2h}$ spatial symmetry and the singlet $^{1}A_g$ ground state reduced this complexity to $9.1 \times 10^{11}$ determinants~\cite{Vogiatzis2017}. Furthermore, exact FCI solutions have been achieved for the propane molecule ($\mathrm{C_3H_8}$) involving over $1.31 \times 10^{12}$ determinants~\cite{Hong2024}, effectively breaching the one-trillion-determinant barrier. Most recently, lossless categorical compression strategies have pushed the frontier to the quadrillion-determinant scale ($10^{15}$) for heavy-element systems such as HBrTe~\cite{Shayit2025}. Yet, despite these heroic classical efforts, the non-polynomial scaling persists; the Hilbert space for a moderate-sized drug molecule or a complex catalytic center remains orders of magnitude beyond even the exascale regime, necessitating a fundamentally different computational paradigm.

    \subsection{Classical Approaches to Many-Electron Correlation}

    To circumvent this barrier, computational chemistry has long relied on different classes of approximations. Wave-function-based correlation methods provide systematically improvable hierarchies~\cite{Szabo1996, Levine2014} starting with mean-field Hartree-Fock (HF)~\cite{Hartree1928 ,Fock1930, Roothaan1951, Hall1951}, while Kohn--Sham Density Functional Theory (DFT)~\cite{Kohn1965, Pribram2015, Jones2015} relying on approximations to the exchange-correlation functional. Both capture a substantial portion of the total energy at modest computational cost~\cite{Goedecker1999, Mardirossian2017}. However, the residual electron correlation energy governs chemically decisive phenomena, including reaction barriers~\cite{Donahue2003, Hait2018, Nitz2025}, weak non-covalent interactions~\cite{Klaus2000, Masumian2024, Khabibrakhmanov2025}, and excited-state processes~\cite{Matsika2018, Jacquemin2023, Peverati2014}. Post-Hartree--Fock methods, most notably Coupled Cluster with Single and Double excitations augmented by perturbative Triples [CCSD(T)]~\cite{Raghavachari1989}, recover a large fraction of this correlation energy with polynomial scaling and have therefore earned the designation of the ``gold standard'' of quantum chemistry~\cite{Mester2024, Nagy2024}. Nonetheless, the steep $\mathcal{O}(N^7)$ scaling of CCSD(T), together with large prefactors, severely restricts its applicability to medium-sized molecules and precludes routine treatment of extended systems or large biomolecules without further fragmentation~\cite{Gordon2012, Raghavachari2015, Broderick2025} or embedding approximations~\cite{Sun2016, Vorwerk2022, Verma2026}.

    Tensor-network methods provide another important route to mitigating the exponential complexity of the many-electron wavefunction by exploiting its entanglement structure. In particular, Matrix Product States (MPS), optimized through the Density Matrix Renormalization Group (DMRG), have become powerful tools for treating strongly correlated molecular systems and large active spaces~\cite{White1992,Chan2002,Chan2011,Wouters2014}. More general tensor-network architectures, including Tree Tensor Network States (TTNS)~\cite{Shi2006,Nakatani2013,Gunst2018}, Projected Entangled-Pair States (PEPS)~\cite{Verstraete2004,Orus2014}, and the Multiscale Entanglement Renormalization Ansatz (MERA)~\cite{Vidal2007,Vidal2008}, extend this framework to alternative network topologies and correlation structures. The accuracy and efficiency of these approaches, however, depend on the entanglement structure of the target state and the resulting tensor ranks and bond dimensions, which can become costly for highly entangled systems~\cite{Szalay2015,Chan2016}.

    \subsection{Quantum Algorithms for Electronic Structure}

    The advent of quantum computing provides a fundamentally different computational paradigm for addressing this challenge~\cite{Feynman1982, McArdle2020, Zhang2025, Patra2025}. By exploiting quantum superposition~\cite{Dirac1958} and entanglement~\cite{Einstein1935, Horodecki2009}, quantum processors can, in principle, represent and manipulate many-body wavefunctions in a Hilbert space whose dimension grows exponentially with the number of qubits, while the physical resources (qubits) scale polynomially~\cite{Nielsen2000, AspuruGuzik2005}. This observation underpins the long-standing notion that quantum computers are natural simulators of quantum systems, capable of addressing regimes that are classically intractable~\cite{Feynman1982, Lloyd1996}. Early algorithmic proposals for quantum chemistry leveraged this insight to formulate polynomial-scaling quantum algorithms for eigenvalue estimation, providing the first concrete evidence of a potential exponential advantage in regimes where the corresponding classical representation or exact computation requires exponentially growing resources~\cite{Abrams1997, AspuruGuzik2005}. The potential advantage of such quantum algorithms does not follow from polynomial scaling alone, since many classical electronic-structure methods also exhibit polynomial scaling; rather, it may arise from more favorable asymptotic scaling with system size or precision~\cite{Abrams1999, Low2017, Low2019, Gilyen2019}, reduced memory requirements for representing many-body states~\cite{Feynman1982, AspuruGuzik2005}, improved treatment of strongly correlated systems~\cite{McArdle2020, Yoo2026, Wang2025}, or access to system sizes or accuracy regimes that are expected to be computationally inaccessible to classical methods~\cite{Lee2023, Beverland2022, Genin2025}.

    A paradigmatic example is Quantum Phase Estimation (QPE), which enables the extraction of eigenvalues of a unitary operator corresponding to the time evolution under a molecular Hamiltonian~\cite{Kitaev1995, AspuruGuzik2005}. When supplied with a sufficiently accurate initial state and implemented using fault-tolerant quantum error correction (QEC), QPE can determine ground- and excited-state energies with asymptotically optimal scaling in the desired precision~\cite{Abrams1999}. While the precise boundary of exponential advantage for ground-state chemistry remains under complexity-theoretic scrutiny~\cite{Lee2023, Dalzell2025, Huang2025}, the fundamental capacity of quantum processors to exceed classical limits has been confirmed through demonstrations of unconditional quantum information supremacy~\cite{Kretschmer2025}. However, the substantial depth and qubit requirements associated with standard QPE protocols firmly place them in the domain of fault-tolerant, application-scale quantum (FASQ) computing~\cite{Eisert2025}. To bridge this gap, recent efforts have pivoted toward early-fault-tolerant architectures, envisioned to operate with 25--100 logical qubits~\cite{Alexeev2025}, employing novel dissipative state preparation protocols~\cite{Lin2025} and partially fault-tolerant schemes~\cite{Toshio2025} to bring practical quantum advantage closer to experimental reality.
    
    In parallel with the pursuit of fault tolerance, the Noisy Intermediate-Scale Quantum (NISQ)~\cite{Preskill2018} era has been defined by hybrid quantum--classical algorithms designed to operate within limited coherence times. The Variational Quantum Eigensolver (VQE) has served as the dominant paradigm~\cite{Peruzzo2014, McClean2016, Fedorov2022}, evolving from static ansätze~\cite{Kandala2017, Barkoutsos2018} to sophisticated adaptive strategies to enhance expressivity. Notable advancements include adaptive derivative-assembled pseudo-trotter VQE (ADAPT-VQE) variants utilizing coupled exchange operators~\cite{Ramoa2025}, parameter-based selection criteria~\cite{He2026}, and automated operator pruning~\cite{Vaquero2025} to drastically reduce circuit depth. Despite these innovations, VQE faces fundamental scaling challenges: optimization landscapes frequently exhibit noise-induced barren plateaus~\cite{McClean2018, Wang2021}, and measurement overheads remain prohibitive even with Hamiltonian partitioning~\cite{Singh2025}. Critically, unmitigated device noise can invalidate the variational bound, complicating the retrieval of accurate molecular properties~\cite{Lima2026} and dynamic Green's functions~\cite{Singh2026}, often preventing meaningful chemical insights for systems as fundamental as benzene on current hardware~\cite{Carreras2026}. These persistent scalability and accuracy constraints have spurred a resurgence of interest in alternative, sample-based paradigms~\cite{Jiang2025} that circumvent the variational optimization loop entirely.

    \subsection{Sample-based Quantum Diagonalization}

    Formalizing this paradigm, sampling-based quantum algorithms such as Quantum-Selected Configuration Interaction (QSCI)~\cite{Kanno2023} decouple wavefunction generation from energy optimization by utilizing the quantum device to identify high-weight determinants from a trial state, subsequently diagonalizing the Hamiltonian within this compressed subspace. To mitigate the burden of ansatz optimization, recent protocols have introduced adaptive input state construction (ADAPT-QSCI)~\cite{Nakagawa2024} and Hamiltonian simulation-based sampling (HSB-QSCI)~\cite{Sugisaki2025, Mikkelsen2025}, which leverage time-evolution to access chemically relevant sectors of the Hilbert space without complex variational loops. This framework has culminated in the concept of quantum-centric supercomputing, demonstrated by large-scale Sample-based Quantum Diagonalization (SQD) simulations of N$_2$ dissociation and impurity models on IBM Heron quantum hardware coupled with classical supercomputers~\cite{Robledo2025}, and most recently by closed-loop workflows integrating 152,064 nodes of the Fugaku supercomputer to tackle electronic structures beyond exact diagonalization~\cite{Shirakawa2025}. The applicability of these sampling-based algorithms has been rapidly expanded to encompass excited states~\cite{Barison2025}, open-shell systems~\cite{Liepuoniute2025}, implicit solvation models~\cite{Kaliakin2025a}, and non-covalent interactions~\cite{Kaliakin2025b}. Furthermore, to address convergence limitations, the paradigm has been extended to Sample-based Krylov Quantum Diagonalization (SKQD)~\cite{Yu2025}, which takes its roots from SQD and Krylov Quantum Diagonalization (KQD)~\cite{Yoshioka2025}, offering a polynomial-time convergence guarantee under assumptions. Further works have shown its efficient implementation via randomized compilation (SqDRIFT)~\cite{Piccinelli2026} or time-independent unitary decomposition (QKUD)~\cite{Asthana2025} to minimize algorithmic error. To capture dynamic correlation missing from the selected subspace, building on the foundational quantum-classical quantum Monte Carlo (QMC)~\cite{Huggins2022}, hybrid workflows merging QSCI with phaseless auxiliary-field quantum Monte Carlo (ph-AFQMC) have been established~\cite{Yoshida2025, Danilov2025}, along with GPU-acceleration support for the classical diagonalization kernels of SQD~\cite{Walkup2026, Doi2026}.
    
    However, despite these algorithmic strides, fundamental limitations remain. Rigorous benchmarks indicate that QSCI and SQD struggle with sampling inefficiency when targeting high-accuracy results, often yielding configuration interaction expansions that are less compact than classical heuristics (e.g., Heat-Bath Configuration Interaction, HCI), where new determinants become exponentially difficult to find~\cite{Reinholdt2025, Raisuddin2025}. Trying to address these scaling bottlenecks, novel variants have emerged. For instance, compact wavefunction-based QSCI leverages stochastic time evolution to predict likely excitations, yielding configuration spaces that are up to 200 times smaller than those obtained with conventional selection criteria, demonstrated on the IQM Emerald superconducting quantum hardware~\cite{Weaving2025}. Similarly, the Handover Iterative VQE (HI-VQE)~\cite{Aidan2025} dynamically exchanges configuration data between quantum samplers and classical solvers; it has been shown to produce wavefunctions substantially more compact than classical HCI for strongly correlated systems like Fe-S clusters~\cite{Yoo2026}, while enabling active spaces as large as (24e, 22o) with superior noise resilience~\cite{Ghasemi2026}. Most recently, half-qubit architectures have been introduced to double the addressable active space size, allowing the qubit count to scale with the number of spatial orbitals rather than spin orbitals~\cite{Smith2025, McFarthing2026, Yoshida2026}. Notable among these are Entanglement Forging SQD (EF-SQD)~\cite{Smith2025}, the independently developed HCI-Half qubit SQD (HCI-HSQD)~\cite{McFarthing2026}, and Doubly Occupied Configuration Interaction-QSCI (DOCI-QSCI)~\cite{Yoshida2026}. EF-SQD utilizes Schmidt decomposition to reduce qubit overhead, enabling the simulation of radical chain reactions on IBM Heron superconducting quantum hardware~\cite{Smith2025}. Addressing sampling efficiency, HCI-HSQD explicitly mitigates the compactness bottleneck by applying classical HCI heuristics to guide sampling, thereby reducing energy errors by up to 76\% in large clusters~\cite{McFarthing2026}. In parallel, DOCI-QSCI leverages seniority-zero schemes to efficiently double the qubit budget, utilizing ph-AFQMC to recover dynamic correlation lost by the seniority restriction across diverse molecular benchmarks~\cite{Yoshida2026}. Complementing these techniques, machine-learning-anchored frameworks such as physics-informed generative SQD (PIGen-SQD) leverage generative models combined with perturbative screening to stochastically recover high-fidelity configurations, improving both sampling efficiency and subspace compactness under noisy conditions~\cite{Chayan2026}. While these methods represent significant progress, optimizing the trade-off between sampling cost and subspace compactness remains an active research direction. A concise comparison of representative quantum algorithms for quantum chemistry is provided in Table~\ref{tab:algo_comparison}.
    
\begin{table*}[htbp]
\centering
\caption{Comparative analysis of representative quantum algorithms for
quantum chemistry. All claims are referenced to primary literature.}
\label{tab:algo_comparison}
\renewcommand{\arraystretch}{1.35}
\setlength{\tabcolsep}{5pt}
\begin{tabularx}{\linewidth}{>{\raggedright\arraybackslash}p{2.15cm}XXX}
\hline
\textbf{Feature}
  & \textbf{QPE (Fault-Tolerant)}
  & \textbf{VQE (Variational)}
  & \textbf{SQD / QSCI (Sampling-Based)} \\
\hline
 
Year introduced
  & 1995~\cite{Kitaev1995}; applied to chemistry in
    2005~\cite{AspuruGuzik2005}
  & 2014~\cite{Peruzzo2014}
  & 2023~\cite{Kanno2023}; SQD variant
    demonstrated in 2025~\cite{Robledo2025} \\
 
Hardware regime
  & Requires full fault-tolerant QEC; target
    of application-scale quantum (FASQ)
    computing~\cite{Kitaev1995, Beverland2022}
  & NISQ -- shallow, noisy circuits
    tolerated~\cite{Preskill2018, Peruzzo2014}
  & NISQ-compatible; noise-tolerant by
    design~\cite{Kanno2023, Robledo2025} \\
 
Role of QPU
  & Phase estimator: extracts the eigenvalue of
    the unitary $e^{-i\hat{H}t}$ via quantum
    Fourier transform~\cite{Kitaev1995,
    AspuruGuzik2005}
  & Energy estimator: evaluates expectation
    values $\langle\psi(\vec{\theta})|\hat{H}|
    \psi(\vec{\theta})\rangle$ for a parametrized
    ansatz~\cite{Peruzzo2014, McClean2016}
  & Configuration sampler: generates
    dominant Slater determinants by
    measuring a parametrized state in the
    computational basis~\cite{Kanno2023,
    Robledo2025} \\
 
Classical role
  & Minimal: state preparation and
    post-measurement classical inference;
    The energy readout is
    single-shot~\cite{AspuruGuzik2005,
    Abrams1999}
  & Moderate: outer optimization loop
    minimizes $E(\vec{\theta})$ over circuit
    parameters using gradient-based or
    gradient-free
    optimizers~\cite{Peruzzo2014, Tilly2022}
  & Dominant: Hamiltonian construction and
    sparse CI diagonalization within the
    sampled subspace~\cite{Kanno2023,
    Robledo2025} \\
 
Ansatz/trial state
  & Arbitrary trial state with non-zero
    overlap $\gamma>0$ with the target
    eigenstate; no parametrized
    optimization~\cite{AspuruGuzik2005,
    Abrams1999}
  & Parametrized quantum circuit
    (e.g., \ hardware-efficient or UCC-type
    ansatz); expressibility determines
    accuracy~\cite{Kandala2017,
    Barkoutsos2018, Grimsley2019}
  & Any pre-initialized parameterized state (VQE-produced,
    Post-HF) or Hamiltonian-simulation-evolved; sampling quality depends on overlap~\cite{Kanno2023, Sugisaki2025,
    Nakagawa2024} \\
 
Measurement cost
  & Low: a single coherent run yields a
    phase estimate; precision $\epsilon$
    requires $\mathcal{O}(1/\epsilon)$ total
    queries (Heisenberg
    scaling)~\cite{Kitaev1995, Abrams1999}
  & High: $\mathcal{O}(N^4)$ Pauli terms
    must be measured for molecular
    Hamiltonians; each requires
    $\mathcal{O}(1/\epsilon^2)$ shots per
    optimization
    step~\cite{Verteletskyi2020,
    McClean2016, Tilly2022}
  & Low-to-moderate: shots define the
    subspace; energy is exact within that
    subspace by classical diagonalization;
    no per-term Pauli measurement
    overhead~\cite{Kanno2023,
    Robledo2025} \\
 
Circuit depth
  & Very deep: requires
    $\mathcal{O}(\mathrm{poly}(N)/\epsilon)$
    controlled-unitary applications and
    inverse QFT; infeasible without
    QEC~\cite{AspuruGuzik2005,
    Beverland2022, Nelson2024}
  & Shallow-to-moderate: depth set by ansatz; compatible with near-term gate
    fidelities~\cite{Peruzzo2014,
    Preskill2018, KSV2025}
  & Shallow-to-moderate: depth set by ansatz; comparable to NISQ VQE circuits~\cite{Kanno2023,
    Robledo2025} \\
 
Variational bound on energy
  & Not variational; returns a statistical
    estimate of the eigenvalue with
    controlled
    precision~\cite{Kitaev1995, Abrams1999}
  & Guaranteed in the noise-free limit;
    Hardware noise can violate the bounds,
    invalidating
    results~\cite{Lima2026, McClean2016}
  & Guaranteed within the selected subspace
    by exact classical diagonalization;
    Energy is an upper bound to
    FCI~\cite{Kanno2023, Robledo2025} \\
 
Noise resilience
  & Low: requires active QEC to suppress
    logical errors; impractical for NISQ
    hardware~\cite{Preskill2018,
    Beverland2022}
  & Moderate: variational bound degrades
    under hardware noise; barren plateaus
    impede optimization for large
    systems~\cite{McClean2018, Wang2021}
  & High: QPU used only for sampling; no
    variational optimisation loop;
    Classical diagonalization is exact and
    noise-free~\cite{Kanno2023,
    Robledo2025} \\
 
Scalability
  & Limited by qubit count, circuit depth,
    and QEC overhead; resource estimates
    for small molecules exceed $10^{10}$
    T-gates~\cite{Beverland2022,
    Nelson2024}
  & Limited by classical optimizer
    convergence, $\mathcal{O}(N^4)$
    measurement overhead, and barren
    plateaus in deep
    circuits~\cite{McClean2018, Tilly2022}
  & Limited by sampling efficiency
    (finding new determinants becomes
    exponentially hard), subspace
    compactness, and classical
    diagonalization
    cost~\cite{Reinholdt2025,
    Raisuddin2025} \\
 
Key limitations
  & Demands full fault tolerance with thousands of logical qubits and deep circuits; far beyond NISQ
    hardware~\cite{Beverland2022,
    Nelson2024}
  & Barren plateaus, noise-induced bias,
    high measurement overhead, and
    difficult optimization landscapes for
    strongly correlated
    systems~\cite{McClean2018, Wang2021,
    Lima2026}
  & Quantum sampling efficiencies; CI expansions can be less compact than classical heuristics (e.g.\ HCI); classical diagonalization becomes a bottleneck for large
    subspaces~\cite{Reinholdt2025,
    Raisuddin2025} \\
 
\hline
\end{tabularx}
\end{table*}

    \subsection{Fragmentation and Embedding for Larger Quantum Simulations}

    To extend the reach of these quantum solvers beyond small molecules to macroscopic systems, fragmentation and embedding strategies have become indispensable~\cite{Yamazaki2018}. Prominent approaches include energy-based fragmentation schemes such as the Fragment Molecular Orbital (FMO) method~\cite{Kitaura1999, Fedorov2012} and the Many-Body Expansion (MBE)~\cite{Xantheas1994, Eriksen2019, Ballesteros2023, Burns2024}, which recover total energies through systematic many-body summation of fragment contributions. Recent adaptations of these methods for quantum hardware have demonstrated FMO-based VQE~\cite{Lim2024} and MBE-VQE workflows~\cite{Xu2024} capable of capturing correlation while reducing the required number of qubits. Similarly, the Divide-and-Conquer (DC) technique~\cite{Yang1995, Nakai2011, Nakai2023} constructs global electronic structures from local density matrices, a strategy that has been mapped onto the VQE class of quantum algorithms to address larger systems~\cite{Fujii2022, Yoshikawa2022}. For systems requiring rigorous self-consistency, the Localized Active Space (LAS) method~\cite{Hermes2019, Hermes2020} and Bootstrap Embedding (BE)~\cite{Welborn2016, Ye2019, Cho2025} enforce consistency constraints between fragments, enabling localized quantum chemistry simulations on quantum hardware~\cite{Otten2022, Liu2023}. Other embedding approaches include self-energy-based methods such as Self-Energy Embedding Theory (SEET)~\cite{Lan2015, Lan2017, Rusakov2019} and related self-energy decomposition and embedding approaches~\cite{Romanova2020}. Coupled-cluster downfolding constructs effective Hamiltonians in reduced active spaces by integrating out external degrees of freedom~\cite{Bauman2019, Kowalski2020, Bauman2022}, while Quantum Flow (QFlow) provides a related reduced-dimensionality approach for many-body simulations~\cite{Kowalski2023}. Recent work has demonstrated coupled-cluster downfolding for quantum simulations involving hundreds of orbitals~\cite{Bauman2026}. Foremost among embedding theories are Dynamical Mean-Field Theory (DMFT)~\cite{Metzner1989, Georges1996, Blumenthal2025}, which treats local impurity problems in lattice systems, and Density Matrix Embedding Theory (DMET)~\cite{Knizia2012, Wouters2016}, which enforces self-consistency between a fragment and its bath via the one-particle reduced density matrix. Recent theoretical analyses have refined the understanding of DMET's convergence~\cite{Cances2025, Negre2025}, and hybrid quantum-classical algorithms such as VQE have been applied within DMET frameworks to model correlated active regions in protein–ligand systems~\cite{Kirsopp2022}.

    \subsection{From SQD to Embedded Quantum Simulations}

    Recent advancements in quantum-centric supercomputing have moved towards integrating these fragmentation and embedding frameworks with the sampling-based quantum solvers such as QSCI/SQD. Notably, SQD algorithm has been coupled with DMET to compute ground-state energies of h-chains, conformers of cyclohexane~\cite{Shajan2025}, and ligand-like molecules~\cite{Patra2026}, effectively treating strongly correlated impurities within a mean-field bath. Parallel efforts have replaced the exact diagonalization step in LAS Self-Consistent Field (LASSCF) theory with SQD (LAS-SQD), enabling the simulation of iron-sulfur clusters with fragment sizes that are intractable for standard LASSCF~\cite{Wang2025}. In the regime of high-accuracy thermochemistry, Quantum Bootstrap Embedding (QBE) has been combined with SQD to recover the correlation energy in large molecular systems by enforcing density-matching constraints across overlapping fragments~\cite{Bierman2026}. Pushing towards macroscopic biological systems, the Embedded Wave Function (EWF) framework has integrated SQD to predict relative conformer energies of the 300-atom Trp-cage miniprotein, explicitly treating difficult fragments on quantum hardware while handling the remainder via classical FCI~\cite{Shajan2026}. Furthermore, layered hybrid protocols such as Our own N-layered Integrated molecular Orbital and molecular Mechanics (ONIOM)~\cite{Chung2015} have been extended to use Time-Evolved QSCI (TE-QSCI) for the quantum-mechanical region, facilitating excited-state calculations of biomolecules on hybrid quantum-HPC architectures using Quantinuum's Reimei trapped-ion quantum hardware~\cite{Yamamoto2026}. Collectively, these results establish QSCI/SQD-driven embedding as a potentially robust and scalable route for near-term quantum chemistry, capable of handling complex electronic structures that defy standalone NISQ algorithms.

    \subsection{The Present Work: SQD and DMET-SQD on Quantum Hardware}
    
    In this work, we build upon these developments by presenting a comprehensive experimental assessment of SQD on the IQM Sirius superconducting Quantum Processing Unit (QPU). The device comprises 24 physical qubits, with up to 16 operational qubits arranged in a star topology that enables all-to-all connectivity via a central resonator. Leveraging this connectivity and the native gate set ($\mathrm{R}$, $\mathrm{CZ}$, $\mathrm{Move}$), we implement SQD using the established Local Unitary Cluster Jastrow (LUCJ)~\cite{Motta2023} ansatz and introduce the Linear-CNOT Unitary Coupled-Cluster Singles and Doubles (LCNot-UCCSD)~\cite{Magoulas2023} ansatz within the SQD framework.
    
    We systematically benchmark these approaches across a range of molecular systems, reporting detailed sampled subspace dimensions and energy comparisons against exact FCI references. Our work includes 1D-PES scans for H$_2$ and HeH$^+$ in both STO-3G and 6-31G basis sets, as well as LiH and BeH$_2$ in the STO-3G basis set. Furthermore, we demonstrate the first experimental realization of a dense $32 \times 32$ 2D-PES scan for the water molecule on superconducting hardware. While previous efforts have achieved chemically accurate ground-state energy estimates for water using VQE on trapped-ion processors~\cite{Nam2020} or constructed 2D surfaces utilizing Daubechies wavelets via simulations~\cite{Hong2022}, our work explicitly captures the global energetic interplay between bond length and bond angle directly on a physical superconducting QPU. The benchmark systems were selected to fit within the QPU's limit of 16 operational qubits, permit comparison with exact FCI reference energies, and probe different correlation regimes. All calculations are performed utilizing 10,000 shots per run and evaluating the impact of post-configuration recovery parameters, comparing high-threshold filtering against symmetry-adaptive thresholds, on the accuracy of the diagonalized subspace.
    
    Finally, to demonstrate applicability beyond small molecules, we integrate the SQD(LUCJ) solver with DMET to treat macroscopic chemical systems. We report active-space energies for a set of eight ligand-like molecules (43–75 Da) and the FDA-approved antiviral drug Amantadine (151 Da), benchmarking the results against DMET-CASCI references. The ligand-like molecules provide complementary evidence as they have also been studied at the same geometries using IBM superconducting hardware, albeit with different active-space treatments \cite{Patra2026}. These results, obtained with total runtimes ranging from minutes for small scans to hours for the full 2D surfaces, validate the hybrid quantum-classical workflow. The remainder of this paper is organized as follows: Section~\ref{sec:methodology} describes the experimental implementation on the IQM Sirius QPU's architecture; Section~\ref{sec:results_and_discussion} presents our results and discussion on the obtained benchmarking data from the QPU; and Section~\ref{sec:conclusion} provides our concluding remarks along with offering an outlook on future scalable workflows. In addition, the necessary theoretical background required for this work is detailed in Section~C of the Supplementary Information.
    
\section{Methodology}
\label{sec:methodology}

    \subsection{Sample-based Quantum Diagonalization (SQD) Implementation}
    \label{subsec:sqd_implementation}

        \subsubsection{Software Stack and Electronic Structure Preprocessing}
        \label{subsubsec:software}

            All experiments were carried out on IQM's Sirius 24-qubit superconducting quantum hardware, with full runtime calibration details, hardware metrics, connectivity map, and qubit utilization provided in Section~E in the Supplementary Information. The experiments were implemented using \texttt{Python v3.11.14}~\cite{Van1995, Python311} and accessed via \texttt{iqm-client v32.1.1}~\cite{IQMClient2025}. Quantum circuits were constructed and managed using \texttt{Qiskit v1.4.2}~\cite{Ali2024}, with ansatz-specific construction relying on \texttt{ffsim v0.0.56}~\cite{ffsim2025} for the LUCJ ansatz and native custom Qiskit gate primitives for the LCNot-UCCSD ansatz. The SQD post-processing layer was performed using \texttt{qiskit-addon-sqd v0.10.0}~\cite{QiskitAddonSQD2025}. Classical electronic structure computations, including mean-field, integral generation, coupled-cluster, and FCI, were performed using \texttt{PySCF v2.11.0}~\cite{Sun2015, Sun2018, Sun2020}. Numerical linear algebra relied on \texttt{NumPy v2.3.5}~\cite{Harris2020} and \texttt{SciPy 1.15.3}~\cite{Virtanen2020}, with \texttt{JAX v0.8.1}~\cite{Jax2018} available as an accelerated backend for internal \texttt{ffsim} operations. Results were aggregated and persisted using \texttt{Pandas v2.3.3}~\cite{Pandas2020}, with each experimental record written incrementally to the output file to guard against runtime interruptions.

            For each molecular system, the molecular Hamiltonian was constructed in the molecular orbital (MO) basis using \texttt{PySCF}. A molecular geometry object was instantiated with the target coordinates and the STO-3G basis set (or 6-31G for selected PES scans as described in Section~\ref{subsec:pes}), and a restricted Hartree–Fock (RHF) solution was computed. The one-electron and two-electron integrals in the MO basis were obtained as
            \begin{equation} 
            h_{pq}^{\mathrm{MO}}
            =
            \sum_{\mu \nu}
            C_{\mu p} \,
            h_{\mu\nu}^{\mathrm{AO}} \,
            C_{\nu q}
            \quad
            \forall\quad
            h_{\mu\nu}^{\mathrm{AO}} = \langle \mu | \hat{T} + \hat{V}_{\mathrm{nuc}} | \nu \rangle,
            \label{eq:h1e_mo}
            \end{equation}
            \begin{equation} 
            h_{pqrs}^{\mathrm{MO}} =
            \sum_{\mu\nu\lambda\sigma}
            C_{\mu p} C_{\nu q} C_{\lambda r} C_{\sigma s}\,
            \langle \mu\nu | \lambda\sigma \rangle^{\mathrm{AO}},
            \label{eq:h2e_mo}
            \end{equation}
            where $C_{\mu p}$ are the Hartree–Fock molecular orbital coefficients, $h_{\mu\nu}^{\mathrm{AO}}$ denotes the $(\mu,\nu)$ element of the one-electron integral matrix in the atomic orbital (AO) basis, and $\langle \mu\nu | \lambda\sigma \rangle^{\mathrm{AO}}$ denotes the $(\mu,\nu,\lambda,\sigma)$ element of the two-electron integral tensor in chemist’s notation. The above expressions therefore represent the explicit two-index and four-index AO-to-MO tensor transformations of the one- and two-electron integrals. The nuclear repulsion energy $E_{\mathrm{nuc}}$ was obtained from the mean-field object and added to the electronic correlation energy from the SQD solver. No frozen-core approximation was applied in the SQD benchmarks and subsequent PES calculations described in Section~\ref{subsec:pes}.

        \subsubsection{Active Space Specification}
        \label{subsubsec:active_space}
        
            The active space for each molecule, defined by the number of active spatial orbitals and active electrons $(N_{\mathrm{orb}}, N_{\mathrm{ele}})$, was specified as follows: H$_2$ (2, 2), LiH (6, 4), BeH$_2$ (7, 6), H$_2$O (7, 10), and NH$_3$ (8, 10). All systems employed the STO-3G basis and a closed-shell singlet reference, for which the total spin is $\mathrm{S} = 0$ and the multiplicity is $2\mathrm{S}+1 = 1$. The qubit register for each molecule maps to $2N_{\mathrm{orb}}$ spin orbitals under the Jordan–Wigner (JW) transformation.
            
            The symmetry-constrained Hilbert space, or simply the symmetry space size $|\mathbb{S}|$, was computed for each active space as
            \begin{equation}
                |\mathbb{S}| = \binom{N_{\mathrm{orb}}}{N_{\mathrm{ele}}/2}^{\!2},
                \label{eq:sym_space}
            \end{equation}
            which represents the number of particle-number-conserving determinants in the spin-restricted subspace and serves as the normalization reference for evaluating subspace coverage.

        \subsubsection{Ansatz Construction and Transpilation}
        \label{subsubsec:ansatz}
        
            \paragraph{Local Unitary Cluster Jastrow (LUCJ) ansatz.}
            The LUCJ circuit was constructed using \texttt{ffsim}, which accepts CCSD $t_1$ and $t_2$ amplitude tensors, to define a spin-balanced UCJ operator restricted to a single repetition layer (LUCJ$_{\mathrm{reps}} = 1$). The initial Hartree-Fock reference state was generated first, which was appended to the UCJ unitary, both employing the JW FTQM. CCSD amplitudes ($t_1$, $t_2$) were obtained using \texttt{PySCF} for the LUCJ ansatz initialization. This initialization strategy fixes the LUCJ parameters at their CCSD-optimal values without further variational optimization on hardware, at a pre-execution classical overhead of a single coupled-cluster calculation.
            
            \paragraph{Linear CNOT Unitary Coupled Cluster Singles and Doubles (LCNot-UCCSD) ansatz.}
            The LCNot-UCCSD circuit was constructed entirely within \texttt{Qiskit} using a custom CNOT-ladder decomposition. The Hartree-Fock reference was prepared by applying $X$ gates on the qubits corresponding to occupied spin orbitals. Single excitations were implemented as a $CX-CR_\mathrm{Y}-CX$ sequence; double excitations employed a three-CNOT ladder (CX on pairs $(b,a)$, $(j,i)$, $(b,j)$) followed by a 3-controlled $R_Y$ gate and its inverse entangling structure.
            Throughout, no approximations of the excitation gates were used, retaining full circuit fidelity at the cost of increased depth. Parameters were initialized with MP2 amplitudes ($t_{ij}^{ab}$) computed. The total parameter count equals the number of single plus double excitations generated for the active space. Further details on the ansatz construction are detailed in Section~C of the Supplementary Information.
            
\begin{figure}[htbp]
                \centering
                \includegraphics[width=0.95\linewidth]{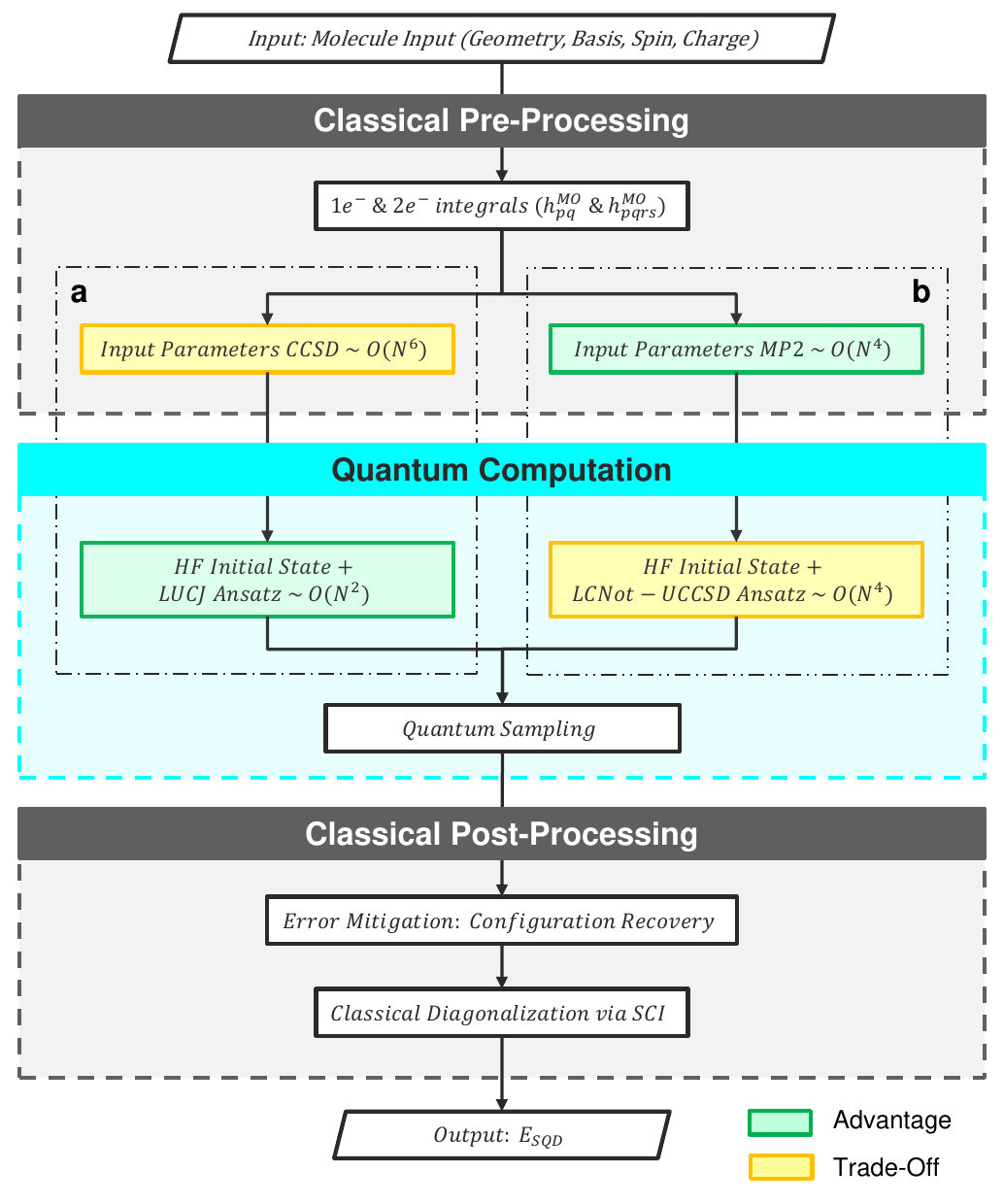}
                \caption{Schematic representation of SQD workflows comparing the (a) LUCJ and (b) LCNot-UCCSD ansätze. The flowchart illustrates the division between Classical Pre-Processing, Quantum Computation, and Classical Post-Processing stages. Green boxes indicate computational advantages (lower algorithmic scaling), while yellow boxes highlight computational trade-offs.}
                \label{fig:sqd_ansatz_workflows}
            \end{figure}

            \paragraph{Transpilation.}
            Both ans\"{a}tze were transpiled to the instruction set architecture (ISA) of the IQM Sirius backend using Qiskit. A custom quantum circuit resource estimator routine computed categorized gate counts (single-qubit, two-qubit, multi-qubit) and circuit depth for each transpiled circuit; these metrics were recorded per experiment in the output CSV, further presented in Section~\ref{sec:results_and_discussion}. The two workflows are schematically represented in Figure~\ref{fig:sqd_ansatz_workflows}. The overall process is divided into Classical Pre-Processing, Quantum Computation, and Classical Post-Processing. After calculating the $1e^-$ and $2e^-$ integrals (Eq.~(\ref{eq:h1e_mo}) and Eq.~(\ref{eq:h2e_mo})), the pipeline splits into two distinct paths. In the LUCJ pipeline (path a), the classically intensive stage is the CCSD pre-computation, representing a classical trade-off with $\mathcal{O}(N^6)$ scaling~\cite{cc_qchem_2007}. However, this is balanced by an advantageous quantum circuit preparation scaling at $\mathcal{O}(N^2)$~\cite{Motta2023}. Conversely, the LCNot-UCCSD pipeline (path b) employs a cheaper, advantageous MP2 initialization at $\mathcal{O}(N^4)$ (considering that the $h_{pq}^{\mathrm{MO}}$ and $h_{pqrs}^{\mathrm{MO}}$ are pre-computed and stored), shifting the complexity to a deeper quantum circuit representing a trade-off scaling at $\mathcal{O}(N^4)$~\cite{Peruzzo2014, Barkoutsos2018}. Both paths converge at quantum sampling, followed by classical error mitigation (configuration recovery) and classical diagonalization via SCI to output the final energy, $E_{SQD}$.

        \subsubsection{Hardware Execution and Sampling Protocol}
        \label{subsubsec:hardware_exec}
        
            Circuits were submitted to the Sirius backend, where the number of shots ($N_{\mathrm{shots}}$) was set to 10{,}000. Each circuit was executed in three independent hardware runs, with separate job submissions and distinct random seeds for the classical post-processing stages. A 0.5~seconds inter-run sleep interval was inserted between submissions to avoid contention on the hardware queue. For each run, a dictionary of observed bitstrings and their frequencies was obtained, which was converted to a full bitstring matrix and probability array for further post-processing.

        \subsubsection{Iterative Configuration Recovery and Classical Diagonalization}
        \label{subsubsec:cr_diag}
        
            The SQD self-consistent loop was executed for 10 iterations per hardware run. On the first iteration, the full bitstring matrix from hardware was used directly without configuration recovery, since no prior estimate of orbital occupancies was available. 
            
            For all subsequent iterations, configurations were recovered using the full bitstring matrix, the empirical probability array, and the running average orbital occupancy vector from the previous iteration, together with target particle counts $N_\alpha = N_\beta = N_{\mathrm{ele}}/2$. This process of configuration recovery ensured a refined bitstring matrix in which configurations inconsistent with the inferred occupancy pattern are replaced by symmetry-recovered alternatives.
            
            At each iteration, the refined bitstring matrix was subjected to Hamming-weight filtering to enforce $N_\alpha$ and $N_\beta$ electrons in the $\alpha$- and $\beta$-spin sectors (corresponding to the right and left halves of each bitstring, respectively). Configurations were then subsampled based on a per-iteration limit $ \epsilon_{\mathrm{s}}$ (samples per batch) for a single batch. The subsampling used a per-run cryptographically seeded RNG to ensure statistical independence across hardware runs. The resulting batch was decomposed into separate $\alpha$- and $\beta$-spin CI string lists.
         
            The subspace Hamiltonian was constructed in the CI basis spanned by all $\alpha$--$\beta$ string pairs from the subsampled batch and diagonalized using a Davidson iterative eigensolver with a maximum of 200 cycles. The solver returns updated average orbital occupancies, which are stored and used to guide configuration recovery in the subsequent iteration as shown in Algorithm~\ref{alg:sqd}.
            
            After 10 iterations, the reported SQD energy is taken to be the minimum over all iterations and their corresponding batches ($n_{\mathrm{batches}} = 1$ throughout) as shown in Algorithm~\ref{alg:sqd}. Each of the three independent hardware runs yields one such energy value, and all results were aggregated in the output file for further analysis. 
            
            The following subspace dimension metrics were recorded per run:
            \begin{itemize}
              \item $|\mathbb{S}_{\mathrm{samp}}|$: number of unique bitstrings in raw measurement counts,
              \item $|\mathbb{S}_{\mathrm{post-cr}}|$: number of configurations after configuration recovery,
              \item $|\mathbb{S}_{\mathrm{sub}}|$: diagonalization subspace dimension,
              \item $|\mathbb{S}|$: symmetry-constrained Hilbert space size from Eq.~(\ref{eq:sym_space}),
              \item $|\mathbb{H}|$: full Hilbert space dimension, $2^{2N_{\mathrm{orb}}}$.
            \end{itemize}

            In this work, the relative sizes of these spaces are characterized by the ratios
            \begin{equation}
            \eta_{\mathrm{sym}} = \frac{|\mathbb{S}|}{|\mathbb{H}|}, 
            \qquad
            \eta_{\mathrm{post\text{-}cr}} = \frac{|\mathbb{S}_{\mathrm{post\text{-}cr}}|}{|\mathbb{S}|},
            \qquad
            \eta_{\mathrm{sub}} = \frac{|\mathbb{S}_{\mathrm{sub}}|}{|\mathbb{S}|},
            \end{equation}
            which quantify the compression achieved at different stages of the SQD workflow. The recovered configurations are then incorporated into the determinant pool prior to classical diagonalization.

            Two sampling strategies were employed for $\epsilon_{\mathrm{s}}$. The first sets $\epsilon_{\mathrm{s}} = 10^8$, which greatly exceeds the total number of sampled configurations ($|\mathbb{S}_{\mathrm{samp}}|$ unique bitstrings from $10^4$ shots) and effectively imposes no subsampling constraint. The second strategy scales $\epsilon_{\mathrm{s}} = \sqrt{|\mathbb{S}|}$, matching the subsample size to the square root of the symmetry-constrained Hilbert space, effectively imposing a symmetry-adapted subsampling constraint.

            For closed-shell systems, it is advantageous to combine the recovered $\alpha$ and $\beta$ configurations into a single set and construct the diagonalization subspace as the tensor product of this set with itself.\cite{Aidan2025} Configurations related by spin exchange often contribute with comparable weight to the ground state; that is, if a configuration of the form $\alpha\beta$ is significant, the corresponding $\beta\alpha$ configuration is typically similarly relevant.\cite{Aidan2025} Consequently, the sets of unique $\alpha$ and $\beta$ configurations obtained during configuration recovery tend to exhibit substantial overlap. Under this commonly observed structure, the effective number of distinct configurations grows approximately with the square root of the full symmetry-constrained Hilbert space, which provides a practical motivation for the $\sqrt{|\mathbb{S}|}$ subsampling heuristic used in this work. The complete SQD execution flow is detailed in Algorithm~\ref{alg:sqd} for further reference.
    
            \begin{algorithm}[htbp]
            \caption{SQD workflow per molecule, per hardware run}
            \label{alg:sqd}
            \begin{algorithmic}[1]
            \Require $h_{pq}^{\mathrm{MO}}$, $h_{pqrs}^{\mathrm{MO}}$, $E_{\mathrm{nuc}}$,
                     \texttt{ansatz\_isa} (transpiled ansatz),
                     $N_\alpha$, $N_\beta$, $N_{\mathrm{shots}}$ (number of shots),
                     $\epsilon_{\mathrm{s}}$ (samples per batch),
                     $I_{\max}$ (iterations),
                     $M_{\mathrm{dav}}$ (Davidson cycles),
                     \texttt{n\_batches} (number of batches)
            \Ensure  $E_{\mathrm{SQD}}$
            
            \State Execute \texttt{ansatz\_isa} on QPU with $N_{\mathrm{shots}}$ shots
                   $\;\rightarrow\;$ \texttt{counts\_dict}
            \State $(\mathbf{B},\,\mathbf{p}) \leftarrow$
                   \Call{counts\_to\_arrays}{\texttt{counts\_dict}}
            \State $\overline{\mathbf{n}} \leftarrow \texttt{None}$
                   \Comment{average orbital occupancy}
            \For{$i = 0,1,\ldots,I_{\max}-1$}
              \If{$\overline{\mathbf{n}}$ is \texttt{None}}
                \State $(\mathbf{B}',\mathbf{p}') \leftarrow (\mathbf{B},\mathbf{p})$
              \Else
                \State $(\mathbf{B}',\mathbf{p}') \leftarrow$
                       \Call{recover\_configurations}{$\mathbf{B},\mathbf{p},
                       \overline{\mathbf{n}},N_\alpha,N_\beta$}
              \EndIf
              \State $\mathrm{batches} \leftarrow$
                     \Call{postselect\_and\_subsample}{$\mathbf{B}',\mathbf{p}',$
                     $N_\alpha, N_\beta, \epsilon_{\mathrm{s}}$, \texttt{n\_batches}}
                     \For{$j = 0,1,\ldots,$\texttt{n\_batches}$-1$}
              \State $(E_j,\,sv_j,\,\overline{n}_j,\,\langle S^2\rangle) \leftarrow$
                     \Call{solve\_fermion}{$\mathrm{batches[j]},
                     h_{pq}^{\mathrm{MO}}, h_{pqrs}^{\mathrm{MO}},
                     \langle S^2\rangle{=}0, M_{\mathrm{dav}}$}
              \State $E_{\mathrm{history}}[i][j] \leftarrow E_j + E_{\mathrm{nuc}}$ 
              \State $\overline{n}_{\mathrm{history}}[i][j] \leftarrow \overline{n}_j$ 
              \EndFor
              \State $\overline{\mathbf{n}} \leftarrow \mathrm{mean}(\overline{n}_{\mathrm{history}}[i][:])$
            \EndFor
            \State \Return $E_{\mathrm{SQD}} \leftarrow \min (E_{\mathrm{history}})$
            \end{algorithmic}
            \smallskip
            \textit{Parameters:}
            $N_{\mathrm{shots}} = 10{,}000$ shots;\;
            $I_{\max} = 10$ iterations;\;
            $\epsilon_{\mathrm{s}} = 10^8$ or $\sqrt{|\mathbb{S}|}$ samples per batch;\;
            $M_{\mathrm{dav}} = 200$ Davidson cycles;\;
            \texttt{n\_batches}$\,= 1$.
            \end{algorithm}

        \subsubsection{Reproducibility}
        \label{subsubsec:repro}
        
            A global \texttt{NumPy} random seed was set at the start of each script to ensure reproducibility of deterministic
            operations. The per-run subsampling seed was drawn from a cryptographically secure source via \texttt{secrets}~\cite{Python311} module, ensuring that the three independent runs operate with distinct stochastic realizations of the configuration recovery and subsampling procedure. Timing information was recorded using the \texttt{timeit}~\cite{Python311} module for the pre-processing, CCSD, FCI, QPU sampling, configuration recovery, and
            post-processing stages independently. All results, including job identifiers, circuit resource estimates, space dimension metrics, energy values, and timings, were written incrementally to the output file after each run to ensure no data loss in the event of a runtime failure.

    \subsection{Potential Energy Surface (PES) Scans Using Sample-based Quantum Diagonalization (SQD)}
    \label{subsec:pes}

        \subsubsection{One-Dimensional PES Implementation}
        \label{subsubsec:pes_1d}

\begin{figure}[H]
                \centering
                \includegraphics[width=\linewidth]{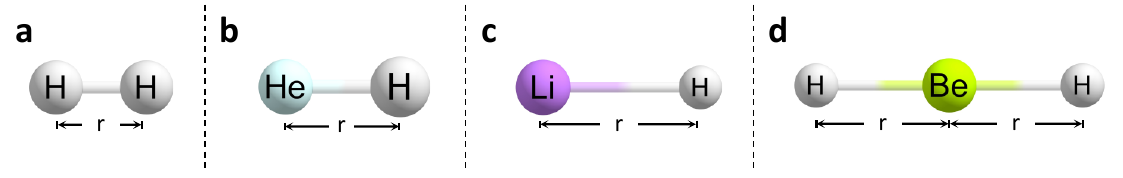}
                \caption{
                    Molecular geometry representations of the one-dimensional potential energy surface (1D-PES) molecules under study: (a) H$_2$, (b) HeH$^+$, (c) LiH, and (d) BeH$_2$. Here, $r$ denotes the varying interatomic distance in ångströms (\AA).
                }
                \label{fig:1d_pes_mol_geom}
            \end{figure}

            The 1D-PES calculations follow the same SQD(LUCJ) workflow described in Section~\ref{subsec:sqd_implementation}, extended by an outer loop over a discrete set of molecular geometries parameterized by the interatomic distance $r$ as shown in Algorithm~\ref{alg:pes}. The molecules studied in the 1D-PES scans are H$_2$ and HeH$^+$ (STO-3G and 6-31G), and LiH and BeH$_2$ (STO-3G), with their coordinate parameterizations illustrated in Figure~\ref{fig:1d_pes_mol_geom}. For all molecules, a non-uniform grid was constructed with closely spaced points covering the bonding region, followed by more widely spaced points extending to capture the dissociation regime. The exact geometries, along with their QPU runtimes, can be found in Section~G in the Supplementary Information for further reference.

            For each grid point, the \texttt{PySCF} workflow (Section~\ref{subsubsec:software}) was re-executed from scratch: the \texttt{gto.Mole} object was rebuilt with the
            updated geometry, a fresh RHF solution was obtained, one- and two-electron integrals were recomputed in the MO basis, and a new CCSD calculation was run
            to provide geometry-specific $t_1$ and $t_2$ amplitudes for the LUCJ circuit. The LUCJ circuit was then transpiled and submitted to hardware with 10{,}000~shots across 3 independent runs. The SQD configuration recovery loop ($I_{\mathrm{max}} = 10$, $\epsilon_{\mathrm{s}} = 10^8$ or $\sqrt{|\mathbb{S}|}$, $M_{\mathrm{dav}}= 200$) was executed independently for each run at each geometry. Per-geometry result records additionally store the $r$ coordinate to enable direct PES reconstruction during analysis. All results were appended incrementally to a single output file per molecule, allowing partial recovery if the scan was interrupted.

        \subsubsection{Two-Dimensional PES Implementation}
        \label{subsubsec:pes_2d}

\begin{figure}[H]
                \centering
                \includegraphics[width=0.33\linewidth]{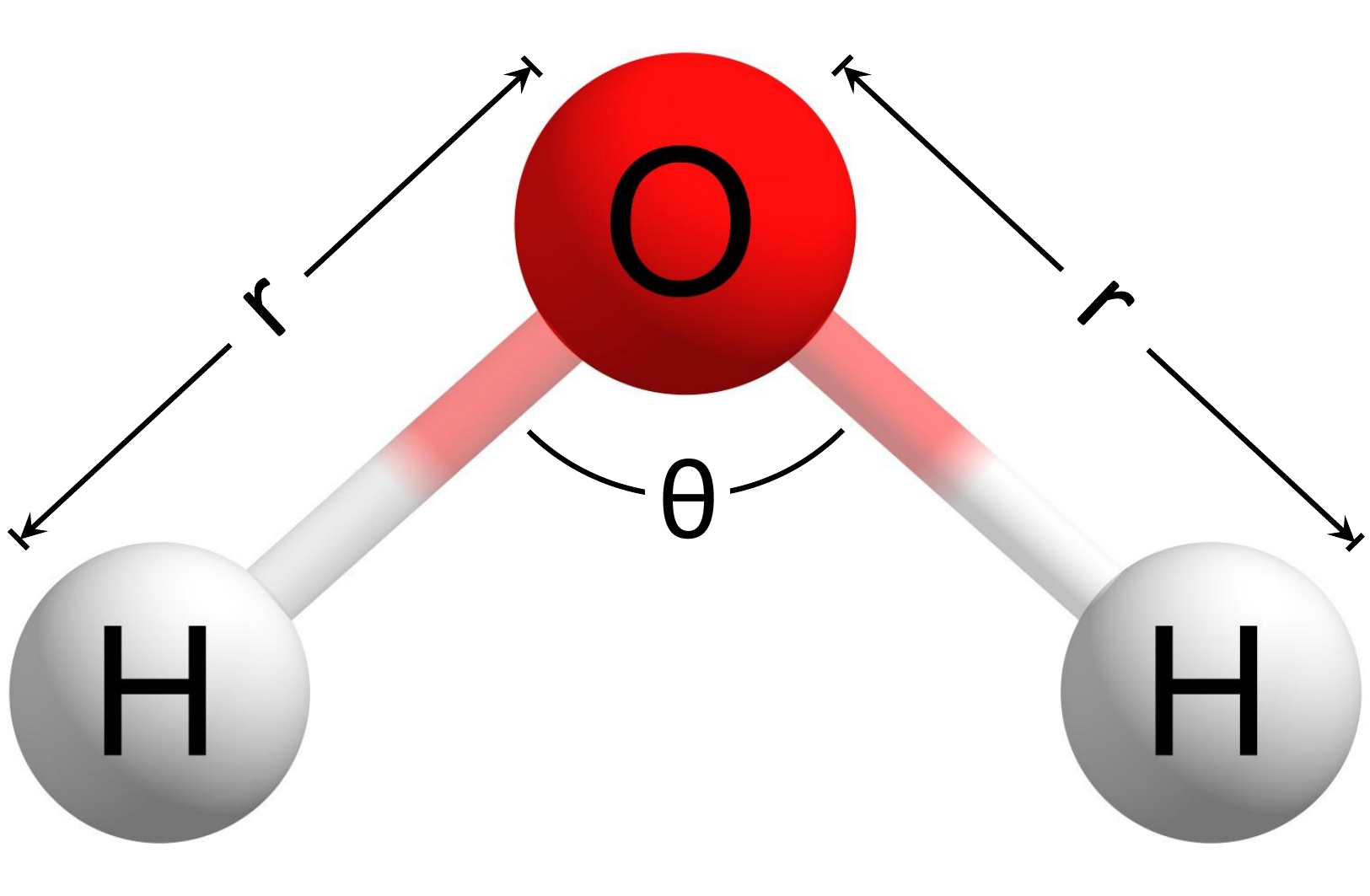}
                \caption{
                    Molecular geometry representation of H$_2$O used for the two-dimensional potential energy surface (2D-PES) scan. Here, $r$ denotes the varying O--H interatomic distance in ångströms (\AA), and $\theta$ denotes the varying H--O--H bond angle in degrees ($^\circ$).
                }
                \label{fig:2d_pes_mol_geom}
            \end{figure}

            The 2D-PES scan for water (H$_2$O, STO-3G) spans a $32 \times 32$ grid in O--H bond length $r$ and H--O--H bond angle $\theta$, as illustrated in Figure~\ref{fig:2d_pes_mol_geom}, for a total of 1{,}024 grid points. The grid geometries were pre-computed and stored as JSON entries, each containing the Cartesian coordinate string, scalar $r$, and scalar $\theta$. The geometry file was loaded at runtime, and the outer loop iterated over all 1{,}024 entries. The LUCJ circuit parameters were re-initialized from geometry-specific CCSD amplitudes at every grid point, following the same procedure as described for the 1D benchmark molecules in Section~\ref{subsubsec:pes_1d}.

           The 2D-PES protocol differs from the 1D case in two respects. The number of independent hardware runs per grid point was set to 1 rather than 3, and the inter-run sleep interval was set to $0.0$~s, reflecting the need to minimize total wait time across 1{,}024 sequential submissions. The SQD configuration recovery loop parameters, such as $I_{\mathrm{max}}$, $\epsilon_{\mathrm{s}}$, and $M_{\mathrm{dav}}$, were identical to those used in the 1D case, as described in Section~\ref{subsubsec:pes_1d}. Each result record stores $r$, $\theta$, the Cartesian geometry string, and the full set of SQD diagnostic metrics, with results appended incrementally to the output file. Details of the construction of the complete set of molecular geometries used in the 2D-PES scan, along with the heatmap of per-geometry QPU runtimes, can be found in Section~G in Supplementary Information. The outer loop of the 2D-PES scan is detailed in Algorithm~\ref{alg:pes}.

            \begin{algorithm}[H]
            \caption{PES scan outer loop (1D shown; 2D is structurally identical
                     with an additional $\theta$ coordinate per entry)}
            \label{alg:pes}
            \begin{algorithmic}[1]
            \Require Geometry dictionary
                     $\mathcal{G} = \{k : (\mathrm{geom}_k, r_k)\}_{k=1}^{K}$;\;
                     SQD parameters: $N_{\mathrm{shots}}$,
                     \textit{runs},
                     $I_{\max}$, $\epsilon_{\mathrm{s}}$, $M_{\mathrm{dav}}$
            \Ensure  $\{E_{\mathrm{SQD}}(r_k)\}_{k=1}^{K}$
            
            \For{each index $k$ in $\mathcal{G}$}
              \State $(\mathrm{geom}_k,\,r_k) \leftarrow \mathcal{G}[k]$
                     \Comment{also read $\theta_k$ for 2D}
              \State Build \texttt{mol} $\leftarrow$
                     \texttt{gto.Mole}($\mathrm{geom}_k$, \textit{basis} = STO-3G)
              \State \texttt{mf} $\leftarrow$ \texttt{scf.RHF(mol).run()}
              \State Compute $h_{pq}^{\mathrm{MO}}$, $h_{pqrs}^{\mathrm{MO}}$,
                     $E_{\mathrm{nuc}}$ from \texttt{mf}
              \State Run \texttt{cc.CCSD(mf)} $\rightarrow$ $t_1$, $t_2$
              \State \texttt{ansatz} $\leftarrow
                     \mathrm{LUCJ}(N_{\mathrm{orb}}, N_{\mathrm{ele}}, t_1, t_2)$
              \State \texttt{ansatz\_isa} $\leftarrow$
                     \texttt{transpile(ansatz, backend)}
              \For{$\mathrm{run} = 1,\ldots,\textit{runs}$}
                \State Execute \texttt{ansatz\_isa} on QPU $\rightarrow$ \texttt{counts}
                \State Run Algorithm~\ref{alg:sqd} $\rightarrow$ $E_{\mathrm{SQD}}$
                \State Append $(k,\,r_k,\,[\theta_k,]\,E_{\mathrm{SQD}},\,
                       \mathrm{diagnostics})$ to output file
              \EndFor
            \EndFor
            \end{algorithmic}
            \smallskip
            \textit{Parameters (1D):}\;
            $K = \{$ H$_2$: $25$, HeH$^+$: $25$, LiH: $30$, BeH$_2$: $27$ $\}$;\; \textit{runs}$\,= 3$;\; $N_{\mathrm{shots}}\,= 10{,}000$.\\
            \textit{Parameters (2D):}\;
            $K = \{$ H$_2$O: $1{,}024$ $\}$ ($32\times32$ grid);\;
            \textit{runs}$\,= 1$;\; $N_{\mathrm{shots}}\,= 10{,}000$.
            \end{algorithm}

    \subsection{Sample-based Quantum Diagonalization within Density Matrix Embedding Framework (DMET-SQD)} \label{sec: dmet_sqd_methodology}

\begin{figure}[H]
            \centering
            \includegraphics[angle=-90,width=0.85\linewidth]{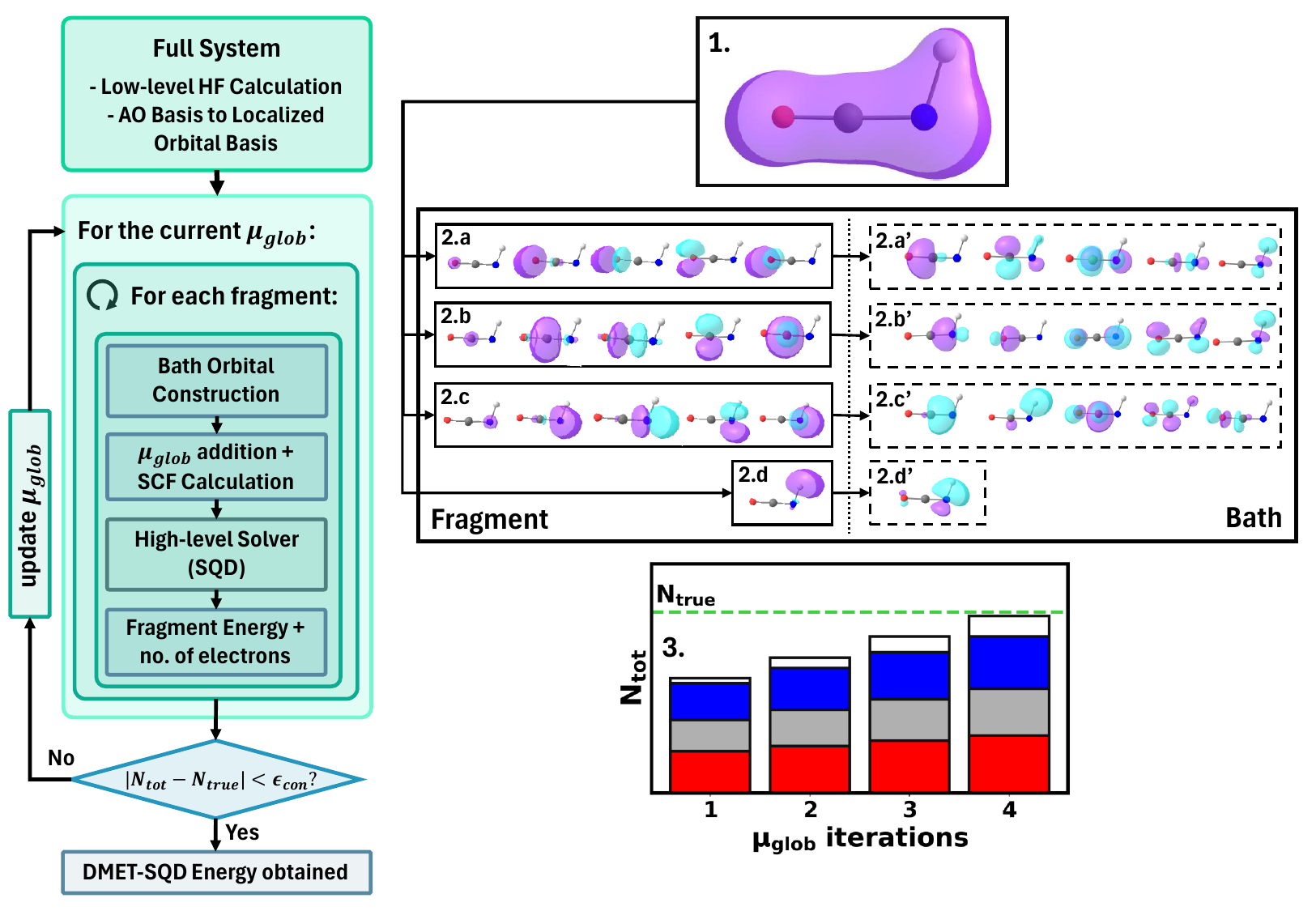}
            \caption{Overall workflow of the DMET-SQD approach. Panel \textbf{1} represents the self-consistent field (SCF) density of the full molecule. Panels \textbf{2.a}/\textbf{2.a$^\prime$}, \textbf{2.b}/\textbf{2.b$^\prime$}, \textbf{2.c}/\textbf{2.c$^\prime$}, and \textbf{2.d}/\textbf{2.d$^\prime$} show the fragment orbitals and corresponding bath orbitals for the O, C, N, and H atom fragments, respectively. Panel \textbf{3} provides a schematic illustration of how successive updates of the global chemical potential $\mu_{\mathrm{glob}}$ drive the total electron number $N_{\mathrm{tot}}$ towards the target value $N_{\mathrm{true}}$.}
            \label{fig:dmet_sqd_workflow}
        \end{figure}

        The current work employs the DMET framework in conjunction with an RHF reference~\cite{Kawashima2021, Kirsopp2022, Shajan2025, Patra2025}, which is appropriate for the closed-shell systems considered here and provides a consistent mean-field starting point for constructing the bath orbitals. The impurity problems are subsequently solved using the SQD approach, which relies on a selected configuration interaction (SCI) procedure for diagonalization. Although SCI is formally neither size-consistent nor size-extensive, the sampling-driven configuration recovery ensures that the dominant determinants in the $\mathbb{H}$ are preferentially captured.  An overall workflow of DMET-SQD is provided in Figure~\ref{fig:dmet_sqd_workflow}. To rigorously benchmark the accuracy of this approach, FCI or Complete Active Space CI (CASCI)~\cite{Olsen1988} is employed as a reference solver within DMET, along with DMET-CCSD, to understand the additional correlation captured by employing DMET-SQD(LUCJ). Furthermore, a one-atom-per-fragment partitioning is adopted as a stringent test of the DMET bath construction, including chemically nontrivial scenarios where covalent bonds are severed across fragments. This aggressive fragmentation minimizes the size of individual impurity problems, thereby reducing quantum resource requirements and sampling overhead, while providing a demanding test of whether inter-fragment correlations are consistently recovered. 
        
        Building on this framework, a set of eight ligand-like molecules is selected~\cite{Patra2026}, several of which are relevant to pharmaceutical and chemical applications~\cite{Listro2022, Ghosh2020, Eads2021, Caimi2017}. The set comprises {HOCN (cyanic acid), CH$_3$NO (formaldehyde oxime), CH$_5$NO (methoxyamine), C$_2$H$_3$NO (methyl isocyanate), C$_2$H$_5$NO (acetaldehyde oxime), CH$_4$N$_2$O (carbamide/urea), NOCl (nitrosyl chloride), and HOSCN (hydroxythiocyanate)}, and is chosen to systematically evaluate the performance and accuracy of the DMET-SQD workflow across chemically diverse yet computationally tractable systems. All systems are treated in the STO-3G basis using a one-atom-per-fragment DMET partitioning. As an illustrative example, Section 2 of Figure~\ref{fig:dmet_sqd_workflow} shows the fragmentation of the CHNO molecule on an atom-by-atom basis. Panels 2.a, 2.b, 2.c, and 2.d depict the isosurfaces of the fragment orbitals, while panels 2.a', 2.b', 2.c', and 2.d' show the corresponding bath orbitals when the selected fragments are [O], [C], [N], and [H], respectively. The fragment orbitals are the localised orbitals associated with the selected fragment. In contrast, the bath orbitals are constructed from the entanglement between the fragment and its environment, and can be interpreted as linear combinations of environment orbitals that are most strongly coupled to the fragment. As a result, they are typically delocalized over the environment, with minimal weight on the fragment itself.
        
        To remain compatible with the available quantum hardware, impurity systems containing more than $16$ spin orbitals were truncated using an active-space selection of four highest occupied molecular orbitals (4 HOMO) and four lowest unoccupied molecular orbitals (4 LUMO), resulting in impurity problems with $(N_{\mathrm{orb}}, N_{\mathrm{ele}})\leq(8, 8)$ and at most $N_Q=16$ qubits. Unless otherwise specified, the SQD sampling experiments were performed using $N_{\mathrm{shots}}=10{,}000$ measurements per fragment, while $\epsilon_s$ was chosen sufficiently large ($10^5$) so that it is a non-factor and to retain all recovered configurations without stochastic truncation.
        
\begin{figure}[H]
            \centering
            \includegraphics[width=0.58\linewidth]{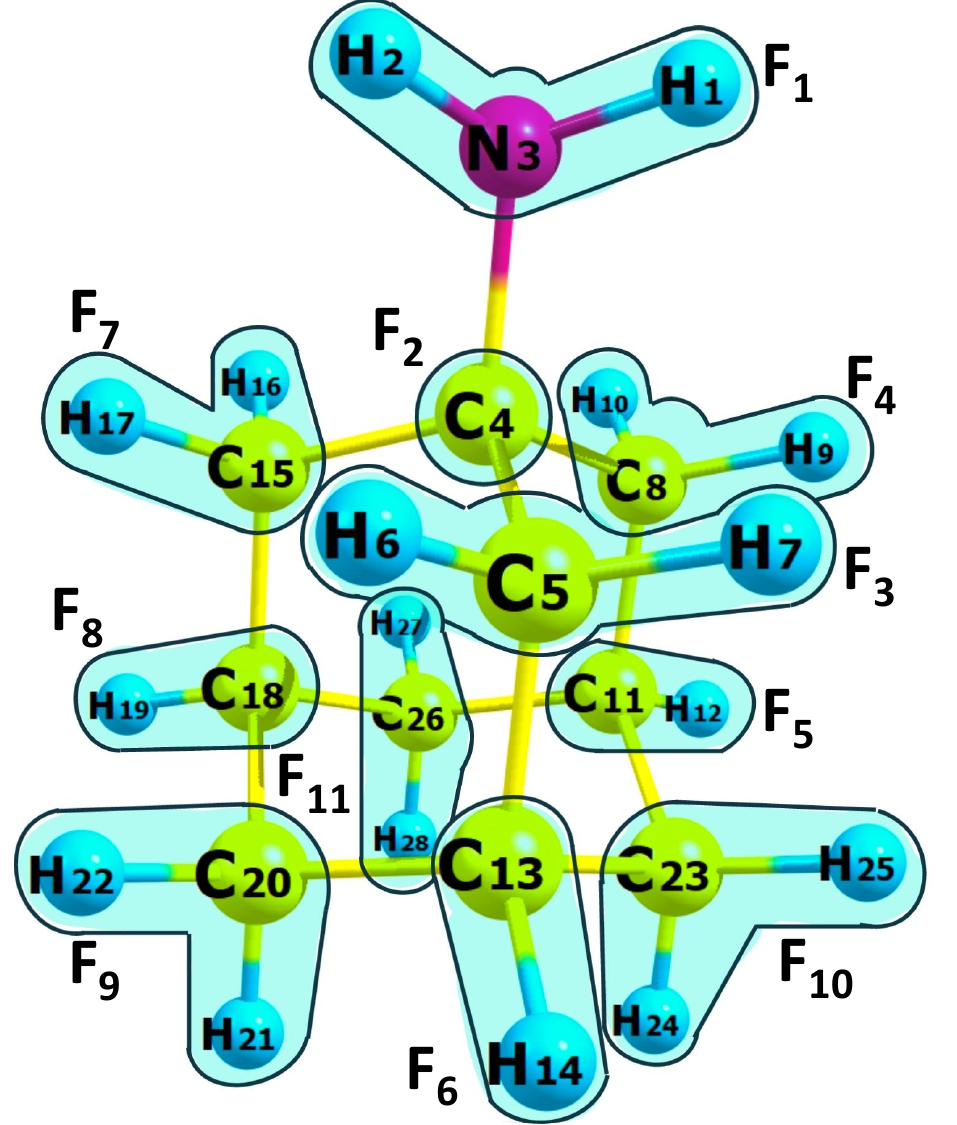}
            \caption{Fragmentation details of Amantadine Molecule: The molecule was fragmented into 11 fragments, such that in STO-3G basis each fragment contains at most 7 molecular spatial orbitals and at least 5 molecular spin orbitals. Using the active space selection method, we pick out the 4 HOMO - 4 LUMO orbitals in each impurity, leading to 16 molecular spin orbitals or 16 qubits per fragment.}
            \label{fig:amantadine_frags}
        \end{figure}

        Next, we extend this work to a significantly larger system, the amantadine molecule ($\mathrm{C_{10}H_{17}N}$)~\cite{Cady2010}, which is an FDA approved~\cite{AmantadineFDA}  antiviral and antiparkinsonian drug, historically used for the treatment of influenza A infections and currently prescribed for Parkinson’s disease. Its molecular weight is $151.25$ Da, and it contains 11 atoms, which are period 2 elements, making it a sufficiently large system suitable for our work. We retain the same minimal basis set (STO-3G), RHF-based DMET formulation, and SQD impurity solver framework, along with DMET-CASCI and DMET-CCSD benchmarking, but adopt a more flexible fragmentation strategy tailored to the molecular structure as shown in Figure~\ref{fig:amantadine_frags}. Specifically, the system is partitioned into 11 fragments containing one (F2), two (F5, F6, F8), or three atoms (F1, F3, F4, F7, F9, F10, F11), moving us towards a fragmentation scheme beyond one-atom per fragment, making larger molecular simulations feasible and a tractable extension. This choice enables a more balanced distribution of correlation within each impurity while maintaining hardware-compatible active spaces. All impurity problems are again truncated to at most $(N_{\mathrm{orb}}, N_{\mathrm{ele}})=(8,8)$, corresponding to $N_Q=16$ qubits, and the SQD sampling parameters are kept consistent with the ligand-like molecule study.  To investigate the influence of the measurement budget on the final DMET-SQD energy, four independent simulations were performed with shot numbers $N_{\mathrm{shots}}=\{500,1000,5000,10000\}$. We also note that a full-scale simulation of the molecule in the same minimal basis set, without any fragmentation, would require 144 qubits, while in the current fragmentation scheme, the quantum simulation of the largest fragment without any active space selection would still only require 28 qubits.
        
\newpage

\section{Results and Discussion} \label{sec:results_and_discussion}

    \subsection{Sample-based Quantum Diagonalization (SQD)} \label{sec:4.1.SQD}

        The SQD framework was applied to five molecular systems of increasing complexity: H$_2$, LiH, BeH$_2$, H$_2$O, and NH$_3$. The relevant electronic structure properties of each system are summarised in Table~\ref{tab:sqd_ccsd_fci_energies}. The QPU execution times and molecular geometries used throughout this subsection are provided in Section~F in the Supplementary Information. As the active space grows from H$_2$ to NH$_3$, the $|\mathbb{H}|$ dimension increases rapidly as a function of the number of qubits $N_Q$, scaling as $2^{N_Q}$ from $2^4 = 16$ to $2^{16} = 65{,}536$, while the $|\mathbb{S}|$ to $|\mathbb{H}|$ dimension ratio $\eta_{\mathrm{sym}}$ decreases considerably, ranging from $0.25$ for H$_2$ down to as low as $0.027$ for H$_2$O. This progressive increase in problem complexity provides a meaningful benchmark for evaluating the performance of the quantum ansätze considered in this work, including LUCJ (Section~\ref{subsec:LUCJ}) and LCNot-UCCSD (Section~\ref{subsec:LCNot-UCCSD}) on IQM Sirius quantum hardware (Section~E in the Supplementary Information provides further information on the QPU's calibration and hardware metrics). For each molecule, the FCI energy serves as the exact reference against which all SQD results are assessed, with CCSD energies additionally reported to contextualise the accuracy of the quantum approaches relative to a well-established classical correlated method, as shown in Table~\ref{tab:sqd_ccsd_fci_energies}.

        \begin{table}[H]
            \centering
            \caption{\label{tab:sqd_ccsd_fci_energies}
            Properties of the molecules under study, including the number of molecular orbitals $N_{\mathrm{orb}}$, electrons $N_{\mathrm{ele}}$, the symmetry space dimension $|\mathbb{S}|$, the Hilbert space dimension $|\mathbb{H}|$, the symmetry space dimension to Hilbert space dimension ratio $\eta_{sub}$, and classical reference ground-state energies obtained using CCSD ($E_{\mathrm{CCSD}}$, Ha) and FCI ($E_{\mathrm{FCI}}$, Ha) for the molecular geometries listed in Section~F in the Supplementary Information.}
    
            \begin{tabular}{|c|c|r|r|r|r|r|}
                \hline
                \multicolumn{1}{|c|}{Molecule} 
                & \multicolumn{1}{c|}{$(N_{\mathrm{orb}}, N_{\mathrm{ele}})$} 
                & \multicolumn{1}{c|}{$|\mathbb{S}|$}
                & \multicolumn{1}{c|}{$|\mathbb{H}|$}
                & \multicolumn{1}{c|}{$\eta_{sym}$}
                & \multicolumn{1}{c|}{$E_{\mathrm{CCSD}}$ (Ha)} 
                & \multicolumn{1}{c|}{$E_{\mathrm{FCI}}$ (Ha)}\\
                \hline
                H$_2$ & (2,2) & 4 & 16 & 0.25 & -1.12784260513579 & -1.12784250431471 \\
                \hline
                LiH  & (6,4) & 225 & 4096 & 0.0549 & -7.88116714390917 & -7.88117835166515 \\
                \hline
                BeH$_2$ & (7,6) & 1125 & 16384 & 0.0748 & -15.54741326429858 & -15.54779792987434 \\
                \hline
                H$_2$O & (7,10) & 441 & 16384 & 0.0269 & -75.01541232667518 & -75.01553351836277 \\
                \hline
                NH$_3$ & (8,10) & 3136 & 65536 & 0.0479 & -55.52092972113001 & -55.52114799754608 \\
                \hline
            \end{tabular}
        \end{table}
    
        \subsubsection{Local Unitary Cluster Jastrow (LUCJ)}
        \label{subsec:LUCJ}

        First, we employ the LUCJ ansatz as the sample-generating circuit within the SQD framework for execution on the IQM Sirius QPU. The corresponding quantum circuit resources for each molecular system are detailed in Section~D of the Supplementary Information.
        
\begin{figure}[H]
                \centering
                \includegraphics[width=0.825\linewidth]{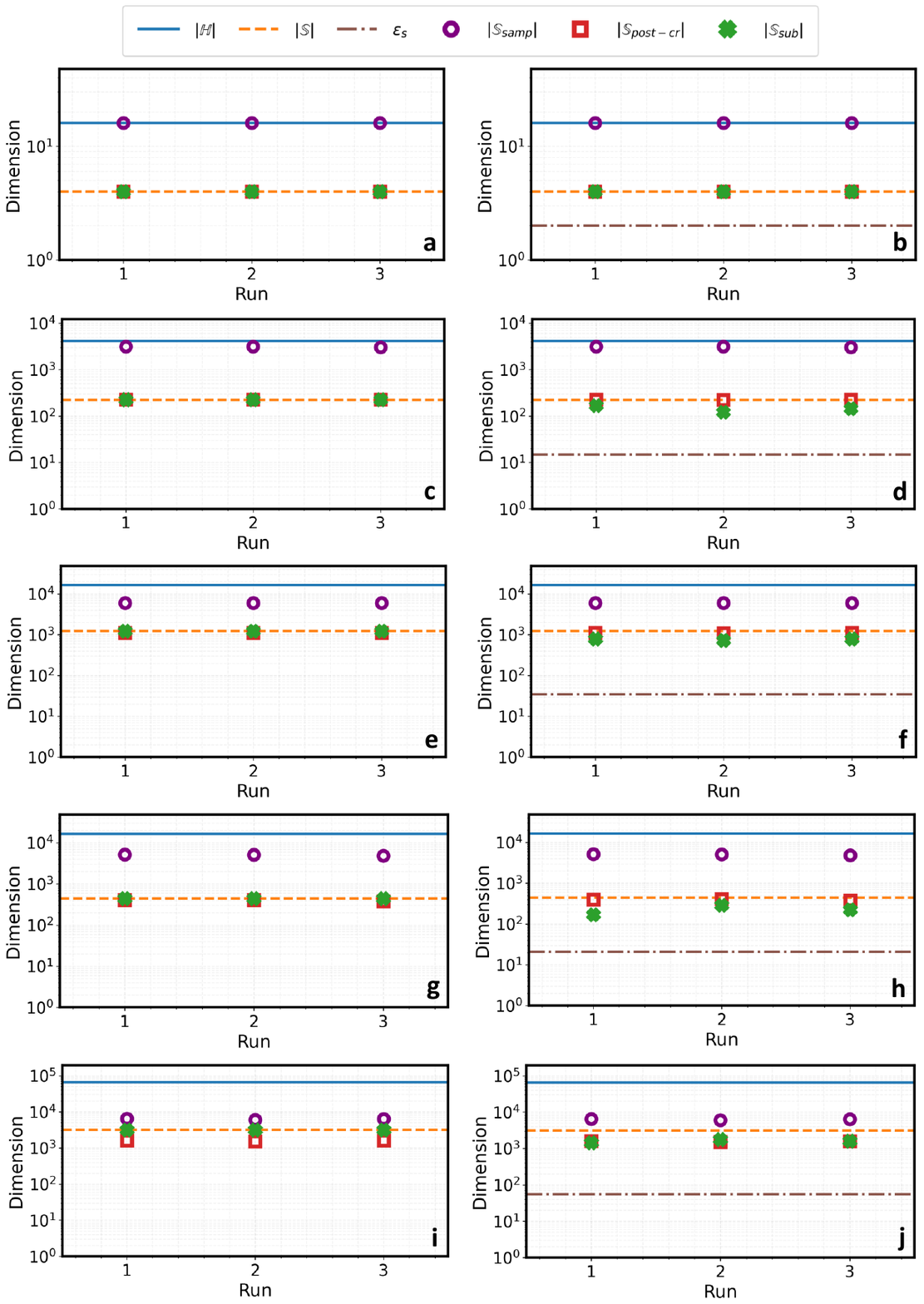}
                \caption{
                Dimensions (log scale) on the y-axis versus runs on the x-axis for SQD(LUCJ) across various molecules. Subplots (a,b) correspond to H$_2$, (c,d) to LiH, (e,f) to BeH$_2$, (g,h) to H$_2$O, and (i,j) to NH$_3$. The quantity $|\mathbb{H}|$ denotes the Hilbert space dimension, $|\mathbb{S}|$ the symmetry space dimension, $|\mathbb{S}_{\mathrm{samp}}|$ the sampling space dimension, $|\mathbb{S}_{\mathrm{post\text{-}cr}}|$ the post–configuration-recovery dimension, $|\mathbb{S}_{\mathrm{sub}}|$ the diagonalization subspace dimension, and $\epsilon_{s}$ the samples per batch, a user-defined SQD parameter. For subplots (a,c,e,g,i), $\epsilon_{s}=10^{8}$ (non-factor), whereas for subplots (b,d,f,h,j), $\epsilon_{s}=\sqrt{|\mathbb{S}|}$ for the respective molecules.
                }
                \label{fig:sqd_lucj_dim}
            \end{figure}

        The dimensional behaviour of the SQD workflow across successive runs is illustrated in Figure~\ref{fig:sqd_lucj_dim} for all five molecules, under two choices of $\epsilon_{\mathrm{s}}$. In the case of the smaller molecules considered, such as H$_2$ and LiH, the post-configuration-recovery dimension $|\mathbb{S}_{\mathrm{post\text{-}cr}}|$ converges rapidly towards the molecule's symmetry space dimension $|\mathbb{S}|$, whereas for BeH$_2$ and H$_2$O we access approximately $0.90$ of $|\mathbb{S}|$, and for NH$_3$ about $0.50$. When $\epsilon_{\mathrm{s}} = 10^8$, the diagonalization subspace dimension $|\mathbb{S}_{\mathrm{sub}}|$ consistently spans the entire $|\mathbb{S}|$, achieving a diagonalization subspace dimension to symmetry space dimension ratio $\eta_{\mathrm{sub}} = 1.0$ across all runs and all molecules. In contrast, with $\epsilon_{\mathrm{s}} = \sqrt{|\mathbb{S}|}$, the subspace coverage relative to $|\mathbb{S}|$ is necessarily reduced, with $\eta_{\mathrm{sub}}$ values ranging from approximately $0.38$ to $0.75$ depending on the molecule and run, reflecting the more compact $|\mathbb{S}_{\mathrm{sub}}|$.

\begin{figure}[H]
                \centering
                \includegraphics[width=\linewidth]{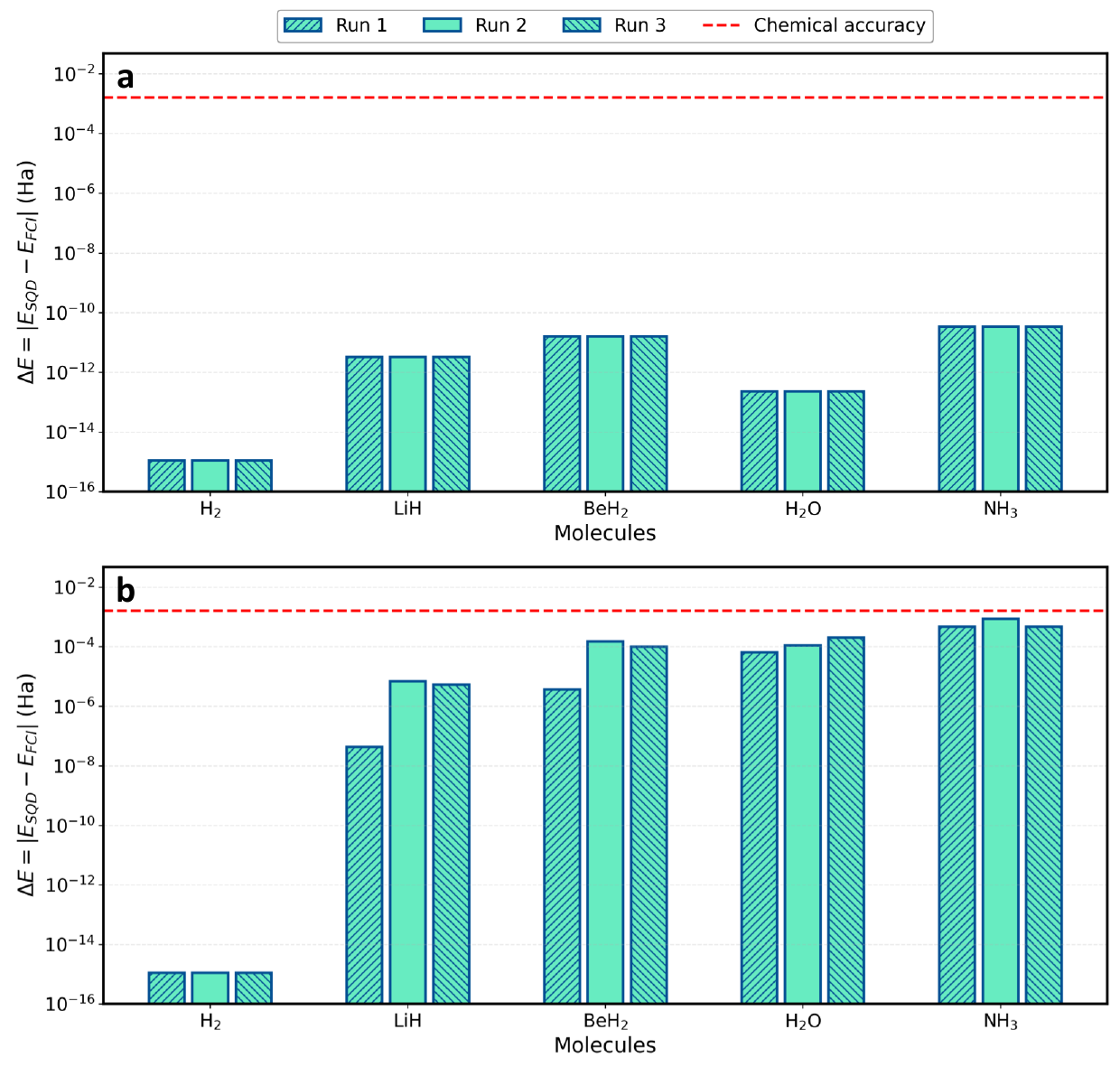}
                \caption{
                Energy difference between SQD(LUCJ) and FCI, in Hartree (Ha), for three independent runs on the IQM Sirius QPU for each molecule. The y-axis represents the energy difference $\Delta E = |E_{\mathrm{SQD}} - E_{\mathrm{FCI}}|$ (log-scale), and the x-axis represents the molecules (H$_2$, LiH, BeH$_2$, H$_2$O, and NH$_3$). Subplots (a) and (b) correspond to $\epsilon_{s} = 10^{8}$ (non-factor) and $\epsilon_{s} = \sqrt{|\mathbb{S}|}$, respectively, where $|\mathbb{S}|$ is the symmetry space dimension of the corresponding molecule.
                }
                \label{fig:sqd_lucj_delta_e}
            \end{figure}

        The energetic consequences of these two subsampling regimes is visualised in Figure~\ref{fig:sqd_lucj_delta_e}. For $\epsilon_{\mathrm{s}} = 10^8$, the SQD(LUCJ) workflow recovers the FCI ground-state energy to within sub-nanohartree precision for all five molecules across all three independent runs, with no run-to-run variance observed. This is primarily due to $|\mathbb{S}_{\mathrm{sub}}|$ spanning the complete $|\mathbb{S}|$, as reflected by $\eta_{\mathrm{sub}}$. When $\epsilon_{\mathrm{s}}$ is reduced to $\sqrt{|\mathbb{S}|}$, the accuracy is seen to degrade depending on the subspace diagonalized and the system size. For H$_2$, $\mathbb{S}$ is small enough that even $\epsilon_{\mathrm{s}} = 2$ suffices to span it completely, and FCI-level accuracy is preserved to numerical precision across all runs. For LiH with $\epsilon_{\mathrm{s}} = 15$, energy deviations from FCI are on the order of $10^{-6}~Ha$. For BeH$_2$, H$_2$O, and NH$_3$, the reduced $|\mathbb{S}_{\mathrm{sub}}|$ leads to a more pronounced degradation, while still providing energetics within the chemical accuracy threshold against the FCI benchmark. The exact energies can be found in Section~D of the Supplementary Information.

        \newpage
        \subsubsection{Linear Scaling CNOT-Unitary Coupled Cluster Singles and Doubles (LCNot-UCCSD)}
        \label{subsec:LCNot-UCCSD}

        The LCNot-UCCSD ansatz was similarly employed as the sample-generating circuit within the SQD framework and executed on the IQM Sirius QPU. The quantum circuit resources associated with each molecular system are detailed in Section~D of the Supplementary Information. The qubit counts are identical to those used in the LUCJ workflows, as the same geometry and complete active space were employed. The transpiled quantum gate depths are, however, substantially larger across all systems. Consequently, it should be noted that the hardware execution of the LCNot-UCCSD circuits resulted in quantum sampling where configuration recovery followed by classical subspace diagonalization was possible for three molecules, H$_2$, LiH, and BeH$_2$, whereas it failed due to the lack of any samples falling in $\mathbb{S}$ for H$_2$O and NH$_3$. This failure may be attributed to the reduced $\eta_{\mathrm{sym}}$ as well as the increased gate depths for these molecules. Detailed analysis of this phenomenon is deferred to future work.

\begin{figure}[H]
            \centering
            \includegraphics[width=0.825\linewidth]{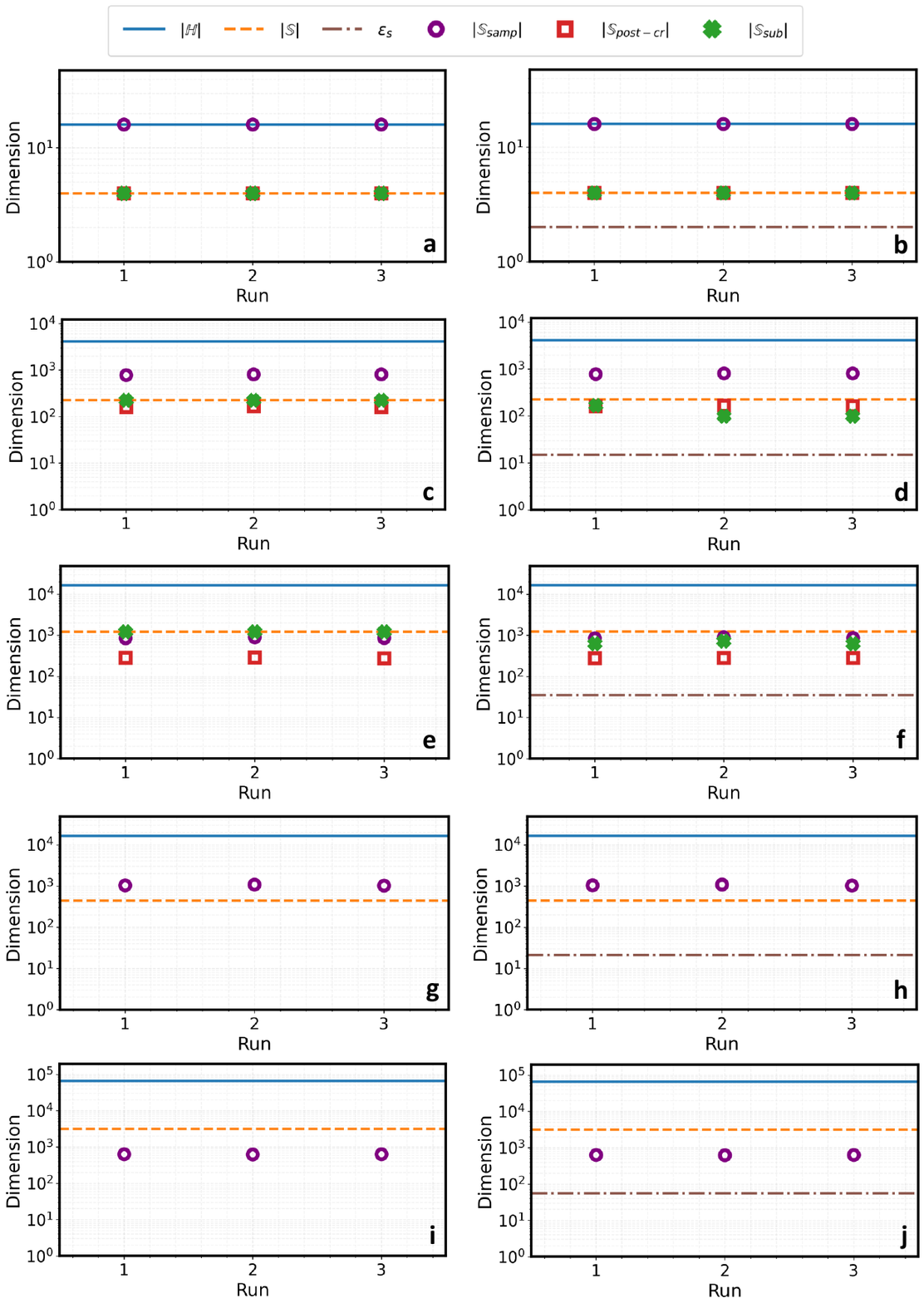}
                \caption{
                Dimensions (log scale) on the y-axis versus runs on the x-axis for SQD(LCNot-UCCSD) across various molecules. Subplots (a,b) correspond to H$_2$, (c,d) to LiH, (e,f) to BeH$_2$, (g,h) to H$_2$O, and (i,j) to NH$_3$. The quantity $|\mathbb{H}|$ denotes the Hilbert space dimension, $|\mathbb{S}|$ the symmetry space dimension, $|\mathbb{S}_{\mathrm{samp}}|$ the sampling space dimension, $|\mathbb{S}_{\mathrm{post\text{-}CR}}|$ the post–configuration-recovery dimension, $|\mathbb{S}_{\mathrm{sub}}|$ the diagonalization subspace dimension, and $\epsilon_{s}$ the samples per batch, a user-defined SQD parameter. For subplots (a,c,e,g,i), $\epsilon_{s}=10^{8}$ (non-factor), whereas for subplots (b,d,f,h,j), $\epsilon_{s}=\sqrt{|\mathbb{S}|}$ for the respective molecules.
                }
            \label{fig:sqd_lcnot-uccsd_dim}
        \end{figure}

        The dimensional convergence behaviour of the SQD(LCNot-UCCSD) workflow is shown in Figure~\ref{fig:sqd_lcnot-uccsd_dim} for both values of $\epsilon_{\mathrm{s}}$. A notable distinction relative to the LUCJ results emerges in the post-configuration-recovery coverage. For LiH, $\eta_{\mathrm{post\text{-}cr}}$ values lie in the range of approximately $0.71$ -- $0.74$, compared to unity for LUCJ, indicating that the deeper LCNot-UCCSD circuits generate bitstring distributions that cover a relatively smaller fraction of $\mathbb{S}$ following configuration recovery. This effect is considerably more pronounced for BeH$_2$, where $\eta_{\mathrm{post\text{-}cr}}$ drops to approximately $0.23$ -- $0.24$, in stark contrast to the $\sim 0.91$ observed for LUCJ. This reduction in post-configuration-recovery coverage is consistent with the expectation that increased $G_D$ leads to greater hardware noise, which in turn distorts the output distribution and reduces the effective overlap of the sampled bitstrings with the ground state. Despite this degraded coverage, the diagonalization subspace under $\epsilon_{\mathrm{s}} = 10^8$ continues to span the full $\mathbb{S}$ ($\eta_{\mathrm{sub}} = 1.0$) for all three molecules across all runs.

\begin{figure}[H]
            \centering
            \includegraphics[width=\linewidth]{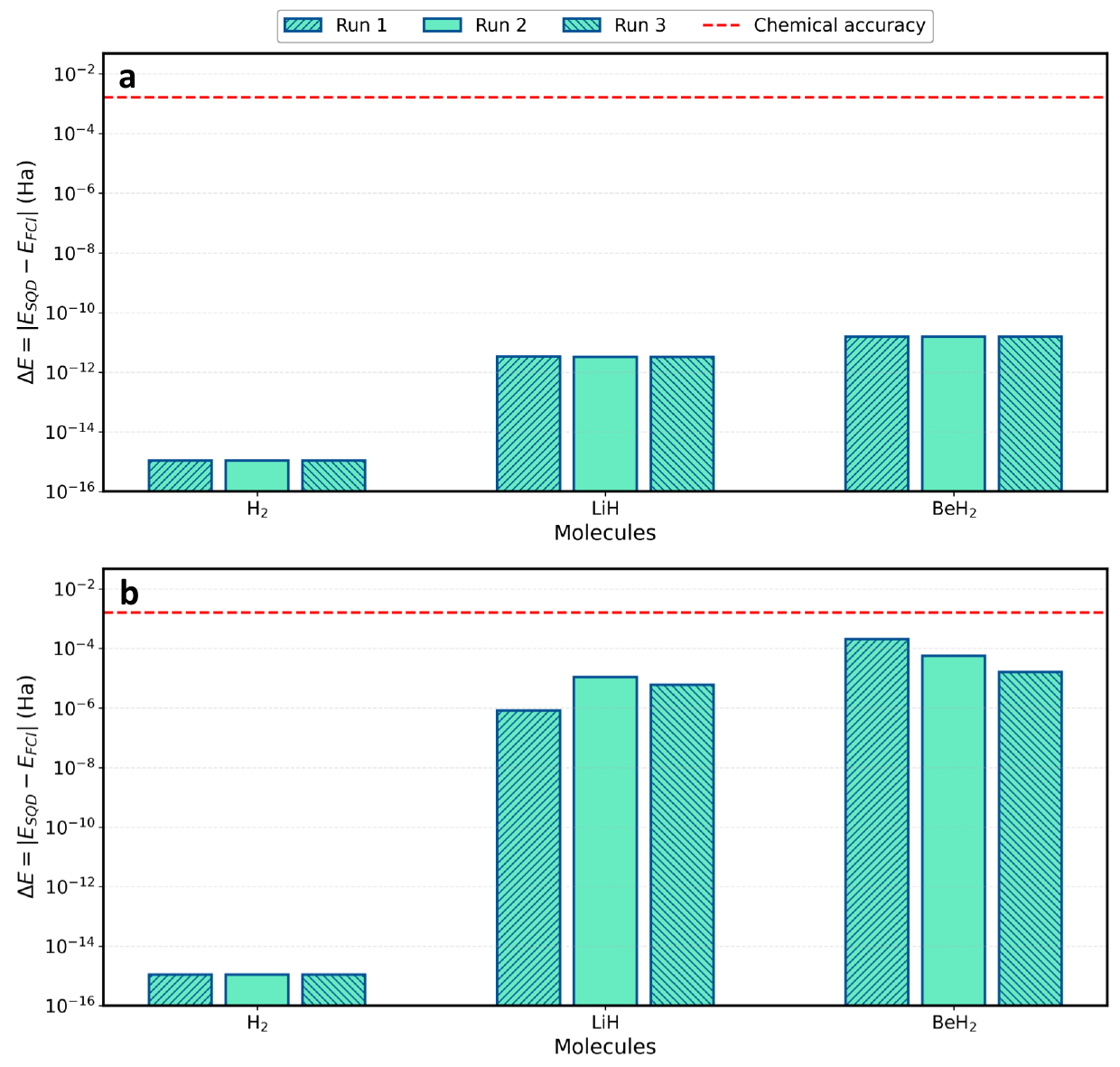}
                \caption{
                Energy difference between SQD(LCNot-UCCSD) and FCI, in Hartree (Ha), for three independent runs on the IQM Sirius QPU for each molecule. The y-axis represents the energy difference $\Delta E = |E_{\mathrm{SQD}} - E_{\mathrm{FCI}}|$ (log-scale), and the x-axis represents the molecules (H$_2$, LiH, BeH$_2$). Subplots (a) and (b) correspond to $\epsilon_{s} = 10^{8}$ (non-factor) and $\epsilon_{s} = \sqrt{|\mathbb{S}|}$, respectively, where $|\mathbb{S}|$ is the symmetry space dimension of the corresponding molecule.
                }
            \label{fig:sqd_lcnot-uccsd_delta_e}
        \end{figure}

        The corresponding ground-state energies are presented in Figure~\ref{fig:sqd_lcnot-uccsd_delta_e}. Under $\epsilon_{\mathrm{s}} = 10^8$, the SQD(LCNot-UCCSD) workflow recovers the FCI energy to within sub-nanohartree accuracy for all three molecules across all three independent runs, mirroring the performance observed for LUCJ in the same regime. This confirms that, provided $\mathbb{S}_{\mathrm{sub}}$ fully spans $\mathbb{S}$, the SQD procedure is robust to the increased noise levels introduced by the deeper LCNot-UCCSD circuits, effectively diagonalizing over the complete $\mathbb{S}$. In the symmetry-adaptive subsampling regime with $\epsilon_{\mathrm{s}} = \sqrt{|\mathbb{S}|}$, the energy accuracy degrades in a manner broadly consistent with the LUCJ trends, though the absolute deviations from FCI are relatively larger. For LiH, the three independent runs yield energy deviations on the order of $10^{-5}$~Ha and $10^{-6}$~Ha, while for BeH$_2$ the deviations are bounded between $10^{-4}$~Ha and $10^{-5}$~Ha, slightly higher when compared to their SQD(LUCJ) counterparts. This marginal degradation may be attributed to the compounding effect of reduced $\eta_{\mathrm{post\text{-}cr}}$ coverage and smaller $\eta_{\mathrm{sub}}$ ratios, both of which are exacerbated by the higher noise levels inherent to the deeper LCNot-UCCSD circuits. Nonetheless, even in this constrained setting, the SQD(LCNot-UCCSD) energies for the molecules H$_2$, LiH, and BeH$_2$ remain within the chemical accuracy threshold when compared against the FCI reference, demonstrating that the method retains meaningful accuracy despite the significant circuit overhead, primarily due to the configuration recovery error mitigation process paired with the SCI proliferation performed on $\mathbb{S}_{\mathrm{post\text{-}cr}}$ to obtain $\mathbb{S}_{\mathrm{sub}}$~\cite{Patra2026}. The exact energies can be found in Seciton~D of the Supplementary Information. Further analysis of hardware noise, $\eta_{\mathrm{sym}}$, and their corresponding effects on $\mathbb{S}_{\mathrm{sub}}$ not falling within $\mathbb{S}$, causing failure of configuration recovery for molecules H$_2$O and NH$_3$, is deferred to future work. 

        \subsubsection{Ansätze Comparison: SQD(LUCJ) vs SQD(LCNot-UCCSD)}
        
\begin{figure}[H]
            \centering
            \includegraphics[width=\linewidth]{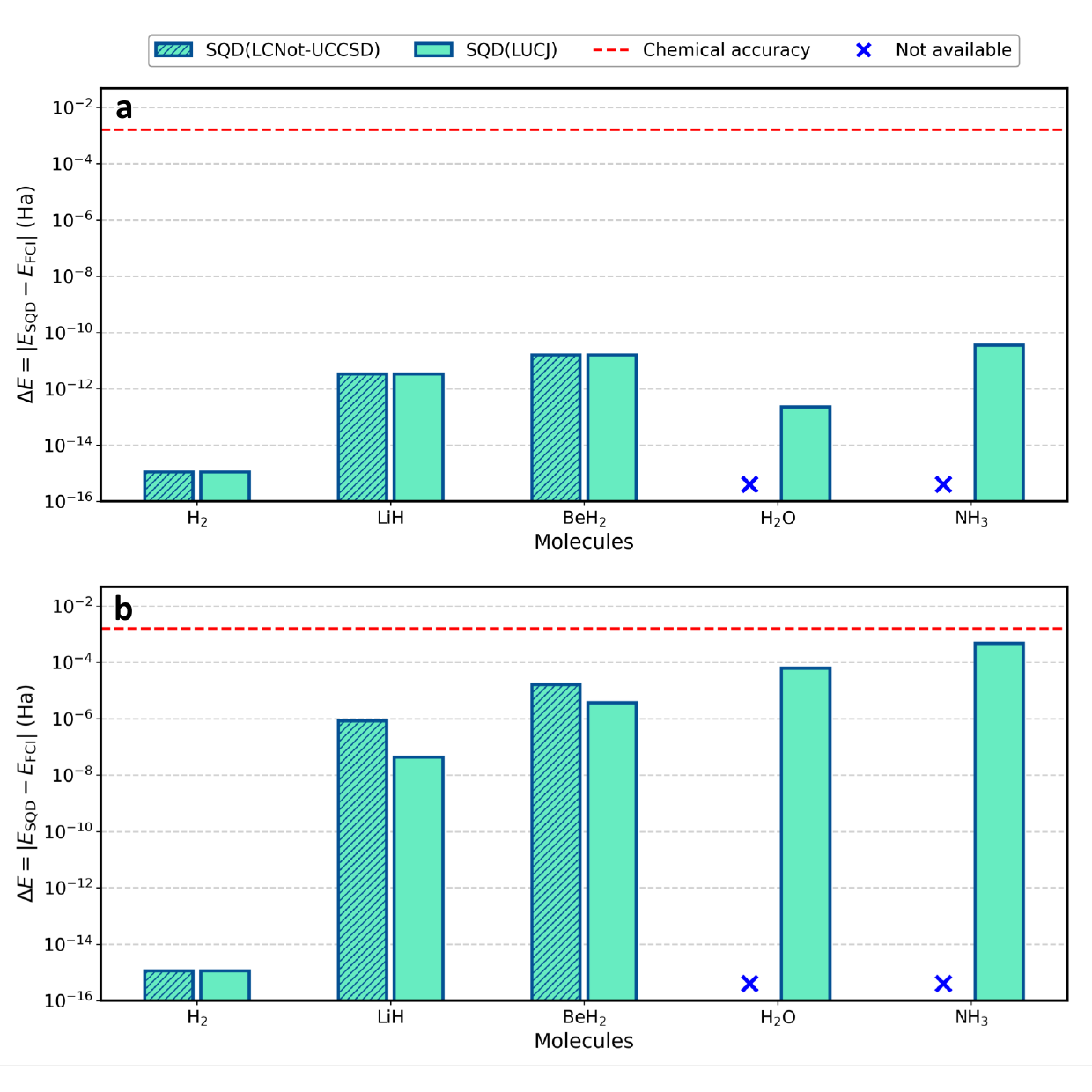}
            \caption{
            Comparison of the energy difference $\Delta E = |E_{\mathrm{SQD}} - E_{\mathrm{FCI}}|$ (Ha) for SQD(LUCJ) and SQD(LCNot-UCCSD) ansatz implementations across molecules. 
            Here, the x-axis represents $\Delta E$ (log scale) and the y-axis represents the molecules for 
            (a) $\epsilon_{s} = 10^{8}$ and 
            (b) $\epsilon_{s} = \sqrt{|\mathbb{S}|}$ for the respective molecule.
            }
            \label{fig:delta_e_sqd_lucj_vs_sqd_lcnot-uccsd}
        \end{figure}

        Having characterised the two ansätze individually, it is instructive to compare them directly. The scaling of the number of variational parameters and the transpiled gate depths for SQD(LUCJ) and SQD(LCNot-UCCSD) across all molecular systems can be found in Section~D of the Supplementary Information. The energetic comparison between the two ansätze is presented in Figure~\ref{fig:delta_e_sqd_lucj_vs_sqd_lcnot-uccsd} in terms of $\Delta E = |E_{\mathrm{SQD}} - E_{\mathrm{FCI}}|$ for both subsampling regimes. For $\epsilon_{\mathrm{s}} = 10^8$, both ansätze achieve essentially identical accuracy across all commonly evaluated molecules, with energy deviations from FCI falling below the sub-nanohartree threshold in every case. This convergence of accuracy underscores a key feature of the SQD framework: provided that $\mathbb{S}_{\mathrm{sub}}$ spans most of $\mathbb{S}$, the quality of the final energy is governed primarily by the SCI proliferation and the subsequent subspace diagonalization rather than the specific distribution generated by the ansatz. As a result, even the considerably noisier output of the deeper LCNot-UCCSD circuits is sufficient to recover FCI-level energies.

        The picture changes meaningfully under $\epsilon_{\mathrm{s}} = \sqrt{|\mathbb{S}|}$, as shown in Figure~\ref{fig:delta_e_sqd_lucj_vs_sqd_lcnot-uccsd}(b). Here, SQD(LUCJ) consistently achieves equal or lower energy deviations than SQD(LCNot-UCCSD) for the H$_2$, LiH, and BeH$_2$ molecules, where both are evaluated. This performance gap may be understood in terms of the interplay between circuit execution noise and $\eta_{\mathrm{sub}}$: the shallower LUCJ circuits produce cleaner bitstring distributions with higher $\eta_{\mathrm{post\text{-}cr}}$, which in turn enables a more representative $\mathbb{S}_{\mathrm{sub}}$ even at reduced subsampling budgets. Conversely, the deeper LCNot-UCCSD circuits suffer from greater hardware-induced noise, which degrades the quality of quantum sampling and the symmetry compliance of the sampled configurations, ultimately limiting the overlap of $\mathbb{S}_{\mathrm{sub}}$ with the ground state when $\epsilon_{\mathrm{s}}$ is symmetry-adapted to $\sqrt{|\mathbb{S}|}$.

        As mentioned earlier in Section~\ref{subsec:LCNot-UCCSD}, for the molecules H$_2$O and NH$_3$, SQD(LCNot-UCCSD) fails because quantum sampling does not produce samples that lie within $\mathbb{S}$ of the molecule to perform configuration recovery; hence, they are marked as not available in Figure~\ref{fig:delta_e_sqd_lucj_vs_sqd_lcnot-uccsd}.

        Taken together, these results point to a clear trade-off between the two ansätze within the SQD framework. SQD(LUCJ) offers a compelling combination of hardware efficiency and robust accuracy, particularly under constrained subsampling conditions, but it requires an expensive CCSD-level parameter initialization, making it well-suited to near-term QPU execution. SQD(LCNot-UCCSD), while chemically expressive and comparable in accuracy to LUCJ when subsampling is unrestricted, incurs a substantially higher circuit cost with less costly MP2-level parameter initialization, but this limits its practical applicability to smaller systems on current hardware. These considerations motivate the use of SQD(LUCJ) as the primary ansatz for the remainder of this work. Further information on the SQD(LUCJ) and SQD(LCNot-UCCSD) output distribution analysis with various post-HF methods is available in Section~D of the Supplementary Information.

    \subsection{One Dimensional Potential Energy Surface (1D-PES) Scans}
    \label{sec:4.2.1D-PES}

        Having established the accuracy and hardware feasibility of the SQD(LUCJ) framework at fixed molecular geometries, we now extend the assessment to one-dimensional potential energy surface (1D-PES) scans, in which the interatomic distance $r$ is varied continuously along the dissociation pathway. This represents a more stringent test of the method, as the electronic structure and, consequently, the optimal ansatz parameterisation change significantly along the scan coordinate, particularly in the strongly correlated regime near bond dissociation~\cite{Helgaker2000}. The 1D-PES scans were performed for four molecular systems, H$_2$, HeH$^+$, LiH, and BeH$_2$, and are reported for two basis sets, STO-3G and 6-31G, in the following subsections. All SQD(LUCJ) energies are assessed against FCI as the exact reference, with CCSD additionally included to benchmark the regime in which single-reference classical methods are known to fail~\cite{Small2012}. The QPU execution times and molecular geometries used throughout this section are provided in Section~G in the Supplementary Information.
    
        \subsubsection{Basis Set: STO-3G}

\begin{figure}[H]
                \centering
                \includegraphics[width=0.98\linewidth]{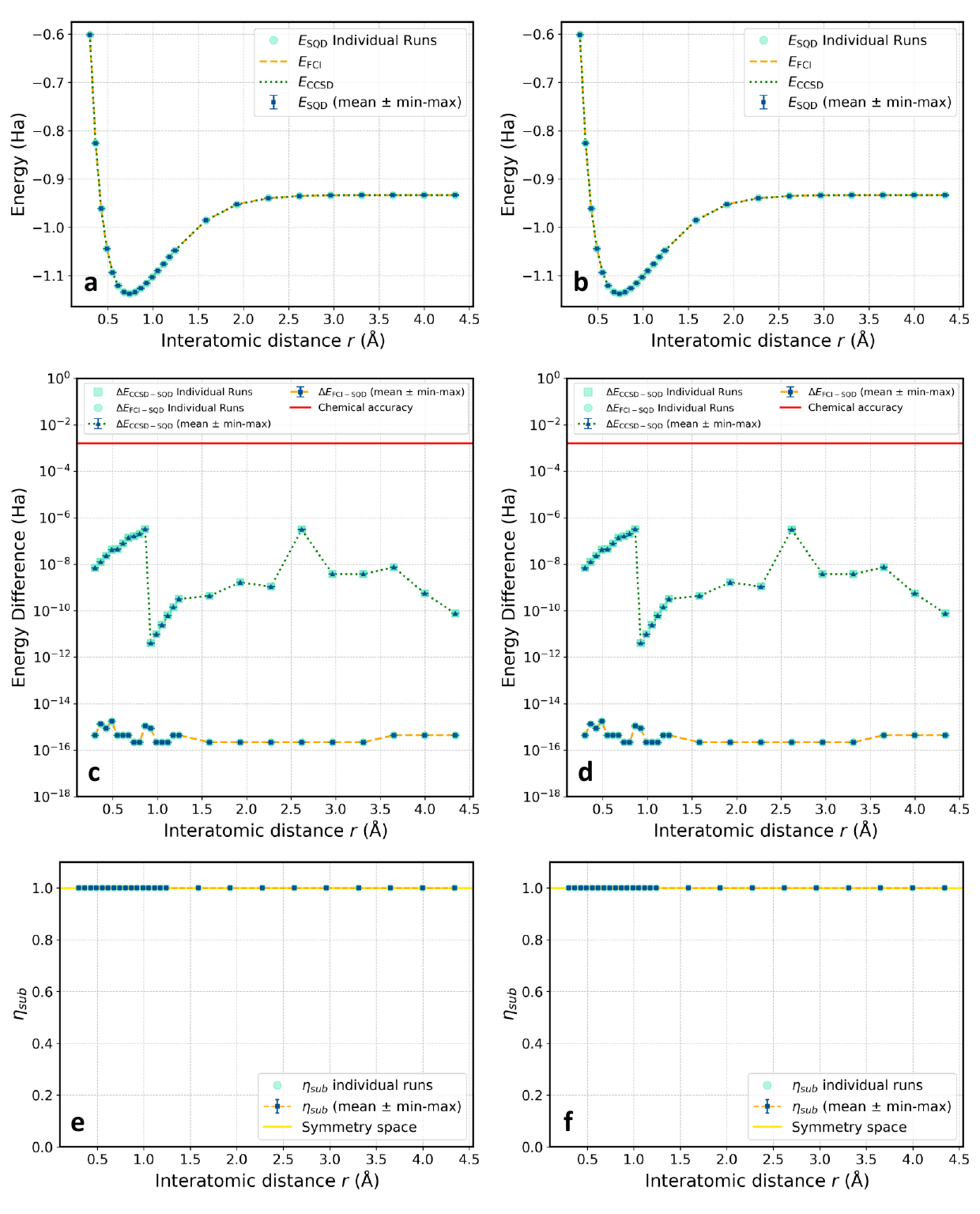}
                \caption{
                One-dimensional potential energy surface (1D-PES) results for the H$_2$ molecule in STO-3G basis. Subplots (a,b) show the PES as a function of interatomic distance $r$ (Å), with energy reported in Hartree. Subplots (c,d) present the energy deviation $\Delta E$ versus $r$, while subplots (e,f) display the ratio of the diagonalization subspace dimension to the symmetry space dimension, $\eta_{\mathrm{sub}}$, as a function of $r$. Results correspond to $\epsilon_s = 10^8$ in (a,c,e) and $\epsilon_s = 2$ in (b,d,f).
                }
                \label{fig:1d_pes_h2_sto-3g}
            \end{figure}

            The 1D-PES results for H$_2$ in the STO-3G basis are presented in Figure~\ref{fig:1d_pes_h2_sto-3g}. The SQD(LUCJ) PES overlaps the FCI curve with perfect visual agreement across the full dissociation pathway under both subsampling conditions, spanning bond lengths from $0.3$ to $4.3$~\AA. The $\Delta E_{\mathrm{FCI-SQD}}$ deviations shown in subplot (c), under $\epsilon_{\mathrm{s}} = 10^8$, lie at the level of numerical precision throughout, falling between $10^{-16}$ and $10^{-14}$~Ha, well below any physically meaningful threshold. By contrast, $\Delta E_{\mathrm{CCSD-SQD}}$, which in this case reflects the CCSD method's own error relative to FCI rather than any deficiency of SQD, sits in the range of $\sim 10^{-8}$~Ha near equilibrium but is itself below the chemical accuracy threshold everywhere, consistent with the near-exactness of CCSD for this two-electron system. Under $\epsilon_{\mathrm{s}} = 2$, shown in subplot (d), the $\Delta E_{\mathrm{FCI-SQD}}$ deviations remain equally flat and at the same numerical precision floor, and $\eta_{\mathrm{sub}} = 1.0$ is maintained throughout the scan in both regimes, as shown in subplots (e) and (f). The complete invariance of H$_2$ with respect to the subsampling budget reflects the trivially small $\mathbb{S}$ of this system, which is fully spanned regardless of the geometry of the molecule.

\begin{figure}[H]
                \centering
                \includegraphics[width=0.98\linewidth]{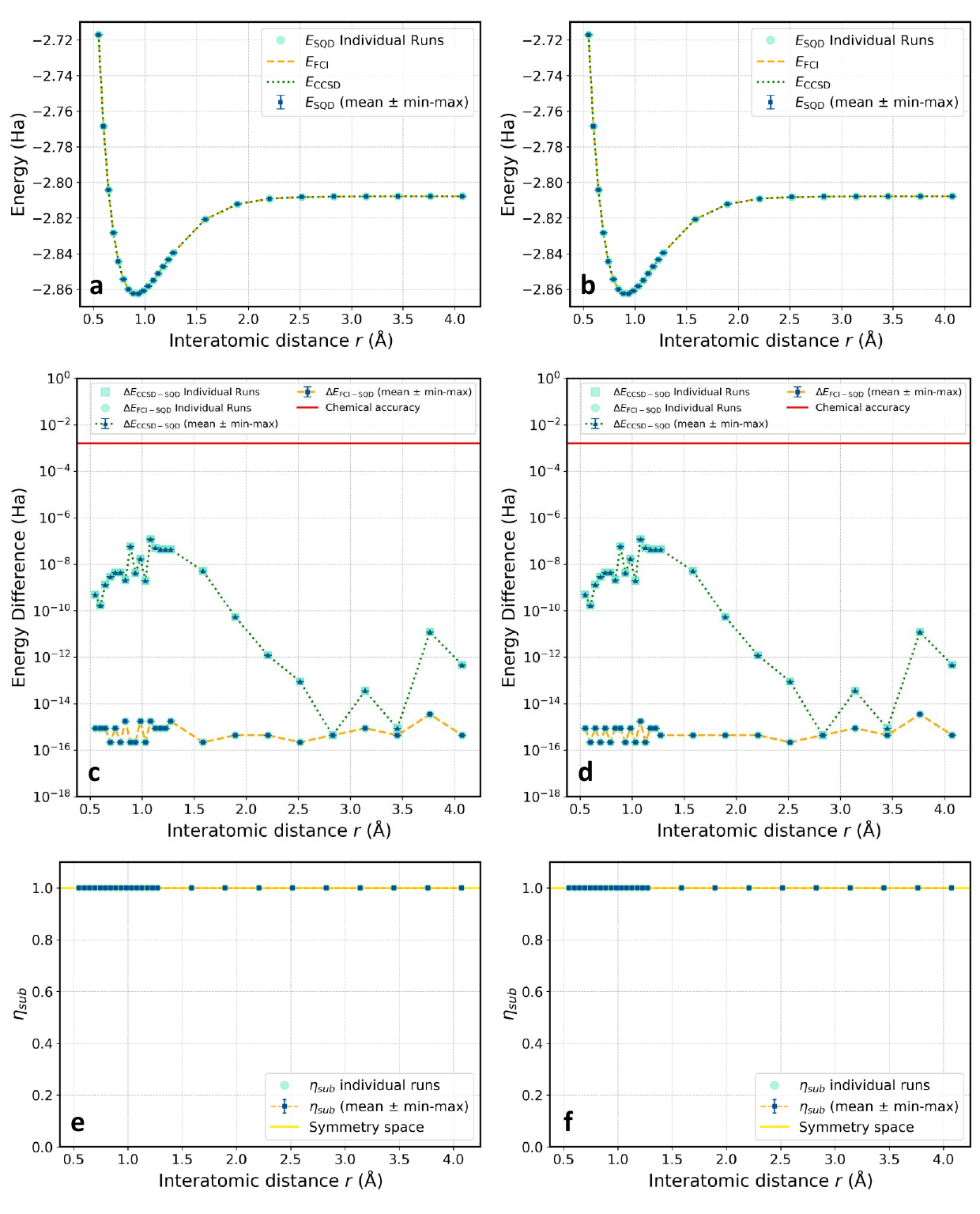}
                \caption{
                One-dimensional potential energy surface (1D-PES) results for the HeH$^+$ molecule in STO-3G basis. Subplots (a,b) show the PES as a function of interatomic distance $r$ (Å), with energy reported in Hartree. Subplots (c,d) present the energy deviation $\Delta E$ versus $r$, while subplots (e,f) display the ratio of the diagonalization subspace dimension to the symmetry space dimension, $\eta_{\mathrm{sub}}$, as a function of $r$. Results correspond to $\epsilon_s = 10^8$ in (a,c,e) and $\epsilon_s = 2$ in (b,d,f).
                }
                \label{fig:1d_pes_heh+_sto-3g}
            \end{figure}

            The corresponding results for HeH$^+$ in STO-3G are shown in Figure~\ref{fig:1d_pes_heh+_sto-3g}. The behaviour closely mirrors that of H$_2$: in both subsampling regimes, SQD(LUCJ) recovers the FCI PES with deviations at the numerical precision floor ($\sim 10^{-15}$~Ha) throughout the scan, and $\eta_{\mathrm{sub}} = 1.0$ is maintained uniformly at all geometries. The CCSD deviations from SQD remain in the range of $\sim 10^{-7}$--$10^{-8}$~Ha, reflecting the intrinsic CCSD error for this two-electron system rather than any SQD inaccuracy. The PES is well resolved across the equilibrium region near $r \approx 0.8$~\AA\ and the full dissociation plateau, confirming that SQD(LUCJ) captures both regimes faithfully. Notably, the discrete reduction in circuit resources observed at $r \approx 2.5$~\AA\ for HeH$^+$ has no energetic consequence, as the accuracy of the method is unaffected by the geometry-localised transpilation.        

\begin{figure}[H]
                \centering
                \includegraphics[width=0.98\linewidth]{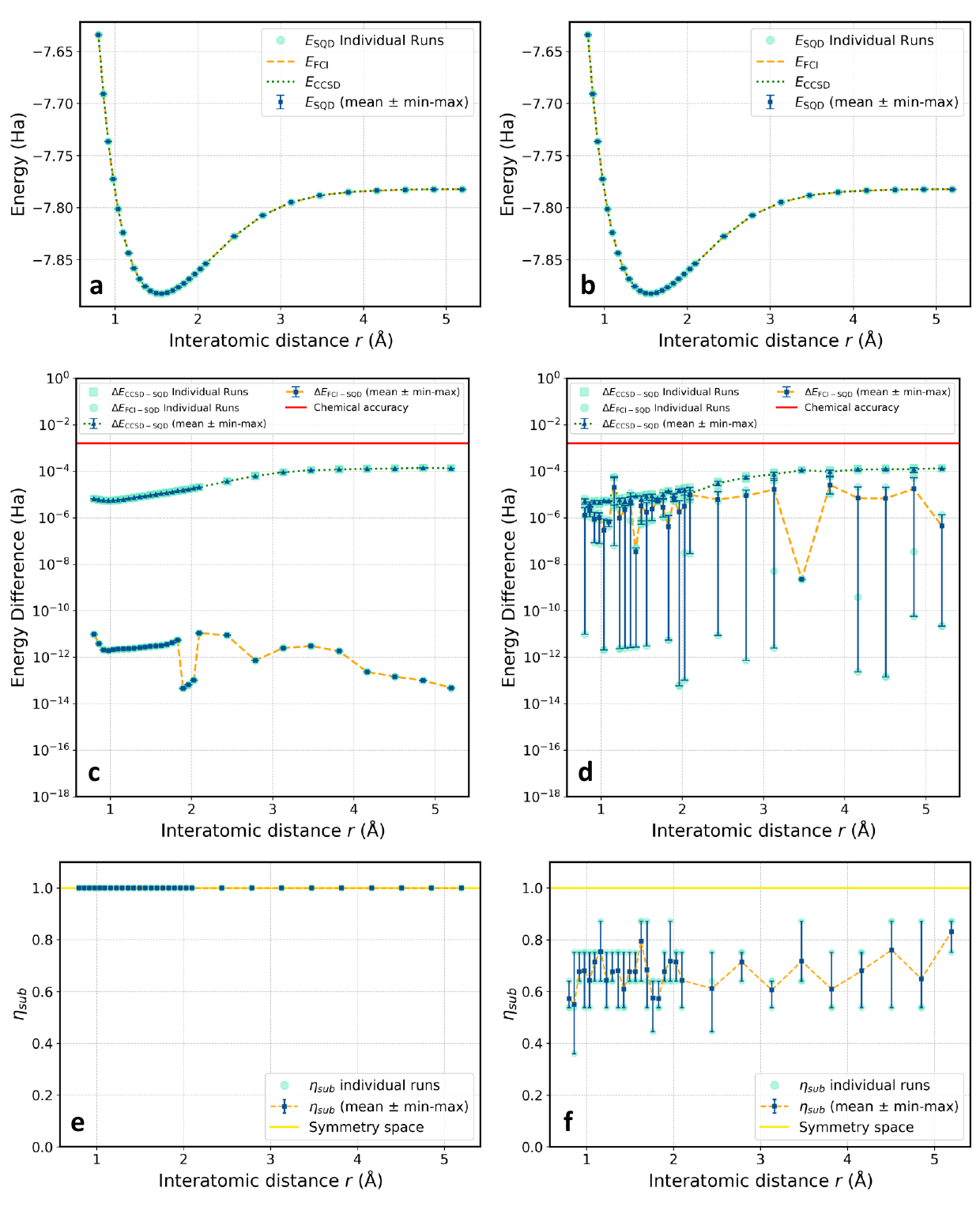}
                \caption{
                One-dimensional potential energy surface (1D-PES) results for the LiH molecule in STO-3G basis. Subplots (a,b) show the PES as a function of interatomic distance $r$ (Å), with energy reported in Hartree. Subplots (c,d) present the energy deviation $\Delta E$ versus $r$, while subplots (e,f) display the ratio of the diagonalization subspace dimension to the symmetry space dimension, $\eta_{\mathrm{sub}}$, as a function of $r$. Results correspond to $\epsilon_s = 10^8$ in (a,c,e) and $\epsilon_s = 15$ in (b,d,f).
                }
                \label{fig:1d_pes_lih_sto-3g}
            \end{figure}

            For LiH, shown in Figure~\ref{fig:1d_pes_lih_sto-3g}, the picture is broadly consistent with the fixed-geometry findings but reveals more physically interesting structure. Under $\epsilon_{\mathrm{s}} = 10^8$, the FCI PES is reproduced with $\Delta E_{\mathrm{FCI-SQD}} \approx 10^{-12}$~Ha uniformly across all geometries, and $\eta_{\mathrm{sub}} = 1.0$ throughout. The CCSD deviation profile shown in subplot (c) is particularly instructive: $\Delta E_{\mathrm{CCSD-SQD}}$ grows monotonically from $\sim 10^{-5}$~Ha near the equilibrium geometry to $\sim 10^{-4}$~Ha at stretched bond lengths beyond $r \approx 3$~\AA. This behaviour highlights the well-known breakdown of the single-reference CCSD approximation in the dissociation regime, which SQD(LUCJ) avoids entirely by diagonalizing the full $\mathbb{S}$. Under $\epsilon_{\mathrm{s}} = 15$, the subspace coverage $\eta_{\mathrm{sub}}$ shown in subplot (f) fluctuates between approximately $0.50$ and $0.85$ across the scan, with considerable run-to-run variability and without any strong systematic trend along $r$. The corresponding $\Delta E_{\mathrm{FCI-SQD}}$ values in subplot (d) range from $\sim 10^{-6}$ to $\sim 10^{-4}$~Ha, correlated with the variations in $\eta_{\mathrm{sub}}$, yet remain below the chemical accuracy threshold across the entire dissociation pathway. The overall shape of the PES is faithfully reproduced even under these constrained subsampling conditions, and the SQD(LUCJ) energies consistently undercut the CCSD reference at all stretched geometries.

            \begin{figure}[H]
                \centering
                \includegraphics[width=0.98\linewidth]{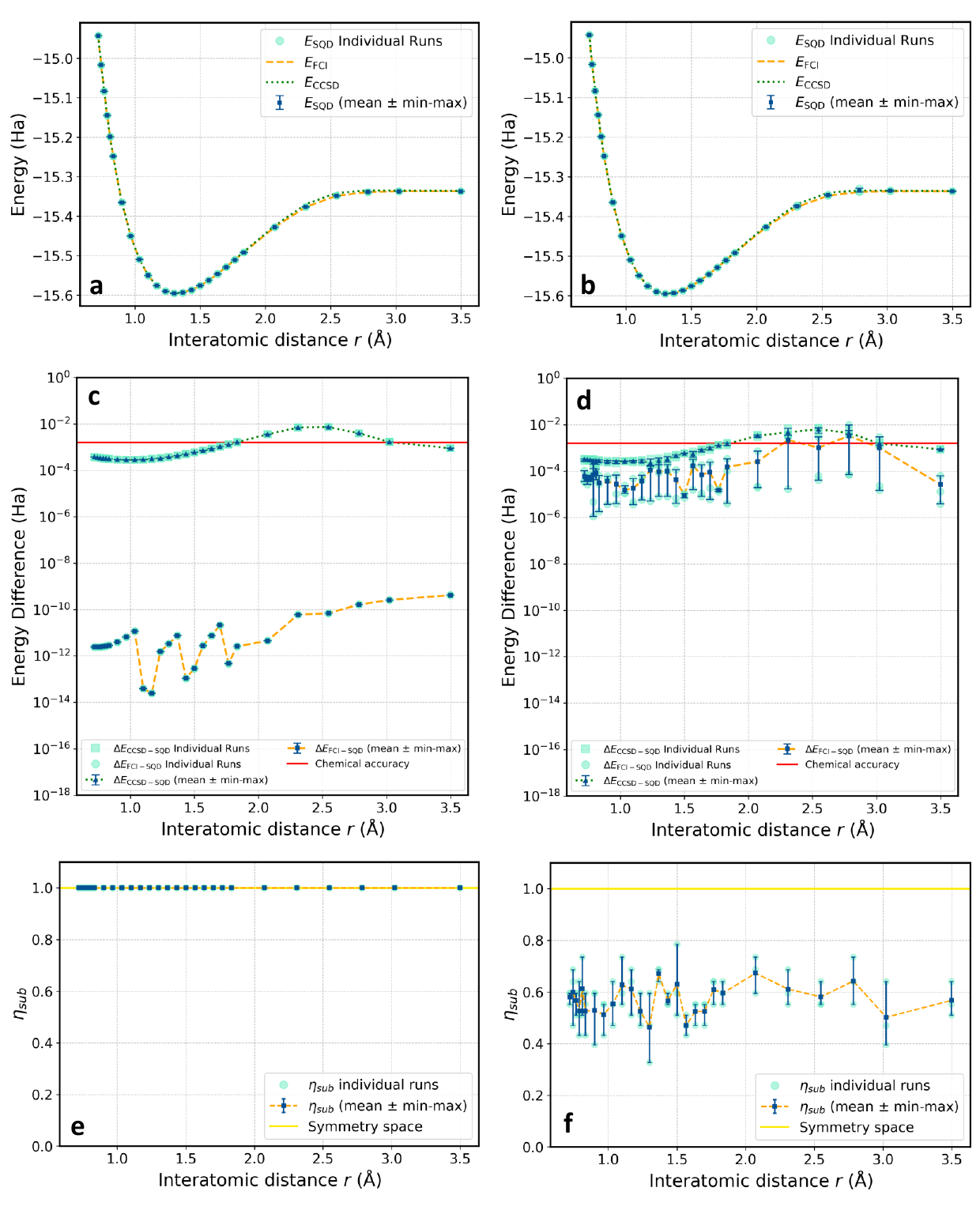}
                \caption{
                One-dimensional potential energy surface (1D-PES) results for the BeH$_2$ molecule in STO-3G basis. Subplots (a,b) show the PES as a function of interatomic distance $r$ (Å), with energy reported in Hartree. Subplots (c,d) present the energy deviation $\Delta E$ versus $r$, while subplots (e,f) display the ratio of the diagonalization subspace dimension to the symmetry space dimension, $\eta_{\mathrm{sub}}$, as a function of $r$. Results correspond to $\epsilon_s = 10^8$ in (a,c,e) and $\epsilon_s = 35$ in (b,d,f).
                }
                \label{fig:1d_pes_beh2_sto-3g}
            \end{figure}

            The BeH$_2$ results in Figure~\ref{fig:1d_pes_beh2_sto-3g} present the most demanding test within the STO-3G set. The equilibrium geometry is located near $r \approx 1.3$~\AA\ at an energy of approximately $-15.6$~Ha, with the PES rising steeply at compressed geometries and plateauing near $-15.30$~Ha in the dissociation limit. Under $\epsilon_{\mathrm{s}} = 10^8$, $\Delta E_{\mathrm{FCI-SQD}}$ remains close to the numerical precision floor around $\sim 10^{-11}$--$10^{-12}$~Ha throughout the full scan, with $\eta_{\mathrm{sub}} = 1.0$ at all geometries. The CCSD deviation profile is strikingly different: $\Delta E_{\mathrm{CCSD-SQD}}$ grows from $\sim 10^{-4}$~Ha near equilibrium and exceeds the chemical accuracy threshold at stretched geometries beyond $r \approx 2.0$~\AA, reaching values approaching $10^{-2}$~Ha at $r \approx 2.5$~\AA. This pronounced CCSD failure in the multireference dissociation regime is cleanly captured and corrected by SQD(LUCJ), which maintains sub-nanohartree accuracy throughout as it diagonalizes over a subspace that spans the complete $\mathbb{S}$. Under $\epsilon_{\mathrm{s}} = 35$, the subspace coverage shown in subplot (f) fluctuates between approximately $0.45$ and $0.70$, with large run-to-run variability across the scan. The associated $\Delta E_{\mathrm{FCI-SQD}}$ values in subplot (d) are correspondingly larger, ranging from $\sim 10^{-6}$ to $\sim 10^{-3}$~Ha and approaching the chemical accuracy threshold at certain geometries. The LUCJ ansatz initialization with CCSD parameters, together with the reduced $\eta_{\mathrm{sub}}$, leads to energies that fall outside the chemical accuracy threshold at the exact geometries where CCSD itself fails, primarily due to the strongly multireference character of BeH$_2$ at stretched bond lengths. A deeper analysis of these effects, including improved multireference ansatz parameter initialization strategies and approaches to reduce the required subspace while maintaining chemical accuracy, is deferred to future work. Despite the reduced subspace coverage, the qualitative shape of the dissociation curve is faithfully reproduced, and the SQD(LUCJ) energies continue to mostly outperform CCSD at stretched geometries where the latter deteriorates. These results collectively demonstrate that the SQD(LUCJ) framework is capable of generating accurate 1D-PES scans in the STO-3G basis, with performance that improves systematically as $\eta_{\mathrm{sub}}$ increases toward unity. Further details on the quantum circuit resource counts and their variation with interatomic distance for all four molecules in STO-3G are provided in Section~D of the Supplementary Information.
            
        \subsubsection{Basis Set: 6-31G}

            The 1D-PES scans were repeated for H$_2$ and HeH$^+$ using the 6-31G basis set, which provides a more flexible description of the electronic structure relative to the minimal STO-3G basis by introducing additional contracted Gaussian functions per atomic shell. This expansion of the orbital basis increases the active space and hence the quantum circuit complexity while offering a more accurate representation of the PES.

\begin{figure}[H]
                \centering
                \includegraphics[width=0.98\linewidth]{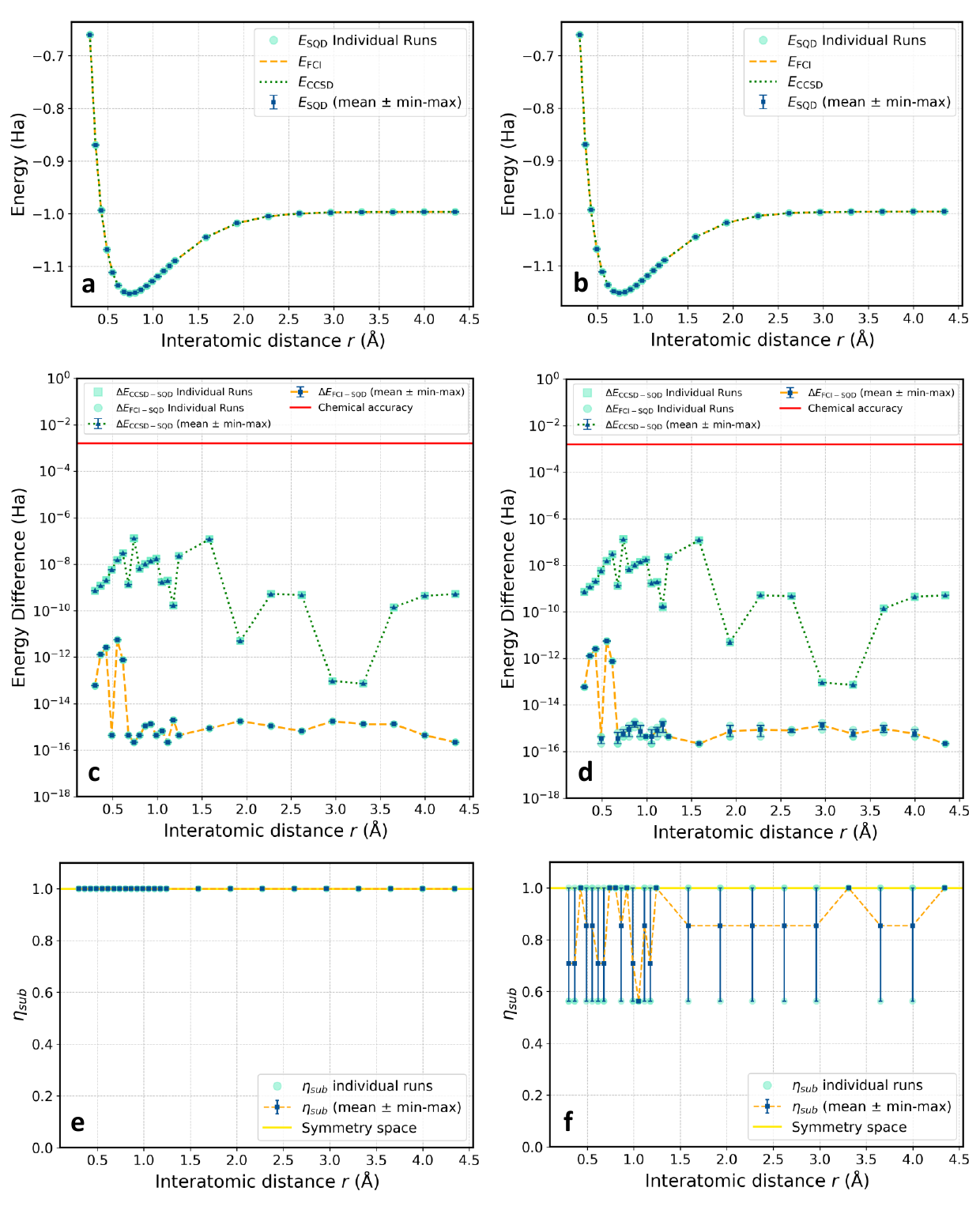}
                \caption{
                One-dimensional potential energy surface (1D-PES) results for the H$_2$ molecule in 6-31G basis. Subplots (a,b) show the PES as a function of interatomic distance $r$ (Å), with energy reported in Hartree. Subplots (c,d) present the energy deviation $\Delta E$ versus $r$, while subplots (e,f) display the ratio of the diagonalization subspace dimension to the symmetry space dimension, $\eta_{\mathrm{sub}}$, as a function of $r$. Results correspond to $\epsilon_s = 10^8$ in (a,c,e) and $\epsilon_s = 2$ in (b,d,f).
                }
                \label{fig:1d_pes_h2_6-31g}
            \end{figure}

            The 1D-PES results for H$_2$ in the 6-31G basis are presented in Figure~\ref{fig:1d_pes_h2_6-31g}. Under $\epsilon_{\mathrm{s}} = 10^8$, the SQD(LUCJ) method faithfully reproduces the FCI PES across the full range of interatomic distances, with $\Delta E_{\mathrm{FCI-SQD}}$ values on average at the numerical precision floor of $\sim 10^{-15}$~Ha throughout, and $\eta_{\mathrm{sub}} = 1.0$ uniformly maintained. The CCSD deviations, which fall in the range $\sim 10^{-7}$--$10^{-10}$~Ha, remain well below the chemical accuracy threshold with sub-microhartree precision, consistent with the essentially exact nature of CCSD for two electrons. In the reduced symmetry-adapted subsampling regime with $\epsilon_{\mathrm{s}} = 2$, the energy deviations in subplot (d) are equally negligible, preserving the same numerical precision floor on average as in the large-subsampling case. However, the $\eta_{\mathrm{sub}}$ profile in subplot (f) reveals a notable feature not observed in the STO-3G counterpart: at most bond lengths $r$ near equilibrium and through the dissociation plateau, $\eta_{\mathrm{sub}}$ drops appreciably from unity, with individual runs showing values as low as $\sim 0.55$.

\begin{figure}[H]
                \centering
                \includegraphics[width=0.98\linewidth]{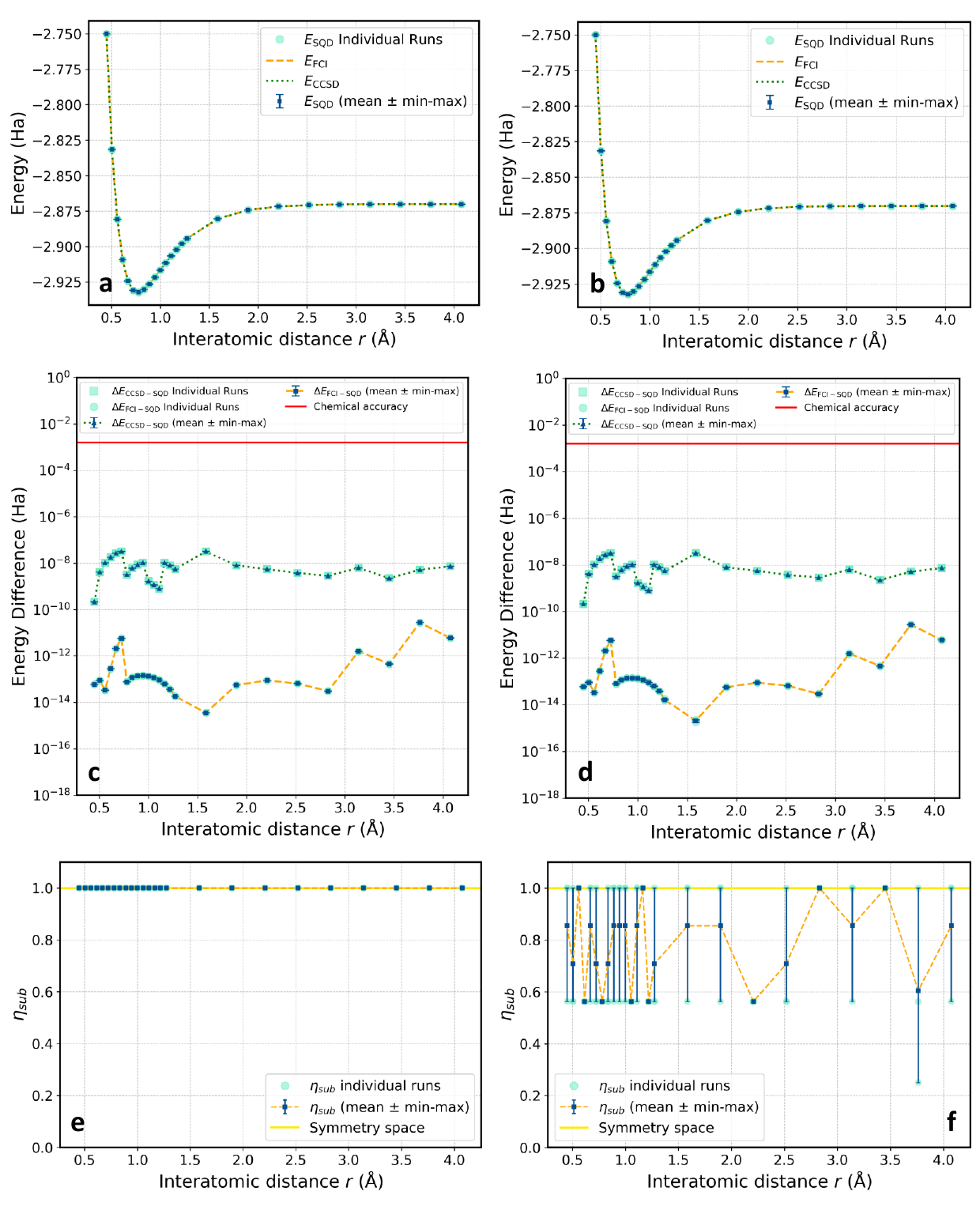}
                \caption{
                One-dimensional potential energy surface (1D-PES) results for the HeH$^+$ molecule in 6-31G basis. Subplots (a,b) show the PES as a function of interatomic distance $r$ (Å), with energy reported in Hartree. Subplots (c,d) present the energy deviation $\Delta E$ versus $r$, while subplots (e,f) display the ratio of the diagonalization subspace dimension to the symmetry space dimension, $\eta_{\mathrm{sub}}$, as a function of $r$. Results correspond to $\epsilon_s = 10^8$ in (a,c,e) and $\epsilon_s = 2$ in (b,d,f).
                }
                \label{fig:1d_pes_heh+_6-31g}
            \end{figure}

            The HeH$^+$ results in the 6-31G basis are shown in Figure~\ref{fig:1d_pes_heh+_6-31g}. Under $\epsilon_{\mathrm{s}} = 10^8$, the performance mirrors the H$_2$ 6-31G case closely, with $\Delta E_{\mathrm{FCI-SQD}}$ values at the numerical precision floor ($\sim 10^{-13}$--$10^{-15}$~Ha) across the entire scan and $\eta_{\mathrm{sub}} = 1.0$ at all geometries. Under $\epsilon_{\mathrm{s}} = 2$, the HeH$^+$ 6-31G system shows a similar $\eta_{\mathrm{sub}}$ profile to that of H$_2$ 6-31G: the subspace coverage drops from unity at most geometries, reaching values as low as $\sim 0.55$. Despite the reduced $\eta_{\mathrm{sub}}$, the energy deviations shown in subplot (d) remain at the same numerical precision floor as in the large-subsampling regime without exception. Additionally, the 6-31G PES for HeH$^+$ shows a well-defined equilibrium near $r \approx 0.77$~\AA\ at an energy of approximately $-2.93$~Ha, deepening the potential well relative to the STO-3G description, while the dissociation plateau converges to approximately $-2.87$~Ha. The robustness of SQD(LUCJ) across both subsampling conditions and both molecules in the 6-31G basis provides a strong foundation for the basis set comparison in the following subsection. Further details on the quantum circuit resource counts and their variation with interatomic distance for both the molecules in 6-31G are provided in Section~D of the Supplementary Information.

        \subsubsection{Basis Set Comparison: STO-3G vs 6-31G}

\begin{figure}[H]
                \centering
                \includegraphics[width=0.98\linewidth]{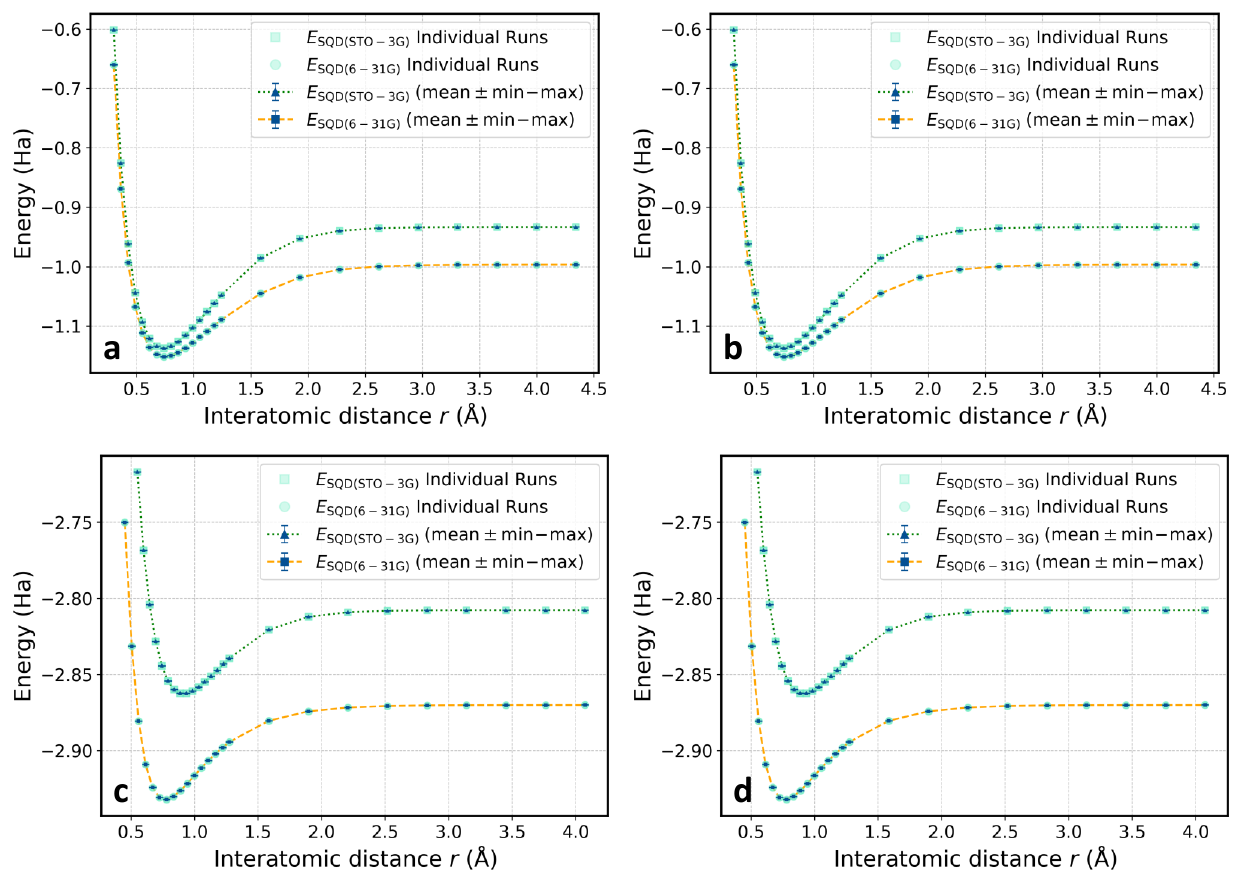}
                \caption{
                One-dimensional potential energy surface (1D-PES) basis set comparison for H$_2$ and HeH$^+$ molecules. Subplots (a,b) show the PES as a function of interatomic distance $r$ (Å), with energy reported in Hartree for the H$_2$ molecule. Similarly, subplots (c,d) present the PES versus $r$ for the HeH$^+$ molecule. Results correspond to $\epsilon_s = 10^8$ in (a,c) and $\epsilon_s = 2$ in (b,d).
                }
                \label{fig:1d_pes_sto-3g_vs_6-31g}
            \end{figure}

            Figure~\ref{fig:1d_pes_sto-3g_vs_6-31g} places the STO-3G and 6-31G potential energy surfaces in direct comparison for both H$_2$ and HeH$^+$ under both subsampling conditions. The figure displays only SQD energies, which, as established in the preceding subsections, coincide closely with the FCI reference to numerical precision in all cases. Therefore, the comparison is directly between the two basis set descriptions for the SQD(LUCJ) method.

            For H$_2$, the STO-3G and 6-31G PES curves converge to the same minimum-energy geometry near $r \approx 0.74$~\AA\ but differ in their absolute energies throughout: the 6-31G minimum lies approximately $0.014$~Ha below the STO-3G minimum, and the two dissociation plateaus differ by a similar margin, with STO-3G approaching $-0.93$~Ha and 6-31G approaching $-1.00$~Ha. For HeH$^+$, the energy separation is more pronounced: the 6-31G PES lies approximately $0.06$--$0.07$~Ha below the STO-3G curve throughout the scan, with the 6-31G minimum at $\sim 0.78$~\AA\ and $-2.93$~Ha compared to $\sim 0.94$~\AA\ and $-2.86$~Ha for STO-3G. In both cases, the larger basis captures a deeper and more accurately located potential well, reflecting the additional variational flexibility afforded by the larger orbital basis of the 6-31G basis set.

            Crucially, across all four panels of Figure~\ref{fig:1d_pes_sto-3g_vs_6-31g}, the SQD(LUCJ) energies in both basis sets are reproduced with equally tight run-to-run consistency, and the error bars on the mean are negligibly small in all cases. The comparison under $\epsilon_{\mathrm{s}} = 2$ in subplots (b) and (d) is equally clean, with no visible degradation in the agreement between basis sets or in the internal self-consistency of the SQD results. This confirms that the reduction in subsampling budget has no differential impact on the accuracy of either basis-set description for these two-electron systems, and that the choice of basis set influences the physical content of the PES---the absolute energies, the depth and location of the potential well, and the dissociation asymptote---but not the reliability of the SQD(LUCJ) method in reproducing it. The superior energetic description offered by 6-31G, combined with the demonstrated ability of SQD(LUCJ) to maintain FCI-level accuracy within it at no additional cost in terms of subsampling requirements, reinforces the case for employing larger basis sets in future quantum simulation efforts where chemical accuracy demands it, provided that the associated circuit overhead remains within the operationally acceptable noise envelope of the available quantum hardware.

    \subsection{Two Dimensional Potential Energy Surface (2D-PES) Scans}
    \label{sec:4.3.2D-PES}

        Building on the one-dimensional scans, the SQD(LUCJ) framework was further applied to map a 2D-PES for the H$_2$O molecule in the STO-3G basis, in which both the O--H bond length $r$ and the H--O--H bond angle $\theta$ are varied simultaneously over a $32 \times 32$ grid with $r \in [0.85, 1.20]$~\AA\ and $\theta \in [85^\circ, 115^\circ]$. This constitutes a considerably more demanding application of the method, as the electronic structure must be accurately resolved across a continuous two-dimensional grid of geometries spanning both stretching and bending degrees of freedom, a setting highly relevant to the practical use of quantum chemistry methods for characterizing vibrational frequencies, molecular force fields and reaction pathways~\cite{Barone2021, Chmiela2017, Qiu2025}. The scan was performed under both $\epsilon_{\mathrm{s}} = 10^8$ and $\epsilon_{\mathrm{s}} = 21$, and assessed against CCSD and FCI reference energies computed at every grid point. The QPU execution time heatmap and the methodology employed to generate the molecular geometries used throughout this subsection are detailed in Section~G in the Supplementary Information.

\begin{figure}[H]
            \centering
            \includegraphics[width=0.9\linewidth]{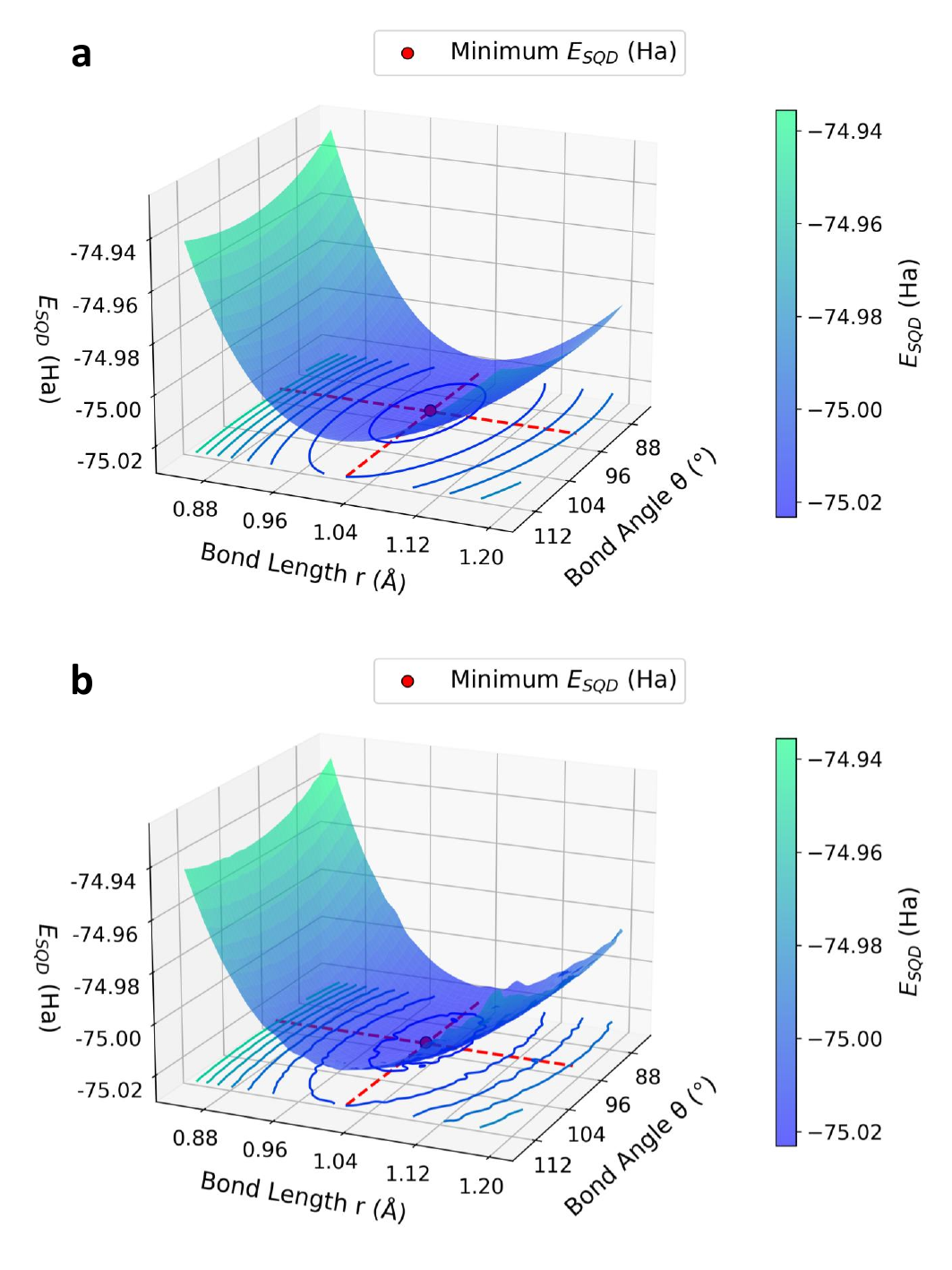}
            \caption{
            Two-dimensional potential energy surface (PES) scan of the H$_2$O molecule in the STO-3G basis, with bond length $r$ (\AA) along the x-axis, bond angle $\theta$ ($^\circ$) along the y-axis, and energy $E_{\mathrm{SQD}}$ (Ha) along the z-axis. Subplots show results for (a) $\epsilon_s = 10^8$ and (b) $\epsilon_s = 21$.
            }
            \label{fig:2d_pes_h2o_sto-3g}
        \end{figure}

\begin{figure}[H]
            \centering
            \includegraphics[width=\linewidth]{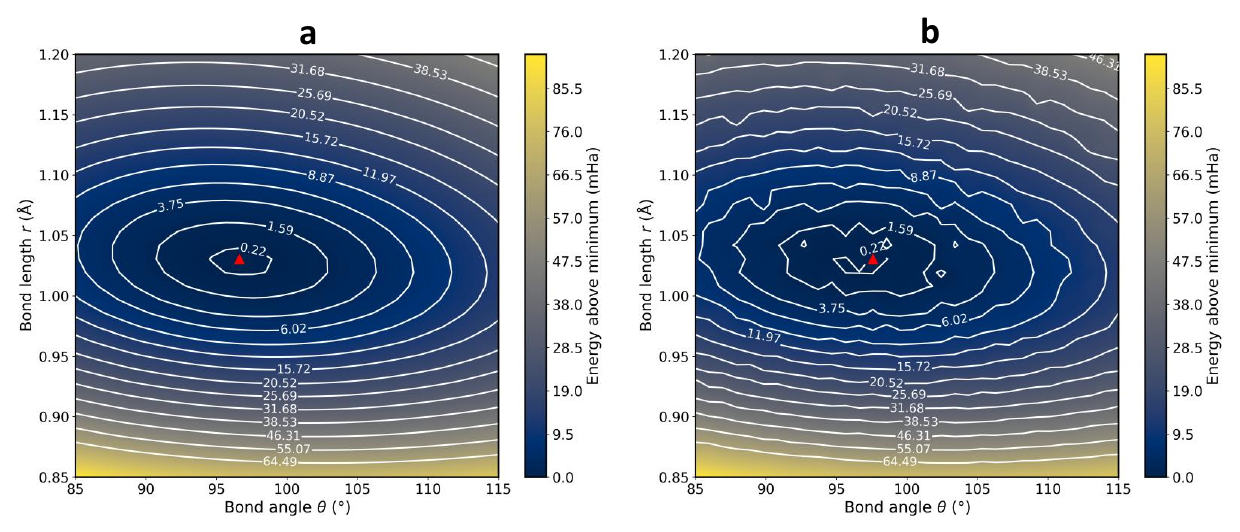}
            \caption{
            Contour plots of the Two-dimensional potential energy surface (PES) scan of the H$_2$O molecule in the STO-3G basis with bond angle $\theta$ ($^\circ$) along the x-axis and bond length $r$ (\AA) along the y-axis; the energy is reported in milli-Hartree (mHa) and encoded by the color bar. The minimum-energy configuration is indicated by a red triangle and set to zero, with all contour levels shifted relative to this minimum. Subplots (a) and (b) correspond to $\epsilon_s = 10^8$ and $\epsilon_s = 21$, respectively.
            }
            \label{fig:2d_contour_h2o_sto-3g}
        \end{figure}

        The 2D-PES is presented as three-dimensional surface plots in Figure~\ref{fig:2d_pes_h2o_sto-3g} and as contour plots in Figure~\ref{fig:2d_contour_h2o_sto-3g}, for both $\epsilon_{\mathrm{s}} = 10^8$ and $\epsilon_{\mathrm{s}} = 21$. Under $\epsilon_{\mathrm{s}} = 10^8$, the SQD(LUCJ) method produces a smooth, well-resolved potential energy surface across the entire $(\theta, r)$ grid. The energy ranges from approximately $-74.94$~Ha at the least stretched and bent geometries down to a well-defined global minimum of approximately $-75.02$~Ha. The contour plot in Figure~\ref{fig:2d_contour_h2o_sto-3g}(a) reveals a well-structured energy landscape with smooth, nearly concentric contour lines surrounding the minimum. The innermost contours at $0.22$ and $1.59$~mHa above the minimum are tightly clustered, indicating a relatively flat basin, while the outer contour levels at $3.75$, $6.02$, $8.87$, $11.97$, $15.72$, $20.52$, $25.69$, $31.68$, $38.53$, $46.31$, $55.07$, and $64.49$~mHa increase in spacing as the energy rises steeply toward the edges of the grid. The obtained surface is notably asymmetric: it rises more steeply along the $r$ axis than along the $\theta$ axis, reflecting the greater stiffness of the O--H stretching mode relative to the bending mode. The minimum-energy configuration of the H$_2$O molecule in the STO-3G basis, indicated by the red triangle, is located at $r_{\mathrm{min}} = 1.03$~\AA\ and $\theta_{\mathrm{min}} = 96.61^\circ$. 

        Under the reduced-subsampling condition of $\epsilon_{\mathrm{s}} = 21$, shown in Figure~\ref{fig:2d_pes_h2o_sto-3g}(b) and Figure~\ref{fig:2d_contour_h2o_sto-3g}(b), the 3D surface retains its qualitative bowl shape but exhibits visible roughness, particularly near the minimum and at larger bond lengths. In the contour plot, the outer contour lines ($> 15$~mHa) remain relatively smooth and well resolved, while the innermost contours at $0.22$ and $1.59$~mHa are noticeably jagged and irregular, indicating that the fine structure of the energy landscape near the minimum is sensitive to the reduced subsampling budget. The minimum under $\epsilon_{\mathrm{s}} = 21$ is displaced slightly to $\theta_{\mathrm{min}} = 97.58^\circ$ at the same bond length of $r_{\mathrm{min}} = 1.03$~\AA, corresponding to a shift of approximately $1^\circ$ relative to the FCI minimum. This deviation arises from stochastic variation in $\eta_{\mathrm{sub}}$ across neighbouring grid points rather than from a systematic bias. Another notable feature is that the practical chemical accuracy bound is represented by the $1.59$~mHa contour in Figure~\ref{fig:2d_contour_h2o_sto-3g}, indicating that all geometries within this contour lie within the chemical accuracy threshold (1 kcal/mol) of the global minimum energy.

\begin{figure}[H]
            \centering
            \includegraphics[width=\linewidth]{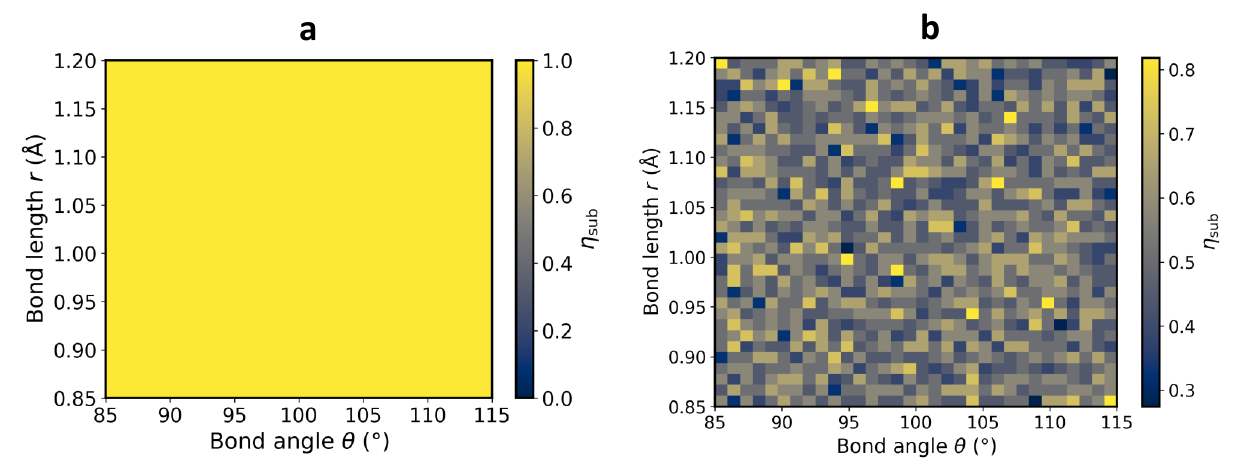}
            \caption{
            Diagonalization subspace dimension to symmetry space dimension ratio $\eta_\mathrm{sub}$ heatmaps with bond angle $\theta$ ($^\circ$) along the x-axis and bond length $r$ (\AA) along the y-axis for: (a) $\epsilon_s = 10^8$ and (b) $\epsilon_s = 21$, respectively.
            }
            \label{fig:2d_eta_sub_h2o_sto-3g}
        \end{figure}

        The subspace coverage heatmaps in Figure~\ref{fig:2d_eta_sub_h2o_sto-3g} reveal a qualitatively distinct behaviour between the two subsampling regimes. Under $\epsilon_{\mathrm{s}} = 10^8$, shown in subplot (a), $\eta_{\mathrm{sub}} = 1.0$ is achieved uniformly across the entire $(\theta, r)$ grid, confirming that $\mathbb{S}_{\mathrm{sub}}$ fully spans $\mathbb{S}$ at every geometry and that the energies in this regime are equivalent to exact diagonalization within the considered active space. Under $\epsilon_{\mathrm{s}} = 21$, shown in subplot (b), the picture changes markedly: $\eta_{\mathrm{sub}}$ varies across the grid in an essentially stochastic fashion, with values ranging from approximately $0.30$ to $0.80$ and no discernible smooth dependence on either $r$ or $\theta$. In this 2D setting, the reduced subsampling budget interacts with the two-dimensional geometry space in a more complex manner, producing a diagonalization subspace coverage that fluctuates quasi-randomly across neighbouring grid points. This stochastic variation in $\eta_{\mathrm{sub}}$ directly underlies the roughness visible in the contour plot under $\epsilon_{\mathrm{s}} = 21$ and is a fundamental characteristic of the SQD framework when operating below the threshold subsampling budget required to fully span $\mathbb{S}$ for the current $N_{\mathrm{shots}} = 10{,}000$.

        \begin{figure}[H]
            \centering
            \includegraphics[width=\linewidth]{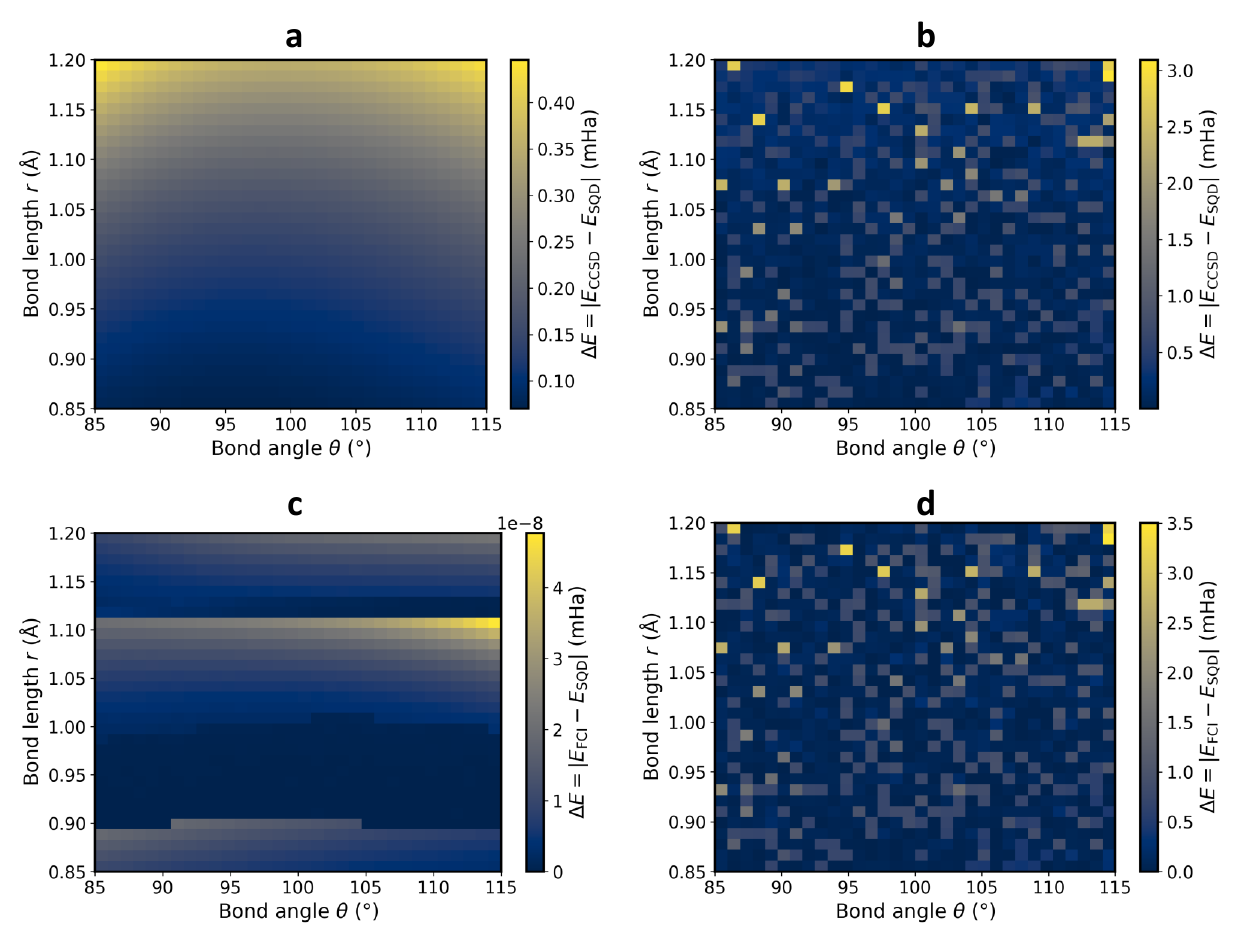}
            \caption{
            Energy-difference ($\Delta E$) heatmaps for the two-dimensional potential energy surface (2D-PES) of the H$_2$O molecule in the STO-3G basis. The bond length $r$ (\AA) and bond angle $\theta$ ($^\circ$) are shown along the x- and y-axes, respectively. Subplots show: (a) $\Delta E = |E_{\mathrm{CCSD}} - E_{\mathrm{SQD}}|$ (mHa) for $\epsilon_s = 10^8$, (b) $\Delta E = |E_{\mathrm{CCSD}} - E_{\mathrm{SQD}}|$ (mHa) for $\epsilon_s = 21$, (c) $\Delta E = |E_{\mathrm{FCI}} - E_{\mathrm{SQD}}|$ (mHa) for $\epsilon_s = 10^8$, and (d) $\Delta E = |E_{\mathrm{FCI}} - E_{\mathrm{SQD}}|$ (mHa) for $\epsilon_s = 21$.
            }
            \label{fig:2d_pes_h2o_delta_heatmaps}
        \end{figure}     

        The accuracy of the 2D-PES relative to both CCSD and FCI is quantified in the energy-difference heatmaps of Figure~\ref{fig:2d_pes_h2o_delta_heatmaps}. Under $\epsilon_{\mathrm{s}} = 10^8$, the deviation from CCSD shown in subplot (a) exhibits a smooth, monotonically increasing gradient along the $r$ axis, ranging from approximately $0.10$~mHa at compressed bond lengths near $r = 0.85$~\AA\ to approximately $0.40$~mHa at the most stretched geometries near $r = 1.20$~\AA, with only weak dependence on $\theta$. This spatially structured pattern reflects the growing difference between the FCI-equivalent SQD energy and the CCSD approximation as the O--H bond is stretched, and multireference character accumulates, and is therefore a consequence of CCSD's increasing inaccuracy rather than any degradation of SQD. The deviation from FCI, shown in subplot (c), is qualitatively different in character: it exhibits a faint but structured horizontal banding pattern, with $\Delta E_{\mathrm{FCI-SQD}}$ values spanning approximately $0$ to $5 \times 10^{-8}$~mHa, a scale that lies entirely within numerical diagonalization precision. The maximum deviations are concentrated near $r = 1.10$--$1.20$~\AA\ at large $\theta$, but remain negligible in any chemically meaningful sense. These results confirm that under $\epsilon_{\mathrm{s}} = 10^8$, the SQD(LUCJ) 2D-PES is numerically indistinguishable from the exact FCI surface at every grid point.

        Under $\epsilon_{\mathrm{s}} = 21$, the spatial character of both deviation heatmaps changes fundamentally. The $\Delta E_{\mathrm{CCSD-SQD}}$ map in subplot (b) and the $\Delta E_{\mathrm{FCI-SQD}}$ map in subplot (d) are both highly heterogeneous, with isolated yellow pixels of elevated deviation reaching up to approximately $3.0$~mHa and $3.5$~mHa, respectively, scattered across the grid in a pattern that mirrors the stochastic $\eta_{\mathrm{sub}}$ distribution. The majority of grid points retain deviations well below the chemical accuracy threshold of $1.59$~mHa, but a non-negligible fraction exceeds this threshold, with the worst-case outliers concentrated at larger bond lengths. Notably, the spatial distribution of high-error points in subplots (b) and (d) is closely correlated, confirming that the SQD energy errors under reduced subsampling are governed primarily by the incompleteness of $\mathbb{S}_{\mathrm{sub}}$ rather than by any systematic bias toward CCSD or FCI. Further analysis of the diversity of bitstrings generated by SQD(LUCJ) with CCSD initialization in these stretched-geometry cases, as well as their interplay with the symmetry-adapted subsampling budget and the overlap of the resulting $\mathbb{S}_{\mathrm{sub}}$ with the ground state, is deferred to future work. Further analysis of the rank and scatter plots, the one-dimensional PES slices, and the corresponding subspace-coverage behaviour under both subsampling regimes is provided in Section~D of the Supplementary Information.

        Taken together, the 2D-PES results demonstrate that the SQD(LUCJ) framework is capable of generating accurate and physically meaningful 2D-PES scans on near-term quantum hardware. Under sufficient quantum sampling and non-factorized subsampling, the method achieves FCI-level accuracy at every point on the grid, producing a smooth and chemically consistent surface whose topological features, such as well depth, anisotropy, minimum location, and contour structure, are all faithfully reproduced. Under reduced symmetry-adapted subsampling, the primary effect is the introduction of stochastic, geometry-uncorrelated noise in the energy landscape, driven by random variation in $\eta_{\mathrm{sub}}$ across the grid. This noise degrades the fine structure of the PES near the minimum and introduces isolated high-error geometries, but preserves the global topology and the relative ordering of energies across the grid. These characteristics suggest a clear path to improvement: increasing the subsampling budget toward $\epsilon_{\mathrm{s}} = \sqrt{|\mathbb{S}|}$ or beyond, together with sufficient quantum sampling, reduces stochastic noise and recovers smooth, chemically accurate surfaces; however, this requires prohibitively large $N_{\mathrm{shots}}$ and leads to a classically intractable $\mathbb{S}_{\mathrm{sub}}$, ultimately encountering a scaling wall. Therefore, efficient $\mathbb{S}_{\mathrm{sub}}$ generation methods beyond configuration recovery are essential. In addition, extending beyond the current single-reference CCSD initialization of the LUCJ ansatz, along with the development of methods better suited for multireference and strongly correlated regimes, is warranted. Nonetheless, the present results establish a well-characterized performance baseline for 2D-PES mapping with SQD(LUCJ) on current hardware, motivating its extension to larger molecules and more complex coordinate spaces, including the incorporation of classical fragmentation techniques such as DMET in the next subsection.
    
    \subsection{SQD(LUCJ) within the Density Matrix Embedding Framework}
    
        To assess the effectiveness of the combined methodology of employing DMET along with SQD as a fragmentation-with-simulation strategy for quantum simulations of molecular systems~\cite{Shajan2025, Patra2026}, we consider two classes of problems of increasing complexity. First, a set of eight ligand-like molecules~\cite{Patra2026}, several of which are relevant to pharmaceutical and chemical applications. These molecules enable a controlled analysis of how fragmentation reduces the complexity of the electronic structure problem while preserving chemically relevant correlations within each impurity. Building on this, we extend the study to a significantly larger system, the amantadine molecule ($\mathrm{C_{10}H_{17}N}$)~\cite{Cady2010}, which represents a substantial increase in both molecule and fragment size. The ability to decompose such a molecule into multiple embedded fragments, each mapped to hardware-compatible active spaces (4 HOMO - 4 LUMO), highlights the key advantage of DMET in enabling simulations that would otherwise be intractable on near-term quantum devices. The details of the molecular geometries used and the QPU execution time are provided in Section~D in the Supplementary Information. 
    
    \subsubsection{Ligand-like Molecules}

        In the present simulations, the impurity Hamiltonians are solved on the IQM Sirius QPU, which provides 16 functional physical qubits. An active-space truncation is applied by selecting the four highest occupied molecular orbitals (4 HOMO) and four lowest unoccupied molecular orbitals (4 LUMO), resulting in a 16-qubit quantum system for each impurity calculation. Detailed resource estimates for the resulting impurity calculations are provided in Section~D of the Supplementary Information.

        \begin{figure}[htbp]
            \centering
            \includegraphics[width=0.9\linewidth]{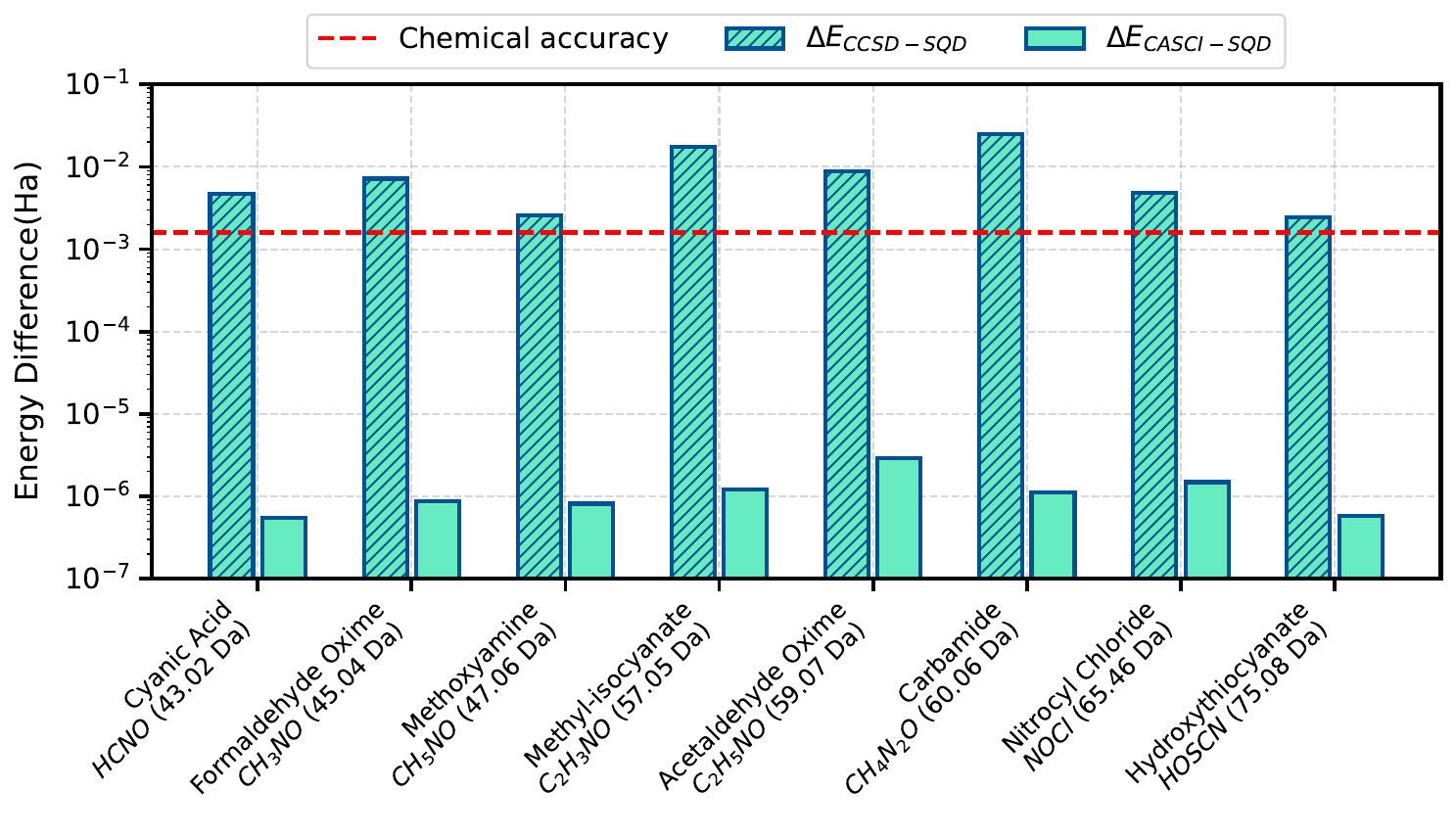}
            \caption{Absolute Energy difference plot obtained with the DMET-SQD workflow for eight ligand-like molecules simulated on the IQM Sirius QPU. The bars show the absolute differences between the converged DMET-SQD energies and the corresponding DMET-CASCI ($\Delta E_{\mathrm{CASCI-SQD}}$) and DMET-CCSD ($\Delta E_{\mathrm{CCSD-SQD}}$) reference energies. The dashed horizontal line indicates the chemical accuracy threshold ($1.594 \times 10^{-3}$ Ha).}
            \label{fig:dmet_sqd_ligand_mols_bars}
        \end{figure}

        The results presented in Figure~\ref{fig:dmet_sqd_ligand_mols_bars} summarize the final energy differences obtained with the DMET-SQD workflow for eight ligand-like molecules. For all molecules considered, the deviation between DMET-SQD and DMET-CASCI remains several orders of magnitude below the chemical accuracy threshold ($1.594 \times 10^{-3}$ Ha), typically reaching the micro-Hartree regime. In contrast, the differences between DMET-SQD and DMET-CCSD are consistently larger and in several cases approach or exceed the chemical accuracy threshold. This systematic behavior indicates that the SQD-based impurity solver is able to reproduce near-FCI-quality energies for the embedded fragments, while CCSD remains limited by its truncated cluster expansion.. The same qualitative trend is observed across all molecules spanning molecular weights from approximately $43$ to $75$ Da, including molecules such as HCNO, CH$_3$NO, CH$_5$NO, C$_2$H$_3$NO, C$_2$H$_5$NO, CH$_4$N$_2$O, NOCl, and HOSCN, demonstrating that the high accuracy of DMET-SQD is maintained consistently across chemically diverse systems.
        
\begin{figure}
            \centering
            \includegraphics[width=0.9\linewidth]{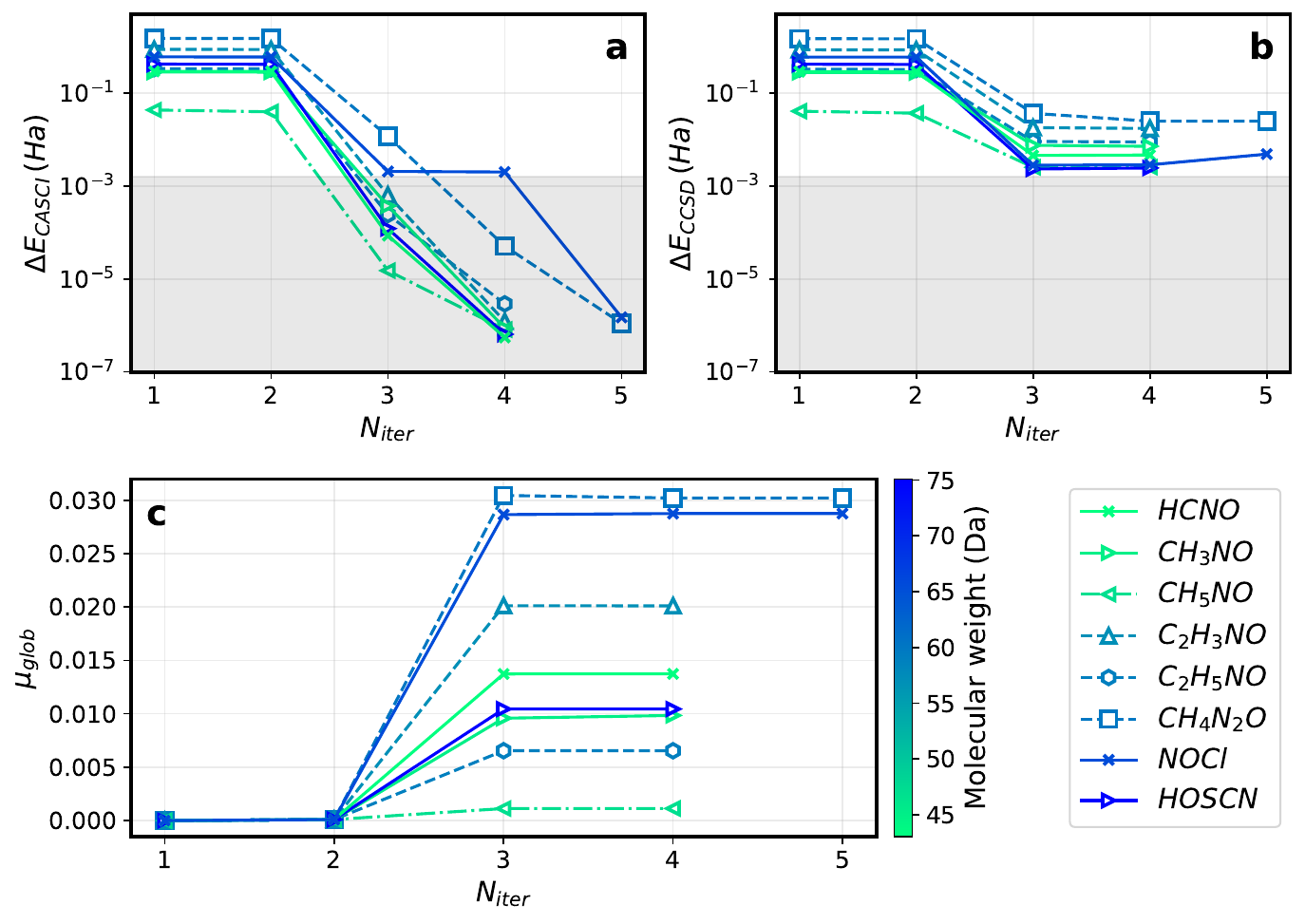}
            \caption{Convergence behavior and accuracy of the DMET–SQD workflow for eight ligand-like molecules simulated on the 16-qubit IQM Sirius quantum processor. (a) Absolute deviation of the DMET-SQD energies from the DMET-CASCI reference, $\Delta E_{\mathrm{FCI}}$, as a function of DMET iteration number $N_{\mathrm{iter}}$. (b) Absolute deviation relative to the DMET-CCSD reference, $\Delta E_{\mathrm{CCSD}}$. (c) Convergence of the global chemical potential $\mu_{\mathrm{glob}}$ during the DMET self-consistency procedure. Molecules are color-coded by molecular weight.}
            \label{fig:dmet_sqd_ligand_mols_full}
        \end{figure}

        To further understand how these highly accurate final energies are obtained, the convergence behavior of the DMET-SQD workflow is analyzed in Figure~\ref{fig:dmet_sqd_ligand_mols_full}. Figure~\ref{fig:dmet_sqd_ligand_mols_full}(c) shows the chemical potential convergence plot, which indicates that the global chemical potential $\mu_{\mathrm{glob}}$ stabilizes rapidly for all systems, reaching convergence within four to five DMET iterations. In particular, six of the eight molecules converge within four iterations, while only $\mathrm{NOCl}$ and $\mathrm{CH_4N_2O}$ require a fifth iteration before the chemical potential stabilizes. The energy convergence plots, shown in Figure~\ref{fig:dmet_sqd_ligand_mols_full}(a) and Figure~\ref{fig:dmet_sqd_ligand_mols_full}(b), further reveal that the DMET-SQD energies approach the DMET-CASCI reference rapidly with increasing DMET iterations. This high accuracy can be attributed to the absence of symmetry-adapted subspace restrictions during configuration recovery: a prohibitively large $\epsilon_{\mathrm{s}}$ was used, ensuring that the recovered subspace was not artificially truncated. Combined with the use of $N_{shots} = 10{,}000$, this allows a large fraction of the relevant $\mathbb{S}$ configurations to be included in the SQD diagonalization for each impurity problem. Consequently, the DMET-SQD energies converge to values well within the chemical accuracy threshold and often approach micro-Hartree agreement with DMET-CASCI, while remaining consistently more accurate than DMET-CCSD for the molecules considered. Further details on the exact corresponding energy results are provided in Section~D of the Supplementary Information.

        \subsubsection{Amantadine Molecule}

\begin{figure}[htbp]
                \centering
                \includegraphics[width=0.88\linewidth]{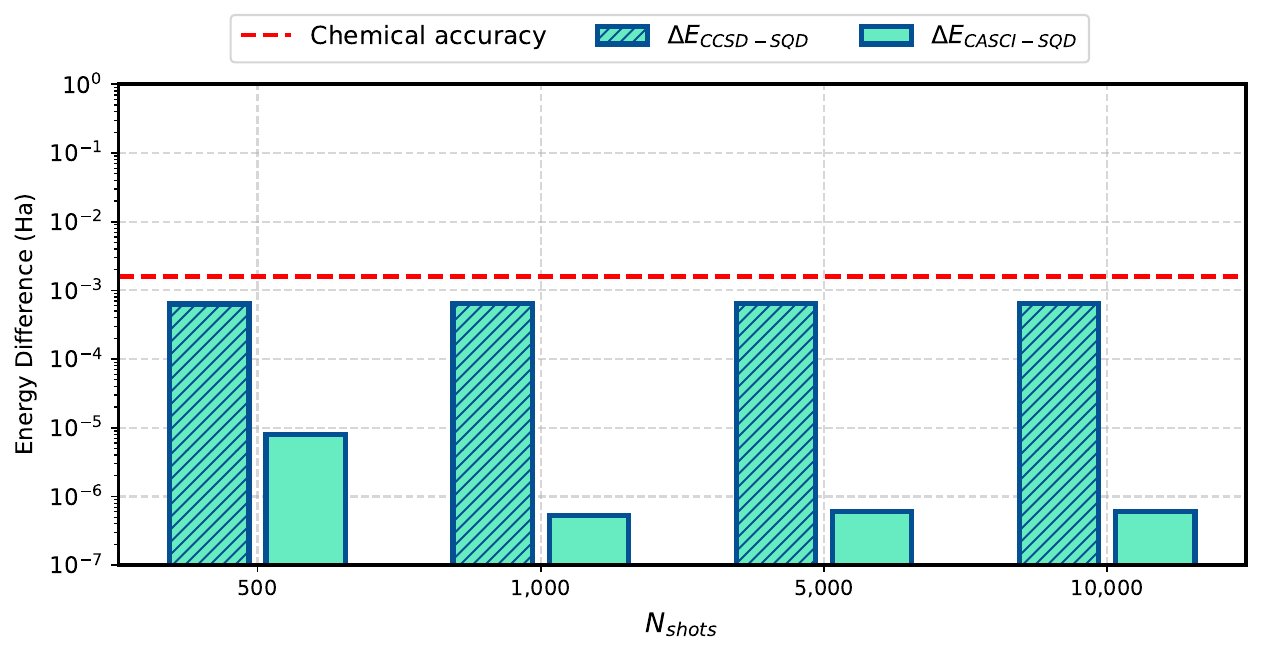}
                \caption{Absolute energy deviations of the final DMET-SQD energies for the amantadine molecule with respect to the DMET-CASCI and DMET-CCSD references for different measurement budgets $N_{\mathrm{shots}}=\{500,1000,5000,10000\}$. The dashed horizontal line indicates the chemical accuracy threshold ($1.594\times10^{-3}$ Ha).
            }
                \label{fig:dmet_sqd_amant_bars}
            \end{figure}

            The DMET-SQD simulation was also performed for the amantadine molecule ($\mathrm{C_{10}H_{17}N}$) as per the partition into eleven subsequent fragments as described in Section~\ref{sec: dmet_sqd_methodology}. For each fragment, the corresponding impurity Hamiltonian was constructed within the DMET framework and subsequently reduced through an active-space selection consisting of the four highest occupied molecular orbitals (4 HOMO) and four lowest unoccupied molecular orbitals (4 LUMO), yielding a 16-qubit problem compatible with the IQM Sirius quantum hardware. The SQD calculations for each impurity were carried out with a prohibitively large threshold on $\epsilon_s$ criterion during configuration recovery in order to avoid artificially restricting the size of the recovered diagonalization subspace. To investigate the influence of the measurement budget on the final DMET-SQD energy, four independent simulations were performed with shot numbers $N_{\mathrm{shots}}=\{500,1000,5000,10000\}$. The resulting bar plot shown in Figure~\ref{fig:dmet_sqd_amant_bars} compares the absolute deviations of the final DMET-SQD energies with respect to DMET-CASCI and DMET-CCSD reference calculations. As shown in the figure~\ref{fig:dmet_sqd_amant_bars}, for all shot counts except $N_{\mathrm{shots}}=500$, the deviation between the DMET-SQD and DMET-CASCI energies lies within the micro-Hartree regime, whereas the case $N_{\mathrm{shots}}=500$ exhibits a slightly larger deviation on the order of $10^{-5}$ Ha. Importantly, only negligible improvements in accuracy are observed when increasing the shot budget from $1{,}000$ to $5{,}000$ or $10{,}000$. This indicates that, for impurity problems of size $16$ qubits, a shot budget of approximately $1{,}000$ measurements is sufficient to saturate $\mathbb{S}_{\mathrm{sub}}$ obtained through configuration recovery. Indeed, for $N_{\mathrm{shots}}=\{1000,5000,10000\}$ the recovered $\mathbb{S}_{\mathrm{sub}}$ for each impurity spans the $\mathbb{S}$, which explains the nearly identical final DMET-SQD energies observed across these shot budgets. In contrast, for $N_{\mathrm{shots}}=500$ the recovered subspace $\mathbb{S}_{\mathrm{sub}}$ typically spans approximately $90\%-100\%$ of $\mathbb{S}$, leading to the slightly larger deviations. Consistent with the previous ligand-like molecular systems, the absolute energy differences with respect to DMET-CCSD remain comparatively larger because the DMET-SQD energies lie much closer to the near-exact DMET-CASCI reference.
            
\begin{figure}[H]
                \centering
                \includegraphics[width=0.88\linewidth]{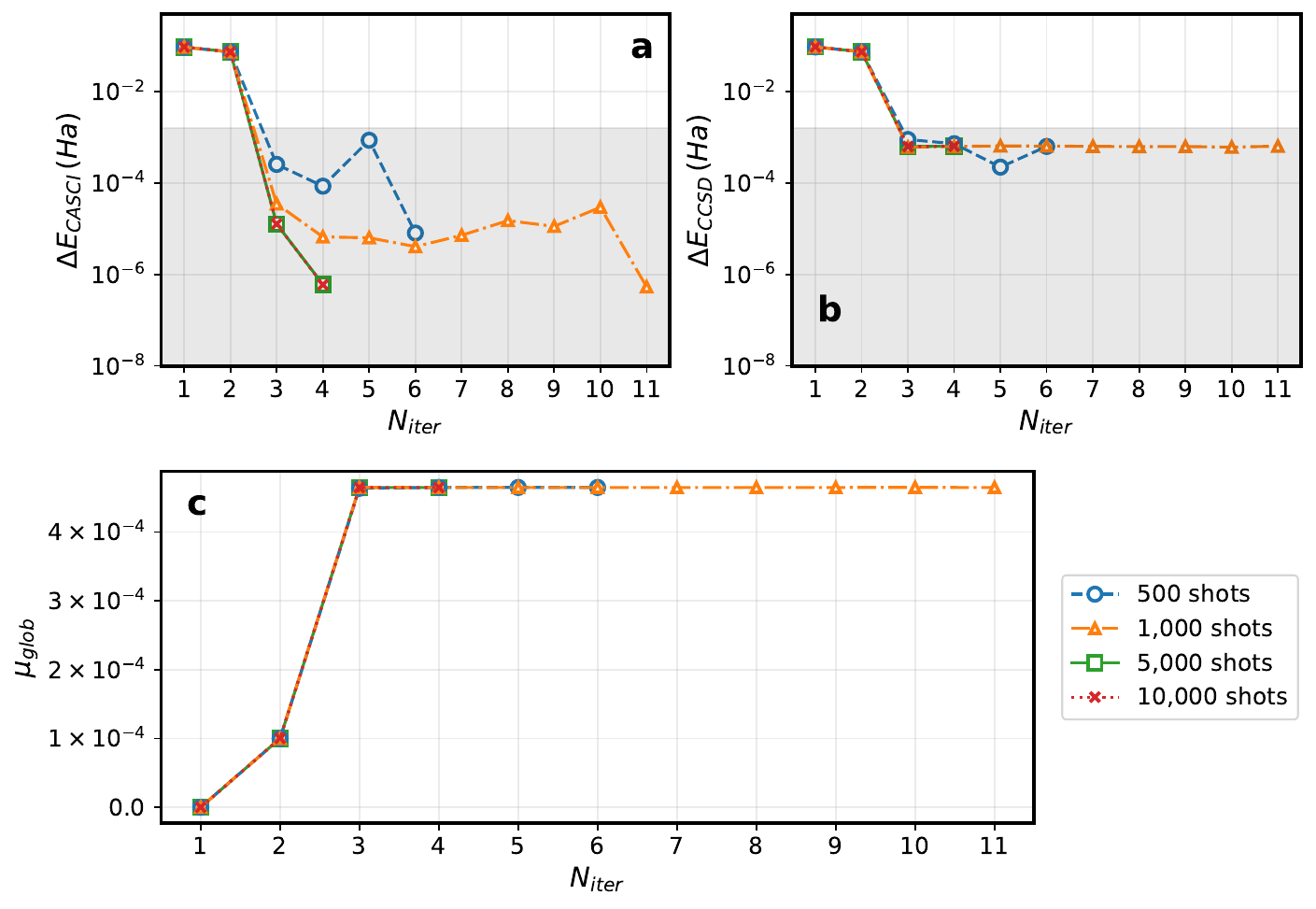}
                \caption{DMET-SQD convergence for the amantadine molecule at different shot budgets $N_{\mathrm{shots}}=\{500,1000,5000,10000\}$. Subplots (a) and (b) show the energy deviations with respect to DMET-CASCI and DMET-CCSD, respectively, while subplot (c) shows the convergence of the global chemical potential $\mu_{\mathrm{glob}}$ during the DMET self-consistency iterations.}
                \label{fig:dmet_amant_full}
            \end{figure}

            Further insight into the behavior of the DMET-SQD workflow can be obtained from the convergence plots of the DMET self-consistency cycle provided in Figure~\ref{fig:dmet_amant_full}. The convergence profiles for the energy errors relative to DMET-CASCI and DMET-CCSD, together with the evolution of the global chemical potential $\mu_{\mathrm{glob}}$, are shown in the accompanying panels Figure~\ref{fig:dmet_amant_full}(a), (b) and (c) respectively. For most shot budgets, the DMET procedure converges rapidly: the cases $N_{\mathrm{shots}}=5{,}000$ and $10{,}000$ reach convergence within four iterations, while $N_{\mathrm{shots}}=500$ converges in six iterations. A notable peculiarity is observed for $N_{\mathrm{shots}}=1{,}000$, where convergence formally required 11 DMET iterations. However, inspection of the chemical potential trajectory reveals that $\mu_{\mathrm{glob}}$ approaches essentially the same value for all shot budgets by the third iteration, with smaller subsequent fluctuations. Because the convergence tolerance for the chemical potential optimization was set to $10^{-8}$, these minute stochastic variations in $\mu_{\mathrm{glob}}$ can lead to additional iterations being counted before the convergence criterion is strictly satisfied. Consequently, the larger iteration count observed for $N_{\mathrm{shots}}=1{,}000$ likely reflects numerical sensitivity to the stringent tolerance rather than a genuine difference in physical convergence behavior.
            
            Importantly, within a fragmentation-based framework such as DMET combined with SQD, the subspace coverage of individual fragments is less critical than the average value of $\eta_{\mathrm{sub}}$ across all fragments. Because the total DMET energy is assembled from contributions of all impurity calculations, the collective coverage of the relevant $\mathbb{S}$ determines the overall accuracy of the simulation. The consistently high values of $\eta_{\mathrm{sub}}$ observed across fragments therefore explain the near-FCI accuracy obtained in the DMET-SQD energies. Moreover, the absence of an explicit symmetry-adapted subspace truncation ensures that the accessible configuration space is governed primarily by the measurement statistics rather than by algorithmic constraints, allowing the SQD procedure to recover near-complete $\mathbb{S}_{\mathrm{sub}}$ for most fragments w.r.to $\mathbb{S}$. These observations further confirm that the DMET-SQD framework remains robust and highly accurate even for larger molecular systems with multiple atoms per fragment, such as amantadine, when implemented on near-term quantum hardware. Further details on the circuit resources, energies, and evolution of $\mathbb{S}_{\mathrm{sub}}$ during the DMET cycle are provided in Section~D of the Supplementary Information.

\section{Conclusion}
\label{sec:conclusion}

    This work presents one of the most comprehensive experimental demonstrations to date of quantum-computing-based molecular simulation on IQM's Sirius 24-qubit superconducting quantum processor, employing up to 16 operational qubits across a diverse range of quantum chemistry tasks, including ground-state energy calculations, 1D- and 2D-PES scans, and embedding-based simulations of ligand-like and pharmacologically relevant molecules. The unifying methodology is the use of sample-based quantum algorithms, in particular SQD, which partially decouple the quality of the computed energy from the fidelity of individual circuit executions by confining the energetic resolution to an exact diagonalization step over a hardware-sampled subspace. Two ansätze were benchmarked within the SQD framework. The hardware efficiency and shallow circuit depth of LUCJ make it the preferred choice for near-term execution, consistently outperforming the newly introduced LCNot-UCCSD for SQD and remaining operational across all five benchmark molecules. LCNot-UCCSD, while offering reduced classical pre-computation overhead, produces circuits of sufficient depth to prevent configuration recovery for H$_2$O and NH$_3$, exposing a noise-induced failure mode whose systematic characterization is left for future work. These complementary strengths and limitations provide a practical basis for ansatz selection in near-term quantum simulation workflows.
    
    The potential energy surface results represent some of the practically significant contributions of this work. SQD(LUCJ) reproduces FCI-quality dissociation profiles across all scanned molecules and basis sets, with accuracy that degrades in a controlled and observable manner as the subsampling budget is reduced. The first experimental 2D-PES for H$_2$O on superconducting quantum hardware to our knowledge, comprising a $32 \times 32$ grid and matching FCI within numerical precision under sufficient subsampling, demonstrates that multi-dimensional surface mapping is now feasible on near-term devices and has the potential to enable improved characterization of transition states, reaction pathways, and force-field parameterization~\cite{Barone2021, Chmiela2017, Qiu2025}. Integrating SQD(LUCJ) with DMET extends the methodology to larger and chemically relevant molecules. For eight ligand-like systems and the pharmacologically relevant amantadine molecule ($\mathrm{C_{10}H_{17}N}$), DMET-SQD recovers energies close to the micro-Hartree regime relative to DMET-CASCI, well within the chemical accuracy threshold and significantly improving upon DMET-CCSD, while using modest measurement budgets and converging reliably within a small number of self-consistency iterations. To our knowledge, the amantadine simulation constitutes one of the first experimental demonstrations of DMET-SQD at this molecular scale on IQM hardware. As the total energy of the system is assembled collectively from these fragmented impurity calculations, the DMET-SQD framework avoids the exponential scaling wall associated with monolithic molecular simulations.
     
    Looking ahead, realizing the full potential of this pipeline requires several natural extensions. On the hardware side, scaling to larger and higher-fidelity QPUs will enable larger active spaces and reduce current circuit depth limitations. On the algorithmic side, establishing the efficiency of quantum sampling beyond current state-of-the-art classical techniques~\cite{Reinholdt2025} and improved error mitigation targeted at configuration recovery~\cite{Weaving2025} would take precedence. Followed by the development of more compact construction of the diagonalization subspace~\cite{Yoo2026, Patra2026ML} and multi-reference ansatz initialization strategies~\cite{Wang2025}, could further enhance accuracy and scalability. Within embedding frameworks, moving beyond the 4 HOMO -- 4 LUMO truncation and exploring alternative fragmentation and bath construction schemes~\cite{Shajan2026} and pushing the sample-based quantum algorithms towards fault-tolerance~\cite{KSV2026} would open the door to economical and efficient simulations of protein-ligand complexes, metal-organic frameworks, and other industrially relevant systems. More broadly, the combination of sampling-based quantum algorithms, chemically inspired hardware-efficient ansätze, and embedding frameworks demonstrated here provides a practical and extensible foundation for chemically accurate quantum simulation, along with a clear quantitative baseline for future hardware and algorithmic advances.

\section*{Acknowledgements}
The authors gratefully acknowledge the credit-free access to the IQM Sirius quantum hardware provided during the IQM Quantum School held from $2^{nd}$ to $4^{th}$ December 2025, presented by Dr. Stefan Seeger, Dr. Daniel Bulmash, and Dr. Nadia Milazzo. All hardware-based quantum chemistry calculations in this work were performed during this program, and the authors thank the organizers for enabling these experiments. The authors would also like to extend their appreciation to the advisors of Qclairvoyance Quantum Labs for their support, constructive discussions, and continuous inspiration throughout the preparation of this work.

 \section*{Funding}
This research received no specific grant from any funding agency in the public, commercial, or not-for-profit sectors.

\section*{Competing Interests}
The authors declare no competing interests.

\section*{Data Availability}
The datasets generated and analysed during the current study are available in the Zenodo repository (DOI: 10.5281/zenodo.19625172).

\newpage
\section*{\Large Supplementary Information}
\addtocontents{toc}{\vspace{0.75em}}
\addcontentsline{toc}{section}{\large Supplementary Information}

\appendix
\setcounter{table}{0}
\renewcommand{\thetable}{A\arabic{table}}

\setcounter{figure}{0}
\renewcommand{\thefigure}{A\arabic{figure}}

\setcounter{equation}{0}
\renewcommand{\theequation}{A\arabic{equation}}

\section{List of Symbols}

    \begin{longtable}{p{0.10\textwidth} p{0.28\textwidth} p{0.10\textwidth} p{0.28\textwidth}}
    \caption{List of symbols used throughout the manuscript.} \\
    \hline
    \textbf{Symbol} & \textbf{Description} & \textbf{Symbol} & \textbf{Description} \\
    \hline
    \endfirsthead
    
    \hline
    \textbf{Symbol} & \textbf{Description} & \textbf{Symbol} & \textbf{Description} \\
    \hline
    \endhead
    
    \hline
    \endfoot
    
    \hline
    \endlastfoot
    
    $\mathbb{H}$ & Hilbert space & 
    $\mathbb{S}$ & Symmetry space \\
    
    $\epsilon_{s}$ & Samples per batch & 
    $\mathbb{S}_{\mathrm{samp}}$ & Sampling space \\
    
    $\mathbb{S}_{\mathrm{post\text{-}sel}}$ & Post-selected subspace & 
    $\mathbb{S}_{\mathrm{post\text{-}cr}}$ & Post-configuration recovery space \\
    
    $\mathbb{S}_{\mathrm{sub}}$ & Diagonalization subspace & 
    $N_{\mathrm{iter}}$ & Number of DMET iterations \\
    
    $N_{\mathrm{shots}}$ & Number of measurement shots & 
    $(N_{\mathrm{orb}},N_{\mathrm{ele}})$ & Number of molecular orbitals and electrons \\
    
    $N_{\alpha}(N_{\beta})$ & Number of electrons in $\alpha(\beta)$-spin & 
    $\vec{\theta}$ & Variational parameters \\
    
    $N_{Q}$ & Number of qubits & 
    $G_{D}$ & Quantum circuit gate depth \\
    
    $Move$ & Number of MOVE gates & 
    $R$ & Number of $R$ gates \\
    
    $CZ$ & Number of controlled-$Z$ gates & 
    $\tau_{\mathrm{qpu}}$ & QPU execution time (s) \\
    
    $G_{q_1}$ & Number of single-qubit gates & 
    $G_{q_2}$ & Number of two-qubit gates \\
    
    $G_{T}$ & Total number of gates & 
    $\eta_{\mathrm{sym}}$ & Symmetry-to-Hilbert space ratio \\
    
    $\eta_{\mathrm{post\text{-}cr}}$ & Post-CR to symmetry space ratio & 
    $\eta_{\mathrm{sub}}$ & Diagonalization Subspace to symmetry space ratio \\
    
    $r_{\min}$ & Bond length at minimum energy & 
    $\theta_{\min}$ & Bond angle at minimum energy \\
    
    $\mu_{\mathrm{glob}}$ & Global chemical potential & $\ket{\psi_{\mathrm{in}}}$ & Input quantum state  \\
    
    $\hat{H}_{ele}$ & Electronic Hamiltonian & 
    $\hat{H}_{\mathrm{sub}}$ & Projected Hamiltonian \\
    
    $\ket{\Phi_0}$ & HF reference determinant & 
    $\ket{\Phi_i^a}$ & Single excitation determinant \\
    
    $\ket{\Phi_{ij}^{ab}}$ & Double excitation determinant & 
    $c_i$ & CI coefficients \\
    
    $\hat{T}$ & Cluster operator & 
    $\hat{T}_1, \hat{T}_2$ & Single and double excitation operators \\
    
    $t_i^a, t_{ij}^{ab}$ & Cluster amplitudes & 
    $\hat{J}$ & Jastrow operator \\
    
    $\hat{n}_{p\sigma}$ & Number operator  & 
    $U(\vec{\theta})$ & Unitary operator \\
    
    $\mu$ & LUCJ layer index & 
    $\hat{a}^\dagger_p, \hat{a}_p$ & Creation and annihilation operators \\
    
    $h_{pq}$ & One-electron integrals & 
    $h_{pqrs} \text{ or } V_{pqrs}$ & Two-electron integrals \\
    
    $D_{\mu\nu}$ & One-particle reduced density matrix & 
    $\Gamma_{pq}^{rs}$ & Two-particle reduced density matrix \\
    
    $\mathbb{H}_{\mathrm{lin}}$ & One-particle Hilbert space & 
    $\mathbb{F}(\mathbb{H}_{\mathrm{lin}})$ & Fermionic Fock space \\
    
    $\mathbf{x}_i$ & Sampled bitstring & 
    $\gamma$ & Overlap with target state \\
    
    \end{longtable}

\newpage

\setcounter{table}{0}
\renewcommand{\thetable}{B\arabic{table}}

\setcounter{figure}{0}
\renewcommand{\thefigure}{B\arabic{figure}}

\setcounter{equation}{0}
\renewcommand{\theequation}{B\arabic{equation}}

\section{List of Abbreviations}

\begin{longtable}{p{0.10\textwidth} p{0.36\textwidth} p{0.10\textwidth} p{0.36\textwidth}}
\caption{List of abbreviations used throughout the manuscript.} \\
\hline
\textbf{Abbreviation} & \textbf{Description} &
\textbf{Abbreviation} & \textbf{Description} \\
\hline
\endfirsthead

\hline
\textbf{Abbreviation} & \textbf{Description} &
\textbf{Abbreviation} & \textbf{Description} \\
\hline
\endhead

\hline
\endfoot

\hline
\endlastfoot

ADAPT-QSCI &  Adaptive Construction of an Input State for Quantum-Selected Configuration Interaction &
AFQMC & Auxiliary-Field Quantum Monte Carlo \\

BE & Bootstrap Embedding &
CASCI & Complete Active Space Configuration Interaction \\

CCSD(T) & Coupled Cluster Singles and Doubles with perturbative Triples &
CI & Configuration Interaction \\

DC & Divide-and-Conquer &
DFT & Density Functional Theory \\

DMET & Density Matrix Embedding Theory &
DMFT & Dynamical Mean-Field Theory \\

DMRG & Density Matrix Renormalization Group &
DOCI-QSCI & Doubly Occupied Configuration Interaction--Quantum-Selected Configuration Interaction \\

EF-SQD & Entanglement Forging Sample-based Quantum Diagonalization &
EWF & Embedded Wave Function \\

FASQ & Fault-Tolerant Application-Scale Quantum &
FCI & Full Configuration Interaction \\

FMO & Fragment Molecular Orbital &
HCI & Heat-Bath Configuration Interaction \\

HI-VQE & Handover Iterative Variational Quantum Eigensolver &
HF & Hartree--Fock \\

HSB-QSCI & Hamiltonian Simulation-Based Quantum-Selected Configuration Interaction &
LAS & Localized Active Space \\

LASSCF & Localized Active Space Self-Consistent Field &
LCNot-UCCSD & Linear-CNOT Unitary Coupled-Cluster Singles and Doubles \\

LCU & Linear Combination of Unitaries &
LUCJ & Local Unitary Cluster Jastrow \\

MBE & Many-Body Expansion &
MERA & Multiscale Entanglement Renormalization Ansatz \\

MPS & Matrix Product State &
NISQ & Noisy Intermediate-Scale Quantum \\

ONIOM & Our own N-layered Integrated molecular Orbital and molecular Mechanics &
PEPS & Projected Entangled-Pair State \\

PES & Potential Energy Surface &
ph-AFQMC & Phaseless Auxiliary-Field Quantum Monte Carlo \\

PIGen-SQD & Physics-Informed Generative Sample-based Quantum Diagonalization &
QBE & Quantum Bootstrap Embedding \\

QEC & Quantum Error Correction &
QKUD & Quantum-Krylov Unitary Decomposition \\

QMC & Quantum Monte Carlo &
QPE & Quantum Phase Estimation \\

QPU & Quantum Processing Unit &
QSCI & Quantum-Selected Configuration Interaction \\

QSP & Quantum Signal Processing &
QSVT & Quantum Singular Value Transformation \\

SEET & Self-Energy Embedding Theory &
SKQD & Sample-based Krylov Quantum Diagonalization \\

SQD & Sample-based Quantum Diagonalization &
TE-QSCI & Time-Evolved Quantum-Selected Configuration Interaction \\

THC & Tensor Hypercontraction &
TN & Tensor Network \\

TTNS & Tree Tensor Network State &
UCCSD & Unitary Coupled-Cluster Singles and Doubles \\

VQE & Variational Quantum Eigensolver &
 &
\\

\end{longtable}
\newpage

\setcounter{table}{0}
\renewcommand{\thetable}{C\arabic{table}}

\setcounter{figure}{0}
\renewcommand{\thefigure}{C\arabic{figure}}

\setcounter{equation}{0}
\renewcommand{\theequation}{C\arabic{equation}}

\section{Theoretical Background}
\label{sec:theoretical_background}

This section outlines the theoretical foundations underlying the methods employed in this work. We briefly review the formulation of the electronic structure problem within the Born--Oppenheimer approximation, the representation of molecular systems in finite basis sets, and the key approximations used to make the many-body Schrödinger equation tractable. These concepts provide the basis for both classical and quantum computational approaches discussed in the subsequent sections.

    \subsection{Ab Initio Quantum Chemistry Methods}
        The fundamental premise of ab initio quantum chemistry is the formulation and exact or approximate solution of the non-relativistic, time-independent Schrödinger equation~\cite{Schrodinger1926} for a system of interacting electrons and nuclei. The total molecular Hamiltonian, which dictates the complete energetic landscape of the system, is defined as a sum of kinetic and potential energy operators:
        \begin{equation}
            \hat{H} = \hat{T}_N + \hat{T}_e + \hat{V}_{Ne} + \hat{V}_{NN} + \hat{V}_{ee}
        \end{equation}
        where $\hat{T}_N$ and $\hat{T}_e$ represent the kinetic energy operators of the atomic nuclei and the electrons, respectively. The potential energy terms $\hat{V}_{Ne}$, $\hat{V}_{NN}$, and $\hat{V}_{ee}$ represent the electron-nucleus Coulombic attraction, the nucleus-nucleus Coulombic repulsion, and the electron-electron Coulombic repulsion, respectively.

        Because the mass of a proton is roughly $1{,}836$ times greater than the mass of an electron~\cite{Lenz1951, Codata2018}, the atomic nuclei move on a significantly slower timescale than the rapidly fluctuating electron cloud. This immense physical disparity justifies the Born-Oppenheimer (BO) approximation~\cite{Born1927}, which mathematically decouples the nuclear and electronic wavefunctions. By treating the nuclear coordinates as fixed parameters rather than dynamic quantum variables, the nuclear kinetic energy term $\hat{T}_N$ is neglected, and the internuclear repulsion $\hat{V}_{NN}$ becomes a strictly constant scalar shift for any given geometry. The remaining electronic Schrödinger equation is expressed as:
        \begin{equation}\label{eqn : ele_schrodinger}
            \hat{H}_{ele} |\Psi_{ele}\rangle = E_{ele} |\Psi_{ele}\rangle
        \end{equation}
        where the electronic Hamiltonian is defined as $\hat{H}_{ele} = \hat{T}_e + \hat{V}_{Ne} + \hat{V}_{ee}$. The total energy of the molecular system at a specific static nuclear geometry is consequently given by $E_{tot} = E_{ele} + \hat{V}_{NN}$.

        \subsubsection{Hartree-Fock (HF) and Restricted HF (RHF)}
        
            The Hartree–Fock~\cite{Hartree1928 ,Fock1930} (HF) method is the foundational mean-field approximation of wavefunction-based quantum chemistry, in which the exact many-electron ground state is approximated by a single Slater determinant, thereby enforcing antisymmetry and the Pauli exclusion principle~\cite{Pauli1925} by construction. The resulting self-consistent field equations are defined by the Fock operator\cite{SzaboOstlund1996}
            \begin{equation}
            \hat{F}_i = \hat{h}_i + \sum_{j=1}^{e} (\hat{J}_j - \hat{K}_j),
            \end{equation}
            where $\hat{h}_i$ contains the one-electron kinetic energy and nuclear attraction terms, $\hat{J}_j$ is the Coulomb operator describing the classical average electron–electron repulsion, and $\hat{K}_j$ is the exchange operator arising from the antisymmetry of the Slater determinant, reducing same-spin repulsion. The HF equations are solved iteratively to self-consistency, yielding a mean-field ground-state reference that neglects dynamical electron correlation~\cite{Roothaan1951}.
            
            In closed-shell systems, Restricted Hartree–Fock (RHF) constrains $\alpha$ and $\beta$ electrons to share identical spatial orbitals, preserving spin symmetry but often failing in bond dissociation and strongly correlated regimes due to the rigid double-occupancy constraint~\cite{Roothaan1951}.

        \subsubsection{Møller–Plesset (MP) Perturbation} 
        
            To systematically recover a portion of the dynamical correlation energy absent in the mean-field HF approximation, Møller–Plesset (MP)~\cite{Moller1934} perturbation theory applies Rayleigh-Schrödinger Perturbation Theory (RSPT)~\cite{SakuraiNapolitano2017_rayleigh_schrodinger} to $\hat{H}_{ele}$. In this framework, the unperturbed zeroth-order Hamiltonian $\hat{H}_0$ is defined exactly as the sum of the one-electron Fock operators, while the perturbation operator $\hat{V}$ is defined as the exact electron-electron repulsion minus the average mean-field HF potential.

            Within RSPT based on the HF reference, the zeroth-order energy equals the sum of canonical orbital energies and the first-order correction recovers the HF energy. The leading correlation contribution therefore arises at second order, yielding the Møller-Plesset second order (MP2) correction. For double excitations from occupied orbitals $i,j$ to virtual orbitals $a,b$, the correlation energy~\cite{jensen2013introduction} is
            
            \begin{equation}
            E_{\mathrm{MP2}} =
            \sum_{i<j}^{\mathrm{occ}}
            \sum_{a<b}^{\mathrm{virt}}
            \frac{\big| \hat{V}_{ijab} - \hat{V}_{ijba} \big|^2}
            {\hat{F}_{ii} + \hat{F}_{jj} - \hat{F}_{aa} - \hat{F}_{bb}}
            =
            \sum_{i<j}^{\mathrm{occ}}
            \sum_{a<b}^{\mathrm{virt}}
            \langle ij||ab\rangle
            \left(
            \frac{\langle ij || ab \rangle}
            {\epsilon_i + \epsilon_j - \epsilon_a - \epsilon_b}
            \right)
            =
            \sum_{i<j}^{\mathrm{occ}}
            \sum_{a<b}^{\mathrm{virt}}
            \langle ij || ab \rangle \, t_{ij}^{ab}.
            \end{equation}
            
            where $\langle ij || ab \rangle = (ij|ab) - (ij|ba)$ are antisymmetrized two-electron integrals in the canonical molecular orbital (MO) basis, with $(pq|rs)$ denoting the Coulomb two-electron integrals and $\hat{V}_{ijab} = (ij|ab)$ in this basis. The quantity $\epsilon_p = \hat{F}_{pp}$ is the diagonal Fock matrix element corresponding to the canonical orbital energy of the $p^{th}$ orbital.
            
            The restricted summations $i<j$ and $a<b$ avoid double counting due to the antisymmetry of the two-electron integrals, the amplitudes
            \begin{equation}\label{eqn : mp2_init}
            t_{ij}^{ab} = \frac{\langle ij || ab \rangle} {(\epsilon_i + \epsilon_j - \epsilon_a - \epsilon_b)}
            \end{equation}
            are the MP2 double-excitation parameters obtained from first-order perturbation theory. These amplitudes constitute the first-order, non-variational approximation to the CCSD~\cite{bartlett_2007_CC} amplitudes and are therefore commonly used to initialize classical CCSD calculations as well as chemically motivated unitary parametrizations such as UCCSD~\cite{Barkoutsos2018} ansatz in variational quantum algorithms.

            When MO two-electron integrals are precomputed and stored, the dominant cost of the MP2 energy \emph{evaluation step} arises from tensor contractions over pairs of occupied and virtual orbitals. The energy contraction itself therefore scales as $\mathcal{O}(N^4)$ with respect to a unified measure of system size $N$, reflecting the double summation over occupied and virtual index pairs. Under these conditions, MP2 remains computationally inexpensive for moderate-to-large molecular systems while recovering approximately 70\%-95\% of the dynamical correlation energy~\cite{Cremer_2011_MP2}. Owing to its favourable cost-to-accuracy ratio and its perturbative foundation on the HF reference, MP2 provides an efficient and physically well-motivated approximation to electron correlation effects.

        \subsubsection{Coupled Cluster (CC)}

            While perturbation theory provides a non-iterative correlation correction, Coupled Cluster (CC) theory~\cite{bartlett_2007_CC, levine2014quantum_ccsd} provides a rigorous, size-consistent, and size-extensive methodology for capturing many-body electron correlation to infinite order within a truncated excitation space. The CC wavefunction~\cite{levine2014quantum_ccsd, jensen2013introduction_ccsd} is formulated via an exponential ansatz applied to the reference RHF determinant:
            \begin{equation}
            |\Psi_{\mathrm{CC}}\rangle = e^{\hat{T}} |\Phi_0\rangle
            \end{equation}
            where the cluster operator $\hat{T}$ generates electronic excitations from the occupied to the virtual space. For the widely utilized CCSD formulation, the operator is truncated at the level of one- and two-body excitations,
            \begin{equation}
            |\Psi_{\mathrm{CCSD}}\rangle 
            = e^{\hat{T}_1 + \hat{T}_2} |\Phi_0\rangle,
            \quad
            \hat{T}_1 = \sum_{i,a} t_i^a \, \hat{a}_a^\dagger \hat{a}_i,
            \quad
            \hat{T}_2 = \frac{1}{4} \sum_{i,j,a,b} t_{ij}^{ab} \, 
            \hat{a}_a^\dagger \hat{a}_b^\dagger \hat{a}_j \hat{a}_i
            \end{equation}
            where $i,j$ denote occupied spin-orbitals and $a,b$ denote virtual spin-orbitals. The amplitudes $t_i^a$ and $t_{ij}^{ab}$ correspond to single and double excitations, respectively.
            
            The defining theoretical advantage of the CC method lies in the exponential structure of $e^{\hat{T}}$. The Taylor series expansion of the exponential generates disconnected higher-order excitations (e.g., the $\hat{T}_2^2 / 2$ term corresponds to disconnected quadruple excitations), thereby guaranteeing size extensivity~\cite{cc_qchem_2007, ccsd_crawford}. This ensures that the computed energy scales properly with system size, in contrast to truncated Configuration Interaction methods, which are neither size-extensive nor size-consistent~\cite{Szabo1996}. Classical CCSD requires the iterative solution of nonlinear amplitude equations and exhibits a computational scaling commonly expressed as $\mathcal{O}(N^6)$~\cite{cc_qchem_2007}.
            
            In the context of the present hybrid quantum-HPC framework, classical CCSD amplitudes are extracted and mapped to initialize the variational parameters of the LUCJ ansatz~\cite{Motta2023}. This informed initialization accelerates convergence of the quantum optimization and places the variational search within a physically meaningful region of parameter space from the outset~\cite{Motta2023}.
            
        \subsubsection{Configuration Interaction (CI)}

            Configuration Interaction (CI)~\cite{Sherrill1999, Knowles1984, slater_configuration_interaction} approaches the electron correlation problem by expressing the exact many-body wavefunction as a linear combination of all possible Slater determinants governed by the number and spin symmetry of the molecular system. These determinants are generated by systematically exciting electrons from the HF reference state $|\Phi_0\rangle$ into all available virtual orbitals:
            \begin{equation}
                |\Psi_{\mathrm{CI}}\rangle = c_0 |\Phi_0\rangle + \sum_{a,r} c_a^r |\Phi_a^r\rangle + \sum_{a<b, r<s} c_{ab}^{rs} |\Phi_{ab}^{rs}\rangle + \dots
            \end{equation}
            
            When all possible excitations are explicitly included, the method is termed Full Configuration Interaction (FCI)~\cite{SzaboOstlund1996_fci_matrix, jensen2007introduction_fci_size}. The FCI wavefunction provides the exact solution to the electronic Schrödinger equation (Eq.~(\ref{eqn : ele_schrodinger})) within the limits of the chosen finite basis set.
            
        Organizing the Slater determinants by excitation rank relative to the HF reference,
        \begin{equation}
        \{ |\Phi_0\rangle,\; |\Phi_i^a\rangle,\; |\Phi_{ij}^{ab}\rangle,\; 
        |\Phi_{ijk}^{abc}\rangle,\; |\Phi_{ijkl}^{abcd}\rangle,\; \dots \}
        \;\equiv\;
        \{ |\Phi_0\rangle,\; |S\rangle,\; |D\rangle,\; |T\rangle,\; |Q\rangle,\; \dots \}.
        \end{equation}

            where, $|S\rangle$, $|D\rangle$, $|T\rangle$, and $|Q\rangle$ denote the manifolds of single, double, triple, and quadruple excitations, respectively.
            The Hamiltonian matrix in this ordered basis acquires a characteristic block-banded structure~\cite{SzaboOstlund1996_fci_matrix}:
            
       \begin{equation} \label{eqn : ci_matrix}
        \renewcommand{\arraystretch}{1.3}
        \begin{array}{c|cccccc}
         & |\Phi_0\rangle & |S\rangle & |D\rangle & |T\rangle & |Q\rangle & \cdots \\
        \hline
        \langle \Phi_0 | 
        & \langle \Phi_0 | \hat H | \Phi_0 \rangle 
        & 0 
        & \langle \Phi_0 | \hat H | D \rangle 
        & 0 
        & 0 
        & \cdots \\
        
        \langle S | 
        & 0 
        & \langle S | \hat H | S \rangle 
        & \langle S | \hat H | D \rangle 
        & \langle S | \hat H | T \rangle 
        & 0 
        & \cdots \\
        
        \langle D | 
        & \langle D | \hat H | \Phi_0 \rangle 
        & \langle D | \hat H | S \rangle 
        & \langle D | \hat H | D \rangle 
        & \langle D | \hat H | T \rangle 
        & \langle D | \hat H | Q \rangle 
        & \cdots \\
        
        \langle T | 
        & 0 
        & \langle T | \hat H | S \rangle 
        & \langle T | \hat H | D \rangle 
        & \langle T | \hat H | T \rangle 
        & \langle T | \hat H | Q \rangle 
        & \cdots \\
        
        \langle Q | 
        & 0 
        & 0 
        & \langle Q | \hat H | D \rangle 
        & \langle Q | \hat H | T \rangle 
        & \langle Q | \hat H | Q \rangle 
        & \cdots \\
        
        \vdots
        & \vdots & \vdots & \vdots & \vdots & \vdots & \ddots
        \end{array}
        \end{equation}
            
             The matrix is Hermitian, and only the upper triangle need be considered.
            
            The block structure of the CI matrix shown in Eq.~(\ref{eqn : ci_matrix}) arises from the two-body form of $\hat{H}_{ele}$, which can change the occupation of at most two spin-orbitals in a Slater determinant. As a consequence, matrix elements between determinants whose excitation ranks differ by more than two vanish, producing the banded coupling pattern between the reference, single, double, and higher excitation manifolds. 
            
            In addition, when canonical HF orbitals are used as the reference, Brillouin’s theorem~\cite{SzaboOstlund1996_brillouins_theorem} states that the HF determinant does not couple to singly excited determinants through the Hamiltonian. This leads to the vanishing reference-single block in the CI Hamiltonian matrix, i.e.
            $\langle \Phi_0 | \hat{H} | \Phi_i^a \rangle = 0$.
            
            The dimension of the FCI space grows combinatorially~\cite{jensen2007introduction_fci_size} with the number of spin-orbitals. For a system with $N_{orb}$ spatial orbitals and $N_{\alpha}$, $N_{\beta}$ electrons of each spin, the symmetry-constrained Hilbert space $\mathbb{S}$ size is

            \begin{equation}
            |\mathbb{S}| = \binom{N_{orb}}{N_{\alpha}} \binom{N_{orb}}{N_{\beta}},
            \end{equation}
            
            which leads to an exponential growth of the CI Hamiltonian dimension with system size.

        \subsection{Basis Sets}

        In \textit{ab initio} quantum chemistry, the electronic wavefunction is represented in a finite basis through the Linear Combination of Atomic Orbitals (LCAO) approximation~\cite{Roothaan1951}, where molecular orbitals are expanded as linear combinations of atom-centered basis functions. Although Slater-type orbitals (STOs)~\cite{slater_1930}, with radial form $e^{-\zeta r}$, provide a physically accurate description of atomic orbitals by reproducing the correct nuclear cusp and exponential decay, evaluating the required multi-center electron repulsion integrals over STOs is computationally demanding. For this reason, practical electronic structure calculations employ Gaussian-type orbitals (GTOs)~\cite{boys_1950, pople_1969}, $e^{-\alpha r^2}$, whose mathematical properties allow efficient analytical evaluation of molecular integrals via the Gaussian product theorem~\cite{slater_1930}. In modern basis sets, contracted Gaussian functions (linear combinations of several primitive Gaussians) are optimized to approximate the radial behavior of STOs while retaining the computational advantages of Gaussian integrals.
        
        The choice of basis set significantly influences the accuracy of computed molecular energetics. Minimal basis sets, such as STO-3G, assign one contracted function to each atomic orbital (AO) and therefore provide a highly compact representation that is computationally inexpensive but lacks the flexibility required to describe polarization and electron correlation effects accurately. Larger split-valence basis sets, such as 6-31G~\cite{hehre_1972}, introduce \textit{additional radial functions for valence orbitals}, allowing a more flexible description of electron density and generally improving energetic predictions. However, the increased number of basis functions enlarges the resulting Hilbert space $\mathbb{H}$ and the number of $\hat{H}_{ele}$ terms, which directly translates to higher qubit requirements and deeper circuits in quantum simulations.

\begin{figure}[htbp]
            \centering
            \includegraphics[width=1.0\linewidth]{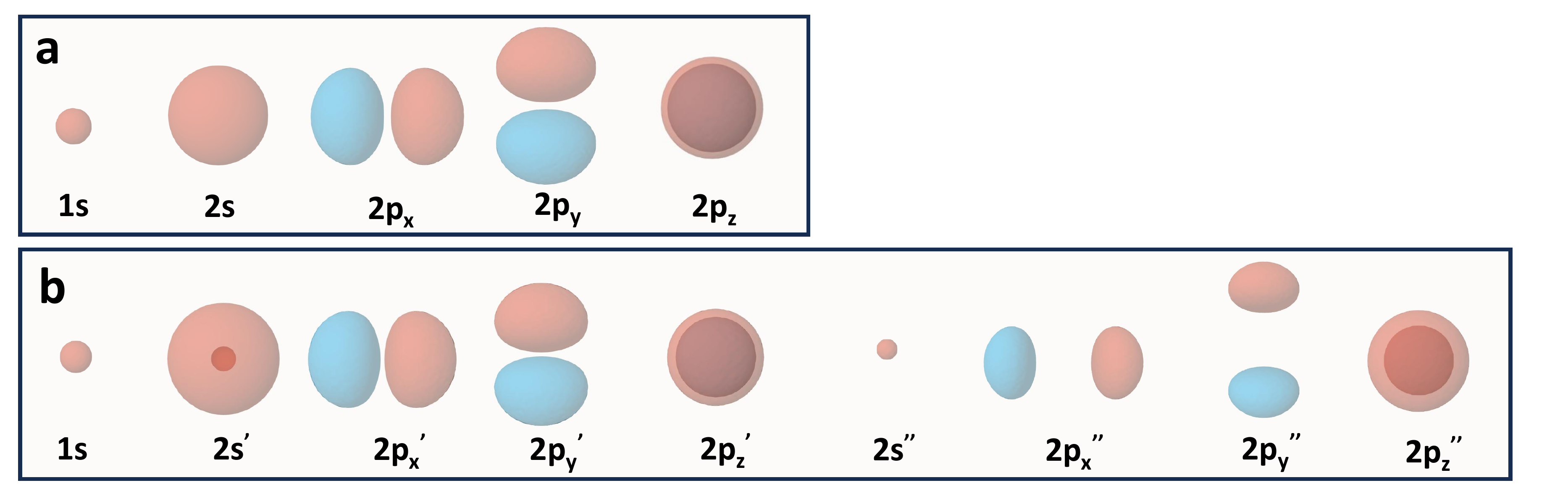}
            \caption{An illustrative depiction of the iso-spheres of the orbitals used for describing the carbon atom in (a) STO-3G basis set requiring 5 orbitals $1s, 2s, 2p_x, 2p_y, 2pz$ and in (b) 6-31G basis set which requires 9 orbitals $1s$, $2s$, $2p_x$, $2p_y$, $2p_z$, $2s'$, $2p_x'$, $2p_y'$, $2p_z'$. In addition to the orbitals which have a possibility of filling in the p-block (carbon atom), an additional orbital from the next block is picked for all the orbitals except the core orbital 1s. Hence, the additional orbitals $2s'$, $2p_x'$, $2p_y'$ and $2p_z'$ are also used. These orbitals were generated using \texttt{PySCF}, and visualized in \texttt{Mol$^*$} (MolStar)~\cite{Sehnal2021MolStar} online software.}
            \label{fig:C_diff_basis_sets}
        \end{figure}

        Figure~\ref{fig:C_diff_basis_sets} shows the basis functions used to describe the carbon atom in (a) the STO-3G basis and (b) the 6-31G basis. In both cases, the $1s$ orbital represents the core orbital. In the 6-31G basis, the valence space is described using a split-valence (double-$\zeta$) representation, in which each valence atomic orbital ($2s$, $2p_x$, $2p_y$, $2p_z$) is expanded using two basis functions with different radial extents: $2s\longrightarrow (2s', 2s'')$, $2p_x\longrightarrow(2p_x', 2p_x'')$, $2p_y\longrightarrow(2p_y', 2p_y'')$, $2p_z\longrightarrow(2p_z', 2p_z'')$. This effectively doubles the number of basis functions associated with the valence shell, leading to four additional basis functions compared to STO-3G. Consequently, while basis sets such as 6-31G provide a more flexible and accurate description of the electronic structure, minimal bases like STO-3G remain a practical choice for current quantum hardware due to their significantly reduced computational cost.

    \subsection{Linear-CNOT Unitary Coupled-Cluster Singles and Doubles (LCNot-UCCSD) Ansatz}

        The Unitary Coupled-Cluster (UCC) framework~\cite{Romero_2019} is one of the most widely studied ansätze in quantum computational chemistry, as it provides a chemically motivated variational form that systematically incorporates electron correlation through excitation operators. In the commonly used singles and doubles truncation (UCCSD), the trial state is generated by exponentiating fermionic excitation operators that promote electrons from occupied to virtual orbitals. To preserve unitarity on a quantum processor, the ansatz is constructed from an anti-Hermitian generator,
        \begin{equation}
            U(\vec{\theta}) = e^{\hat{T}(\vec{\theta}) - \hat{T}^\dagger(\vec{\theta})},
        \end{equation}
        where $\hat{T}(\vec{\theta})$ contains the single and double excitation operators. In practice, the exponential operator is factorized using a Trotter–Suzuki decomposition~\cite{Suzuki1976}, producing a sequence of exponentiated Pauli strings acting on qubits. When fermionic operators are mapped to qubits using transformations such as Jordan-Wigner (JW)~\cite{Jordan1928}, each excitation decomposes into multiple Pauli strings whose number grows rapidly with excitation rank: a \textit{single excitation} produces \textit{two Pauli strings}, a \textit{double excitation} produces \textit{eight}, and higher-rank excitations generate increasingly large operator sets. These Pauli strings contain long parity chains required to enforce fermionic statistics, which in turn lead to substantial control (C-NOT) overhead in the resulting quantum circuits.
                
        To mitigate this overhead, alternative formulations have been proposed that reinterpret fermionic excitations directly in the qubit space. In particular, the qubitized excitation operators~\cite{Magoulas2023} reformulate the excitation generators so that they can be implemented using simplified Pauli operators with reduced control (C-NOT) structure~\cite{Magoulas2023_LCNOT_first}. In this formulation, fermionic excitations are replaced by qubit excitation operators that act directly on pairs or quartets of qubits corresponding to occupied and virtual orbitals. The corresponding circuits for implementing the single and double excitation operators are shown in Fig.~\ref{fig:lcnot_uccsd_ansatz_excitation_gates}. Building on this idea, the UCCSD ansatz with excitation gates constructed with linearly scaling C-NOT gates~\cite{Magoulas2023} (henceforth abbreviated as LCNot-UCCSD) modifies the implementation of excitation operators by relaxing the fermionic parity structure introduced by mappings such as JW. In the standard construction, the resulting Pauli strings contain long $Z$-parity chains that are implemented using C-NOT ladders to propagate parity across the qubit register, leading to circuit depths that grow with orbital separation. 
        
        LCNot-UCCSD, on the other hand, removes selected parity-control links, thereby shortening these ladders and reducing the overall C-NOT gate cost while preserving the variational structure of the ansatz. As a result, the implementation of each excitation operator requires only a linear number of C-NOT gates, rather than the exponential growth characteristic of standard constructions~\cite{Barkoutsos2018}. This modification significantly reduces circuit depth, making practical implementations feasible on near-term quantum hardware. Importantly, while the circuit depth is reduced, the overall operator and parameter complexity remains unchanged, scaling as $\mathcal{O}(N^4)$ with the number of spin orbitals $2N_{\mathrm{orb}}$ (equivalently $N_Q$ qubits), consistent with standard UCCSD formulations~\cite{Peruzzo2014, Barkoutsos2018}. Some LCNot-UCCSD approximations allow for the exact fermionic parity constraints to be partially relaxed, resulting in operators that no longer strictly conserve particle number and spin projection, thereby introducing a small degree of symmetry leakage~\cite{Magoulas2023}. However, in the present work, no such approximations are considered, as the goal is to obtain a one-to-one correspondence with the standard UCCSD ansatz.

        \begin{figure}[H]
            \centering
            \begin{subfigure}[c]{0.38\textwidth}
                \centering
                \begin{quantikz}
                    \lstick{$|q_a\rangle$} & \ctrl{1} & \gate{R_y(\theta)} & \ctrl{1} & \qw \\
                    \lstick{$|q_i\rangle$} & \targ{}  & \ctrl{-1}          & \targ{}  & \qw
                \end{quantikz}
            \end{subfigure}
            \hfill
            \begin{subfigure}[c]{0.58\textwidth}
                \centering
                \begin{quantikz}
                    \lstick{$|q_b\rangle$} & \ctrl{1} & \ctrl{2}  & \gate{R_y(\theta)} & \ctrl{2} & \ctrl{1} & \qw \\
                    \lstick{$|q_a\rangle$} & \targ{}  & \qw       & \octrl{-1}          & \qw      & \targ{}  & \qw \\
                    \lstick{$|q_j\rangle$} & \ctrl{1} & \targ{}   & \ctrl{-2}           & \targ{}  & \ctrl{1} & \qw \\
                    \lstick{$|q_i\rangle$} & \targ{}  & \qw       & \octrl{-3}          & \qw      & \targ{}  & \qw
                \end{quantikz}
            \end{subfigure}
            \caption{The single excitation gate \textit{(left)} and double excitation gate \textit{(right)} 
                     represented in the Qubit Excitation Based (QEB) method~\cite{Magoulas2023}.}
            \label{fig:lcnot_uccsd_ansatz_excitation_gates}
        \end{figure}

    \subsection{Local Unitary Cluster Jastrow (LUCJ) Ansatz}
        
        The LUCJ ansatz provides a highly localized, Hubbard-inspired alternative that avoids the steep CNOT scaling associated with JW transformed UCC methods~\cite{Motta2023}. The canonical Unitary Cluster Jastrow (UCJ) ansatz is constructed as an $L$-layer unitary acting on the HF reference state $|\Phi_0\rangle$, where each layer alternates between orbital rotations and a Jastrow correlation operator. The LUCJ ansatz is written as
        \begin{equation}
            U(\vec{\theta}) = \prod_{\mu=1}^{L} e^{\hat{T}_\mu} e^{\hat{J}_\mu} e^{-\hat{T}_\mu} .
        \end{equation}
        A representative quantum circuit implementation of this layered structure is shown in Figure~\ref{fig:lucj_circuit}. This structure separates the wavefunction transformation into two physically meaningful components: a kinetic/cluster operator $\hat{T}_{\mu}$ generating generalized one-body orbital rotations and a Jastrow correlator $\hat{J}_{\mu}$ introducing explicit electron–electron interactions.
        
        The cluster operator $\hat{T}_\mu$ generates orbital rotations within the active space and can be expressed as a generalized one-body fermionic operator,
        \begin{equation}
            \hat{T}_\mu = \sum_{p,q,\sigma} T_{pq}^{\mu}\,\hat{a}_{p\sigma}^\dagger \hat{a}_{q\sigma},
        \end{equation}
        where $T_{pq}^{\mu}$ is an anti-Hermitian matrix of variational parameters, $\hat{a}_{p\sigma}^\dagger$ and $\hat{a}_{q\sigma}$ denote fermionic creation and annihilation operators, and $\sigma \in \{\alpha,\beta\}$ labels the spin degree of freedom. Exponentiation of this operator, $e^{\hat{T}_\mu}$ implements a unitary orbital rotation acting within the active space, dynamically reconfiguring the single-particle basis without introducing the higher-order fermionic excitation operators characteristic of conventional UCC expansions.
        
        The Jastrow factor $\hat{J}_{\mu}$ complements this transformation by encoding explicit two-body correlations through number–number interactions,
        \begin{equation}
            \hat{J}_\mu = i \sum_{p,q,\sigma,\tau} J_{pq,\sigma\tau}^{\mu}\,\hat{n}_{p\sigma}\hat{n}_{q\tau},
        \end{equation}
        where $\hat{n}_{p\sigma}=\hat{a}_{p\sigma}^\dagger \hat{a}_{p\sigma}$ is the number operator and $J_{pq,\sigma\tau}^{\mu}$ is a real symmetric tensor of variational parameters. This operator penalizes energetically unfavorable electron configurations and introduces conditional dependence between occupations, thereby introducing correlation effects beyond the single-reference HF description.
        
        In the LUCJ variant, the operator pool is strongly sparsified by restricting correlations to physically motivated local interactions, such as on-site opposite-spin terms and nearest-neighbor couplings. This locality assumption reflects the spatially short-ranged nature of electronic correlation and allows the ansatz to map naturally onto hardware connectivity graphs, eliminating the need for deep fermionic SWAP networks. Under all-to-all qubit connectivity, this sparsification further ensures that the operator and parameter complexity scales as $\mathcal{O}(N^2)$ in the number of spin orbitals~\cite{Motta2023}. As a result, the number of variational parameters and the resulting circuit depth remain tightly controlled, enabling hardware-efficient implementations while retaining sufficient flexibility to capture strongly correlated electronic structure~\cite{Motta2023}.

\definecolor{myblue}{HTML}{005dc7}
\definecolor{myteal}{HTML}{005dc7}
\definecolor{myaqua}{HTML}{00f2ea}
\definecolor{mygreen}{HTML}{3ef7ba}

\begin{figure}[H]
  \centering
  \begin{quantikz}[row sep=0.30cm, column sep=0.70cm]
    %-- alpha register --
    \lstick{$\alpha_1$} &
      \gate[4, nwires=3, style={fill=mygreen!40, draw=mygreen}]{e^{-\hat{T}_{1\alpha}}} &
      \gate[4, nwires=3, style={fill=myblue!40, draw=myblue}]{e^{i\hat{J}_{1\alpha\alpha}}} &
      \gate[8, nwires={3,7}, style={fill=myaqua!60, draw=myteal}]{e^{i\hat{J}_{1\alpha\beta}}} &
      \gate[4, nwires=3, style={fill=mygreen!40, draw=mygreen}]{e^{-\hat{T}_{2\alpha}}e^{\hat{T}_{1\alpha}}} &
      \meter{} \\
    \lstick{$\alpha_2$}
      & & & & & \meter{} \\
    \lstick{$\vdots$}\setwiretype{n}
      & \setwiretype{n} & \setwiretype{n} & \setwiretype{n} & \setwiretype{n} & \\
    \lstick{$\alpha_{N_{\mathrm{orb}}}$}
      & & & & & \meter{} \\
    %-- beta register --
    \lstick{$\beta_1$} &
      \gate[4, nwires=3, style={fill=mygreen!40, draw=mygreen}]{e^{-\hat{T}_{1\beta}}} &
      \gate[4, nwires=3, style={fill=myteal!40, draw=myteal}]{e^{i\hat{J}_{1\beta\beta}}} &
      &
      \gate[4, nwires=3, style={fill=mygreen!40, draw=mygreen}]{e^{-\hat{T}_{2\beta}}e^{\hat{T}_{1\beta}}} &
      \meter{} \\
    \lstick{$\beta_2$}
      & & & & & \meter{} \\
    \lstick{$\vdots$}\setwiretype{n}
      & \setwiretype{n} & \setwiretype{n} & \setwiretype{n} & \setwiretype{n} & \\
    \lstick{$\beta_{N_{\mathrm{orb}}}$}
      & & & & & \meter{}
  \end{quantikz}
  \caption{Quantum circuit for a single layer of the LUCJ ansatz acting on $\alpha$ and $\beta$ spin registers. Each layer implements a sequence of localized orbital rotations $e^{\pm \hat{T}}$ and spin-resolved Jastrow correlators $e^{i\hat{J}}$, including inter-spin coupling $e^{i\hat{J}_{\alpha\beta}}$, enabling efficient encoding of one- and two-body correlations within a locality-constrained framework.}
  \label{fig:lucj_circuit}
\end{figure}

\subsection{Quantum-Selected Configuration Interaction (QSCI) / Sample-Based Quantum Diagonalization (SQD)}

Quantum-Selected Configuration Interaction (QSCI)~\cite{Kanno2023} is a hybrid quantum–classical algorithm designed to exploit the sampling capability of QPUs while avoiding the large measurement overheads associated with VQEs. In conventional VQE~\cite{Peruzzo2014}, the quantum device repeatedly evaluates expectation values of many Pauli terms in the Hamiltonian, often requiring millions of measurement shots to suppress statistical noise. QSCI instead uses the QPU purely as a sampler of the wavefunction in the computational basis. An approximate state $|\psi_{\mathrm{in}}\rangle$, prepared using a shallow or heuristic ansatz with $|\vec{\theta}|$ variational parameters and non-zero overlap with the ground state, is repeatedly measured in the Slater-determinant basis using $N_{\mathrm{shots}}$ measurements. The resulting bitstrings $\{x_i\}$ reveal the dominant electronic configurations of the wavefunction, enabling the construction of a truncated but chemically relevant diagonalization subspace $\mathbb{S}_{\mathrm{sub}} \subset \mathbb{S}$ with dimension $|\mathbb{S}_{\mathrm{sub}}| \ll |\mathbb{S}|$, where $|\mathbb{H}|$ denotes the full Hilbert-space dimension. Because the QPU is used only to identify important configurations rather than to evaluate Hamiltonian expectation values, the algorithm avoids the severe measurement bottleneck that limits many NISQ-era variational algorithms.

More specifically, measurements of the prepared state $\ket{\psi_{\mathrm{in}}}$ generate a sampling space of determinants
\begin{equation}
\mathbb{S}_{\mathrm{samp}} = \{\, |x_i\rangle \,\}_{i=1}^{|\mathbb{S}_{\mathrm{samp}}|},
\end{equation}
whose size $|\mathbb{S}_{\mathrm{samp}}|$ depends on the number of shots $N_{\mathrm{shots}}$. From this space, symmetry filtering is applied to retain only determinants belonging to the physically allowed symmetry sector $\mathbb{S} \subset \mathbb{H}$ with dimension $|\mathbb{S}|$, corresponding to the correct electron number and spin quantum numbers for a molecular system with $(N_{orb}, N_{ele})$, to obtain the post-selected subspace $\mathbb{S}_{\mathrm{post-sel}}$.

The distinct $\alpha$ and $\beta$-spin configurations in $\mathbb{S}_{\mathrm{post-sel}}$ are identified and combined to form a unified set, which is subsequently tensored with itself to construct the final subspace for Hamiltonian diagonalization~\cite{Robledo2025, Patra2026}. Once $\mathbb{S}_{\mathrm{sub}}$ has been identified, all subsequent steps are performed classically. The Hamiltonian is projected into the selected determinant basis using Slater–Condon rules,
\begin{equation}
(\hat{H}_{\mathrm{sub}})_{ij} = \langle x_i | \hat{H} | x_j \rangle,
\qquad |x_i\rangle, |x_j\rangle \in \mathbb{S}_{\mathrm{sub}},
\end{equation}
producing a noise-free matrix representation of the Hamiltonian in $\mathbb{S}_{\mathrm{sub}}$. The resulting matrix is then diagonalized using classical iterative eigensolvers such as Davidson method~\cite{Davidson1975}. Because all matrix elements are evaluated deterministically on classical hardware, the resulting energy obeys a strict variational bound~\cite{Sakurai1993Variational}
\begin{equation}
E_{\mathrm{QSCI}} = \min \operatorname{eig}(\hat{H}_{\mathrm{sub}}) \ge E_{\mathrm{exact}},
\end{equation}
ensuring that the estimated ground-state energy cannot fall below the true value. In this framework, quantum errors do not corrupt the Hamiltonian evaluation itself but only affect the quality of $\mathbb{S}_{\mathrm{sub}}$ by potentially missing important configurations.

Sample-Based Quantum Diagonalization (SQD)~\cite{Robledo2025} extends QSCI by incorporating classical error-mitigation procedures that improve robustness to noise on NISQ devices. In practice, hardware errors frequently produce bitstrings that violate physical symmetries such as particle-number or spin conservation. Rather than discarding these samples, SQD introduces symmetry-aware configuration recovery (CR) methods that probabilistically correct corrupted bitstrings toward a reference orbital-occupancy vector derived from valid samples or mean-field estimates. This recovery procedure generates a corrected determinant pool $\mathbb{S}_{\mathrm{post\text{-}cr}}$, with dimension $|\mathbb{S}_{\mathrm{post\text{-}cr}}|$, which is subsequently used to construct the final diagonalization subspace $\mathbb{S}_{\mathrm{sub}}$, by means of proliferation~\cite{Robledo2025, Patra2026}.

This recovery process enables SQD to retain useful information even when a large fraction of raw samples are noisy, significantly improving performance on current quantum hardware. As a result, SQD represents a particularly promising architecture for near-term quantum chemistry simulations, where the QPU is used to perform quantum sampling, while classical supercomputers perform deterministic Hamiltonian construction and diagonalization within $\mathbb{S}_{\mathrm{sub}}$.

\subsection{Density-Matrix Embedding Theory (DMET)}

DMET~\cite{Knizia2012, Wouters2016} originates from the recognition that conventional embedding and fragmentation approaches~\cite{Wesolowski1993_frozen_density_functional, MFCC_Zhang_2003, Gadre1994_MTA} fail for fragments strongly coupled to their environment, such as through covalent bonds. In such cases, the fragment constitutes an open quantum system entangled with the rest of the system. Formally, an exact embedding can be achieved by replacing the environment with a finite quantum bath that reproduces the fragment–environment entanglement of the full many-body state. This follows from the Schmidt decomposition~\cite{ekert_schmidt_1995}, which guarantees that this entanglement can be represented using a finite number of bath degrees of freedom.

A practical formulation of DMET~\cite{Negre2025} proceeds at the level of the one-particle reduced density matrix (1-RDM). Let $\mathbb H_{\mathrm{lin}}$ denote the one-particle Hilbert space of dimension $L$ (where $L$ corresponds to the number of spin orbitals), with associated fermionic Fock space $\mathbb F(\mathbb {H}_{\mathrm{lin}})=\bigoplus_{n=0}^{L}\wedge^n\mathbb{H}_{\mathrm{lin}}$, corresponding to $\mathbb{H}$. For a many-body state described by a density operator $\Gamma$, the 1-RDM $D$, a Hermitian operator acting on $\mathbb{H}_{\mathrm{lin}}$ is defined as
\begin{equation}
D_{\mu\nu}
=
\mathrm{Tr}
\left(
\Gamma \, \hat a^\dagger_\nu \hat a_\mu
\right),
\end{equation}
with $0 \le D \le 1$ and $\mathrm{Tr}(D)=N$, and $\hat{a}^{\dagger}_{\nu}$($\hat{a}_{\mu}$) correspond to the creation(annihilation) operator acting on the $\nu^{th}$ ($\mu^{th}$) spin orbital. For a pure state $\Gamma=\ket{\Psi}\bra{\Psi}$, this reduces to
\begin{equation}
D_{\mu\nu}
=
\bra{\Psi}
\hat a^\dagger_\nu \hat a_\mu
\ket{\Psi},
\end{equation}
making explicit that the 1-RDM encodes all one-body observables. In the special case of a single Slater determinant, the 1-RDM reduces to $D = C_{\mathrm{occ}} C_{\mathrm{occ}}^\dagger$, where $C_{\mathrm{occ}}$ contains the occupied molecular orbital coefficients.

In conventional DMET~\cite{Wouters2016}, the low-level description is restricted to idempotent one-particle density matrices ($D^2 = D$). Let $\mathbb{H}_{\mathrm{lin}} = \mathbb H_S \oplus \mathbb H_E$ denote a partition of the one-particle Hilbert space, where $\mathbb{H}_S$ and $\mathbb{H}_E$ are the Hilbert spaces of the system and the environment, with $|\mathbb{H}_S| = \ell$. The idempotency of $D$ implies that the fragment-environment coupling is of rank at most $\ell$, or equivalently that the environment block of $D$ has no more than $\ell$ eigenvalues in the interval $(0,1)$~\cite{MacDonald1933} in the molecular spin orbital basis and $(0, 2)$ in the molecular spatial orbital basis. The corresponding eigenvectors define the bath orbitals, which, together with the fragment orbitals, span the impurity subspace.

In practice, the bath is constructed from an approximate mean-field state. The Schmidt decomposition then reduces to a single-particle problem involving the diagonalization of the fragment-environment block of the 1-RDM. The resulting impurity Hamiltonian, defined on the fragment and bath orbitals, is solved using a high-level correlated method (like SQD, in our case) to obtain embedded 1- and 2- RDMs.

The contribution of fragment $A$ to the total energy~\cite{Wouters2016} is expressed as
\begin{equation}
E_A
=
\sum_{i \in A}\sum_{p}
D_{i}^{\;p} \, h_{p i}
+
\frac{1}{2}
\sum_{i \in A}\sum_{p q r}
\Gamma_{i p}^{\;\;q r} \, V_{q r i p},
\end{equation}
where $i,j$ label fragment orbitals and $p,q,r$ span the impurity space. The two-particle reduced density matrix is defined as
\begin{equation}
\Gamma_{i p}^{\;\;q r}
=
\langle a_q^\dagger a_r^\dagger a_p a_i \rangle,
\end{equation}
and $h_{pi}$ and $V_{qrip}$ denote one- and two-electron integrals in the impurity basis. Restricting at least one index to fragment $A$ ensures each interaction term is counted once while allowing exchange of particles and correlations with the bath. The total energy is then
\begin{equation}
E_{\mathrm{tot}} = \sum_A E_A,
\end{equation}
yielding a size-extensive partitioning within the embedding framework~\cite{Wouters2016}.

To ensure global particle number consistency, a chemical potential $\mu_{\mathrm{glob}}$ is introduced and iteratively optimized such that the total electron number obtained from the embedded fragments matches the target electron number $N_{\mathrm{ele}}$. This is achieved by solving
\begin{equation}
\sum_A \mathrm{Tr}\big(D^{(A)}(\mu_{\mathrm{glob}})\big) = N_{\mathrm{ele}},
\end{equation}
where $D^{(A)}$ denotes the embedded 1-RDM of fragment $A$. The chemical potential $\mu_{\mathrm{glob}}$ is adjusted self-consistently until this condition is satisfied.

\newpage

\setcounter{table}{0}
\renewcommand{\thetable}{D\arabic{table}}

\setcounter{figure}{0}
\renewcommand{\thefigure}{D\arabic{figure}}

\setcounter{equation}{0}
\renewcommand{\theequation}{D\arabic{equation}}

\section{Supplementary Results} \label{si_sec:results_and_discussion}

    \subsection{Sample-based Quantum Diagonalization (SQD)} \label{si_sec:4.1.SQD}
    
        \subsubsection{Local Unitary Cluster Jastrow (LUCJ)}
        \label{si_subsec:LUCJ}     

            \begin{table}[H]
                \centering
                \caption{\label{tab:sqd_resource_summary}
                Quantum circuit resource summary for the SQD(LUCJ) workflows for the molecular systems executed on IQM Sirius quantum hardware. Here, $N_{Q}$ denotes the number of qubits, $|\vec{\theta}|$ the number of ansatz parameters, $G_{D}$ the circuit depth, $R$ the number of single-qubit rotation gates, $CZ$ the number of two-qubit Controlled-Z gates, $Move$ the number of two-qubit Move gates, $G_{q_1}$ the number of single-qubit gates, $G_{q_2}$ the number of two-qubit gates, and $G_{T}$ the total number of quantum gates.}
                \begin{tabular}{|c|r|r|r|r|r|r|r|r|r|}
                \hline
                \multicolumn{1}{|c|}{Molecule}
                & \multicolumn{1}{c|}{$N_{Q}$}
                & \multicolumn{1}{c|}{$|\vec{\theta}|$}
                & \multicolumn{1}{c|}{$G_{D}$}
                & \multicolumn{1}{c|}{$R$}
                & \multicolumn{1}{c|}{$CZ$}
                & \multicolumn{1}{c|}{$Move$}
                & \multicolumn{1}{c|}{$G_{q_1}$}
                & \multicolumn{1}{c|}{$G_{q_2}$}
                & \multicolumn{1}{c|}{$G_{T}$} \\
                \hline
                H$_2$ & 4 & 2 & 64 & 35 & 18 & 32 & 35 & 50 & 85 \\
                % \hline
                LiH & 12 & 72 & 638 & 358 & 202 & 348 & 358 & 550 & 908 \\
                % \hline
                BeH$_2$ & 14 & 156 & 1025 & 558 & 342 & 524 & 558 & 866 & 1424 \\
                % \hline
                H$_2$O & 14 & 110 & 892 & 492 & 270 & 486 & 492 & 756 & 1248 \\
                % \hline
                NH$_3$ & 16 & 240 & 1371 & 746 & 450 & 702 & 746 & 1152 & 1898 \\
                \hline
                \end{tabular}
            \end{table}

            \begin{table}[H]
                \centering
                \caption{\label{tab:sqd_lucj_energy_summary}
                SQD(LUCJ) ground-state energies obtained using different values of $\epsilon_{s}$ (samples per batch). For each molecular system, three independent runs are reported. Here, $\eta_{\mathrm{post\text{-}cr}}$ denotes the ratio of the post-configuration-recovery space dimension to the symmetry space dimension, $\eta_{\mathrm{sub}}$ denotes the ratio of the diagonalization subspace dimension to the symmetry space dimension, and $E_{\mathrm{SQD}}$ is the final SQD energy reported in Hartree for the corresponding molecule.}
                \begin{tabular}{|c|c|c|c|c|r|}
                \hline
                \multicolumn{1}{|c|}{Molecule}
                & \multicolumn{1}{c|}{$\epsilon_{s}$}
                & \multicolumn{1}{c|}{Run}
                & \multicolumn{1}{c|}{$\eta_{\mathrm{post\text{-}cr}}$}
                & \multicolumn{1}{c|}{$\eta_{\mathrm{sub}}$}
                & \multicolumn{1}{c|}{$E_{\mathrm{SQD}}$ (Ha)} \\
                \hline
                \multirow{6}{*}{\centering H$_2$}
                & \multirow{3}{*}{10$^8$} & 1 & 1.0000 & 1.0000 & -1.127842504 \\
                &                               & 2 & 1.0000 & 1.0000 & -1.127842504 \\
                &                               & 3 & 1.0000 & 1.0000 & -1.127842504 \\
                \cline{2-6}
                & \multirow{3}{*}{2}            & 1 & 1.0000 & 1.0000 & -1.127842504 \\
                &                               & 2 & 1.0000 & 1.0000 & -1.127842504 \\
                &                               & 3 & 1.0000 & 1.0000 & -1.127842504 \\
                \hline
                \multirow{6}{*}{\centering LiH}
                & \multirow{3}{*}{10$^8$} & 1 & 1.0000 & 1.0000 & -7.881178352 \\
                &                               & 2 & 1.0000 & 1.0000 & -7.881178352 \\
                &                               & 3 & 1.0000 & 1.0000 & -7.881178352 \\
                \cline{2-6}
                & \multirow{3}{*}{15}         & 1 & 1.0000 & 0.7511 & -7.881178308 \\
                &                               & 2 & 0.9956 & 0.5378 & -7.881171560 \\
                &                               & 3 & 1.0000 & 0.6400 & -7.881173091 \\
                \hline
                \multirow{6}{*}{\centering BeH$_2$}
                & \multirow{3}{*}{10$^8$} & 1 & 0.9127 & 1.0000 & -15.54779793 \\
                &                               & 2 & 0.9069 & 1.0000 & -15.54779793 \\
                &                               & 3 & 0.9078 & 1.0000 & -15.54779793 \\
                \cline{2-6}
                & \multirow{3}{*}{35}         & 1 & 0.9159 & 0.6400 & -15.54779418 \\
                &                               & 2 & 0.9020 & 0.5951 & -15.54765172 \\
                &                               & 3 & 0.9159 & 0.6400 & -15.54769945 \\
                \hline
                \multirow{6}{*}{\centering H$_2$O}
                & \multirow{3}{*}{10$^8$} & 1 & 0.9229 & 1.0000 & -75.01553352 \\
                &                               & 2 & 0.9252 & 1.0000 & -75.01553352 \\
                &                               & 3 & 0.8617 & 1.0000 & -75.01553352 \\
                \cline{2-6}
                & \multirow{3}{*}{21}         & 1 & 0.9048 & 0.3832 & -75.01547008 \\
                &                               & 2 & 0.9274 & 0.6553 & -75.01542294 \\
                &                               & 3 & 0.8707 & 0.5102 & -75.01533280 \\
                \hline
                \multirow{6}{*}{\centering NH$_3$}
                & \multirow{3}{*}{10$^8$} & 1 & 0.5080 & 1.0000 & -55.52114800 \\
                &                               & 2 & 0.4837 & 1.0000 & -55.52114800 \\
                &                               & 3 & 0.5124 & 1.0000 & -55.52114800 \\
                \cline{2-6}
                & \multirow{3}{*}{56}         & 1 & 0.5102 & 0.4605 & -55.52067640 \\
                &                               & 2 & 0.4786 & 0.5625 & -55.52029544 \\
                &                               & 3 & 0.5003 & 0.5102 & -55.52067337 \\
                \hline
                \end{tabular}
            \end{table}

        \subsubsection{Linear Scaling CNOT-Unitary Coupled Cluster Singles and Doubles (LCNot-UCCSD)}
        \label{si_subsec:LCNot-UCCSD}

        \begin{table}[H]
            \centering
            \caption{\label{tab:sqd_lcnot-uccsd_resource_summary}
            Quantum circuit resource summary for the SQD(LCNot-UCCSD) workflows for the molecular systems executed on IQM Sirius quantum hardware. Here, $N_{Q}$ denotes the number of qubits, $|\vec{\theta}|$ the number of ansatz parameters, $G_{D}$ the circuit depth, $R$ the number of single-qubit rotation gates, $CZ$ the number of two-qubit Controlled-Z gates, $Move$ the number of two-qubit Move gates, $G_{q_1}$ the number of single-qubit gates, $G_{q_2}$ the number of two-qubit gates, and $G_{T}$ the total number of quantum gates.}
            \begin{tabular}{|c|r|r|r|r|r|r|r|r|r|}
            \hline
            \multicolumn{1}{|c|}{Molecule}
            & \multicolumn{1}{c|}{$N_{Q}$}
            & \multicolumn{1}{c|}{$|\vec{\theta}|$}
            & \multicolumn{1}{c|}{$G_{D}$}
            & \multicolumn{1}{c|}{$R$}
            & \multicolumn{1}{c|}{$CZ$}
            & \multicolumn{1}{c|}{$Move$}
            & \multicolumn{1}{c|}{$G_{q_1}$}
            & \multicolumn{1}{c|}{$G_{q_2}$}
            & \multicolumn{1}{c|}{$G_{T}$} \\
            \hline
            H$_2$ & 4 & 3 & 117 & 82 & 34 & 46 & 82 & 80 & 162 \\
            % \hline
            LiH & 12 & 92 & 6389 & 4642 & 1928 & 2486 & 4642 & 4414 & 9056 \\
            % \hline
            BeH$_2$ & 14 & 204 & 14964 & 10868 & 4512 & 5836 & 10868 & 10348 & 21216 \\
            % \hline
            H$_2$O & 14 & 140 & 9991 & 7238 & 3046 & 3856 & 7238 & 6902 & 14140 \\
            % \hline
            NH$_3$ & 16 & 315 & 23623 & 17128 & 7140 & 9198 & 17128 & 16338 & 33466 \\
            \hline
            \end{tabular}
        \end{table}

        \begin{table}[H]
            \centering
            \caption{\label{tab:sqd_lcnot-uccsd_energy_summary}
            SQD(LCNot-UCCSD) ground-state energies obtained using different values of $\epsilon_{s}$ (samples per batch). For each molecular system, three independent runs are reported. Here, $\eta_{\mathrm{post\text{-}cr}}$ denotes the ratio of the post-configuration-recovery space dimension to the symmetry space dimension, $\eta_{\mathrm{sub}}$ denotes the ratio of the diagonalization subspace dimension to the symmetry space dimension, and $E_{\mathrm{SQD}}$ is the final SQD energy reported in Hartree for the corresponding molecule.}
            \begin{tabular}{|c|c|c|c|c|r|}
            \hline
            \multicolumn{1}{|c|}{Molecule}
            & \multicolumn{1}{c|}{$\epsilon_{s}$}
            & \multicolumn{1}{c|}{Run}
            & \multicolumn{1}{c|}{$\eta_{\mathrm{post\text{-}cr}}$}
            & \multicolumn{1}{c|}{$\eta_{\mathrm{sub}}$}
            & \multicolumn{1}{c|}{$E_{\mathrm{SQD}}$ (Ha)} \\
            \hline
            \multirow{6}{*}{H$_2$}
             & \multirow{3}{*}{10$^8$} & 1 & 1.0000 & 1.0000 & -1.127842504 \\
             &                             & 2 & 1.0000 & 1.0000 & -1.127842504 \\
             &                             & 3 & 1.0000 & 1.0000 & -1.127842504 \\
            \cline{2-6}
             & \multirow{3}{*}{2}          & 1 & 1.0000 & 1.0000 & -1.127842504 \\
             &                             & 2 & 1.0000 & 1.0000 & -1.127842504 \\
             &                             & 3 & 1.0000 & 1.0000 & -1.127842504 \\
            \hline
            \multirow{6}{*}{LiH}
             & \multirow{3}{*}{10$^8$} & 1 & 0.7111 & 1.0000 & -7.881178352 \\
             &                             & 2 & 0.7378 & 1.0000 & -7.881178352 \\
             &                             & 3 & 0.7156 & 1.0000 & -7.881178352 \\
            \cline{2-6}
             & \multirow{3}{*}{15}        & 1 & 0.7244 & 0.7511 & -7.881177519 \\
             &                            & 2 & 0.7378 & 0.4444 & -7.881167474 \\
             &                            & 3 & 0.7200 & 0.4444 & -7.881172350 \\
            \hline
            \multirow{6}{*}{BeH$_2$}
             & \multirow{3}{*}{10$^8$} & 1 & 0.2335 & 1.0000 & -15.54779793 \\
             &                             & 2 & 0.2359 & 1.0000 & -15.54779793 \\
             &                             & 3 & 0.2278 & 1.0000 & -15.54779793 \\
            \cline{2-6}
             & \multirow{3}{*}{35}        & 1 & 0.2269 & 0.5102 & -15.54758854 \\
             &                            & 2 & 0.2294 & 0.5951 & -15.54774085 \\
             &                            & 3 & 0.2294 & 0.5102 & -15.54778165 \\
            \hline
            \end{tabular}
        \end{table}

        \subsubsection{Ansätze Comparison: SQD(LUCJ) vs SQD(LCNot-UCCSD)}

         \begin{figure}[H]
            \centering
            \includegraphics[width=\linewidth]{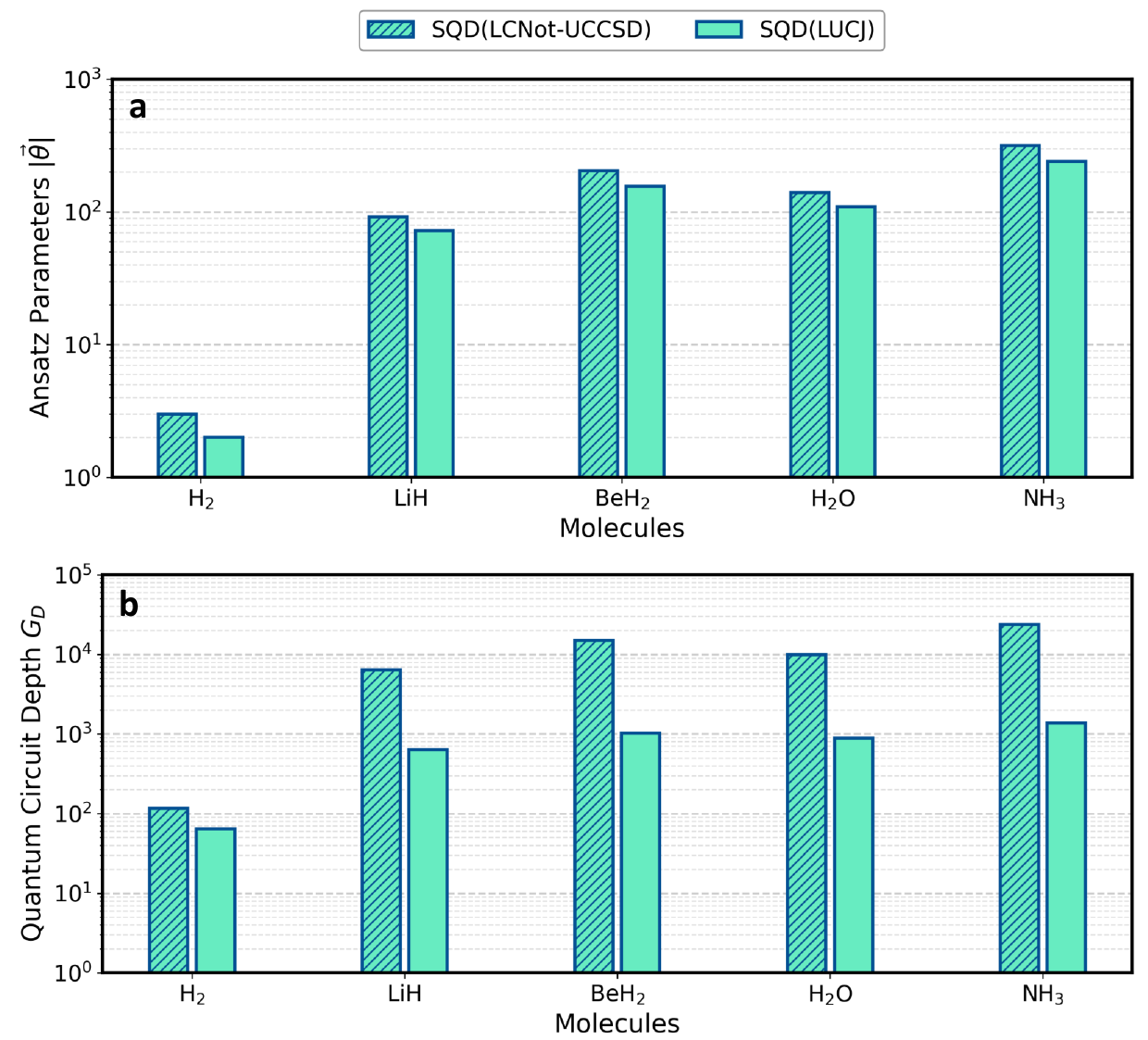}
            \caption{
            Quantum circuit resource comparison metrics for SQD(LUCJ) and SQD(LCNot-UCCSD) ansatz implementations across molecules. 
            (a) Ansatz parameters $|\vec{\theta}|$ (log scale) on the y-axis versus molecules on the x-axis, and 
            (b) quantum circuit depth $G_{D}$ (log scale) on the y-axis versus molecules on the x-axis.
            }
            \label{fig:qc_res_sqd_lucj_vs_sqd_lcnot-uccsd}
        \end{figure}

        \subsubsection{SQD Distribution Comparison with Various Post-HF Methods}

\begin{figure}[H]
            \centering
            \includegraphics[width=0.98\linewidth]{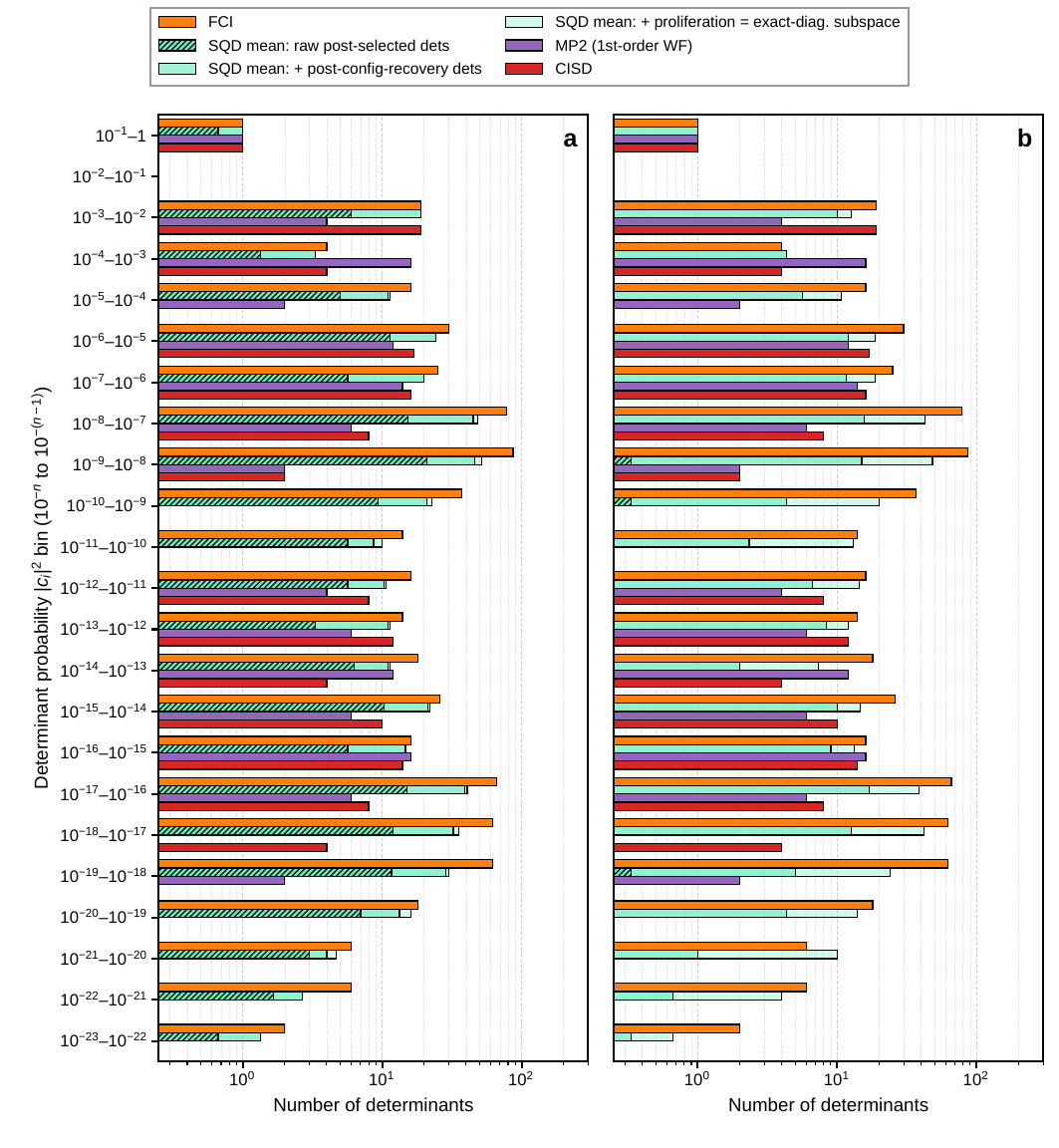}
           \caption{Probability distribution $P_i=|c_i|^2$ of the determinants of BeH$_2$ (STO-3G) for (a)~LUCJ and (b)~LCNot-UCCSD. Determinants are grouped by $P_i$ (vertical axis, $10^{-1}$ to $10^{-23}$); bars give their number (horizontal axis, log scale) for FCI, SQD (layers: raw, $+$recovered, $+$subspace; mean of three runs, $\epsilon_{s} = \sqrt{|\mathbb{S}|} = 35$), MP2 and CISD.}
            \label{fig:sqd_wf_dist}
        \end{figure}

        Figure~\ref{fig:sqd_wf_dist} compares the probability distribution $P_i=|c_i|^2$ of the determinants at the successive stages of SQD with that of the classical wavefunctions for BeH$_2$ (STO-3G). The first-order MP2 wavefunction contains only the HF determinant and the doubles, so it doesn't span the complete probability range, and CISD adds the singles and extends this. FCI, the gold-standard reference, populates the entire range. The SQD distribution is shown at three stages: the raw post-selected sampled determinants, the determinants added by configuration recovery, and the full proliferated subspace used in the diagonalization. Each stage enlarges the set, and the final subspace covers most of the FCI determinants, well beyond MP2 and CISD, though not all of them. The two circuits differ mainly in the raw post-selected sampled counts. For LUCJ, many sampled determinants survive post-selection and populate most of the probability range, whereas for LCNot-UCCSD almost none do, so its final subspace comes almost entirely from configuration recovery and the subsequent proliferation. The raw LUCJ distribution is therefore better in coverage.

\begin{table}[t]
  \centering
  \caption{Total energies of BeH$_2$ (STO-3G) from HF, MP2, CISD, CCSD, SQD,
  and FCI. SQD values are the best of the three runs for each circuit and
  setting. $\Delta E=E-E_{\mathrm{FCI}}$.}
  \label{tab:energies}
  \begin{tabular}{lccc}
    \toprule
    Method & Samples per batch & $E$ (Ha) & $\Delta E$ (mHa) \\
    \midrule
    HF                & ---       & $-15.508656$ & 39.142 \\
    MP2               & ---       & $-15.533964$ & 13.833 \\
    CISD              & ---       & $-15.546894$ & 0.904  \\
    CCSD              & ---       & $-15.547413$ & 0.385  \\
    SQD (LUCJ)        & 35        & $-15.547794$ & 0.0037 \\
    SQD (LUCJ)        & $10^{8}$  & $-15.547798$ & $1.6\times10^{-8}$ \\
    SQD (LCNOT-UCCSD) & 35        & $-15.547782$ & 0.0163 \\
    SQD (LCNOT-UCCSD) & $10^{8}$  & $-15.547798$ & $1.6\times10^{-8}$ \\
    FCI               & ---       & $-15.547798$ & 0      \\
    \bottomrule
  \end{tabular}
\end{table}

Table~\ref{tab:energies} lists the total energies, with FCI as the accuracy
reference. With $10^{8}$ samples per batch all recovered configurations are
used, so the SQD subspace is the full FCI space and SQD reproduces FCI; with 35
samples per batch the subspace is smaller, and the best-run error is still below
$0.02$~mHa. To compare the determinants retained by SQD with those of the
classical wavefunctions, Figure~\ref{fig:sqd_wf_dist} (computed with 35 samples per
batch) shows their probability distribution $P_i=|c_i|^2$.

    \subsection{One Dimensional Potential Energy Surface (1D-PES) Scans}
    \label{si_sec:4.2.1D-PES}

        \subsubsection{Basis Set: 6-31G}

            \begin{figure}[H]
                \centering
                \includegraphics[width=0.95\linewidth]{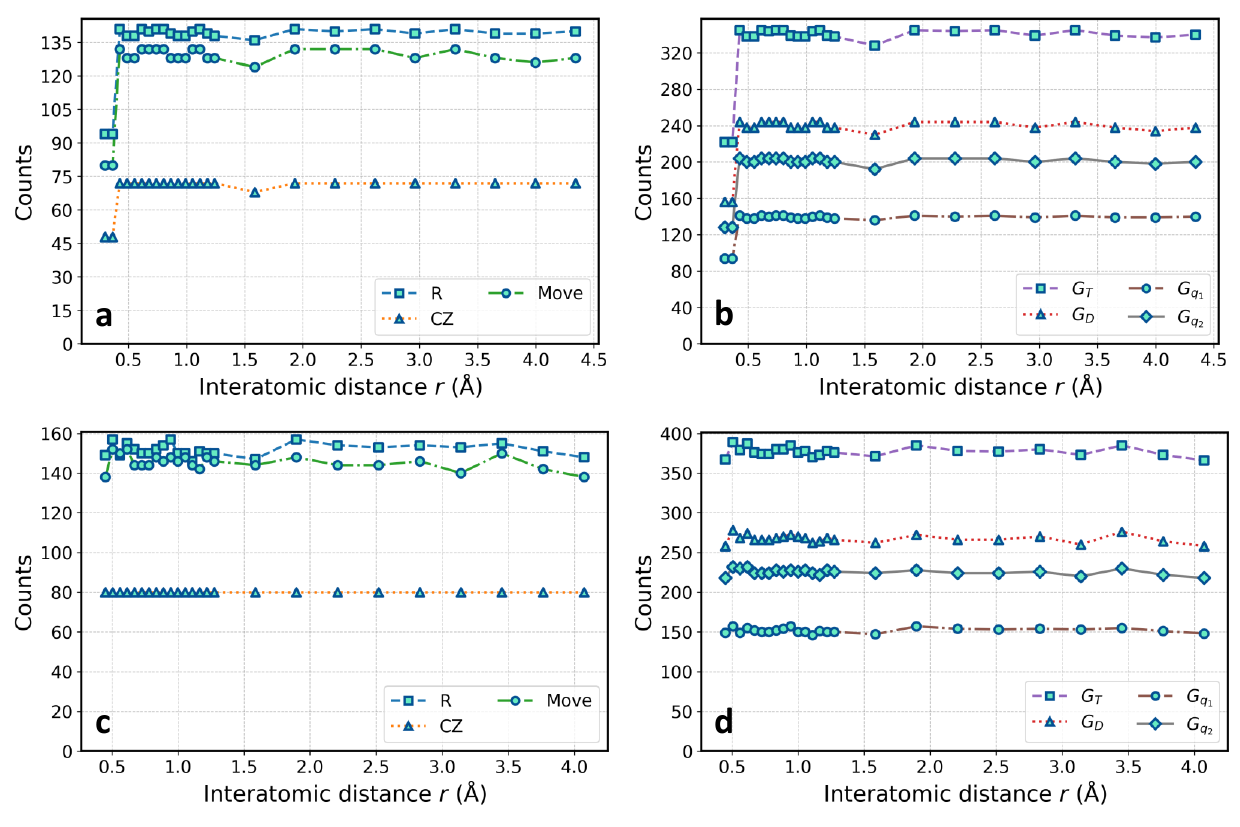}
                    \caption{
                    Quantum circuit resource counts on the y-axis versus interatomic distance $r$ (\AA) on the x-axis for SQD(LUCJ) across various molecules in 6-31G basis. Subplots (a,b) correspond to H$_2$, and (c,d) to HeH$^+$. Here, $\mathrm{R}$ is the number of single-qubit rotation gates, $\mathrm{CZ}$ is the number of two-qubit Controlled-Z gates, $\mathrm{Move}$ is the number of two-qubit Move gates, $G_{T}$ is the total number of quantum gates, $G_{D}$ is the circuit depth, $G_{q_1}$ is the number of single-qubit gates, and $G_{q_2}$ is the number of two-qubit gates.
                    }
                \label{fig:1d_pes_res_est_6-31g}
            \end{figure}
    
        \subsubsection{Basis Set: STO-3G}

            \begin{figure}[H]
                \centering
                \includegraphics[width=0.825\linewidth]{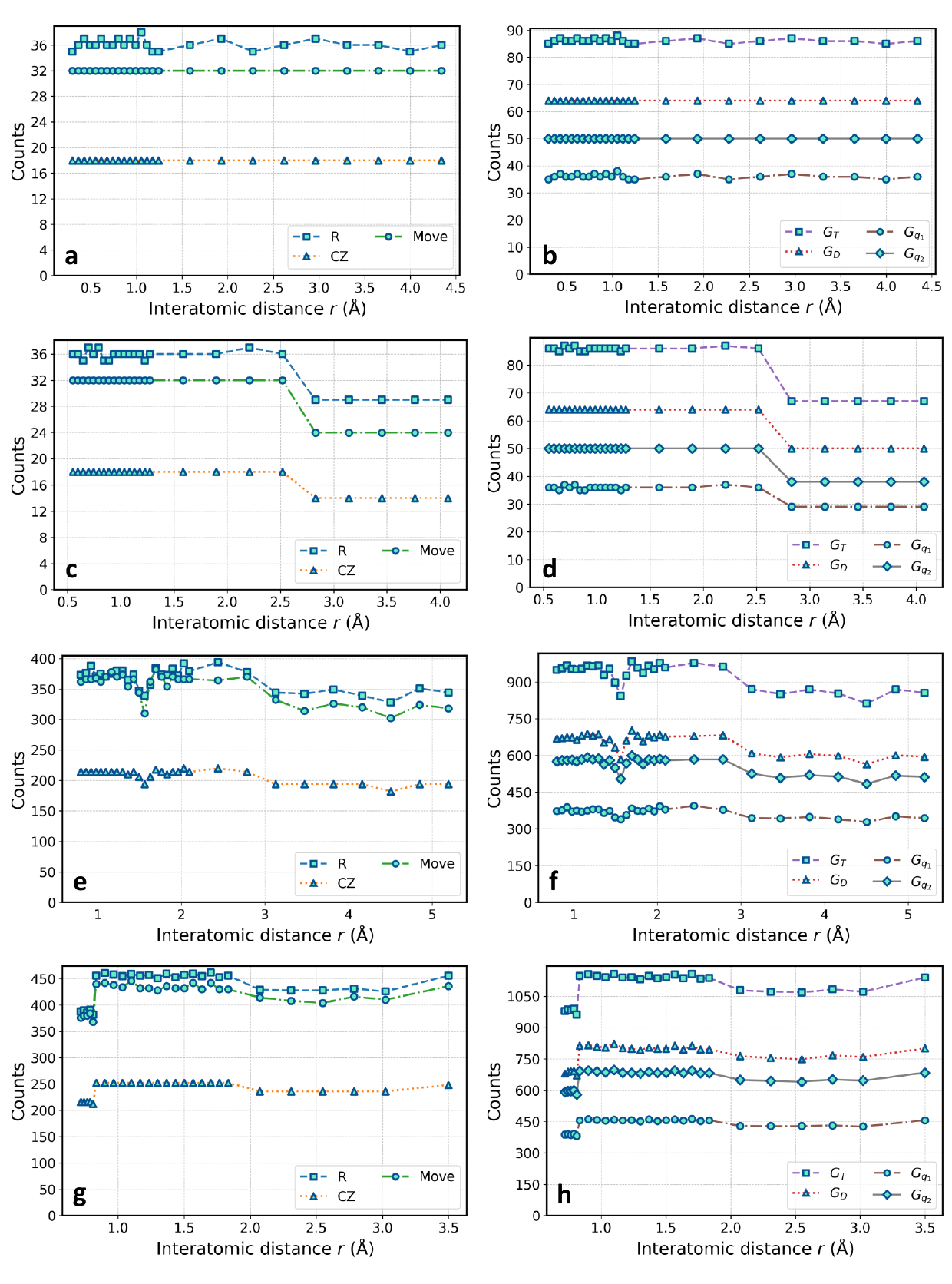}
                    \caption{
                    Quantum circuit resource counts on the y-axis versus interatomic distance $r$ (\AA) on the x-axis for SQD(LUCJ) across various molecules in STO-3G basis. Subplots (a,b) correspond to H$_2$, (c,d) to HeH$^+$, (e,f) to LiH, and (g,h) to BeH$_2$. Here, $\mathrm{R}$ is the number of single-qubit rotation gates, $\mathrm{CZ}$ is the number of two-qubit Controlled-Z gates, $\mathrm{Move}$ is the number of two-qubit Move gates, $G_{T}$ is the total number of quantum gates, $G_{D}$ is the circuit depth, $G_{q_1}$ is the number of single-qubit gates, and $G_{q_2}$ is the number of two-qubit gates.
                    }
                \label{fig:1d_pes_res_est}
            \end{figure}            

    \subsection{Two Dimensional Potential Energy Surface (2D-PES) Scans}
    \label{si_sec:4.3.2D-PES}

        \begin{figure}[H]
            \centering
            \includegraphics[width=0.95\linewidth]{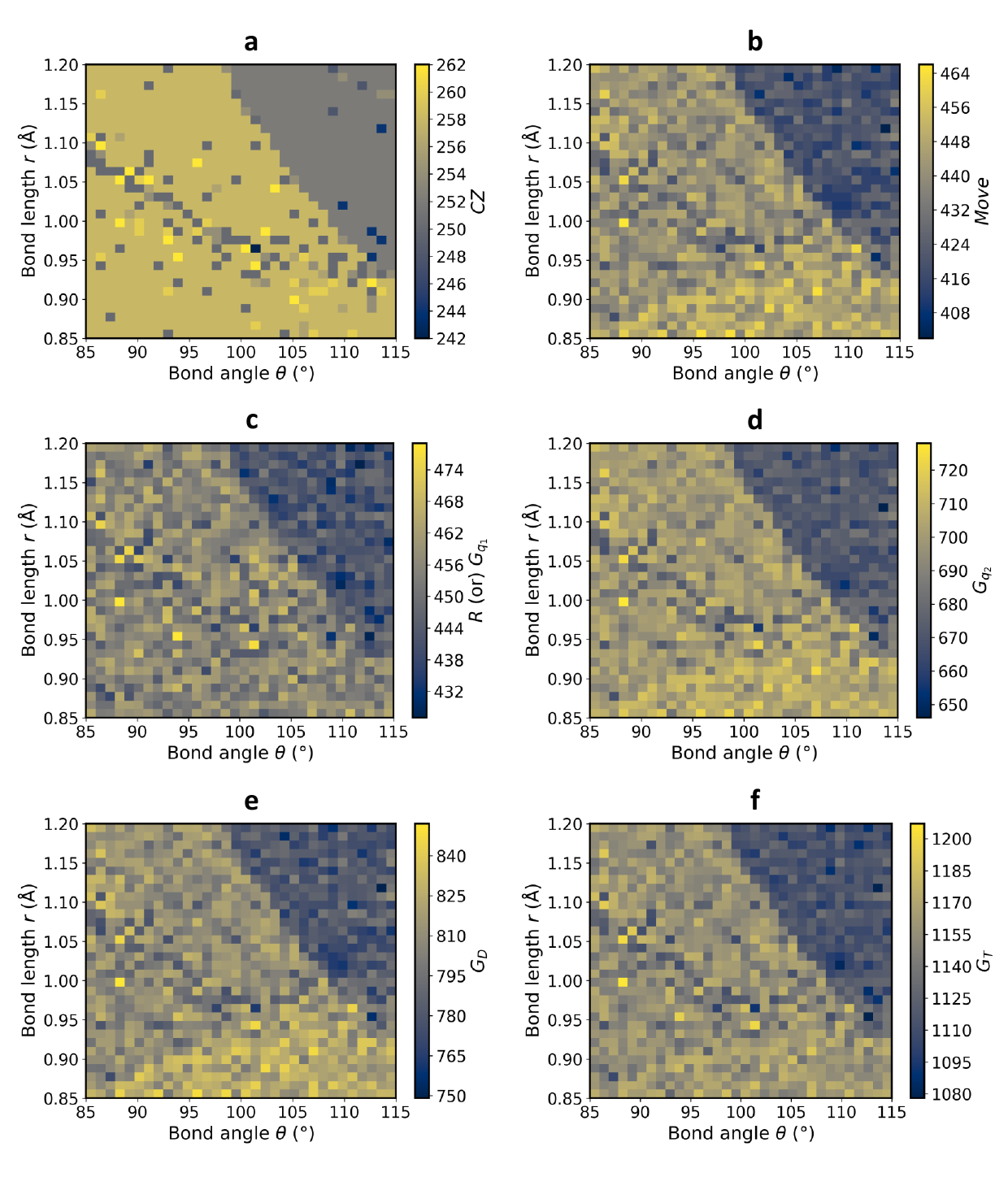}
            \caption{
                Quantum circuit resource summary for a two-dimensional potential energy surface (PES) scan of the H$_2$O molecule in the STO-3G basis. 
                All heatmaps are plotted as a function of the H--O--H bond angle $\theta$ ($^\circ$) along the x-axis and the O--H bond length $r$ (\AA) along the y-axis. 
                Subplots show: (a) the number of two-qubit controlled-$Z$ ($CZ$) gates; 
                (b) the number of two-qubit Move gates; 
                (c) the number of rotation ($R$) gates, equivalent to the total number of one-qubit gates ($G_{q_1}$); 
                (d) the total number of two-qubit gates ($G_{q_2}$); 
                (e) the quantum circuit depth ($G_D$); and 
                (f) the total number of quantum gates in the circuit ($G_T$).
                }
            \label{fig:2d_pes_res_est}
        \end{figure}

        \begin{figure}[H]
            \centering
            \includegraphics[width=0.95\linewidth]{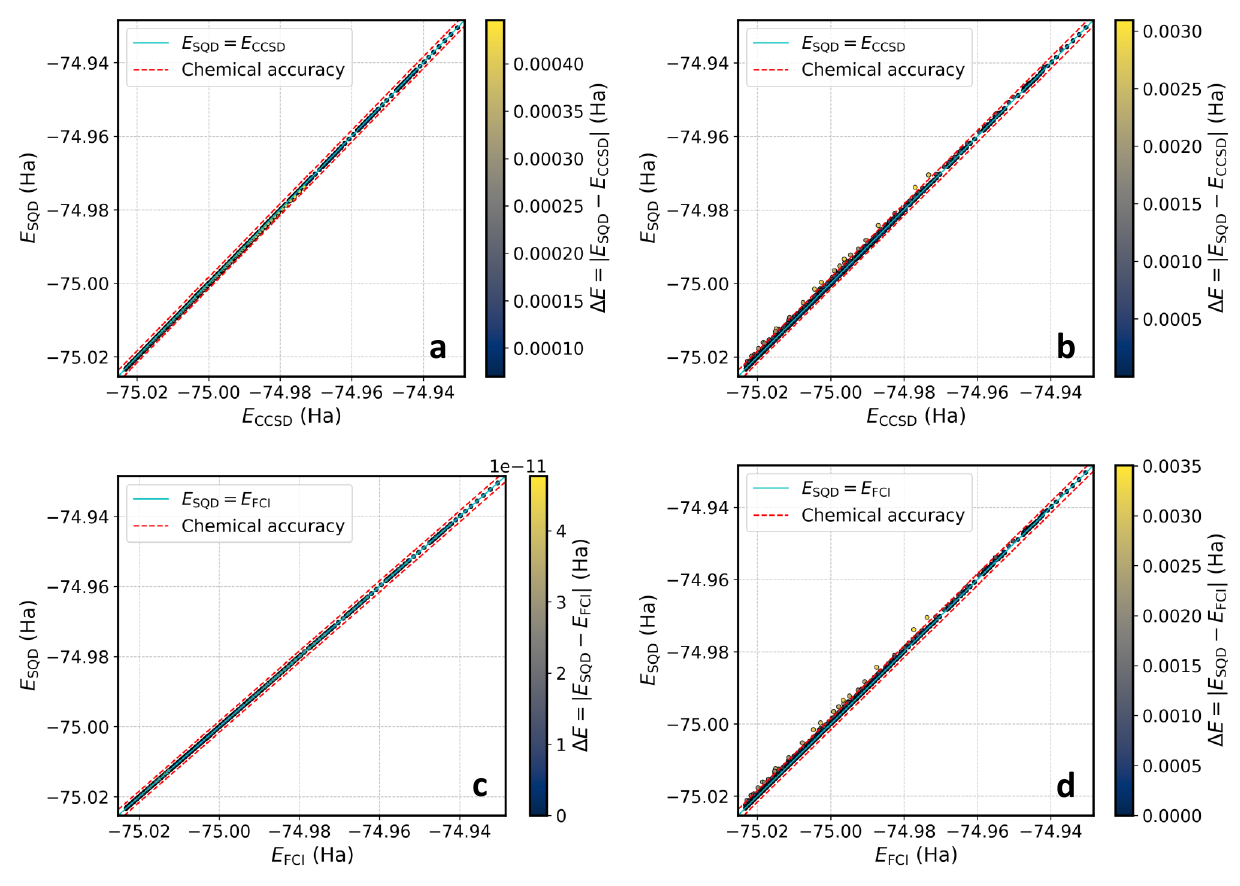}
            \caption{
            Rank plots of absolute energies for the H$_2$O molecule in the STO-3G basis. Classically computed reference energies $E_{\mathrm{CCSD}}$ (a, b) and $E_{\mathrm{FCI}}$ (c, d) in Hartree (Ha) are shown on the x-axis, while the corresponding quantum-computed energies $E_{\mathrm{SQD}}$ in Hartree (Ha) are shown on the y-axis. Subplots correspond to: (a, c) $\epsilon_s = 10^8$ and (b, d) $\epsilon_s = 21$.
            }
            \label{fig:2d_pes_h2o_rank_plots}
        \end{figure}

        \begin{figure}[H]
            \centering
            \includegraphics[width=0.95\linewidth]{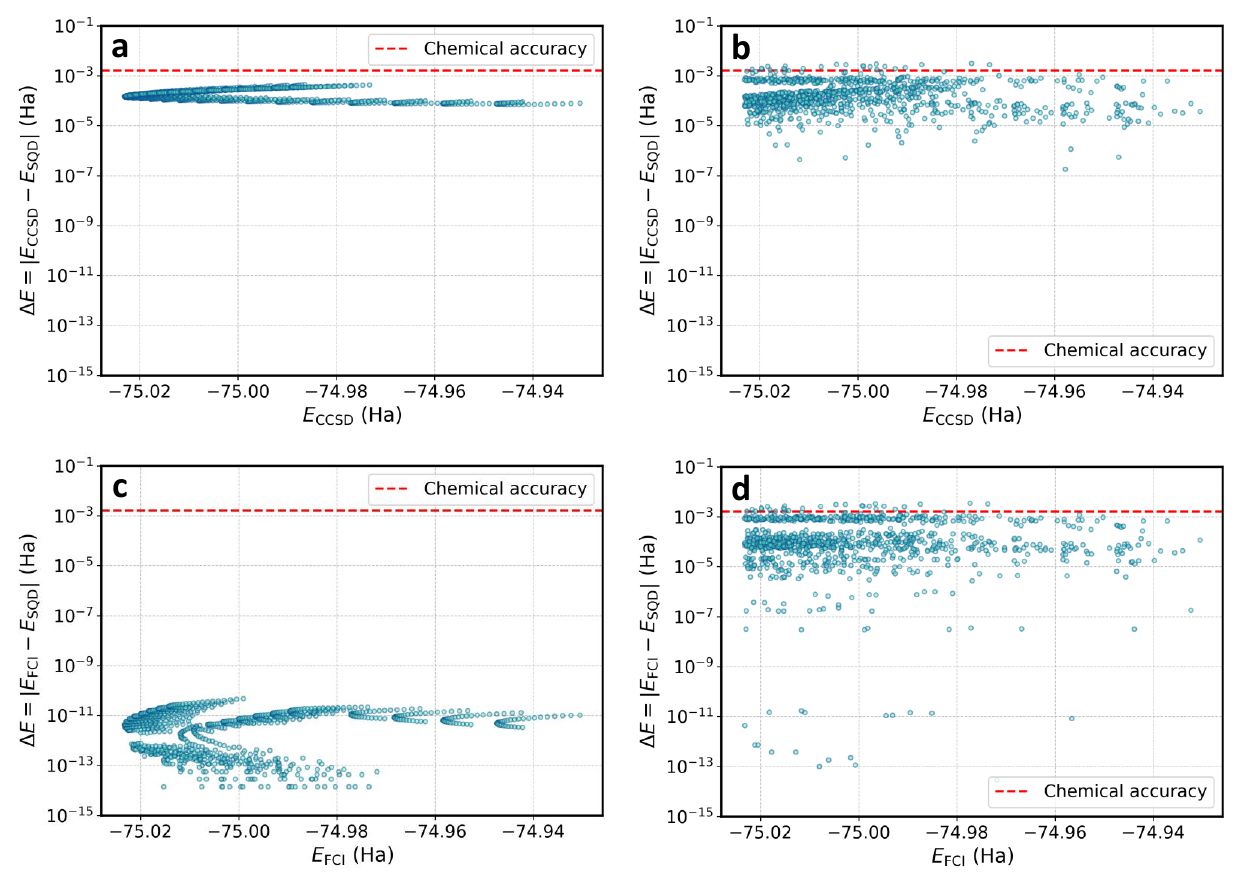}
            \caption{
            Scatter plots of absolute energies to energy differences $\Delta E$ for the H$_2$O molecule in the STO-3G basis. Classically computed reference energies $E_{\mathrm{CCSD}}$ (a, b) and $E_{\mathrm{FCI}}$ (c, d) in Hartree (Ha) are shown on the x-axis, while the corresponding energy differences with quantum-computed energies $E_{\mathrm{SQD}}$ in Hartree (Ha) are shown on the y-axis. Subplots correspond to: (a, c) $\epsilon_s = 10^8$ and (b, d) $\epsilon_s = 21$.
            }
            \label{fig:2d_pes_h2o_scatter_plots}
        \end{figure}

        \begin{figure}[H]
            \centering
            \includegraphics[width=0.95\linewidth]{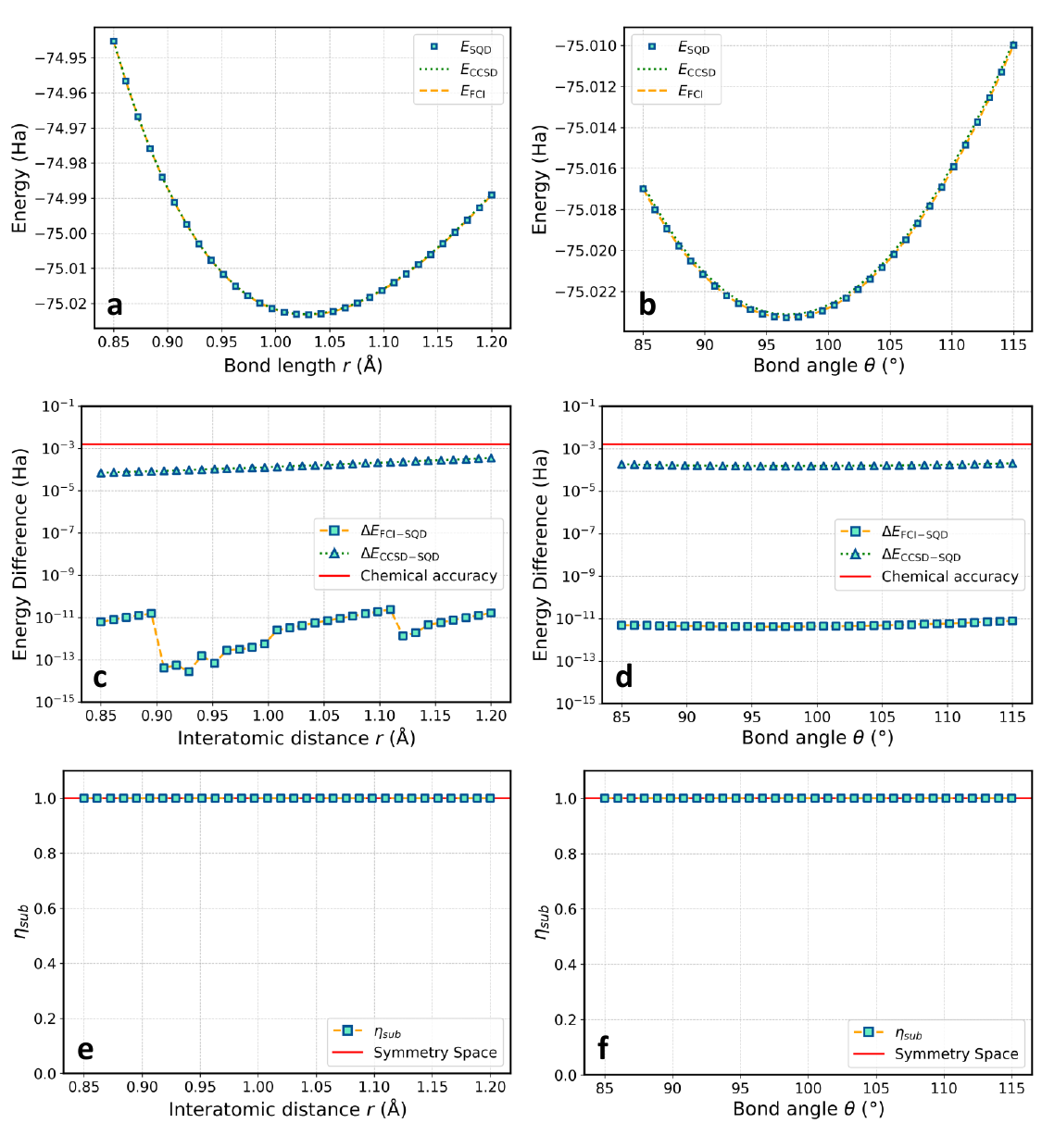}
            \caption{
            One-dimensional potential energy surface (1D-PES) slices extracted for $\epsilon_s = 10^8$ at $E_{\mathrm{SQD}}(r_{\mathrm{min}}, \theta_{\mathrm{min}})$ and $E_{\mathrm{FCI}}(r_{\mathrm{min}}, \theta_{\mathrm{min}})$ where
            $r_{\mathrm{min}} = 1.03$ \AA$, \theta_{\mathrm{min}} = 96.61^\circ$ from the two-dimensional PES (2D-PES) of the H$_2$O molecule in the STO-3G basis. Subplots (a, b) show the absolute energy in Hartree (Ha) as a function of the interatomic distance $r$ (\AA) at $\theta_{\mathrm{min}}$ and the H--O--H bond angle $\theta$ ($^\circ$) at $r_{\mathrm{min}}$, respectively. Subplots (c, d) show the corresponding energy differences in Hartree (Ha) as functions of $r$ at $\theta_{\mathrm{min}}$ and $\theta$ at $r_{\mathrm{min}}$, respectively. Subplots (e, f) display the ratio of the diagonalization subspace dimension to the symmetry space dimension, $\eta_{\mathrm{sub}}$, as functions of $r$ at $\theta_{\mathrm{min}}$ and $\theta$ at $r_{\mathrm{min}}$, respectively.
            }
            \label{fig:1d_slice_sqd_fci_min_final1}
        \end{figure}        

        \begin{figure}[H]
            \centering
            \includegraphics[width=0.95\linewidth]{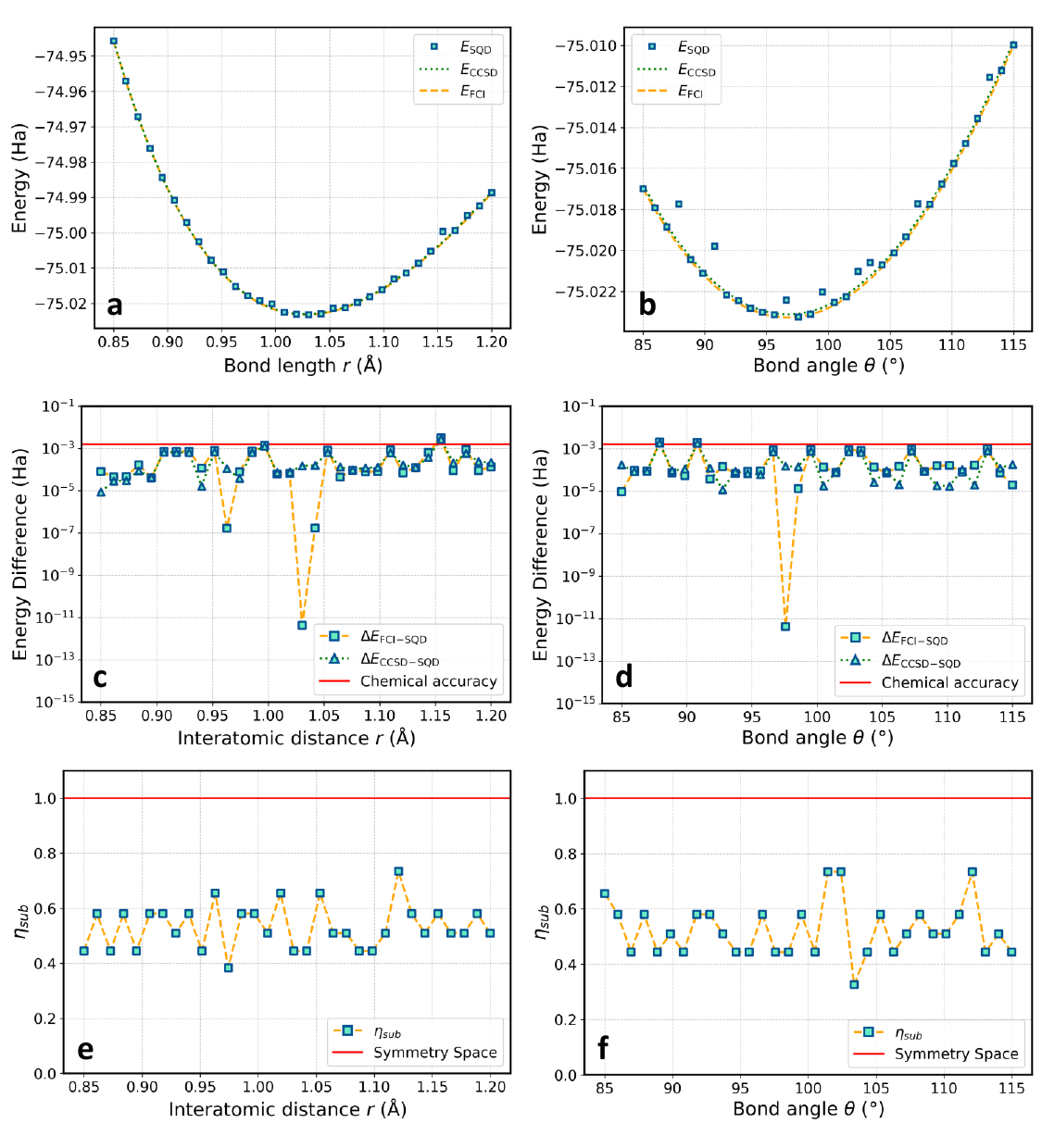}
            \caption{
             One-dimensional potential energy surface (1D-PES) slices extracted for $\epsilon_s = 21$ at $E_{\mathrm{SQD}}(r_{\mathrm{min}}, \theta_{\mathrm{min}})$ where
            $r_{\mathrm{min}} = 1.03$ \AA$, \theta_{\mathrm{min}} = 97.58^\circ$ from the two-dimensional PES (2D-PES) of the H$_2$O molecule in the STO-3G basis. Subplots (a, b) show the absolute energy in Hartree (Ha) as a function of the interatomic distance $r$ (\AA) at $\theta_{\mathrm{min}}$ and the H--O--H bond angle $\theta$ ($^\circ$) at $r_{\mathrm{min}}$, respectively. Subplots (c, d) show the corresponding energy differences in Hartree (Ha) as functions of $r$ at $\theta_{\mathrm{min}}$ and $\theta$ at $r_{\mathrm{min}}$, respectively. Subplots (e, f) display the ratio of the diagonalization subspace dimension to the symmetry space dimension, $\eta_{\mathrm{sub}}$, as functions of $r$ at $\theta_{\mathrm{min}}$ and $\theta$ at $r_{\mathrm{min}}$, respectively.
            }
            \label{fig:1d_slice_sqd_min_final2}
        \end{figure}

        \begin{figure}[H]
            \centering
            \includegraphics[width=0.95\linewidth]{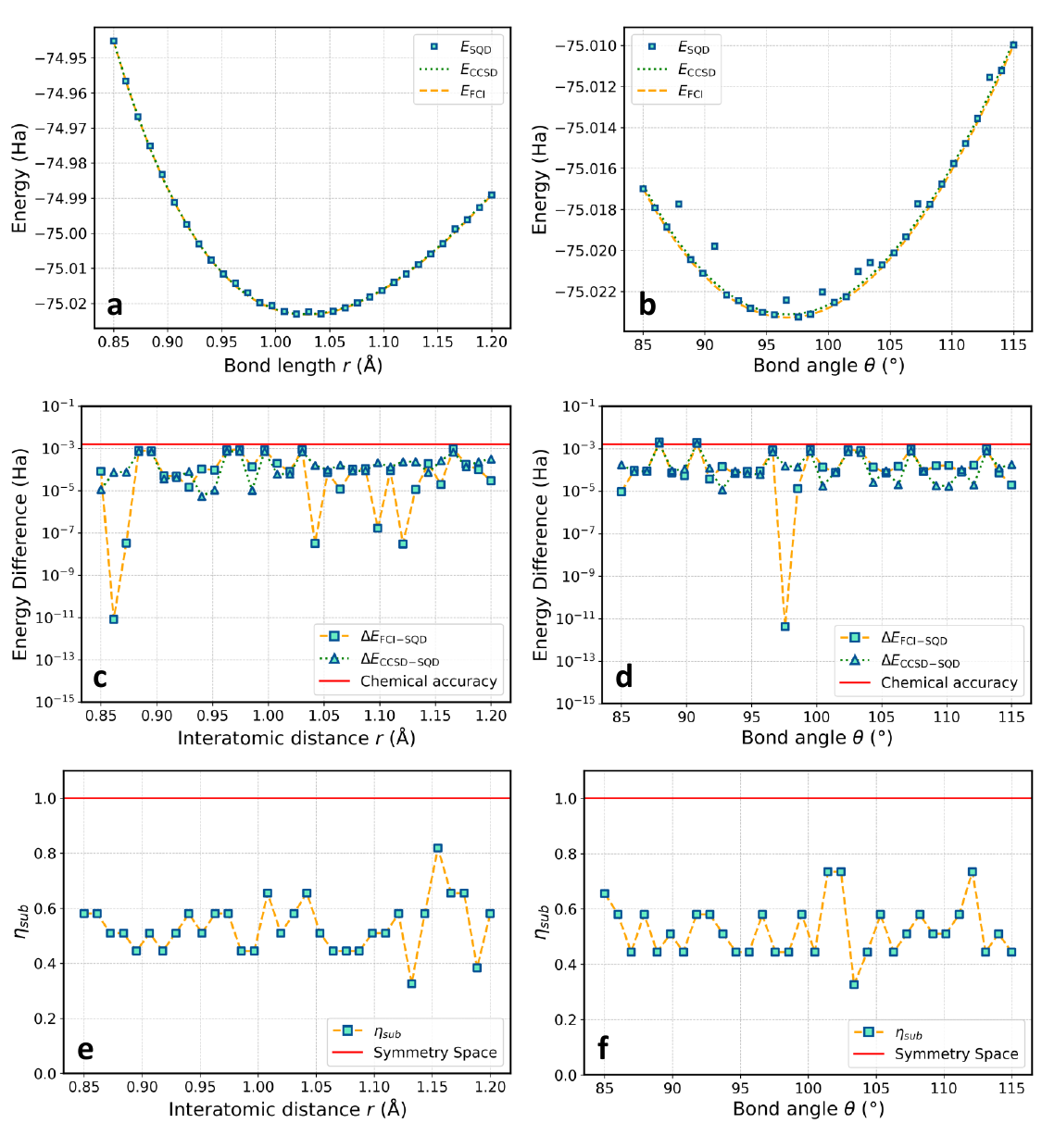}
            \caption{
             One-dimensional potential energy surface (1D-PES) slices extracted for $\epsilon_s = 21$ at $E_{\mathrm{FCI}}(r_{\mathrm{min}}, \theta_{\mathrm{min}})$ where
            $r_{\mathrm{min}} = 1.03$ \AA$, \theta_{\mathrm{min}} = 96.61^\circ$ from the two-dimensional PES (2D-PES) of the H$_2$O molecule in the STO-3G basis. Subplots (a, b) show the absolute energy in Hartree (Ha) as a function of the interatomic distance $r$ (\AA) at $\theta_{\mathrm{min}}$ and the H--O--H bond angle $\theta$ ($^\circ$) at $r_{\mathrm{min}}$, respectively. Subplots (c, d) show the corresponding energy differences in Hartree (Ha) as functions of $r$ at $\theta_{\mathrm{min}}$ and $\theta$ at $r_{\mathrm{min}}$, respectively. Subplots (e, f) display the ratio of the diagonalization subspace dimension to the symmetry space dimension, $\eta_{\mathrm{sub}}$, as functions of $r$ at $\theta_{\mathrm{min}}$ and $\theta$ at $r_{\mathrm{min}}$, respectively.
            }
            \label{fig:1d_slice_fci_min_final2}
        \end{figure}
    
    \subsection{SQD(LUCJ) within the Density Matrix Embedding Framework}

        \subsubsection{Ligand-like Molecules}
    
        \begingroup
            \begin{table}[H]
            \centering
            \small
            \caption{Quantum circuit resource summary for the ligand-like molecules' corresponding molecular fragments, utilizing $10{,}000$ shots per fragment. The depth and the gates shown here correspond to the final iteration of DMET chemical potential convergence.}
            \label{tab1:resource_dmet_sqd_ligands}
                \begin{tabular}{|c|c|c|c|c|c|c|c|c|c|}
                \hline
                Molecule & Frag & $(N_{\mathrm{orb}}, N_{\mathrm{ele}})$ & $N_{Q}$ & $G_{D}$ & $R$ & $CZ$ & $Move$ & $|\mathbb{S}|$ & $|\mathbb{H}|$ \\
                \hline
                \multirow{4}{*}{$\mathrm{HOCN}$}
                 & [H] & (2, 2) & 4 & 46 & 24 & 12 & 28 & 4 & 16 \\
                 & [O] & (8, 8) & 16 & 1{,}154 & 626 & 366 & 660 & 4{,}900 & 65{,}536 \\
                 & [C] & (8, 8) & 16 & 1{,}148 & 623 & 366 & 656 & 4{,}900 & 65{,}536 \\
                 & [N] & (8, 8) & 16 & 1{,}151 & 614 & 366 & 652 & 4{,}900 & 65{,}536 \\
                \hline
                \multirow{6}{*}{$\mathrm{CH_3NO}$}
                 & [C] & (8, 8) & 16 & 1{,}162 & 605 & 366 & 660 & 4{,}900 & 65{,}536 \\
                 & [H] & (2, 2) & 4 & 46 & 24 & 12 & 28 & 4 & 16 \\
                 & [H] & (2, 2) & 4 & 46 & 24 & 12 & 28 & 4 & 16 \\
                 & [H] & (2, 2) & 4 & 46 & 24 & 12 & 28 & 4 & 16 \\
                 & [N] & (8, 8) & 16 & 1{,}155 & 632 & 366 & 656 & 4{,}900 & 65{,}536 \\
                 & [O] & (8, 8) & 16 & 1{,}190 & 637 & 366 & 680 & 4{,}900 & 65{,}536 \\
                \hline
                \multirow{8}{*}{$\mathrm{CH_5NO}$}
                 & [C] & (8, 8) & 16 & 1{,}132 & 617 & 366 & 640 & 4{,}900 & 65{,}536 \\
                 & [H] & (2, 2) & 4 & 46 & 24 & 12 & 28 & 4 & 16 \\
                 & [H] & (2, 2) & 4 & 46 & 24 & 12 & 28 & 4 & 16 \\
                 & [H] & (2, 2) & 4 & 46 & 24 & 12 & 28 & 4 & 16 \\
                 & [H] & (2, 2) & 4 & 46 & 24 & 12 & 28 & 4 & 16 \\
                 & [H] & (2, 2) & 4 & 46 & 24 & 12 & 28 & 4 & 16 \\
                 & [O] & (8, 8) & 16 & 1{,}170 & 614 & 366 & 664 & 4{,}900 & 65{,}536 \\
                 & [N] & (8, 8) & 16 & 1{,}147 & 608 & 366 & 656 & 4{,}900 & 65{,}536 \\
                \hline
                \multirow{7}{*}{$\mathrm{C_2H_3NO}$}
                 & [C] & (8, 8) & 16 & 1{,}176 & 619 & 366 & 668 & 4{,}900 & 65{,}536 \\
                 & [C] & (8, 8) & 16 & 1{,}137 & 612 & 366 & 652 & 4{,}900 & 65{,}536 \\
                 & [H] & (2, 2) & 4 & 46 & 24 & 12 & 28 & 4 & 16 \\
                 & [H] & (2, 2) & 4 & 46 & 24 & 12 & 28 & 4 & 16 \\
                 & [H] & (2, 2) & 4 & 46 & 24 & 12 & 28 & 4 & 16 \\
                 & [N] & (8, 8) & 16 & 1{,}165 & 614 & 366 & 660 & 4{,}900 & 65{,}536 \\
                 & [O] & (8, 8) & 16 & 1{,}132 & 606 & 366 & 644 & 4{,}900 & 65{,}536 \\
                \hline
                \multirow{9}{*}{$\mathrm{C_2H_5NO}$}
                 & [C] & (8, 8) & 16 & 1{,}166 & 635 & 366 & 664 & 4{,}900 & 65{,}536 \\
                 & [C] & (8, 8) & 16 & 1{,}160 & 599 & 366 & 660 & 4{,}900 & 65{,}536 \\
                 & [H] & (2, 2) & 4 & 46 & 24 & 12 & 28 & 4 & 16 \\
                 & [H] & (2, 2) & 4 & 46 & 24 & 12 & 28 & 4 & 16 \\
                 & [H] & (2, 2) & 4 & 46 & 24 & 12 & 28 & 4 & 16 \\
                 & [H] & (2, 2) & 4 & 46 & 24 & 12 & 28 & 4 & 16 \\
                 & [H] & (2, 2) & 4 & 46 & 24 & 12 & 28 & 4 & 16 \\
                 & [N] & (8, 8) & 16 & 1{,}142 & 613 & 366 & 652 & 4{,}900 & 65{,}536 \\
                 & [O] & (8, 8) & 16 & 1{,}171 & 628 & 366 & 668 & 4{,}900 & 65{,}536 \\
                \hline
                \multirow{8}{*}{$\mathrm{CH_4N_2O}$}
                 & [C] & (8, 8) & 16 & 1{,}149 & 603 & 366 & 652 & 4{,}900 & 65{,}536 \\
                 & [H] & (2, 2) & 4 & 46 & 24 & 12 & 28 & 4 & 16 \\
                 & [H] & (2, 2) & 4 & 46 & 24 & 12 & 28 & 4 & 16 \\
                 & [H] & (2, 2) & 4 & 46 & 24 & 12 & 28 & 4 & 16 \\
                 & [H] & (2, 2) & 4 & 46 & 24 & 12 & 28 & 4 & 16 \\
                 & [N] & (8, 8) & 16 & 1{,}162 & 626 & 366 & 668 & 4{,}900 & 65{,}536 \\
                 & [N] & (8, 8) & 16 & 1{,}113 & 618 & 366 & 628 & 4{,}900 & 65{,}536 \\
                 & [O] & (8, 8) & 16 & 1{,}171 & 608 & 366 & 672 & 4{,}900 & 65{,}536 \\
                \hline
                \multirow{3}{*}{$\mathrm{NOCl}$}
                 & [N] & (7, 8) & 14 & 818 & 434 & 256 & 464 & 1{,}225 & 16{,}384 \\
                 & [O] & (7, 8) & 14 & 806 & 422 & 256 & 454 & 1{,}225 & 16{,}384 \\
                 & [Cl] & (7, 8) & 14 & 811 & 447 & 256 & 458 & 1{,}225 & 16{,}384 \\
                \hline
                \multirow{5}{*}{$\mathrm{HOSCN}$}
                 & [H] & (2, 2) & 4 & 46 & 24 & 12 & 28 & 4 & 16 \\
                 & [O] & (8, 8) & 16 & 1{,}119 & 611 & 366 & 624 & 4{,}900 & 65{,}536 \\
                 & [S] & (8, 8) & 16 & 1{,}163 & 626 & 366 & 672 & 4{,}900 & 65{,}536 \\
                 & [C] & (8, 8) & 16 & 1{,}146 & 602 & 366 & 648 & 4{,}900 & 65{,}536 \\
                 & [N] & (8, 8) & 16 & 1{,}115 & 618 & 366 & 640 & 4{,}900 & 65{,}536 \\
                \hline
                \end{tabular}
            \end{table}
        \endgroup

        \begin{figure}[htbp]
            \centering
            \includegraphics[width=1.0\linewidth]{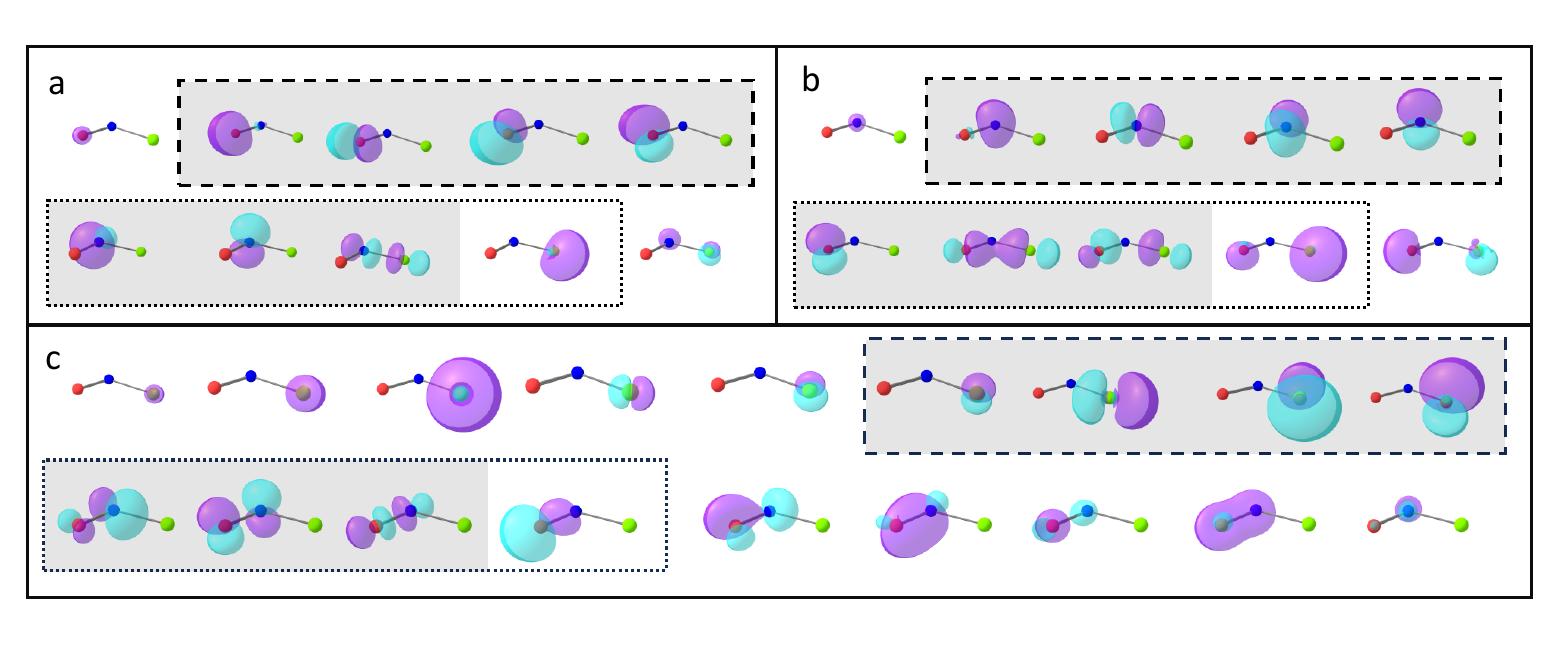}
            \caption{DMET impurity orbital construction for NOCl with atom-centered fragments [O], [N], and [Cl] shown in panels (a)–(c). The isospheres of the fragment orbitals (top row) and corresponding bath orbitals (bottom row) are displayed. An active space of 4 HOMO – 4 LUMO orbitals is selected to restrict each impurity to 16 qubits. Unlike other systems, only a subset of bath orbitals (3 orbitals) contributes significantly to the fragment–environment correlation in NOCl.}
            \label{fig:NOCl_bath}
        \end{figure}

        \begin{table}[htbp]
        \centering
        \caption{Energy results obtained from the DMET--SQD simulations for the studied ligand-like molecules. The DMET--CCSD column reports reference coupled-cluster energies evaluated on the same fragments (values rounded to 8 decimal places).}
        \label{tab:nat_mols_energy}
            \begin{tabular}{|p{4.2cm}|p{3.8cm}|p{3.8cm}|p{3.8cm}|}
            \hline
            \textbf{Molecule} 
            & \textbf{$E_{\mathrm{DMET\text{-}CCSD}}$ (Ha)}
            & \textbf{$E_{\mathrm{DMET\text{-}CASCI}}$ (Ha)}
            & \textbf{$E_{\mathrm{DMET\text{-}SQD}}$ (Ha)} \\
            \hline
            
            \makecell[l]{Cyanic Acid \\ ($\mathrm{HOCN}$)} 
            & -165.76903095
            & -165.77369080 
            & -165.77369135 \\
            \hline
            
            \makecell[l]{Formaldehyde Oxime \\ ($\mathrm{CH_3NO}$)} 
            & -166.83748626
            & -166.84466291 
            & -166.84466203 \\
            \hline
            
            \makecell[l]{Methoxyamine \\ ($\mathrm{CH_5NO}$)} 
            & -167.98584469
            & -167.98839994 
            & -167.98840075 \\
            \hline
            
            \makecell[l]{Methyl Isocyanate \\ ($\mathrm{C_2H_3NO}$)} 
            & -204.46392213
            & -204.48135661 
            & -204.48135783 \\
            \hline
            
            \makecell[l]{Acetaldehyde Oxime \\ ($\mathrm{C_2H_5NO}$)} 
            & -205.49795846
            & -205.50685229 
            & -205.50684937 \\
            \hline
            
            \makecell[l]{Carbamide / Urea \\ ($\mathrm{CH_4N_2O}$)} 
            & -221.26440756
            & -221.28940628 
            & -221.28940740 \\
            \hline
            
            \makecell[l]{Nitrosyl Chloride \\ ($\mathrm{NOCl}$)} 
            & -582.34549362
            & -582.38437288 
            & -582.38437139 \\
            \hline
            
            \makecell[l]{Hydroxythiocyanate \\ ($\mathrm{HOSCN}$)} 
            & -559.01368538
            & -559.01613855 
            & -559.01613914 \\
            \hline
            
            \end{tabular}
        \end{table}

        \subsubsection{Amantadine Molecule}
        
            \begingroup
            \renewcommand{\arraystretch}{1.3}
   
                \begin{table}[H]
                \centering
                \caption{Quantum resources summary for the Amantadine molecule. Here, the resources required for the final iteration of the chemical potential convergence are shown, for the experiment performed with 10{,}000 shots.}
                \label{tab1:resource_dmet_sqd}
                    \begin{tabular}{|c|c|c|c|c|c|c|c|c|c|}
                    \hline
                    Molecule & Frag & $(N_{\mathrm{orb}}, N_{\mathrm{ele}})$ & $N_Q$ & $G_D$ & $R$ & $CZ$ & $Move$ & $|\mathbb{S}|$ & $|\mathbb{H}|$ \\
                    \hline
                    
                    \multirow{11}{*}{\centering \makecell{Amantadine \\ ($\mathrm{C_{10}H_{17}N}$)}}
                     & [NH$_2$] & (8, 8) & 16 & 1{,}145 & 602 & 366 & 664 & 4{,}900 & 65{,}536 \\
                     & [C]     & (8, 8) & 16 & 1{,}127 & 590 & 366 & 636 & 4{,}900 & 65{,}536 \\
                     & [CH$_2$]  & (8, 8) & 16 & 1{,}139 & 604 & 366 & 648 & 4{,}900 & 65{,}536 \\
                     & [CH$_2$]  & (8, 8) & 16 & 1{,}140 & 623 & 366 & 648 & 4{,}900 & 65{,}536 \\
                     & [CH]    & (8, 8) & 16 & 1{,}132 & 612 & 366 & 636 & 4{,}900 & 65{,}536 \\ 
                     & [CH]    & (8, 8) & 16 & 1{,}161 & 617 & 366 & 672 & 4{,}900 & 65{,}536 \\ 
                     & [CH$_2$]  & (8, 8) & 16 & 1{,}174 & 610 & 366 & 674 & 4{,}900 & 65{,}536 \\ 
                     & [CH]    & (8, 8) & 16 & 1{,}144 & 622 & 366 & 656 & 4{,}900 & 65{,}536 \\
                     & [CH$_2$]  & (8, 8) & 16 & 1{,}182 & 611 & 366 & 676 & 4{,}900 & 65{,}536 \\
                     & [CH$_2$]  & (8, 8) & 16 & 1{,}155 & 608 & 366 & 660 & 4{,}900 & 65{,}536 \\
                     & [CH$_2$]  & (8, 8) & 16 & 1{,}192 & 617 & 366 & 688 & 4{,}900 & 65{,}536 \\
                    \hline
                    \end{tabular}
                \end{table}
            \endgroup

            \begin{table}[htbp]
            \centering
            \caption{DMET energies using SQD with different shot counts, compared with CCSD and CASCI.}
            \label{tab:dmet_energies}
                \begin{tabular}{|c|c|c|c|c|c|c|}
                \hline
                \multirow{2}{*}{\textbf{Method}} 
                & \multicolumn{4}{c|}{\textbf{DMET-SQD}} 
                & \multirow{2}{*}{\textbf{DMET-CCSD}} 
                & \multirow{2}{*}{\textbf{DMET-CASCI}} \\
                \cline{2-5}
                & $N_{\text{shots}}=500$ 
                & $N_{\text{shots}}=1\text{k}$ 
                & $N_{\text{shots}}=5\text{k}$ 
                & $N_{\text{shots}}=10\text{k}$ 
                &  &  \\
                \hline
                \textbf{Energy (Ha)} 
                & -438.09785800
                & -438.09785047
                & -438.09785054
                & -438.09785054
                & -438.09849129
                & -438.09784993 \\
                \hline
                \end{tabular}
            \end{table}

            \begin{figure}[H]
                \centering
                \includegraphics[width=0.9\linewidth]{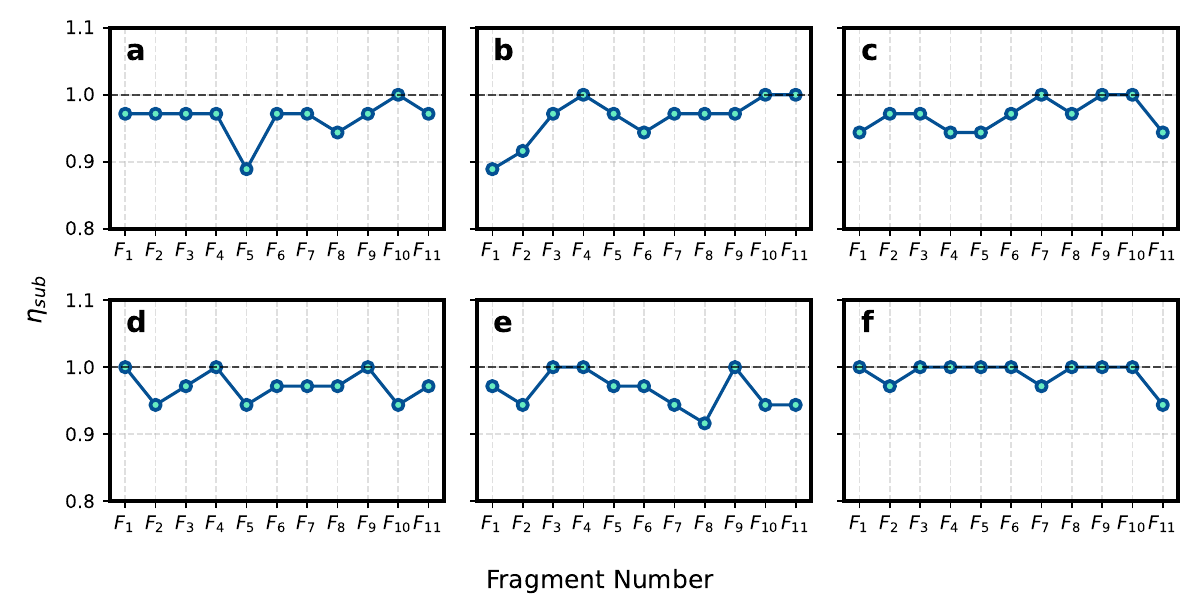}
                \caption{Evolution of the diagonalization subspace fraction $\eta_{\mathrm{sub}}$ across the eleven fragments ($F_1$–$F_{11}$) of the amantadine molecule during the DMET-SQD self-consistency cycle for the amantadine molecule with $N_{shots} = 500$. Subplots (a)-(f) correspond to successive values of the global chemical potential $\mu_{\mathrm{glob}}$ obtained during the DMET iterations, with subplot (f) representing the converged iteration. The results show that the SQD procedure accesses a large fraction of the symmetry-constrained space for most fragments throughout the cycle, with eight fragments reaching $\eta_{\mathrm{sub}} = 1$ in the final iteration.}
                \label{fig: dmet_amant_500}
            \end{figure}
            
\newpage

\setcounter{table}{0}
\renewcommand{\thetable}{E\arabic{table}}

\setcounter{figure}{0}
\renewcommand{\thefigure}{E\arabic{figure}}

\setcounter{equation}{0}
\renewcommand{\theequation}{E\arabic{equation}}

\section{IQM Sirius QPU Details}
\label{appendix:iqm_sirius_qpu}

    All quantum computations reported in this work were performed on the IQM Sirius quantum processing unit (QPU), accessed via the IQM Resonance cloud platform. IQM Sirius is a superconducting transmon qubit-based QPU supporting up to 24~qubits arranged in a star lattice topology (STAR\,24). All qubits are coupled to a single shared central computational resonator (COMPR1), forming a hub-and-spoke connectivity in which any qubit can interact with any other qubit through resonator-mediated operations, providing effective all-to-all connectivity without the need for SWAP gates. The device was calibrated on 4~December~2025 at 10:34~IST (GMT+5:30); all calculations in this work were executed within the window 4~Dec~2025 14:49~IST -- 6~Dec~2025 05:48~IST, which falls within the validity period of this single calibration. The active qubit set comprised 16~qubits: QB1, QB2, QB3, QB4, QB5, QB8, QB9, QB10, QB11, QB13, QB15,
    QB16, QB17, QB19, QB20, and QB21.

    \subsection*{Hardware Topology}
    
\begin{figure}[H]
            \centering
            \includegraphics[width=0.90\textwidth]{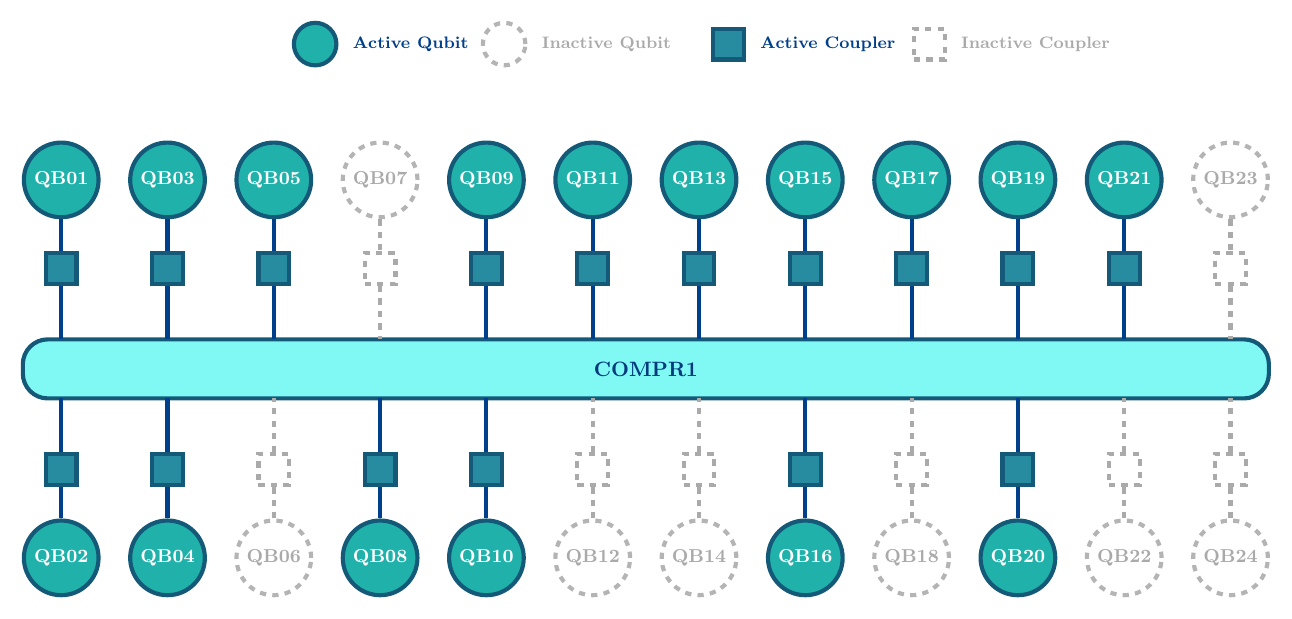}
            \caption{%
                Hardware topology of the IQM Sirius QPU (STAR\,24). All 16~active qubits (filled circles) are connected to the central computational resonator COMPR1 (horizontal bar) through individual couplers (filled squares), providing effective all-to-all qubit connectivity. Dashed outlines denote inactive couplers and qubits.
            }
            \label{fig:iqm_topology}
        \end{figure}

        The physical topology of the IQM Sirius QPU is shown in Figure~\ref{fig:iqm_topology}. Each active qubit (filled circle) is connected through a coupler element (filled square) to the central computational resonator COMPR1 (horizontal bar). Dashed outlines indicate the inactive couplers and qubits (QB6, QB7, QB12, QB14, QB18, QB22--QB24) in this configuration.
        
    \subsection*{Native Gate Set}
    
        The native gate set of IQM Sirius consists of the following operations,
        as defined in the IQM Client API~\cite{IQMInstructionAPI2025}:
        
        \begin{itemize}
            \item \textbf{PRX}($\theta$, $\phi$) --- \emph{phased rotation} (universal
                  single-qubit gate):
                  \begin{equation}
                      \mathrm{PRX}(\theta,\phi)
                      = e^{-i\,\pi\theta\,
                        (X\cos 2\pi\phi + Y\sin 2\pi\phi)}
                      = R_Z(2\pi\phi)\,R_X(2\pi\theta)\,R_Z^\dagger(2\pi\phi),
                      \label{eq:prx}
                  \end{equation}
                  where $\theta$ is the rotation angle and $\phi$ is the phase
                  (azimuthal) angle, both in units of full turns;
                  $X$ and $Y$ are the Pauli matrices, as the $\textbf{\textrm{R}}$ gate in the main text. The default pulse implementation uses a DRAG Cosine Rise--Fall envelope.
            \item \textbf{CZ} --- native two-qubit controlled-$Z$ gate, represented as $\textbf{\textrm{CZ}}$ gate in the main text:
            \[
                \mathrm{CZ} =
                \begin{bmatrix}
                1 & 0 & 0 & 0 \\
                0 & 1 & 0 & 0 \\
                0 & 0 & 1 & 0 \\
                0 & 0 & 0 & -1
                \end{bmatrix}
            \]
            \item \textbf{MOVE} --- Resonator-mediated excitation-exchange gate that
              swaps the populations of the qubit--resonator states
              $|01\rangle \leftrightarrow |10\rangle$ within the invariant
              subspace $\mathcal{S} = \mathrm{span}\{|00\rangle, |01\rangle, |10\rangle\}$.
              Its action can be written as:
              \[
              \mathrm{MOVE}_{\mathcal{S}} = \frac{1}{\mathcal{N}}(
              |00\rangle\langle 00|
              + a\,|10\rangle\langle 01|
              + a^{-1}\,|01\rangle\langle 10|),
              \]
              where $\mathcal{N}$ is the normalization constant and $a$ is an unknown complex phase that cancels when the gate
              is applied twice. The action of MOVE on the orthogonal subspace
              (in particular $|11\rangle$) is undefined. This operation enables
              excitation shuttling between a qubit and the resonator COMPR1,
              providing effective all-to-all connectivity, represented as $\textbf{\textrm{Move}}$ gate in the main text.
            \item \textbf{MEASURE} --- Measurement in the computational ($Z$) basis.
        \end{itemize}
    
    \subsection*{Hardware Characterisation Metrics}
    
        Table~\ref{tab:iqm_summary} presents device-level summary statistics and Table~\ref{tab:iqm_perqubit} lists the full per-qubit metrics; both were derived from the Calibration data obtained as a JSON export on 4~December~2025 at 10:34~IST (GMT+5:30).
    
        \begin{table}[H]
        \centering
        \caption{%
            IQM Sirius QPU benchmarking summary (Calibration: 4~December~2025, 10:34~IST). Values are computed from the IQM Resonance metrics export and
            cross-validated against the IQM Resonance dashboard
            (all 14 entries, including 7 averages and 7 medians, agree to within
            dashboard rounding).
        }
        \label{tab:iqm_summary}
            \begin{tabular}{lcc}
            \hline
            \textbf{Property} & \textbf{Average} & \textbf{Median} \\
            \hline
            $T_1$ time                      & 28.69\,$\mu$s & 28.71\,$\mu$s \\
            $T_2^{*}$ time (Ramsey)         & 17.11\,$\mu$s & 18.62\,$\mu$s \\
            $T_2^{\rm echo}$ time           & 27.68\,$\mu$s & 27.98\,$\mu$s \\
            PRX gate fidelity               & 99.79\%       & 99.78\%       \\
            Single-qubit readout fidelity   & 97.84\%       & 98.49\%       \\
            CZ gate fidelity                & 98.42\%       & 98.54\%       \\
            MOVE-MOVE fidelity              & 98.79\%       & 98.93\%       \\
            \hline
            \end{tabular}
        \end{table}
        
        \medskip
        \noindent\textbf{Metric definitions.}
        \begin{itemize}
            \item $T_1$: is the energy relaxation time.
            \item $T_2^{*}$ and $T_2^{\rm echo}$: are the dephasing times measured by Ramsey and spin-echo sequences, respectively.
            \item PRX gate fidelity: is obtained from DRAG-calibrated randomised benchmarking.
            \item Single-qubit readout fidelity: is the symmetric average state-assignment fidelity from single-shot readout.
            \item CZ gate fidelity: is derived from interleaved randomised benchmarking (\texttt{crf\_crf} for 13 pairs; \texttt{tgss} for QB2, QB10, and QB16, indicated by $\dagger$).
            \item MOVE-MOVE fidelity: is the total fidelity of a MOVE--MOVE sequence obtained from interleaved RB (\texttt{crf\_crf}, $n_{\rm interleaved}=2$).
        \end{itemize}
        
        \noindent
        Summary averages for $T_1$ and $T_2^{*}$ include the COMPR1 resonator as the 17th entry; $T_2^{\rm echo}$, CZ, and MOVE statistics are over the 16 active qubits/pairs only.
        
        The majority of qubits show stable single-qubit performance, with the PRX gate fidelities uniformly above 99.5\% and readout fidelities above 97\%. Two qubits, QB13 and QB15, exhibit notably reduced performance: QB13 has the shortest $T_2^{*}$ ($2.29\,\mu$s) and QB15 the lowest readout fidelity ($91.05\%$), with correspondingly reduced CZ gate fidelities
        ($97.28\%$ and $97.17\%$, respectively). These qubits were retained in all circuits; their reduced coherence and
        readout performance represent a conservative upper bound on hardware noise in the reported results. The two-qubit gate performance is otherwise uniformly stable, with CZ and MOVE-MOVE fidelities exceeding $98\%$ for 14 of 16 qubit--resonator
        pairs.
        
        \begin{table}[H]
        \centering
        \small
        \setlength{\tabcolsep}{5pt}
        \caption{%
            Per-qubit hardware metrics for the 16~active qubits of IQM Sirius
            (Calibration: 4~December~2025, 10:34~IST).
            The COMPR1 resonator row lists its independently measured $T_1$ and
            $T_2^{*}$ and is included in the device-level averages.
            CZ fidelities marked~$\dagger$ are obtained from \texttt{tgss}
            protocol rather than \texttt{crf\_crf}.
            All values are rounded to two decimal places.
        }
        \label{tab:iqm_perqubit}
            \begin{tabular}{lrrrrrrr}
            \hline
            \textbf{Qubit} &
            \boldsymbol$T_1$ &
            \boldsymbol$T_2^{*}$ &
            \boldsymbol$T_2^{\rm echo}$ &
            \textbf{PRX} &
            \textbf{Readout} &
            \textbf{CZ} &
            \textbf{MOVE} \\
            & ($\mu$s) & ($\mu$s) & ($\mu$s) & (\%) & (\%) & (\%) & (\%) \\
            \hline
            COMPR1 & 16.60 & 29.24 & ---   & ---   & ---   & ---   & ---   \\
            \hline
            QB1    & 22.45 & 18.62 & 28.99 & 99.85 & 98.22 & 98.14 & 98.64 \\
            QB2    & 26.79 & 28.22 & 43.85 & 99.76 & 98.62 & 98.11$^{\dagger}$ & 99.24 \\
            QB3    & 27.17 &  5.93 &  9.24 & 99.84 & 98.83 & 98.87 & 98.93 \\
            QB4    & 19.87 & 25.86 & 31.08 & 99.91 & 98.60 & 98.86 & 99.21 \\
            QB5    & 39.42 & 23.77 & 29.50 & 99.94 & 98.75 & 98.96 & 99.19 \\
            QB8    & 31.02 &  6.38 & 19.66 & 99.91 & 98.25 & 98.64 & 98.01 \\
            QB9    & 30.12 &  4.93 & 16.13 & 99.77 & 99.12 & 98.96 & 98.87 \\
            QB10   & 28.71 &  8.03 & 24.50 & 99.76 & 97.72 & 98.32$^{\dagger}$ & 99.04 \\
            QB11   & 38.30 & 29.57 & 35.10 & 99.77 & 98.72 & 98.71 & 98.75 \\
            QB13   & 18.95 &  2.29 &  6.28 & 99.53 & 95.58 & 97.28 & 98.49 \\
            QB15   & 18.19 & 14.04 & 26.98 & 99.64 & 91.05 & 97.17 & 97.96 \\
            QB16   & 33.96 & 33.05 & 39.76 & 99.77 & 98.55 & 98.32$^{\dagger}$ & 98.93 \\
            QB17   & 35.83 &  4.85 & 16.38 & 99.78 & 98.80 & 98.76 & 98.96 \\
            QB19   & 20.48 &  9.22 & 23.87 & 99.76 & 98.42 & 98.44 & 98.95 \\
            QB20   & 32.79 & 19.55 & 40.63 & 99.82 & 98.22 & 98.98 & 99.11 \\
            QB21   & 47.13 & 27.28 & 50.90 & 99.81 & 98.02 & 98.24 & 98.42 \\
            \hline
            \textbf{Mean}   & 28.69 & 17.11 & 27.68 & 99.79 & 97.84 & 98.42 & 98.79 \\
            \textbf{Median} & 28.71 & 18.62 & 27.98 & 99.78 & 98.49 & 98.54 & 98.93 \\
            \hline
            \end{tabular}
        \end{table}

\newpage

\setcounter{table}{0}
\renewcommand{\thetable}{F\arabic{table}}

\setcounter{figure}{0}
\renewcommand{\thefigure}{F\arabic{figure}}

\setcounter{equation}{0}
\renewcommand{\theequation}{F\arabic{equation}}

\section{Sample-based Quantum Diagonalization: Molecule Geometries \& QPU Runtimes}
\label{appendix:sqd_geometries_qpu}

    This appendix reports, in Table~\ref{tab:sqd_geom_qpu_final}, the Cartesian geometric coordinates (optimized at the CCSD level) and the corresponding quantum processing unit (QPU) execution runtimes for all molecular systems studied using sample-based quantum diagonalization (SQD) workflows in Section~4.1 of the main text. The tabulated geometries represent fixed nuclear configurations consistently employed across SQD(LUCJ) and SQD(LCNot--UCCSD) energy evaluations, as well as in the associated post-Hartree--Fock classical benchmark calculations at the CCSD and FCI levels. For each molecule, QPU runtimes are reported from three independent executions performed at a fixed number of measurement shots, enabling transparent comparison of execution-time variability across workflows. No additional structural optimization was performed within either the quantum or classical computational pipelines.

    \begin{table}[H]
    \centering
    \caption{
    Cartesian geometries and QPU execution runtimes for SQD workflows on IQM Sirius quantum hardware.
    All QPU measurements were performed using $N_{\mathrm{shots}} = 10{,}000$. Here, $\boldsymbol{\tau_{\mathrm{sqd}}^{\prime}}$ denotes the QPU execution time for SQD using the LUCJ ansatz, and $\boldsymbol{\tau_{\mathrm{sqd}}^{\prime\prime}}$ denotes the QPU execution time for SQD using the LCNot–UCCSD ansatz, both reported in seconds (s).
    }
    \label{tab:sqd_geom_qpu_final}
    
    \renewcommand{\arraystretch}{1.15}
    \setlength{\tabcolsep}{6pt}
    
        \begin{tabular}{|
        p{2.8cm}|
        c|
        c|c|c|
        c|
        c|
        c|
        }
        \hline
        \textbf{Molecule} &
        \textbf{Atom} &
        \multicolumn{3}{c|}{\textbf{Geometry (\AA)}} &
        \textbf{Run} &
        $\boldsymbol{\tau_{\mathrm{sqd}}^{\prime}}$ \textbf{(s)} &
        $\boldsymbol{\tau_{\mathrm{sqd}}^{\prime\prime}}$ \textbf{(s)} \\
        \cline{3-5}
         & & \textbf{x} & \textbf{y} & \textbf{z} & & & \\
        \hline
        
        \multirow{3}{*}{\makecell[l]{Hydrogen\\(H$_2$)}}
        & H & -8.0563 & 4.7154 & 0.0000  & 1 & 5.9399 & 6.0922 \\
        & H & -8.5983 & 4.7845 & -0.3333 & 2 & 5.8913 & 5.9191 \\
        &   &         &        &         & 3 & 5.8319 & 6.3319 \\
        \hline
        
        \multirow{3}{*}{\makecell[l]{Lithium Hydride\\(LiH)}}
        & Li & -1.3452 & 3.1070 & 0.0000 & 1 & 6.5436 & 10.5604 \\
        & H  & -0.5971 & 2.1184 & 1.0889 & 2 & 6.5822 & 10.8045 \\
        &    &         &        &        & 3 & 6.5957 & 10.6617 \\
        \hline
        
        \multirow{3}{*}{\makecell[l]{Beryllium Hydride\\(BeH$_2$)}}
        & Be & -0.6141 & 3.0632 & 0.0000 & 1 & 6.9852 & 18.7704 \\
        & H  & -1.7489 & 3.2078 & -0.6978 & 2 & 7.0202 & 18.5576 \\
        & H  & -0.5883 & 3.9542 & 1.0005  & 3 & 6.6800 & 18.5296 \\
        \hline
        
        \multirow{3}{*}{\makecell[l]{Water\\(H$_2$O)}}
        & O & -8.6252 & 5.1579 & -0.0147 & 1 & 6.9381 & 23.4240 \\
        & H & -7.7020 & 5.4505 & -0.0474 & 2 & 6.8318 & 14.9879 \\
        & H & -9.0559 & 5.8062 & 0.5626  & 3 & 6.9861 & 14.8362 \\
        \hline
        
        \multirow{4}{*}{\makecell[l]{Ammonia\\(NH$_3$)}}
        & N & -9.0297 & 4.0906 & 0.0167  & 1 & 7.1387 & 28.1490 \\
        & H & -8.5642 & 3.4395 & 0.6473  & 2 & 7.2257 & 27.2112 \\
        & H & -8.4095 & 4.1973 & -0.7848 & 3 & 7.2158 & 27.1455 \\
        & H & -9.8553 & 3.6042 & -0.3299 &   &        &         \\
        \hline
        
        \end{tabular}
    \end{table}

\newpage

\setcounter{table}{0}
\renewcommand{\thetable}{G\arabic{table}}

\setcounter{figure}{0}
\renewcommand{\thefigure}{G\arabic{figure}}

\setcounter{equation}{0}
\renewcommand{\theequation}{G\arabic{equation}}

\section{Potential Energy Surface Scans: Molecule Geometries \& QPU Runtimes}
\label{appendix:pes_geometries_qpu}

    This appendix reports the Cartesian geometric coordinates and corresponding QPU execution runtimes for all molecular systems used in the one-dimensional potential energy surface (1D-PES) scans discussed in Section~4.2 of the main text, as well as the geometry construction and runtime characterization for the two-dimensional PES (2D-PES) of the H$_2$O molecule discussed in Section~4.3 of the main text. 
    
    For each 1D-PES system, the tabulated geometries represent fixed nuclear configurations employed uniformly across all quantum computing–based energy evaluations using SQD(LUCJ), as well as in the associated post-Hartree–Fock classical benchmark calculations at the CCSD and FCI levels. In addition, each table includes the measured QPU runtimes, $\tau_{\mathrm{qpu}}$, obtained from multiple independent executions at a fixed shot count, providing a transparent record of execution variability across geometries. No further structural optimization was performed in either the quantum or classical workflows. Specifically, Table~\ref{tab:h2_1d_pes_sto-3g_geometries_qpu} reports H$_2$ geometries and QPU runtimes in the STO-3G basis, Table~\ref{tab:h2_1d_pes_6-31g_geometries_qpu} reports H$_2$ in the 6-31G basis, Table~\ref{tab:hehp_1d_pes_sto-3g_geometries_qpu} reports HeH$^+$ in the STO-3G basis, Table~\ref{tab:hehp_1d_pes_6-31g_geometries_qpu} reports HeH$^+$ in the 6-31G basis, Table~\ref{tab:lih_1d_pes_sto-3g_geometries_qpu} reports LiH in the STO-3G basis, and Table~\ref{tab:beh2_1d_pes_sto-3g_geometries_qpu} reports BeH$_2$ in the STO-3G basis.

    \begin{table}[H]
    \centering
    \caption{Cartesian geometries used for the 1D bond-length PES scan of H$_2$ in STO-3G basis, along with corresponding QPU runtimes, $\tau_{\mathrm{qpu}}$ (s), from three independent executions at $N_{shots} = 10,000$. The interatomic distance $r$ is given in angstrom (\AA). All geometries are aligned along the $z$-axis with one hydrogen atom fixed at the origin.}
    \label{tab:h2_1d_pes_sto-3g_geometries_qpu}
    \renewcommand{\arraystretch}{1.25}
    \setlength{\tabcolsep}{8pt}
        \begin{tabular}{|c|c|p{6.2cm}|c|c|c|}
        \hline
        \textbf{S.No.} & \textbf{$r$ (\AA)} & \textbf{Geometry (Cartesian coordinates in \AA)} 
        & $\boldsymbol{\tau_{\mathrm{qpu}}^{(1)}}$ \textbf{(s)} & $\boldsymbol{\tau_{\mathrm{qpu}}^{(2)}}$ \textbf{(s)} & $\boldsymbol{\tau_{\mathrm{qpu}}^{(3)}}$ \textbf{(s)} \\
        \hline
        1  & 0.3000 & H (0.0, 0.0, 0.0); H (0.0, 0.0, 0.3000) & 7.2979 & 6.1516 & 5.9550 \\ \hline
        2  & 0.3627 & H (0.0, 0.0, 0.0); H (0.0, 0.0, 0.3627) & 5.9553 & 6.0366 & 6.4405 \\ \hline
        3  & 0.4255 & H (0.0, 0.0, 0.0); H (0.0, 0.0, 0.4255) & 5.9228 & 6.1361 & 6.1373 \\ \hline
        4  & 0.4882 & H (0.0, 0.0, 0.0); H (0.0, 0.0, 0.4882) & 5.8478 & 6.3280 & 6.1624 \\ \hline
        5  & 0.5509 & H (0.0, 0.0, 0.0); H (0.0, 0.0, 0.5509) & 6.0579 & 5.9691 & 5.9320 \\ \hline
        6  & 0.6137 & H (0.0, 0.0, 0.0); H (0.0, 0.0, 0.6137) & 5.9747 & 6.0296 & 5.9973 \\ \hline
        7  & 0.6764 & H (0.0, 0.0, 0.0); H (0.0, 0.0, 0.6764) & 5.9981 & 5.9824 & 5.9425 \\ \hline
        8  & 0.7391 & H (0.0, 0.0, 0.0); H (0.0, 0.0, 0.7391) & 5.7185 & 6.1526 & 6.0083 \\ \hline
        9  & 0.8019 & H (0.0, 0.0, 0.0); H (0.0, 0.0, 0.8019) & 6.1911 & 6.6158 & 6.0033 \\ \hline
        10 & 0.8646 & H (0.0, 0.0, 0.0); H (0.0, 0.0, 0.8646) & 5.9217 & 6.6960 & 6.1015 \\ \hline
        11 & 0.9273 & H (0.0, 0.0, 0.0); H (0.0, 0.0, 0.9273) & 5.9897 & 6.1654 & 5.9972 \\ \hline
        12 & 0.9901 & H (0.0, 0.0, 0.0); H (0.0, 0.0, 0.9901) & 6.0618 & 6.0720 & 5.9927 \\ \hline
        13 & 1.0528 & H (0.0, 0.0, 0.0); H (0.0, 0.0, 1.0528) & 6.1565 & 6.0630 & 5.9105 \\ \hline
        14 & 1.1155 & H (0.0, 0.0, 0.0); H (0.0, 0.0, 1.1155) & 5.9083 & 5.9779 & 6.1222 \\ \hline
        15 & 1.1783 & H (0.0, 0.0, 0.0); H (0.0, 0.0, 1.1783) & 7.0812 & 6.1627 & 6.0876 \\ \hline
        16 & 1.2410 & H (0.0, 0.0, 0.0); H (0.0, 0.0, 1.2410) & 6.1679 & 5.9223 & 6.2450 \\ \hline
        17 & 1.5854 & H (0.0, 0.0, 0.0); H (0.0, 0.0, 1.5854) & 6.0374 & 5.9551 & 5.7936 \\ \hline
        18 & 1.9299 & H (0.0, 0.0, 0.0); H (0.0, 0.0, 1.9299) & 5.9870 & 5.9465 & 5.8130 \\ \hline
        19 & 2.2743 & H (0.0, 0.0, 0.0); H (0.0, 0.0, 2.2743) & 6.1093 & 6.1022 & 6.2906 \\ \hline
        20 & 2.6188 & H (0.0, 0.0, 0.0); H (0.0, 0.0, 2.6188) & 5.9763 & 6.3109 & 6.1536 \\ \hline
        21 & 2.9632 & H (0.0, 0.0, 0.0); H (0.0, 0.0, 2.9632) & 5.9698 & 6.0650 & 5.9278 \\ \hline
        22 & 3.3077 & H (0.0, 0.0, 0.0); H (0.0, 0.0, 3.3077) & 6.0132 & 5.8522 & 6.0788 \\ \hline
        23 & 3.6521 & H (0.0, 0.0, 0.0); H (0.0, 0.0, 3.6521) & 6.0385 & 6.1290 & 5.9444 \\ \hline
        24 & 3.9966 & H (0.0, 0.0, 0.0); H (0.0, 0.0, 3.9966) & 5.9143 & 6.0414 & 6.0460 \\ \hline
        25 & 4.3410 & H (0.0, 0.0, 0.0); H (0.0, 0.0, 4.3410) & 6.0281 & 6.1387 & 6.0042 \\ \hline
        \end{tabular}
    \end{table}

    \begin{table}[H]
    \centering
    \caption{Cartesian geometries used for the 1D bond-length PES scan of H$_2$ in 6-31G basis, along with corresponding QPU runtimes, $\tau_{\mathrm{qpu}}$ (s), from three independent executions at $N_{shots} = 10,000$. The interatomic distance $r$ is given in angstrom (\AA). All geometries are aligned along the $z$-axis with one hydrogen atom fixed at the origin.}
    \label{tab:h2_1d_pes_6-31g_geometries_qpu}
    \renewcommand{\arraystretch}{1.25}
    \setlength{\tabcolsep}{8pt}
        \begin{tabular}{|c|c|p{6.2cm}|c|c|c|}
        \hline
        \textbf{S.No.} & \textbf{$r$ (\AA)} & \textbf{Geometry (Cartesian coordinates in \AA)} 
        & $\boldsymbol{\tau_{\mathrm{qpu}}^{(1)}}$ \textbf{(s)} & $\boldsymbol{\tau_{\mathrm{qpu}}^{(2)}}$ \textbf{(s)} & $\boldsymbol{\tau_{\mathrm{qpu}}^{(3)}}$ \textbf{(s)} \\
        \hline
        1  & 0.3000 & H (0.0, 0.0, 0.0); H (0.0, 0.0, 0.3000) & 6.1700 & 5.8312 & 6.4196 \\ \hline
        2  & 0.3627 & H (0.0, 0.0, 0.0); H (0.0, 0.0, 0.3627) & 6.1612 & 5.9671 & 6.0942 \\ \hline
        3  & 0.4255 & H (0.0, 0.0, 0.0); H (0.0, 0.0, 0.4255) & 6.2658 & 6.3419 & 5.9834 \\ \hline
        4  & 0.4882 & H (0.0, 0.0, 0.0); H (0.0, 0.0, 0.4882) & 6.1994 & 6.2379 & 15.9657 \\ \hline
        5  & 0.5509 & H (0.0, 0.0, 0.0); H (0.0, 0.0, 0.5509) & 6.2950 & 6.0310 & 6.1382 \\ \hline
        6  & 0.6137 & H (0.0, 0.0, 0.0); H (0.0, 0.0, 0.6137) & 6.1667 & 6.4015 & 6.0511 \\ \hline
        7  & 0.6764 & H (0.0, 0.0, 0.0); H (0.0, 0.0, 0.6764) & 6.6754 & 6.0823 & 6.0986 \\ \hline
        8  & 0.7391 & H (0.0, 0.0, 0.0); H (0.0, 0.0, 0.7391) & 6.1571 & 6.2088 & 6.1755 \\ \hline
        9  & 0.8019 & H (0.0, 0.0, 0.0); H (0.0, 0.0, 0.8019) & 6.0316 & 6.2185 & 6.2606 \\ \hline
        10 & 0.8646 & H (0.0, 0.0, 0.0); H (0.0, 0.0, 0.8646) & 5.9971 & 6.2708 & 6.2557 \\ \hline
        11 & 0.9273 & H (0.0, 0.0, 0.0); H (0.0, 0.0, 0.9273) & 6.0303 & 6.3036 & 6.0418 \\ \hline
        12 & 0.9901 & H (0.0, 0.0, 0.0); H (0.0, 0.0, 0.9901) & 6.0956 & 6.1590 & 6.0282 \\ \hline
        13 & 1.0528 & H (0.0, 0.0, 0.0); H (0.0, 0.0, 1.0528) & 6.0287 & 6.2573 & 6.2267 \\ \hline
        14 & 1.1155 & H (0.0, 0.0, 0.0); H (0.0, 0.0, 1.1155) & 6.0695 & 6.4471 & 6.0187 \\ \hline
        15 & 1.1783 & H (0.0, 0.0, 0.0); H (0.0, 0.0, 1.1783) & 6.3106 & 6.1846 & 6.2552 \\ \hline
        16 & 1.2410 & H (0.0, 0.0, 0.0); H (0.0, 0.0, 1.2410) & 6.0697 & 6.0810 & 6.2068 \\ \hline
        17 & 1.5854 & H (0.0, 0.0, 0.0); H (0.0, 0.0, 1.5854) & 6.2628 & 5.9745 & 6.1503 \\ \hline
        18 & 1.9299 & H (0.0, 0.0, 0.0); H (0.0, 0.0, 1.9299) & 6.1822 & 6.6125 & 6.2142 \\ \hline
        19 & 2.2743 & H (0.0, 0.0, 0.0); H (0.0, 0.0, 2.2743) & 6.0858 & 6.1836 & 6.8559 \\ \hline
        20 & 2.6188 & H (0.0, 0.0, 0.0); H (0.0, 0.0, 2.6188) & 6.2074 & 6.0729 & 6.1786 \\ \hline
        21 & 2.9632 & H (0.0, 0.0, 0.0); H (0.0, 0.0, 2.9632) & 6.2356 & 6.2927 & 6.3299 \\ \hline
        22 & 3.3077 & H (0.0, 0.0, 0.0); H (0.0, 0.0, 3.3077) & 6.6367 & 6.2286 & 6.4128 \\ \hline
        23 & 3.6521 & H (0.0, 0.0, 0.0); H (0.0, 0.0, 3.6521) & 7.2147 & 6.2212 & 6.2125 \\ \hline
        24 & 3.9966 & H (0.0, 0.0, 0.0); H (0.0, 0.0, 3.9966) & 6.1331 & 6.1968 & 6.3045 \\ \hline
        25 & 4.3410 & H (0.0, 0.0, 0.0); H (0.0, 0.0, 4.3410) & 6.2003 & 6.2773 & 6.5061 \\ \hline
        \end{tabular}
    \end{table}

    \begin{table}[H]
    \centering
    \caption{Cartesian geometries used for the 1D bond-length PES scan of HeH$^+$ in STO-3G basis, along with corresponding QPU runtimes, $\tau_{\mathrm{qpu}}$ (s), from three independent executions at $N_{shots} = 10,000$. The interatomic distance $r$ is given in angstrom (\AA). All geometries are aligned along the $z$-axis with one hydrogen atom fixed at the origin.}
    \label{tab:hehp_1d_pes_sto-3g_geometries_qpu}
    \renewcommand{\arraystretch}{1.25}
    \setlength{\tabcolsep}{8pt}
        \begin{tabular}{|c|c|p{6.2cm}|c|c|c|}
        \hline
        \textbf{S.No.} & \textbf{$r$ (\AA)} & \textbf{Geometry (Cartesian coordinates in \AA)} 
        & $\boldsymbol{\tau_{\mathrm{qpu}}^{(1)}}$ \textbf{(s)} & $\boldsymbol{\tau_{\mathrm{qpu}}^{(2)}}$ \textbf{(s)} & $\boldsymbol{\tau_{\mathrm{qpu}}^{(3)}}$ \textbf{(s)} \\
        \hline
        1  & 0.5500 & He (0.0, 0.0, 0.0); H (0.0, 0.0, 0.5500) & 5.9937 & 6.9913 & 6.1094 \\ \hline
        2  & 0.5983 & He (0.0, 0.0, 0.0); H (0.0, 0.0, 0.5983) & 5.9140 & 6.0741 & 6.1593 \\ \hline
        3  & 0.6465 & He (0.0, 0.0, 0.0); H (0.0, 0.0, 0.6465) & 5.9487 & 6.1472 & 6.0793 \\ \hline
        4  & 0.6948 & He (0.0, 0.0, 0.0); H (0.0, 0.0, 0.6948) & 6.3038 & 6.2373 & 5.8171 \\ \hline
        5  & 0.7431 & He (0.0, 0.0, 0.0); H (0.0, 0.0, 0.7431) & 5.9789 & 5.9495 & 5.8637 \\ \hline
        6  & 0.7913 & He (0.0, 0.0, 0.0); H (0.0, 0.0, 0.7913) & 6.0862 & 6.1246 & 6.0152 \\ \hline
        7  & 0.8396 & He (0.0, 0.0, 0.0); H (0.0, 0.0, 0.8396) & 5.8899 & 6.0294 & 5.8034 \\ \hline
        8  & 0.8879 & He (0.0, 0.0, 0.0); H (0.0, 0.0, 0.8879) & 6.0519 & 5.8819 & 5.9775 \\ \hline
        9  & 0.9361 & He (0.0, 0.0, 0.0); H (0.0, 0.0, 0.9361) & 6.1266 & 5.8488 & 6.0615 \\ \hline
        10 & 0.9844 & He (0.0, 0.0, 0.0); H (0.0, 0.0, 0.9844) & 5.9444 & 6.0185 & 6.0479 \\ \hline
        11 & 1.0327 & He (0.0, 0.0, 0.0); H (0.0, 0.0, 1.0327) & 6.1281 & 6.1791 & 6.1898 \\ \hline
        12 & 1.0809 & He (0.0, 0.0, 0.0); H (0.0, 0.0, 1.0809) & 6.0941 & 5.9533 & 5.9359 \\ \hline
        13 & 1.1292 & He (0.0, 0.0, 0.0); H (0.0, 0.0, 1.1292) & 15.4878 & 6.0701 & 6.0763 \\ \hline
        14 & 1.1775 & He (0.0, 0.0, 0.0); H (0.0, 0.0, 1.1775) & 6.2186 & 6.0398 & 5.9692 \\ \hline
        15 & 1.2257 & He (0.0, 0.0, 0.0); H (0.0, 0.0, 1.2257) & 5.9709 & 5.9988 & 6.0263 \\ \hline
        16 & 1.2740 & He (0.0, 0.0, 0.0); H (0.0, 0.0, 1.2740) & 6.1320 & 6.1994 & 5.9911 \\ \hline
        17 & 1.5851 & He (0.0, 0.0, 0.0); H (0.0, 0.0, 1.5851) & 6.0623 & 5.9470 & 5.8928 \\ \hline
        18 & 1.8962 & He (0.0, 0.0, 0.0); H (0.0, 0.0, 1.8962) & 6.1877 & 6.0365 & 6.0054 \\ \hline
        19 & 2.2073 & He (0.0, 0.0, 0.0); H (0.0, 0.0, 2.2073) & 11.9733 & 6.2745 & 5.9660 \\ \hline
        20 & 2.5184 & He (0.0, 0.0, 0.0); H (0.0, 0.0, 2.5184) & 5.8456 & 5.9821 & 6.0214 \\ \hline
        21 & 2.8296 & He (0.0, 0.0, 0.0); H (0.0, 0.0, 2.8296) & 5.9289 & 5.8350 & 6.0853 \\ \hline
        22 & 3.1407 & He (0.0, 0.0, 0.0); H (0.0, 0.0, 3.1407) & 5.9151 & 6.1138 & 6.1205 \\ \hline
        23 & 3.4518 & He (0.0, 0.0, 0.0); H (0.0, 0.0, 3.4518) & 6.0054 & 6.0066 & 5.9501 \\ \hline
        24 & 3.7629 & He (0.0, 0.0, 0.0); H (0.0, 0.0, 3.7629) & 5.9995 & 6.0248 & 5.9608 \\ \hline
        25 & 4.0740 & He (0.0, 0.0, 0.0); H (0.0, 0.0, 4.0740) & 5.8320 & 5.9449 & 5.8420 \\ \hline
        \end{tabular}
    \end{table}

    \begin{table}[H]
    \centering
    \caption{Cartesian geometries used for the 1D bond-length PES scan of HeH$^+$ in 6-31G basis, along with corresponding QPU runtimes, $\tau_{\mathrm{qpu}}$ (s), from three independent executions at $N_{shots} = 10,000$. The interatomic distance $r$ is given in angstrom (\AA). All geometries are aligned along the $z$-axis with the helium atom fixed at the origin.}
    \label{tab:hehp_1d_pes_6-31g_geometries_qpu}
    \renewcommand{\arraystretch}{1.25}
    \setlength{\tabcolsep}{8pt}
        \begin{tabular}{|c|c|p{6.2cm}|c|c|c|}
        \hline
        \textbf{S.No.} & \textbf{$r$ (\AA)} & \textbf{Geometry (Cartesian coordinates in \AA)} 
        & $\boldsymbol{\tau_{\mathrm{qpu}}^{(1)}}$ \textbf{(s)} & $\boldsymbol{\tau_{\mathrm{qpu}}^{(2)}}$ \textbf{(s)} & $\boldsymbol{\tau_{\mathrm{qpu}}^{(3)}}$ \textbf{(s)} \\
        \hline
        1  & 0.4500 & He (0.0, 0.0, 0.0); H (0.0, 0.0, 0.4500) & 6.2737 & 6.2738 & 6.2271 \\ \hline
        2  & 0.5049 & He (0.0, 0.0, 0.0); H (0.0, 0.0, 0.5049) & 6.0166 & 6.3871 & 6.3318 \\ \hline
        3  & 0.5599 & He (0.0, 0.0, 0.0); H (0.0, 0.0, 0.5599) & 6.0435 & 6.9888 & 6.5851 \\ \hline
        4  & 0.6148 & He (0.0, 0.0, 0.0); H (0.0, 0.0, 0.6148) & 6.2107 & 6.1794 & 6.4662 \\ \hline
        5  & 0.6697 & He (0.0, 0.0, 0.0); H (0.0, 0.0, 0.6697) & 6.6714 & 6.5610 & 6.1933 \\ \hline
        6  & 0.7247 & He (0.0, 0.0, 0.0); H (0.0, 0.0, 0.7247) & 6.8343 & 6.5450 & 6.2113 \\ \hline
        7  & 0.7796 & He (0.0, 0.0, 0.0); H (0.0, 0.0, 0.7796) & 6.1873 & 6.2182 & 6.0013 \\ \hline
        8  & 0.8345 & He (0.0, 0.0, 0.0); H (0.0, 0.0, 0.8345) & 6.1636 & 6.2252 & 6.2059 \\ \hline
        9  & 0.8895 & He (0.0, 0.0, 0.0); H (0.0, 0.0, 0.8895) & 6.7252 & 6.6528 & 6.3301 \\ \hline
        10 & 0.9444 & He (0.0, 0.0, 0.0); H (0.0, 0.0, 0.9444) & 6.1561 & 6.5367 & 6.2345 \\ \hline
        11 & 0.9993 & He (0.0, 0.0, 0.0); H (0.0, 0.0, 0.9993) & 6.0686 & 7.0564 & 7.1170 \\ \hline
        12 & 1.0543 & He (0.0, 0.0, 0.0); H (0.0, 0.0, 1.0543) & 7.3921 & 6.0140 & 6.0151 \\ \hline
        13 & 1.1092 & He (0.0, 0.0, 0.0); H (0.0, 0.0, 1.1092) & 6.4106 & 6.4244 & 6.3907 \\ \hline
        14 & 1.1641 & He (0.0, 0.0, 0.0); H (0.0, 0.0, 1.1641) & 6.1680 & 6.1359 & 6.3287 \\ \hline
        15 & 1.2191 & He (0.0, 0.0, 0.0); H (0.0, 0.0, 1.2191) & 6.2126 & 6.1561 & 6.5339 \\ \hline
        16 & 1.2740 & He (0.0, 0.0, 0.0); H (0.0, 0.0, 1.2740) & 6.2698 & 7.3350 & 7.6462 \\ \hline
        17 & 1.5851 & He (0.0, 0.0, 0.0); H (0.0, 0.0, 1.5851) & 6.2266 & 6.0591 & 6.1860 \\ \hline
        18 & 1.8962 & He (0.0, 0.0, 0.0); H (0.0, 0.0, 1.8962) & 7.3691 & 6.0313 & 6.2810 \\ \hline
        19 & 2.2073 & He (0.0, 0.0, 0.0); H (0.0, 0.0, 2.2073) & 6.0623 & 6.2442 & 6.3034 \\ \hline
        20 & 2.5184 & He (0.0, 0.0, 0.0); H (0.0, 0.0, 2.5184) & 6.1897 & 6.2835 & 6.3677 \\ \hline
        21 & 2.8296 & He (0.0, 0.0, 0.0); H (0.0, 0.0, 2.8296) & 6.2115 & 6.2355 & 6.2503 \\ \hline
        22 & 3.1407 & He (0.0, 0.0, 0.0); H (0.0, 0.0, 3.1407) & 6.4736 & 6.1903 & 6.0395 \\ \hline
        23 & 3.4518 & He (0.0, 0.0, 0.0); H (0.0, 0.0, 3.4518) & 6.3867 & 6.4235 & 6.6657 \\ \hline
        24 & 3.7629 & He (0.0, 0.0, 0.0); H (0.0, 0.0, 3.7629) & 6.1007 & 6.2881 & 6.2068 \\ \hline
        25 & 4.0740 & He (0.0, 0.0, 0.0); H (0.0, 0.0, 4.0740) & 6.1294 & 6.3392 & 6.1096 \\ \hline
        \end{tabular}
    \end{table}

    \begin{table}[H]
    \centering
    \caption{Cartesian geometries used for the 1D bond-length PES scan of LiH in STO-3G basis, along with corresponding QPU runtimes, $\tau_{\mathrm{qpu}}$ (s), from three independent executions at $N_{shots} = 10,000$. The interatomic distance $r$ is given in angstrom (\AA). All geometries are aligned along the $z$-axis with the lithium atom fixed at the origin.}
    \label{tab:lih_1d_pes_sto-3g_geometries_qpu}
    \renewcommand{\arraystretch}{1.25}
    \setlength{\tabcolsep}{8pt}
        \begin{tabular}{|c|c|p{6.2cm}|c|c|c|}
        \hline
        \textbf{S.No.} & \textbf{$r$ (\AA)} & \textbf{Geometry (Cartesian coordinates in \AA)}
        & $\boldsymbol{\tau_{\mathrm{qpu}}^{(1)}}$ \textbf{(s)} & $\boldsymbol{\tau_{\mathrm{qpu}}^{(2)}}$ \textbf{(s)} & $\boldsymbol{\tau_{\mathrm{qpu}}^{(3)}}$ \textbf{(s)} \\
        \hline
        1  & 0.8000 & Li (0.0, 0.0, 0.0); H (0.0, 0.0, 0.8000) & 6.6105 & 6.7314 & 6.5239 \\ \hline
        2  & 0.8592 & Li (0.0, 0.0, 0.0); H (0.0, 0.0, 0.8592) & 6.3240 & 6.4909 & 6.5194 \\ \hline
        3  & 0.9184 & Li (0.0, 0.0, 0.0); H (0.0, 0.0, 0.9184) & 6.6621 & 6.6183 & 6.2601 \\ \hline
        4  & 0.9776 & Li (0.0, 0.0, 0.0); H (0.0, 0.0, 0.9776) & 7.4973 & 6.4484 & 6.6277 \\ \hline
        5  & 1.0368 & Li (0.0, 0.0, 0.0); H (0.0, 0.0, 1.0368) & 7.1162 & 6.5821 & 6.4737 \\ \hline
        6  & 1.0960 & Li (0.0, 0.0, 0.0); H (0.0, 0.0, 1.0960) & 6.5866 & 6.4544 & 6.6322 \\ \hline
        7  & 1.1627 & Li (0.0, 0.0, 0.0); H (0.0, 0.0, 1.1627) & 6.6017 & 6.3138 & 7.8663 \\ \hline
        8  & 1.2293 & Li (0.0, 0.0, 0.0); H (0.0, 0.0, 1.2293) & 6.8034 & 6.5531 & 6.5790 \\ \hline
        9  & 1.2960 & Li (0.0, 0.0, 0.0); H (0.0, 0.0, 1.2960) & 6.7020 & 6.6111 & 6.8629 \\ \hline
        10 & 1.3627 & Li (0.0, 0.0, 0.0); H (0.0, 0.0, 1.3627) & 6.4236 & 6.6864 & 6.4323 \\ \hline
        11 & 1.4293 & Li (0.0, 0.0, 0.0); H (0.0, 0.0, 1.4293) & 6.4246 & 6.4935 & 6.6092 \\ \hline
        12 & 1.4960 & Li (0.0, 0.0, 0.0); H (0.0, 0.0, 1.4960) & 6.7005 & 6.4013 & 6.4722 \\ \hline
        13 & 1.5627 & Li (0.0, 0.0, 0.0); H (0.0, 0.0, 1.5627) & 6.2868 & 6.4638 & 6.5032 \\ \hline
        14 & 1.6293 & Li (0.0, 0.0, 0.0); H (0.0, 0.0, 1.6293) & 6.8374 & 6.6009 & 6.8490 \\ \hline
        15 & 1.6960 & Li (0.0, 0.0, 0.0); H (0.0, 0.0, 1.6960) & 6.9978 & 6.6726 & 6.6939 \\ \hline
        16 & 1.7627 & Li (0.0, 0.0, 0.0); H (0.0, 0.0, 1.7627) & 6.7075 & 6.6723 & 6.5056 \\ \hline
        17 & 1.8293 & Li (0.0, 0.0, 0.0); H (0.0, 0.0, 1.8293) & 6.7507 & 7.4605 & 6.4880 \\ \hline
        18 & 1.8960 & Li (0.0, 0.0, 0.0); H (0.0, 0.0, 1.8960) & 6.4749 & 6.5708 & 6.7451 \\ \hline
        19 & 1.9627 & Li (0.0, 0.0, 0.0); H (0.0, 0.0, 1.9627) & 6.5895 & 6.5987 & 6.5162 \\ \hline
        20 & 2.0293 & Li (0.0, 0.0, 0.0); H (0.0, 0.0, 2.0293) & 6.7619 & 8.1165 & 16.0304 \\ \hline
        21 & 2.0960 & Li (0.0, 0.0, 0.0); H (0.0, 0.0, 2.0960) & 6.7832 & 6.7063 & 6.4066 \\ \hline
        22 & 2.4404 & Li (0.0, 0.0, 0.0); H (0.0, 0.0, 2.4404) & 7.1057 & 6.6864 & 6.6921 \\ \hline
        23 & 2.7849 & Li (0.0, 0.0, 0.0); H (0.0, 0.0, 2.7849) & 6.7119 & 6.3962 & 6.7159 \\ \hline
        24 & 3.1293 & Li (0.0, 0.0, 0.0); H (0.0, 0.0, 3.1293) & 6.5403 & 6.6544 & 7.0440 \\ \hline
        25 & 3.4738 & Li (0.0, 0.0, 0.0); H (0.0, 0.0, 3.4738) & 6.7604 & 6.4909 & 6.6951 \\ \hline
        26 & 3.8182 & Li (0.0, 0.0, 0.0); H (0.0, 0.0, 3.8182) & 6.7622 & 6.6986 & 6.4735 \\ \hline
        27 & 4.1627 & Li (0.0, 0.0, 0.0); H (0.0, 0.0, 4.1627) & 6.4291 & 6.5989 & 6.5565 \\ \hline
        28 & 4.5071 & Li (0.0, 0.0, 0.0); H (0.0, 0.0, 4.5071) & 6.4096 & 6.4240 & 6.7371 \\ \hline
        29 & 4.8516 & Li (0.0, 0.0, 0.0); H (0.0, 0.0, 4.8516) & 6.8221 & 6.7164 & 6.5276 \\ \hline
        30 & 5.1960 & Li (0.0, 0.0, 0.0); H (0.0, 0.0, 5.1960) & 6.4271 & 6.4395 & 6.2613 \\ \hline
        \end{tabular}
    \end{table}

    \begin{table}[H]
    \centering
    \caption{Cartesian geometries used for the 1D bond-length PES scan of BeH$_2$ in STO-3G basis, along with corresponding QPU runtimes, $\tau_{\mathrm{qpu}}$ (s), from three independent executions at $N_{shots} = 10,000$. The interatomic distance $r$ is given in angstrom (\AA). All geometries are linear along the $z$-axis with the beryllium atom fixed at the origin.}
    \label{tab:beh2_1d_pes_sto-3g_geometries_qpu}
    \renewcommand{\arraystretch}{1.25}
    \setlength{\tabcolsep}{8pt}
        \begin{tabular}{|c|c|p{8.5cm}|c|c|c|}
        \hline
        \textbf{S.No.} & \textbf{$r$ (\AA)} & \textbf{Geometry (Cartesian coordinates in \AA)}
        & $\boldsymbol{\tau_{\mathrm{qpu}}^{(1)}}$ \textbf{(s)} & $\boldsymbol{\tau_{\mathrm{qpu}}^{(2)}}$ \textbf{(s)} & $\boldsymbol{\tau_{\mathrm{qpu}}^{(3)}}$ \textbf{(s)} \\
        \hline
        1  & 0.7200 & Be (0.0, 0.0, 0.0); H (0.0, 0.0, 0.7200); H (0.0, 0.0, -0.7200) & 6.6106 & 6.6678 & 6.4738 \\ \hline
        2  & 0.7428 & Be (0.0, 0.0, 0.0); H (0.0, 0.0, 0.7428); H (0.0, 0.0, -0.7428) & 6.7844 & 6.5851 & 6.3720 \\ \hline
        3  & 0.7656 & Be (0.0, 0.0, 0.0); H (0.0, 0.0, 0.7656); H (0.0, 0.0, -0.7656) & 6.6972 & 6.7438 & 6.8278 \\ \hline
        4  & 0.7884 & Be (0.0, 0.0, 0.0); H (0.0, 0.0, 0.7884); H (0.0, 0.0, -0.7884) & 6.3312 & 6.5289 & 6.5240 \\ \hline
        5  & 0.8112 & Be (0.0, 0.0, 0.0); H (0.0, 0.0, 0.8112); H (0.0, 0.0, -0.8112) & 6.8564 & 6.7196 & 6.6531 \\ \hline
        6  & 0.8340 & Be (0.0, 0.0, 0.0); H (0.0, 0.0, 0.8340); H (0.0, 0.0, -0.8340) & 6.7562 & 6.8221 & 6.7377 \\ \hline
        7  & 0.9007 & Be (0.0, 0.0, 0.0); H (0.0, 0.0, 0.9007); H (0.0, 0.0, -0.9007) & 6.5101 & 6.7169 & 6.6753 \\ \hline
        8  & 0.9673 & Be (0.0, 0.0, 0.0); H (0.0, 0.0, 0.9673); H (0.0, 0.0, -0.9673) & 6.6361 & 6.9130 & 6.5877 \\ \hline
        9  & 1.0340 & Be (0.0, 0.0, 0.0); H (0.0, 0.0, 1.0340); H (0.0, 0.0, -1.0340) & 7.2480 & 6.6442 & 6.9104 \\ \hline
        10 & 1.1007 & Be (0.0, 0.0, 0.0); H (0.0, 0.0, 1.1007); H (0.0, 0.0, -1.1007) & 6.7268 & 6.6135 & 6.6615 \\ \hline
        11 & 1.1673 & Be (0.0, 0.0, 0.0); H (0.0, 0.0, 1.1673); H (0.0, 0.0, -1.1673) & 7.7520 & 7.4272 & 6.3733 \\ \hline
        12 & 1.2340 & Be (0.0, 0.0, 0.0); H (0.0, 0.0, 1.2340); H (0.0, 0.0, -1.2340) & 6.5072 & 6.6901 & 6.6654 \\ \hline
        13 & 1.3007 & Be (0.0, 0.0, 0.0); H (0.0, 0.0, 1.3007); H (0.0, 0.0, -1.3007) & 6.8089 & 6.6751 & 6.4889 \\ \hline
        14 & 1.3673 & Be (0.0, 0.0, 0.0); H (0.0, 0.0, 1.3673); H (0.0, 0.0, -1.3673) & 7.1670 & 6.5617 & 6.5469 \\ \hline
        15 & 1.4340 & Be (0.0, 0.0, 0.0); H (0.0, 0.0, 1.4340); H (0.0, 0.0, -1.4340) & 6.8466 & 6.6929 & 6.7522 \\ \hline
        16 & 1.5007 & Be (0.0, 0.0, 0.0); H (0.0, 0.0, 1.5007); H (0.0, 0.0, -1.5007) & 6.6260 & 6.7341 & 6.5788 \\ \hline
        17 & 1.5673 & Be (0.0, 0.0, 0.0); H (0.0, 0.0, 1.5673); H (0.0, 0.0, -1.5673) & 6.7361 & 6.5224 & 6.7584 \\ \hline
        18 & 1.6340 & Be (0.0, 0.0, 0.0); H (0.0, 0.0, 1.6340); H (0.0, 0.0, -1.6340) & 6.8055 & 6.6261 & 6.8842 \\ \hline
        19 & 1.7007 & Be (0.0, 0.0, 0.0); H (0.0, 0.0, 1.7007); H (0.0, 0.0, -1.7007) & 6.6008 & 7.0599 & 6.5021 \\ \hline
        20 & 1.7673 & Be (0.0, 0.0, 0.0); H (0.0, 0.0, 1.7673); H (0.0, 0.0, -1.7673) & 7.3756 & 6.7832 & 6.5648 \\ \hline
        21 & 1.8340 & Be (0.0, 0.0, 0.0); H (0.0, 0.0, 1.8340); H (0.0, 0.0, -1.8340) & 6.5205 & 6.6326 & 6.8316 \\ \hline
        22 & 2.0718 & Be (0.0, 0.0, 0.0); H (0.0, 0.0, 2.0718); H (0.0, 0.0, -2.0718) & 6.6125 & 6.6399 & 6.4728 \\ \hline
        23 & 2.3096 & Be (0.0, 0.0, 0.0); H (0.0, 0.0, 2.3096); H (0.0, 0.0, -2.3096) & 6.6628 & 6.5029 & 6.5046 \\ \hline
        24 & 2.5473 & Be (0.0, 0.0, 0.0); H (0.0, 0.0, 2.5473); H (0.0, 0.0, -2.5473) & 6.7331 & 7.0441 & 7.2652 \\ \hline
        25 & 2.7851 & Be (0.0, 0.0, 0.0); H (0.0, 0.0, 2.7851); H (0.0, 0.0, -2.7851) & 7.2212 & 7.5278 & 7.0331 \\ \hline
        26 & 3.0229 & Be (0.0, 0.0, 0.0); H (0.0, 0.0, 3.0229); H (0.0, 0.0, -3.0229) & 7.1399 & 7.8295 & 6.5676 \\ \hline
        27 & 3.4984 & Be (0.0, 0.0, 0.0); H (0.0, 0.0, 3.4984); H (0.0, 0.0, -3.4984) & 6.6436 & 6.6932 & 7.1953 \\ \hline
        \end{tabular}
    \end{table}

    \newpage 
    
\begin{figure}[H]
            \centering
            \includegraphics[width=0.5\linewidth]{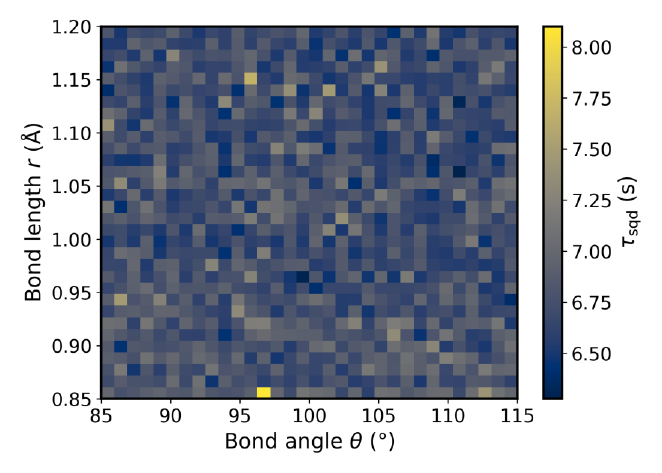}
            \caption{
            QPU execution runtime $\tau_{\mathrm{qpu}}$ heatmap for the two-dimensional potential energy surface (2D-PES) of the H$_2$O molecule in the STO-3G basis, with H--O--H bond angle $\theta$ ($^\circ$) along the x-axis and O--H bond length $r$ (\AA) along the y-axis. Each pixel corresponds to a single $(r,\theta)$ nuclear configuration on the $32 \times 32$ grid at a fixed shot count of $N_{shots} = 10,000$.
            }
            \label{fig:2d_pes_qpu_runtime_h2o_sto-3g}
        \end{figure}

    In addition to the one-dimensional PES scans summarized above, a two-dimensional potential energy surface (2D-PES) was constructed for the H$_2$O molecule in the STO-3G basis by simultaneously varying the O--H bond length and the H--O--H bond angle. Owing to the size of the resulting configuration space, individual Cartesian geometries are not tabulated for the 2D-PES. Instead, all nuclear configurations are specified analytically through a deterministic geometry-generation function, and QPU execution times are reported in aggregated form via a runtime heatmap, as shown in Figure~\ref{fig:2d_pes_qpu_runtime_h2o_sto-3g}.
    
    The geometries were generated using a symmetric internal-coordinate parameterization with the oxygen atom fixed at the origin. For each pair of internal coordinates $(r,\theta)$, the Cartesian positions of the nuclei are given by
        \begin{equation}
            \mathbf{R}_{\mathrm{O}} = (0.0, \,0.0, \,0.0),
        \end{equation}
        \begin{equation}
            \mathbf{R}_{\mathrm{H}_1} = \left( r \sin\frac{\theta}{2},\; 0.0,\; r \cos\frac{\theta}{2} \right),
        \end{equation}
        \begin{equation}
            \mathbf{R}_{\mathrm{H}_2} = \left( -r \sin\frac{\theta}{2},\; 0.0,\; r \cos\frac{\theta}{2} \right),
        \end{equation}
    where $r$ denotes the O--H bond length and $\theta$ denotes the H--O--H bond angle. This construction preserves molecular symmetry and ensures that all variations in the PES arise exclusively from controlled distortions of the internal coordinates.
    
    The internal coordinates were sampled uniformly on a rectangular grid defined by
        \begin{equation}
            r \in [0.85, 1.20]\ \text{\AA}, \qquad \theta \in [85^\circ, 115^\circ],
        \end{equation}
        using $N_r = 32$ points along the bond-length dimension and $N_\theta = 32$ points along the angular dimension, yielding a total of $N_r \times N_\theta = 1024$ distinct nuclear configurations. The corresponding grid spacings are
        \begin{equation}
            \Delta r = \frac{1.20 - 0.85}{31}, \qquad
            \Delta \theta = \frac{115^\circ - 85^\circ}{31}.
        \end{equation}
    All geometries were used directly, without further optimization, in the SQD(LUCJ) quantum workflow as well as in the associated CCSD and FCI classical benchmark calculations. QPU execution runtimes $\tau_{\mathrm{qpu}}(r,\theta)$ in seconds$(s)$ are reported graphically in the form of a heatmap as shown in Figure~\ref{fig:2d_pes_qpu_runtime_h2o_sto-3g}. 
    
\setcounter{table}{0}
\renewcommand{\thetable}{H\arabic{table}}

\setcounter{figure}{0}
\renewcommand{\thefigure}{H\arabic{figure}}

\setcounter{equation}{0}
\renewcommand{\theequation}{H\arabic{equation}}

\section{DMET-SQD: Molecule Geometries \& QPU Runtimes}
\label{sec:coords_supp}

    \begin{table}[H]
    \centering
    \caption{Geometric coordinates of the molecules used for simulation, along with SQD runtime and DMET convergence iterations.}
    
        \begin{tabular}{|p{3.8cm}|c|c|c|c|c|c|}
        \hline
        \multirow{2}{*}{\textbf{Molecule}} 
        & \multicolumn{4}{c|}{\textbf{Geometric Coordinates}} 
        & \multirow{2}{*}{\textbf{$\boldsymbol{\tau_{\mathrm{qpu}}}$ (s)}} 
        & \multirow{2}{*}{$\boldsymbol{N_{it}}$} \\ 
        \cline{2-5}
         & \textbf{Atom} & \textbf{x} & \textbf{y} & \textbf{z} &  &  \\
        \hline
        
        \multirow{4}{*}{\makecell[l]{Cyanic Acid \\ ($\mathrm{HOCN}$)}} 
         & H & -1.4586 & -0.2728 &  0.0655 & \multirow{4}{*}{112.9437} & \multirow{4}{*}{4} \\
         & O & -0.6006 &  0.0996 & -0.3212 &  &  \\
         & C &  0.5444 &  0.0905 &  0.4227 &  &  \\
         & N &  1.5148 &  0.0828 &  1.0531 &  &  \\
        \hline
        
        \multirow{6}{*}{\makecell[l]{Formaldehyde Oxime \\ ($\mathrm{CH_3NO}$)}} 
         & C & -0.9105 & -0.0209 & -0.3325 & \multirow{6}{*}{157.8839} & \multirow{6}{*}{4} \\
         & H &  2.1707 & -0.1169 &  0.6124 &  &  \\
         & H & -1.0055 &  1.0435 & -0.5074 &  &  \\
         & H & -1.7610 & -0.6672 & -0.5009 &  &  \\
         & N &  0.2079 & -0.5236 &  0.0747 &  &  \\
         & O &  1.2984 &  0.2851 &  0.2934 &  &  \\
        \hline
        
        \multirow{8}{*}{\makecell[l]{Methoxyamine \\ ($\mathrm{CH_5NO}$)}} 
         & C & -0.9661 &  0.2013 &  0.1500 & \multirow{8}{*}{204.0350} & \multirow{8}{*}{4} \\
         & H & -1.0003 &  0.9655 & -0.6585 &  &  \\
         & H & -1.6672 &  0.5167 &  0.9498 &  &  \\
         & H & -1.3069 & -0.7850 & -0.2370 &  &  \\
         & H &  1.8825 &  0.4411 & -0.4096 &  &  \\
         & H &  1.6013 & -1.1638 & -0.0232 &  &  \\
         & N &  1.1321 & -0.2814 & -0.3282 &  &  \\
         & O &  0.3246 &  0.1057 &  0.6921 &  &  \\
        \hline
        
        \multirow{7}{*}{\makecell[l]{Methyl Isocyanate \\ ($\mathrm{C_2H_3NO}$)}} 
         & C & -0.8727 & -0.0486 & -0.0127 & \multirow{7}{*}{187.3237} & \multirow{7}{*}{4} \\
         & C &  1.4988 &  0.2006 & -0.1212 &  &  \\
         & H & -0.6585 & -0.7319 &  0.8393 &  &  \\
         & H & -1.4198 & -0.6122 & -0.7965 &  &  \\
         & H & -1.5141 &  0.7828 &  0.3460 &  &  \\
         & N &  0.3512 &  0.5000 & -0.5815 &  &  \\
         & O &  2.6150 & -0.0907 &  0.3266 &  &  \\
        \hline
        
        \multirow{9}{*}{\makecell[l]{Acetaldehyde Oxime \\ ($\mathrm{C_2H_5NO}$)}} 
         & C & -1.3220 &  0.0798 & -0.0599 & \multirow{9}{*}{237.6822} & \multirow{9}{*}{4} \\
         & C &  0.1251 &  0.1946 &  0.2783 &  &  \\
         & H & -1.8488 & -0.4928 &  0.7319 &  &  \\
         & H & -1.7722 &  1.0918 & -0.1346 &  &  \\
         & H & -1.4404 & -0.4413 & -1.0330 &  &  \\
         & H &  0.5670 &  1.1752 &  0.4092 &  &  \\
         & H &  2.6590 &  0.0768 &  0.8612 &  &  \\
         & N &  0.8453 & -0.8736 &  0.4138 &  &  \\
         & O &  2.1868 & -0.8104 &  0.7261 &  &  \\
        \hline
        
        \multirow{8}{*}{\makecell[l]{Carbamide / Urea \\ ($\mathrm{CH_4N_2O}$)}} 
         & C &  0.0236 &  0.1748 &  0.4750 & \multirow{8}{*}{206.0734} & \multirow{8}{*}{5} \\
         & H &  2.1523 & -0.0418 &  0.2442 &  &  \\
         & H &  1.1878 & -0.5180 & -1.1992 &  &  \\
         & H & -2.1229 &  0.2590 &  0.3460 &  &  \\
         & H & -1.3040 & -0.3426 & -1.1399 &  &  \\
         & N & -1.2432 &  0.0166 & -0.1616 &  &  \\
         & N &  1.2242 & -0.1571 & -0.2203 &  &  \\
         & O &  0.0823 &  0.6092 &  1.6557 &  &  \\
        \hline
        
        \multirow{3}{*}{\makecell[l]{Nitrosyl Chloride \\ ($\mathrm{NOCl}$)}} 
         & N  & -0.2226 &  0.4586 &  0.0000 & \multirow{3}{*}{102.4765} & \multirow{3}{*}{5} \\
         & O  & -1.1576 & -0.2817 &  0.0000 &  &  \\
         & Cl &  1.3803 & -0.1769 &  0.0000 &  &  \\
        \hline
        
        \multirow{5}{*}{\makecell[l]{Hydroxythiocyanate \\ ($\mathrm{HOSCN}$)}} 
         & H & -1.5762 &  0.5676 & -0.3044 & \multirow{5}{*}{140.3231} & \multirow{5}{*}{4} \\
         & O & -1.2722 &  0.4432 &  0.6302 &  &  \\
         & S & -0.4213 & -1.0482 &  0.6165 &  &  \\
         & C &  1.1270 & -0.2448 &  0.3537 &  &  \\
         & N &  2.1426 &  0.2822 &  0.1812 &  &  \\
        \hline
        
        \end{tabular}
    \end{table}

    \begin{table}[H]
    \centering
    
    \caption{Geometric coordinates of Amantadine used for simulation (geometry-optimized up to force-field level).}
    
        \begin{tabular}{|p{4.2cm}|c|c|c|c|}
        \hline
        \multirow{2}{*}{\textbf{Molecule}} & \multicolumn{4}{c|}{\textbf{Geometric Coordinates}} \\ \cline{2-5}
         & \textbf{Atom} & \textbf{x} & \textbf{y} & \textbf{z} \\
        \hline
        
        \multirow{30}{*}{\makecell[l]{ Amantadine \\ ($\mathrm{C_{10}H_{17}N}$)}} 
         & H & -0.72057 &  1.14346 &  0.93144 \\
         & H & -1.53537 &  1.21683 & -0.47761 \\
         & N & -0.73805 &  0.77679 & -0.01956 \\
         & C & -0.92076 & -0.67321 &  0.01059 \\
         & C & -2.22915 & -1.03654 &  0.74833 \\
         & H & -3.09165 & -0.57160 &  0.25272 \\
         & H & -2.21056 & -0.65093 &  1.77628 \\
         & C &  0.26844 & -1.33324 &  0.74228 \\
         & H &  0.34582 & -0.95489 &  1.77025 \\
         & H &  1.21362 & -1.08313 &  0.24234 \\
         & C &  0.09005 & -2.86228 &  0.76987 \\
         & H &  0.93676 & -3.32380 &  1.29009 \\
         & C & -0.98906 & -1.22001 & -1.43216 \\
         & H & -0.07298 & -0.96728 & -1.98243 \\
         & H & -1.82253 & -0.75964 & -1.97920 \\
         & C & -1.17095 & -2.74873 & -1.41062 \\
         & H & -1.21856 & -3.12973 & -2.43684 \\
         & C & -2.47313 & -3.09321 & -0.66778 \\
         & H & -2.62462 & -4.17964 & -0.66246 \\
         & H & -3.33092 & -2.65475 & -1.19237 \\
         & C & -2.41390 & -2.56473 &  0.77559 \\
         & H & -3.34286 & -2.81538 &  1.29977 \\
         & C & -1.21881 & -3.20616 &  1.50117 \\
         & H & -1.34806 & -4.29459 &  1.54495 \\
         & H & -1.17399 & -2.84898 &  2.53734 \\
         & C &  0.01730 & -3.38948 & -0.67334 \\
         & H &  0.95170 & -3.16434 & -1.20199 \\
         & H & -0.09033 & -4.48112 & -0.66794 \\
        \hline
        
        \end{tabular}
    \end{table}

    \begin{table}[H]
    \centering
    \caption{Dependence of SQD runtime and DMET convergence on the number of measurement shots.}
    \label{tab:shots_time_nit}
    
        \begin{tabular}{|c|c|c|}
        \hline
        \textbf{$\boldsymbol{N_{\mathrm{shots}}}$} & \textbf{$\boldsymbol{\tau_{\mathrm{qpu}}}$ (s)} & $\boldsymbol{N_{it}}$ \\
        \hline
        500      & 238.6378 & 6  \\
        1{,}000  & 459.7616 & 11 \\
        5{,}000  & 235.1656 & 4  \\
        10{,}000 & 324.2494 & 4  \\
        \hline
        \end{tabular}
    \end{table}

\bibliography{references}

\end{document}